\documentclass{aastex631}
\usepackage{booktabs}
\usepackage{graphicx}
\usepackage{makecell}
\usepackage{hyperref}
\usepackage{threeparttable}
\usepackage{xcolor}

\begin{document}

\title{Nascent Embedded-protostar Survey in Taurus (NEST) II: Measuring Dust Mass, Disk Size, and Gas Mass}

\correspondingauthor{Noshin Yesmin}
\author[0009-0009-4618-8049]{Noshin Yesmin}
\affiliation{Department of Astronomy, University of Virginia, 530 McCormick Road, Charlottesville, VA 22904, USA}
\email{pqt7tv@virginia.edu}

\author[0000-0002-9209-8708]{Patrick Sheehan}
\affiliation{National Radio Astronomy Observatory, 520 Edgemont Rd., Charlottesville, VA 22901, USA}

\author[0000-0002-6195-0152]{John Tobin}
\affiliation{National Radio Astronomy Observatory, 520 Edgemont Rd., Charlottesville, VA 22901, USA}

\author[0009-0003-3073-9148] {Aislinn Coleman-Plante}
\affiliation{National Radio Astronomy Observatory, 520 Edgemont Rd., Charlottesville, VA 22901, USA}

\author[0000-0002-4276-3730]{Nicholas P. Ballering}
\affiliation{Space Science Institute, Boulder, CO 80301, USA}
\affiliation{Department of Astronomy, University of Wisconsin-Madison, Madison, WI 53706, USA}

\author[0000-0001-7491-0048]{Tyler L. Bourke}
\affiliation{SKA Observatory, Jodrell Bank, Macclesfield SK11 9FT, United Kingdom}

\author[0000-0002-1031-4199]{Josh Eisner}
\affiliation{University of Arizona Department of Astronomy and Steward Observatory, 933 North Cherry Ave., Tucson, AZ 85721, USA}

\author[0000-0003-2300-2626]{Hauyu Baobab Liu}
\affiliation{Department of Physics, National Sun Yat-Sen University, No. 70, Lien-Hai Road, Kaohsiung City 80424, Taiwan}
\affiliation{Center of Astronomy and Gravitation, National Taiwan Normal University, Taipei 116, Taiwan}

\author[0000-0002-7402-6487]{Zhi-Yun Li}
\affiliation{Department of Astronomy, University of Virginia, 530 McCormick Road, Charlottesville, VA 22904, USA}




\begin{abstract}
Envelope-embedded protostellar disks represent the earliest stage of protoplanetary disk evolution, but their masses and sizes are difficult to measure because disk emission is entangled with the envelope. We analyze 26 protostellar disk systems in Taurus using ALMA Band~7 (345~GHz; $\sim$0.3\arcsec) and VLA Ka-band (33~GHz; $\sim$0.2\arcsec) observations that trace compact disk emission, together with molecular line data to constrain the disk gas masses. At 345~GHz, the median flux density, dust mass, and radius are 71~mJy, 5.5~$M_\oplus$, and 28~AU, with 68\% ranges of 54--107~mJy, 3.9--9.4~$M_\oplus$, and 25--39~AU. At 33~GHz, the corresponding medians are 0.43~mJy, 39~$M_\oplus$, and 32~AU, with ranges of 0.41-0.80~mJy, 34--52~$M_\oplus$, and 29--33~AU. Taurus Class~I disks are fainter and less massive than those in Orion, comparable to Perseus Class~I disks but fainter than Perseus Class~0 disks, and brighter and more massive than those in Ophiuchus. Within Taurus, Class~0/I disks are brighter than Class~II disks at both frequencies, although their inferred dust masses are comparable at 345~GHz and slightly higher at~33 GHz. Radiative-transfer modeling of CO isotopologue emission yields a median gas mass of $M_{\rm gas,med}=6.7\times10^{-4}\,M_\odot$. The resulting CO-inferred gas-to-dust ratios span a broad range, with a mean of $147 \pm 75$, a median of 26, and a 16th--84th percentile range of 8--147. This distribution overlaps the Taurus Class~II population at the low end and ISM-like or higher values, including the AGE-PRO Ophiuchus Class~0/I population, at the high end.
\end{abstract}


\section{Introduction} \label{sec:intro}
Protoplanetary disks—rotating disks of gas and dust surrounding young stars—are the birthplaces of planets. Although planet formation is known to occur within these disks \citep[e.g.,][]{PDS70_Keppler18, PDS70_Haffert19}, the physical pathways and timescales remain uncertain. To better understand the physical processes and their timescales that lead from disk formation to planetary systems, it is essential to study the youngest disks ($<1$ Myr; Class~0/I), which provide insight into the initial conditions of disks that influence the properties of the planetary systems that form within them \citep[e.g.,][]{EPPSIV_Modarsini12, PPVI-planet-population_Benz14}. Some of the key properties of young disks include the total disk mass (dust and gas) and the spatial extent of the dust disk, since the total mass determines the initial mass budget for planet formation and the dust extent sets the initial spatial scale over which planetary systems can form. The dust mass provides the reservoir for the formation of solid cores, while the gas mass influences whether those cores can accrete substantial gaseous atmospheres and grow into gas giant planets \citep[e.g.,][]{core-accretion_Pollack96}. Additionally, the gas–to–dust mass ratio regulates grain growth, radial drift, and the conditions for the streaming instability that produces planetesimals \citep[e.g.,][]{streaminginstability_Youdin05, streaming-instability_Simon16, streaming-instability_Li19}. 

\par Empirical results from Class~II disks emphasize why measuring the dust mass in Class~0/I systems is crucial. In older ($>$1 Myr) Class~II disks, the observed dust masses are systematically lower than the solid material required to build giant–planet cores, suggesting that a significant fraction of solids may already be locked into larger bodies or otherwise hidden from millimeter observations by the time disks reach the Class~II stage \citep{mass-budget_Greaves10, massbudget_Najita14, mass-discrepancy_Manara18, VANDAMPerseus_Tychoniec18} (with some notable exceptions; e.g., \citealt{ALMAPerseus_Tychoniec20}, \citealt{mass-budget_Mulders21}, \citealt{mass-budget_Savvidou25}). We note, however, that optically thick millimeter continuum emission may contribute in many systems (e.g., \citealt{beta_Li17, alpha_Liu19, SMA-taurus-classII_Chung24, rt_Ballering19}) and is not always explicitly accounted for in these analyses, which would lead to underestimates of the dust mass. In addition, high–resolution ALMA imaging of Class~II disks shows widespread substructures—rings, gaps, cavities, and spirals—that are typically interpreted as signs of planet–disk interactions, indicating that substantial solid evolution is already underway by these ages \citep[e.g.,][]{ALMAHLTau_ALMA15, DSHARP_Andrews18, substructuresTaurus_Long18, substructures_Dong15, disk-structure-review_Andrews20}. To determine when this evolution begins—and whether disks start out with enough solids to form planets—we must directly measure the dust masses of the youngest disks, before substantial growth, drift, and planet formation have reshaped their solid reservoirs.

\par Complementary to the dust component, the gas reservoir comprises most of a disk’s mass, yet it is typically inferred indirectly by adopting the canonical ISM gas–to–dust ratio of 100 \citep{ISM_Bohlin78}, since direct molecular–line detections are observationally challenging and interpreting those observations as gas masses is perhaps even more challenging \citep[e.g.,][]{CO-abundance_Yu17, CO-abundance_Krijt18, CO-abundance_Zhang19}. However, relying on this assumption introduces substantial uncertainty: disk chemistry and dynamics evolve over time, and processes such as grain growth, dust–gas decoupling, and chemical evolution can cause the gas–to–dust ratio to diverge significantly from ISM values \citep{gastemp_Kamp04, dustgrowth_Alessio06}. Therefore, independently measuring the gas and dust masses is essential for establishing the true mass budget of young disks and for accurately constraining their gas–to–dust mass ratios. 

\par Despite the importance of measuring the total (dust + gas) mass and size of young disks, embedded Class 0/I systems remain observationally challenging. These objects are deeply embedded within their natal envelopes, making it difficult to disentangle disk emission from extended envelope structure, which requires high angular resolution, and tracing disk kinematics requires high sensitivity. As a result, previous work on embedded disks has often been restricted to single objects or small samples \citep{CARMAClassI_Eisner12, VLACARMA_binary_Tobin13, ClassI_Taurus_Sheehan17}, limiting our ability to generalize their properties. Larger surveys, such as The Mass Assembly of Stellar Systems and their Evolution with the SMA (MASSES) in Perseus \citep{MASSES_Andersen19}, The VLA/ALMA Nascent Disk and Multiplicity (VANDAM) in Perseus and Orion \citep{VANDAMPerseus_Tobin16, VANDAMPerseus_Tychoniec18, VANDAMOrionI_2020}, The Ophiuchus DIsc Survey Employing ALMA (ODISEA) in Ophiuchus \citep{ODISEA_Williams19}, and the CAMPOS survey \citep{CAMPOSI_Hseih24}, have significantly expanded the available samples and improved constraints on dust emission, but none of these programs were sensitive enough in molecular lines to trace the bulk disk gas across the full sample. On the gas side, the ALMA Large Program eDisk \citep{eDisk_Ohashi23} surveyed 19 Class 0/I protostars at $\sim$5 AU resolution, using molecular-line observations to characterize Keplerian disk kinematics, but the sample was biased toward bright, well-studied sources rather than a region-complete census. More recently, efforts such as the AGE-PRO survey \citep{AGE-PRO_Zhang25}, which measured gas masses for 10 Ophiuchus disks ($<$1 Myr), represent an important step forward, but the sample is small and not a region-complete census. 

\par In this paper, we present the first dust+gas study of all consistently classified sample of Class~0/I systems in a single star-forming region: the Nascent Embedded-protostar Survey in Taurus (NEST), which includes 26 embedded systems in Taurus (23 Class~I and 3 Class~0) and excludes Flat Spectrum sources. NEST I \cite{NESTI_Plante26} focused on a multiplicity study based on this sample of embedded protostellar systems, along with an extended sample. Here we combine ALMA Band 7 (345 GHz; 870 $\mu m$) continuum and $^{13}$CO and C$^{18}$O $J{=}3$–2 line observations with VLA Ka-band (33 GHz; 9 mm) continuum, with angular resolutions of $\sim$0.3$''$ (42 AU) at 345 GHz and $\sim$0.2$''$ (30 AU) at 33 GHz. From the 0.87 mm continuum we measure dust masses and dust radii for the full sample. For the gas component of the disk, we infer CO-based gas masses using Keplerian-masked $^{13}$CO and C$^{18}$O $J{=}3$–2 fluxes and a radiative–transfer model grid, following approaches previously used for more evolved disks \citep{gasmodelgrid_Williams14, co-chamaeleon_Long19, Lupus-CO_Miotello17}. Together, these measurements provide the first unbiased census of the solid and gas reservoirs in the youngest disks in Taurus.

\par The remainder of this paper is organized as follows. In Section~\ref{sec:obs}, we describe the sample selection and the ALMA 345 GHz continuum and line observations, as well as the VLA 33 GHz continuum data. In Section~\ref{subsec:analysis-results}, we present our continuum imaging and dust disk measurements, describe the CO line imaging and Keplerian masking procedure, and outline our approach for estimating gas masses using the radiative–transfer model grid. In Section~\ref{sec:discussions}, we discuss the continuum properties of Taurus Class~0/I disks in the broader context of other star-forming regions, compare the dust mass of Taurus Class~0/I and Class~II disks, and discuss the resulting gas-to-dust mass ratios. Finally, Section~\ref{sec:conclusions} summarizes our main findings.
\section{Observations and Data Reduction} \label{sec:obs}
\subsection{Sample Selection}\label{subsec:sample} 
\begin{table*}[ht] 
\centering 
\caption{Source properties for the Taurus Class 0/I sample.} 
\label{tab:source_properties} 
\resizebox{\textwidth}{!}{ 
\begin{threeparttable}
\begin{tabular}{llllllllc} 
\toprule Source & Other Name & RA (J2000)\tnote{c} & Dec (J2000)\tnote{c} & Class\tnote{d} & Distance (pc)\tnote{e} & $L_{\mathrm{bol}}$ ($L_\odot$)\tnote{d} & $T_{\mathrm{bol}}$ (K)\tnote{d} & $\alpha_{33-345}$\tnote{f}\\ 
\midrule
IRAS 04016+2610 & L1489 IRS & 04:04:43.08 & +26:18:56.11 & I & 138 & 3.8 & 204 & 2.98 $\pm$ 0.10\\
IRAS 04108+2803B & - & 04:13:54.73 & +28:11:32.25 & I & 138 & 0.52 & 203 & 2.08 $\pm$ 0.07\\
IRAS 04158+2805\tnote{a} & - & 04:18:58.15 & +28:12:22.74 & I & 129 & 0.14 & 427 & 2.48 $\pm$ 0.06 \\
IRAS 04166+2706 & & 04:19:42.51 & +27:13:35.79 & 0 & 130 & 0.45 & 56 & 2.01 $\pm$ 0.06\\
IRAS 04169+2702 & - & 04:19:58.48 & +27:09:56.80 & I & 130 & 1.6 & 161 & 2.26 $\pm$ 0.07\\
IRAS 04181+2654A & - & 04:21:11.49 & +27:01:08.95 & I & 130 & 0.73 & 252 & 1.62 $\pm$ 0.09\\
IRAS 04181+2654B\tnote{b} & - & 04:21:10.39 & +27:01:37.27 & I & 130 & 0.30 & 306 & -\\
IRAS 04191+1523A & - & 04:22:00.09 & +15:30:24.59 & I & 159 & 0.52 & 89 & $>$3.88\\
IRAS 04191+1523B & - &  04:22:00.43 & +15:30:21.18 & I & 159 & 0.52 & 89 & 2.64 $\pm$ 0.10 \\
IRAS 04239+2436A & - & 04:26:56.26 & +24:43:34.76 & I & 149 & 1.3 & 257 & 1.82 $\pm$ 0.06\\
IRAS 04239+2436B & - & 04:26:56.28 & +24:43:34.75 & I & 149 & 1.3 & 257 & 1.82 $\pm$ 0.06\\
IRAS 04248+2612A & HH 31 IRS & 04:27:57.34 & +26:19:17.89 & I & 130 & 0.33 & 224 & $>$2.42\\
IRAS 04248+2612B & HH 31 IRS & 04:27:57.32 & +26:19:17.79 & I & 130 & 0.33 & 224 & $>$2.32\\
IRAS 04248+2612C & HH 31 IRS & 04:27:56.37 & +26:19:17.61 & I & 130 & 0.33 & 224 & $>$2.04\\
IRAS 04260+2642 & - & 04:29:05.00 & +26:49:06.80 & I & 137 & 0.08 & 353 & 2.49 $\pm$ 0.10\\
IRAS 04263+2426A & Haro 6-10, GV Tau & 04:29:23.73 & +24:33:01.01 & I & 159 & 7.7 & 351 & 2.01 $\pm$ 0.06\\
IRAS 04263+2426B & Haro 6-10, GV Tau & 04:29:23.75 & +24:32:59.66 & I & 159 & 7.7 & 351 & 1.60 $\pm$ 0.06\\
IRAS 04264+2433A & - & 04:29:30.10 & +24:39:54.55 & I & 159 & 0.39 & 209 & 1.66 $\pm$ 0.07\\
IRAS 04264+2433B & - & 04:29:30.10 & +24:39:54.90 & I & 159 & 0.39 & 209 & 1.68 $\pm$ 0.10\\
IRAS 04287+1801\tnote{a} & L1551 IRS & 04:31:34.16 & +18:08:04.58 & I & 159 & 27 & 111 & 2.42 $\pm$ 0.06\\
IRAS 04288+1802\tnote{a} & L1551 NE & 04:31:44.51 & +18:08:31.34 & I & 159 & 3.2 & 101 & 2.46 $\pm$ 0.06\\
IRAS 04295+2251\tnote{a} & L1536 IRS & 04:32:32.08 & +22:57:26.10 & I & 159 & 0.68 & 337 & 2.42 $\pm$ 0.06\\
IRAS 04302+2247 & - & 04:33:16.50 & +22:53:20.23 & I & 159 & 0.41 & 181 & 2.57 $\pm$ 0.09\\
IRAS 04325+2402A & - & 04:35:35.42 & +24:08:18.78 & I & 159 & 0.85 & 112 & 2.08 $\pm$ 0.09\\
IRAS 04325+2402B & - & 04:35:35.32 & +24:08:26.80 & I & 159 & 1.6 & 77 & 1.93 $\pm$ 0.10\\
IRAS 04361+2547 & TMR 1 & 04:39:13.91 & +25:53:20.33 & I & 147 & 3.3 & 138 & 1.95 $\pm$ 0.06\\
IRAS 04365+2535 & TMC-1A & 04:39:35.21 & +25:41:44.08 & I & 147 & 2.5 & 183 & 2.41 $\pm$ 0.06\\
IRAS 04368+2557 & L1527 IRS & 04:39:53.88 & +26:03:09.40 & 0 & 147 & 1.5 & 40 & 2.20 $\pm$ 0.06\\
IRAS 04381+2540A & TMC 1 & 04:41:12.69 & +25:46:34.63 & I & 147 & 0.62 & 174 & 1.83 $\pm$ 0.10\\
IRAS 04381+2540B & TMC 1 & 04:41:12.73 & +25:46:34.65 & I & 147 & 0.62 & 174 & 1.62 $\pm$ 0.10\\
IRAS 04385+2550 & - & 04:41:38.84 & +25:56:26.30 & I & 147 & 0.45 & 607 & 2.32 $\pm$ 0.10\\
IRAS 04489+3042A & - & 04:52:06.69 & +30:47:16.93 & I & 170 & 0.39 & 424 & 2.58 $\pm$ 0.13\\
IRAS 04489+3042B & - & 04:52:06.75 & +30:47:19.74 & I & 170 & 0.39 & 424 & $>$1.82\\
DG TauB & IRAS 04240+2559 & 04:27:02.58 & +26:05:30.08 & I & 130 & 0.91 & 194 & 2.50 $\pm$ 0.07\\
HH 30 & - &  04:31:37.49 & +18:12:23.76 & I & 159 & 0.02 & 320 & $>$ 3.40\\
IRAM 04191 & IRAM 04191+1522 & 04:21:56.90 & +15:29:46.05 & 0 & 159 & 0.10 & 28 & $>$2.89\\
\bottomrule 
\end{tabular} 
\begin{tablenotes}
\footnotesize
\item[a] Binary system with a circumbinary disk; treated as a single source in our analysis. For IRAS 04295+2251, our ALMA continuum data show a disk with a large central cavity. \citet{NESTI_Plante26} suggests that this structure may be consistent with the presence of a companion and a circumbinary disk.
\item[b] Not detected with ALMA or the VLA; excluded from the analysis.
\item[c] RA and Dec measured from 2D Gaussian fits to the ALMA continuum images using CASA \texttt{imfit}.
\item[d] Adopted from \citet{NESTI_Plante26}. 
\item[e] Assigned by matching each source position to the nearest of the 10 Taurus subregion centers from \cite{taurus-distance-zucker20} using \texttt{astropy.coordinates.match\_to\_catalog\_sky}, and taking that subregion’s mean Gaia-based distance.
\item[f] Spectral index between 33 and 345~GHz, computed from the integrated flux densities; lower limits are reported for sources with 33~GHz upper limits.
\end{tablenotes}
\end{threeparttable}
}
\end{table*}

\begin{table}
\centering
\caption{ALMA 345~GHz CO line observations.}
\label{tab:alma_line_obs}
\begin{tabular}{lccccc}
\hline
Line & Center (GHz) & \multicolumn{2}{c}{Bandwidth (MHz)} & Chan. width (MHz) & $\Delta v$ (km s$^{-1}$) \\
\cline{3-4}
 & & 12 m & 7 m & & \\
\hline
C$^{18}$O(3--2) & 329.3305525 & 234 & 250 & 0.244 & 0.222 \\
$^{13}$CO(3--2) & 330.5879653 & 234 & 250 & 0.244 & 0.221 \\
\hline
\end{tabular}
\end{table}
\par Our sample is comprised of all of 26 protostellar systems in the Taurus molecular cloud that are consistently identified as Class 0 or Class I sources across multiple independent studies (e.g., \citealt{sample_Motte01, sample_Andrews05, sample_Furlan08}). Taurus, at a distance of $\sim$140 pc \citep{taurus-distance_Galli18, taurus-distance-zucker20}, is the nearest large star-forming region with a significant embedded protostar population, and its sparse stellar density and absence of massive stars \citep{taurus-cloud_Kenyon08} reduce crowding and external irradiation, providing a clean laboratory for studying early disk evolution. For each system in our sample, we analyze ALMA 345~GHz continuum observations and CO isotopologue ($^{13}$CO and C$^{18}$O) line data, together with VLA 33~GHz continuum observations. For systems with multiple disks, we treat each disk as a separate source, whereas for systems with circumbinary disks (04287+1801, 04158+2805, 04288+1802, and possibly 04295+2251; see \citet{NESTI_Plante26} for details on 04295+2251), we treat the system as a single source. This yields 35 individual disks in our sample; IRAS 04181+2654B is excluded from further analysis because it is not detected in either the ALMA or VLA continuum data. Table \ref{tab:source_properties} lists the sources along with their coordinates, evolutionary class, bolometric luminosity ($L_{\rm bol}$), and bolometric temperature ($T_{\rm bol}$). We adopt $L_{\rm bol}$ and $T_{\rm bol}$ from \citet{NESTI_Plante26}, who constructed spectral energy distributions (SEDs) using VizieR photometry \citep{vizier}. Distances are assigned using the ten Taurus subregions defined by \citet{taurus-distance-zucker20}; for each source, we matched its sky position to the nearest subregion center with \texttt{astropy.coordinates.match\_to\_catalog\_sky} and adopted that subregion’s mean \textit{Gaia}-based distance.

\subsection{ALMA 345 GHz Continuum Observations}\label{subsec:ALMA_obs_cont} The ALMA Band 7 observations (centered at 345 GHz, or 870 $\mu$m) were conducted under project code 2019.1.00847.S in Cycle 7, using configurations C43-4, C43-1, and the Atacama Compact Array (ACA). The array design was chosen to provide both the $\sim$0.3$''$ spatial resolution necessary to resolve disk structures down to $\sim$21 AU, and sensitivity to spatial scales up to $\sim$20$''$, corresponding to $\sim$3000 AU, to recover emission from the surrounding envelope. Calibration was performed using the standard ALMA pipeline \citep{alma-pipeline_Hunter23}. Continuum emission was extracted from the line-free channels identified by the pipeline across all spectral windows. The final continuum maps have synthesized beam sizes of $\sim$0.3$''$, with rms noise levels ranging from 0.091–0.466~mJy~beam$^{-1}$. 

\par We used pipeline-calibrated ALMA Band~7 measurement sets to produce continuum images. We first identified line-free channels and averaged them in frequency to create continuum measurement sets. We then imaged the data in CASA with \texttt{tclean}, applying Briggs weighting with robust=0.5 and using auto-multithresh masking during deconvolution. We applied primary-beam correction to all final images. Since our ALMA datasets includ both 12~m and ACA 7~m coverage, we used the \texttt{mosaic} gridder in \texttt{tclean} to jointly image the data and recover both compact and extended emission. 

\subsection{VLA 33 GHz Observations}\label{subsec:VLA_obs} The VLA Ka-band (33 GHz; 9 mm) continuum observations were obtained under project code \texttt{VLA/18B-179} in both the B and C configurations. The combined dataset delivers an angular resolution of $\sim$0.2\arcsec\ ($\sim$30 AU at 140 pc) and a typical rms sensitivity of $\sim$0.02 mJybeam$^{-1}$ in the continuum images. Calibration was performed using the standard VLA pipeline. Of the 35 disks in the our sample, 7 disks are not detected in the VLA data, and the 04239+2436AB binary, unlike in the ALMA data, is unresolved at 33 GHz and is therefore treated as a single source.

\par We produced the VLA continuum images in CASA using \texttt{tclean}, applying Briggs weighting with robust=2.0 and using auto-multithresh masking during deconvolution. We applied primary-beam correction to all final images. The synthesized beam sizes are $\sim$0.2$^{\prime\prime}$ for the VLA images. 
\par
\begin{figure*}
\centering

\begin{minipage}{0.48\textwidth}
  \centering
  \includegraphics[width=\linewidth]{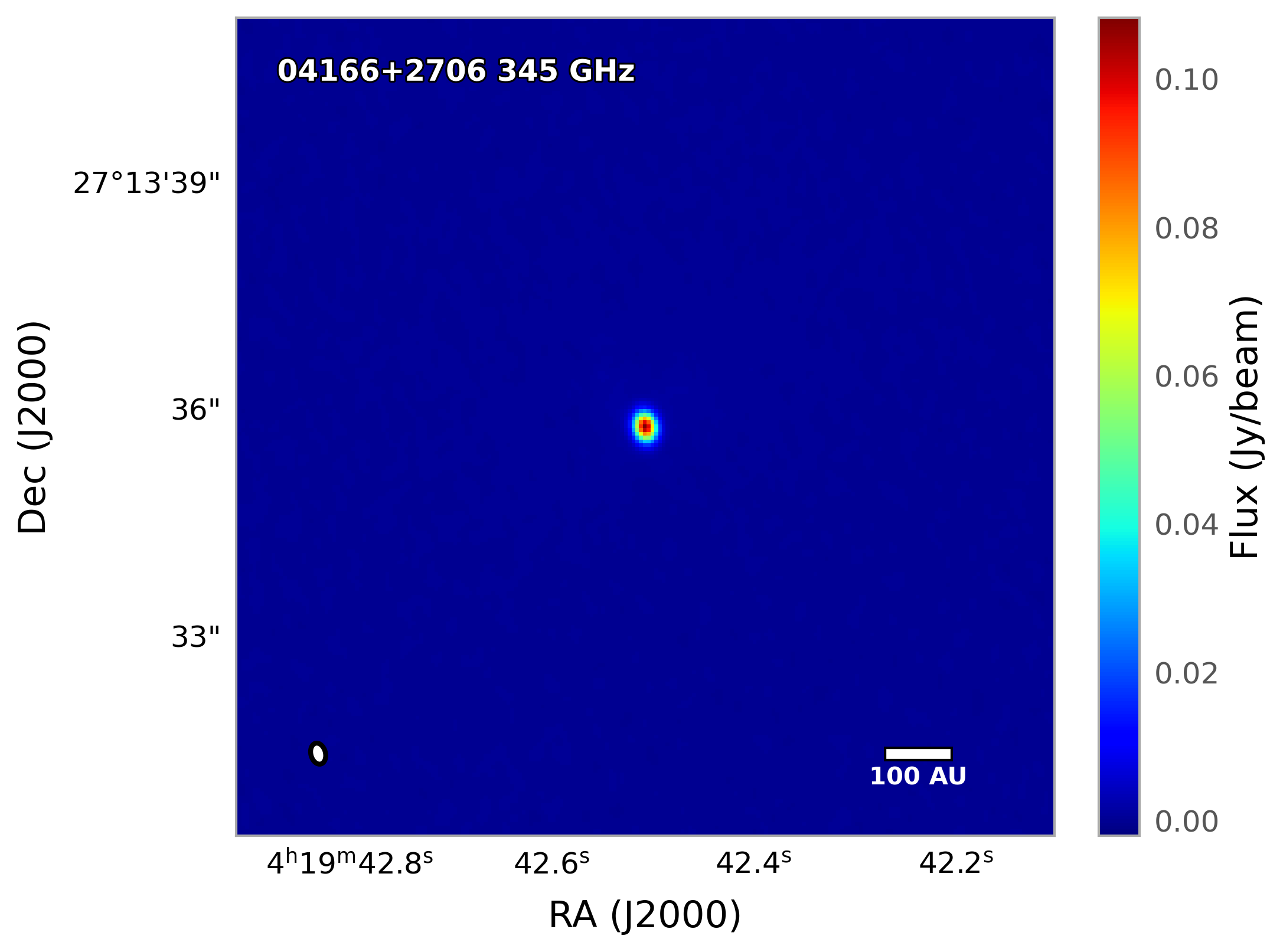}
\end{minipage}\hfill
\begin{minipage}{0.48\textwidth}
  \centering
  \includegraphics[width=\linewidth]{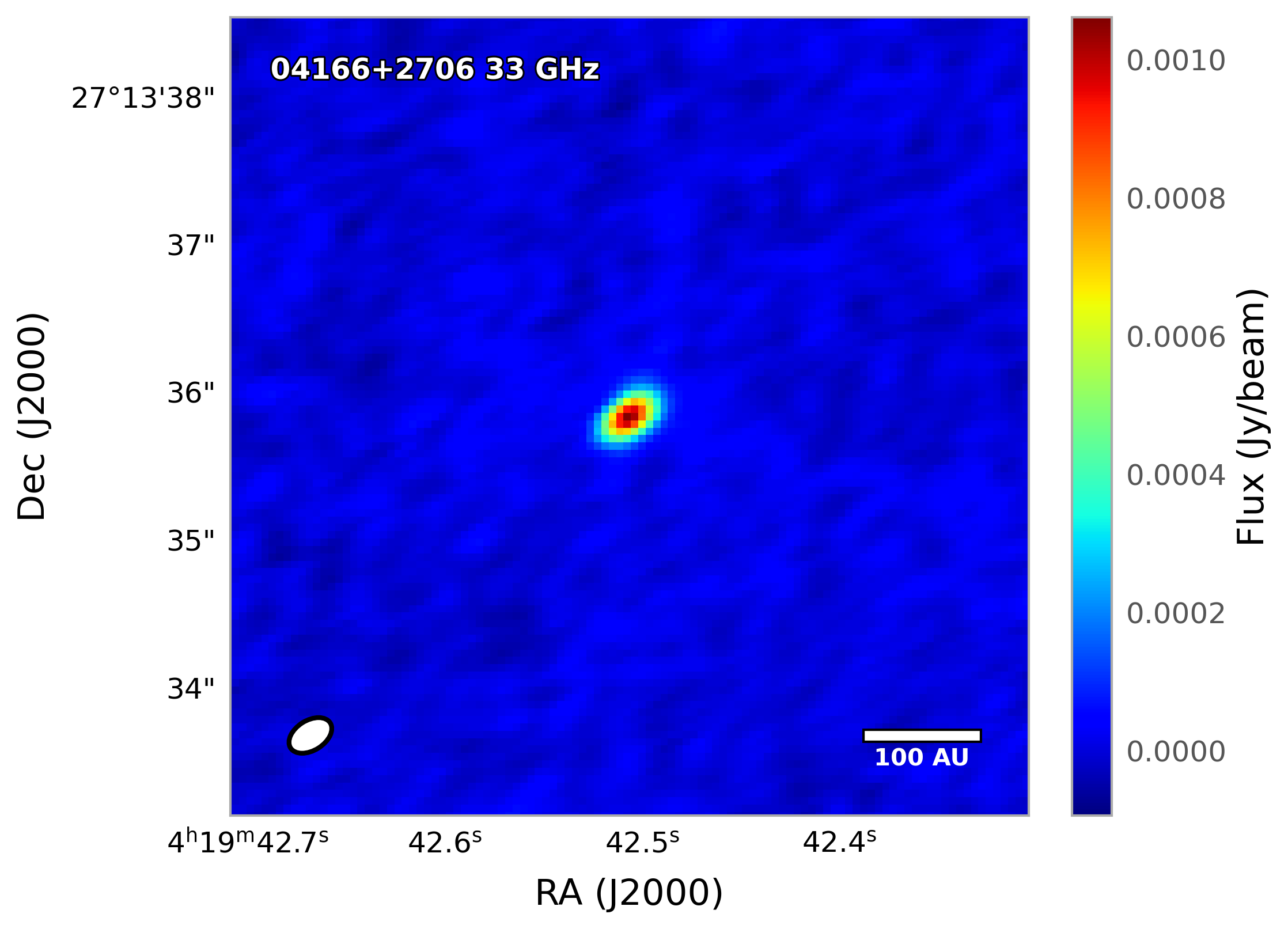}
\end{minipage}

\vspace{0.02\textheight}

\begin{minipage}{0.48\textwidth}
  \centering
  \includegraphics[width=\linewidth]{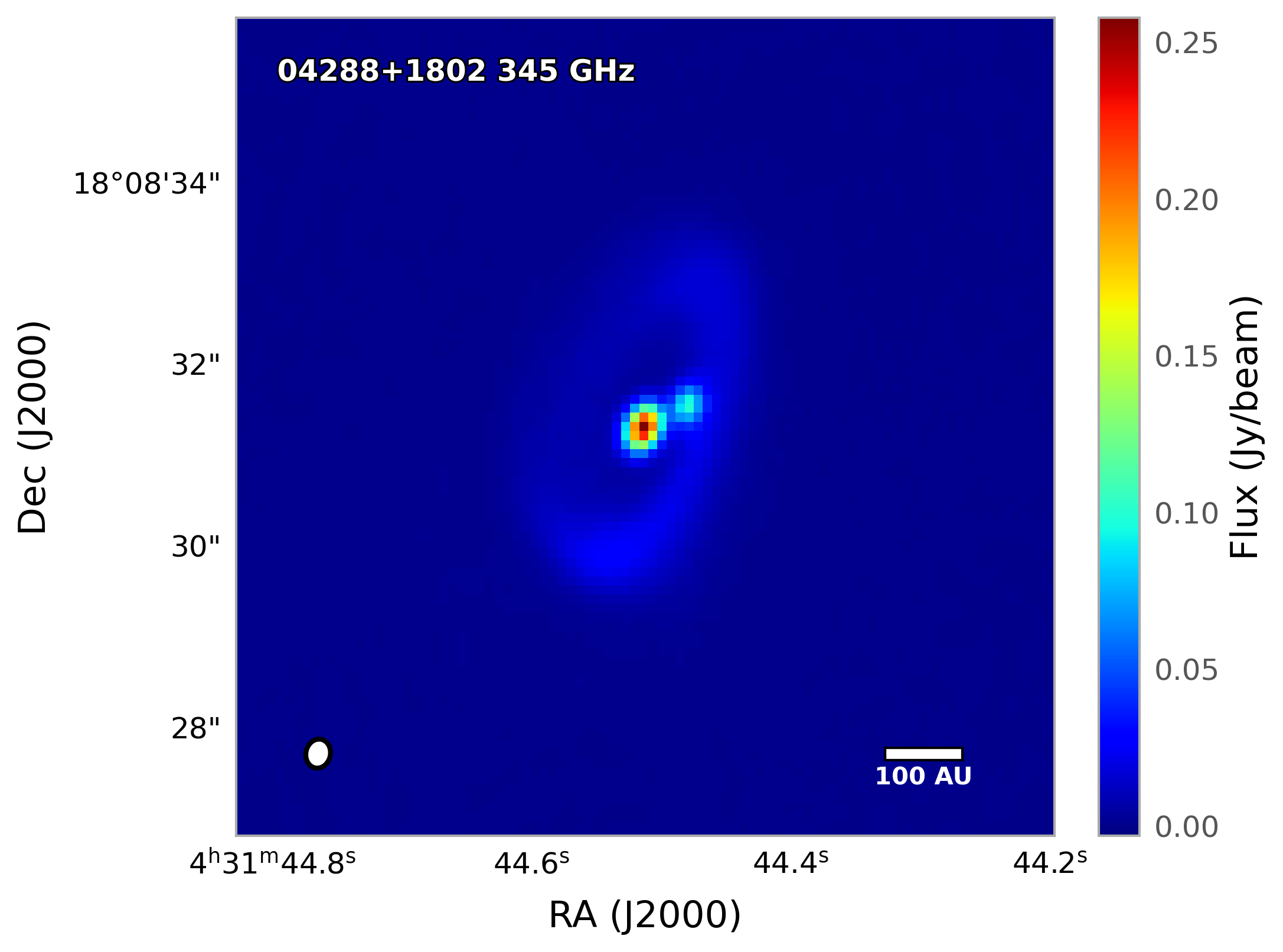}
\end{minipage}\hfill
\begin{minipage}{0.48\textwidth}
  \centering
  \includegraphics[width=\linewidth]{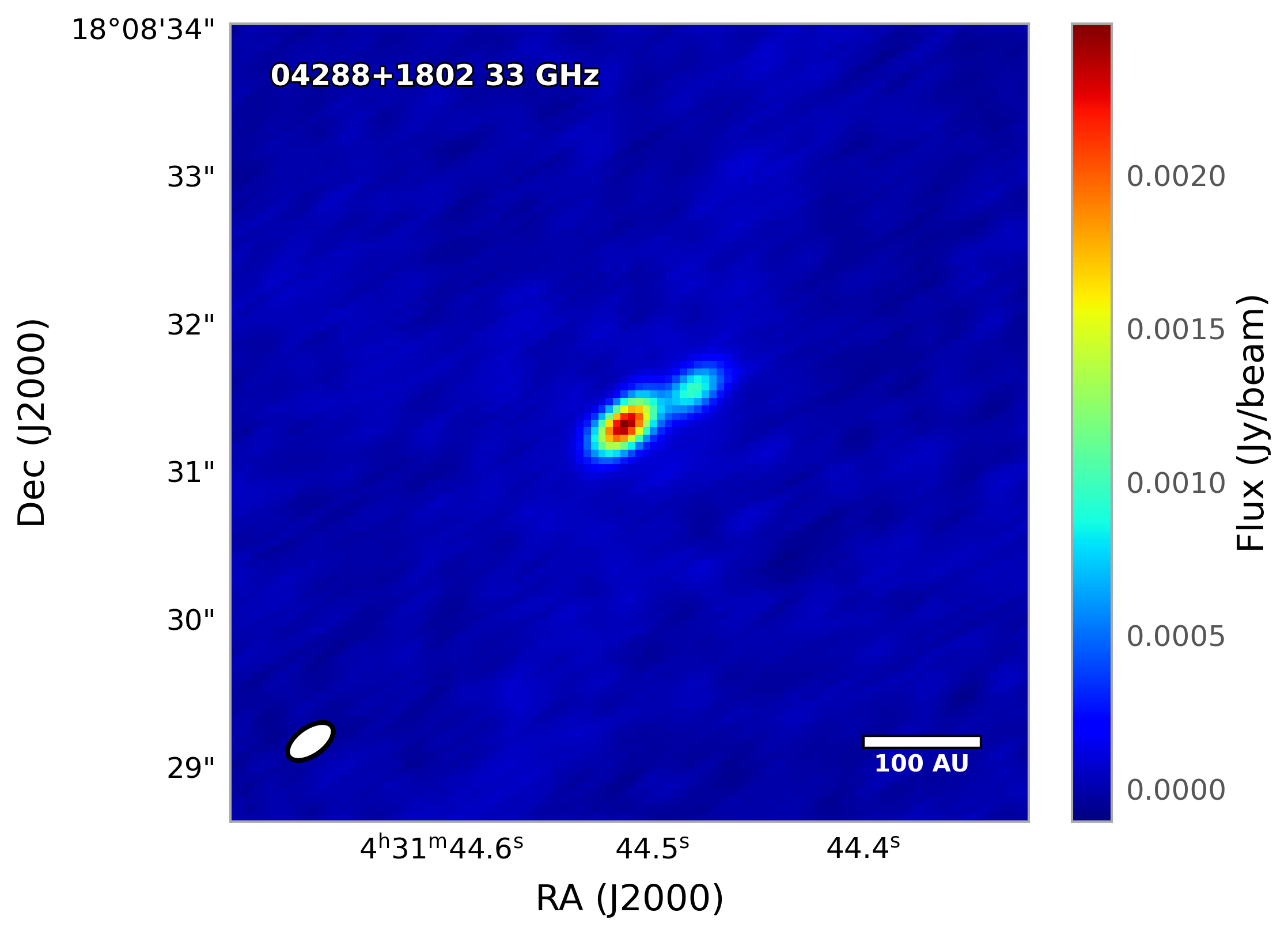}
\end{minipage}

\caption{Representative continuum maps. ALMA 345 GHz (12 m + 7 m arrays) and VLA 33 GHz images are primary--beam corrected; restoring beams are shown at lower left and 100~AU scale bars at lower right. Source 04166+2706 illustrates a compact, marginally resolved disk, while 04288+1802 shows a binary system with extended circumbinary emission. Continuum maps for the full sample are provided in Appendix~\ref{app:disk_images}.}
\label{fig:rep_cont}
\end{figure*}

 In Figure~\ref{fig:rep_cont}, we show ALMA (345~GHz) and VLA (33~GHz) continuum images for two representative targets: 04166+2706, which is compact and marginally resolved, and 04288+1802, which shows extended circumbinary emission. Appendix~\ref{app:disk_images} presents continuum maps for the full sample.

\subsection{ALMA 345 GHz Line Observations}\label{subsec:ALMA_obs_line} The ALMA Band~7 setup targeted the CO isotopologue transitions $^{13}$CO~($J{=}3{-}2$) and C$^{18}$O~($J{=}3{-}2$), with rest frequencies of 330.5879653 GHz and 329.3305525 GHz, respectively, along with additional molecular lines; however, we focus exclusively on $^{13}$CO and C$^{18}$O in this paper. The complete spectral configuration, such as, spectral window centers, bandwidths, channel spacings, and the corresponding velocity resolutions, is summarized in Table~\ref{tab:alma_line_obs}.

\par We use the $^{13}$CO and C$^{18}$O line observations because their differing optical depths provide complementary sensitivity to molecular gas in embedded protostellar disks. Compared to the most abundant CO isotopologue ($^{12}$CO), these rarer isotopologues are less optically thick and are therefore better suited for probing disk-scale emission in the presence of surrounding envelope material. To extract these CO isotopologue lines from our observations, we first subtracted the continuum from the visibilities with \texttt{uvcontsub}, selecting line–free channels per source. We then imaged the $^{13}$CO and C$^{18}$O $J\!=\!3\text{--}2$ transitions with \texttt{tclean} using the \texttt{mosaic} gridder (joint 12 m + ACA 7 m), Briggs weighting (\texttt{robust}=0.5), and the \texttt{auto-multithresh} algorithm (\texttt{usemask=auto-multithresh}), which automatically generates clean masks based on emission signal-to-noise. The image cubes were produced at 0.25 kms$^{-1}$ spectral resolution in the Local Standard of Rest kinematic (LSRK) frame, covering $\pm20$ kms$^{-1}$ around each source’s systemic velocity. Primary–beam correction was applied to all cubes, and the typical common restoring beam is $\sim0.3''$.

\par See \citet{NESTI_Plante26} for further details on the observing setup and calibration. 

\section{Analysis and Results}\label{subsec:analysis-results}
\subsection{Continuum Analysis and Dust Disk Results}\label{subsec:continuum}

\begin{table*}[t]
\caption{ALMA Band 7 (345 GHz) dust continuum properties of Taurus Class 0/I disks.}
\label{tab:alma_continuum}
\small
\centering

\resizebox{\textwidth}{!}{%
\begin{tabular}{lcccccccc}
\toprule
Source & \shortstack{$F_\nu^{\mathrm{b}}$ \\ (mJy)} & \shortstack{Peak $I_\nu^{\mathrm{b}}$ \\ (mJy/beam)} & \shortstack{rms \\ (mJy/beam)} & \shortstack{Deconv. Size \\ ($''$)} & \shortstack{PA$^{\mathrm{b}}$ \\ (deg)} & \shortstack{Incl.$^{\mathrm{b}}$ \\ (deg)} & \shortstack{$M_{\mathrm{dust}}^{\mathrm{c}}$ \\ ($M_\oplus$)} & \shortstack{Radius$^{\mathrm{b}}$ \\ (AU)} \\ 
\midrule
04016+2610 & 205.27 ± 7.13 & 3.57 ± 0.12 & 0.142 & 3.13 $\times$ 0.99 & 68.5 ± 1.0 & 71.6 ± 2.2 & $9.5 \pm 2.1$ & $367 \pm 39$ \\
04108+2803B & 70.49 ± 0.68 & 50.52 ± 0.31 & 0.096 & 0.18 $\times$ 0.08 & 75.5 ± 2.4 & 62.4 ± 4.0 & $5.9 \pm 1.3$ & $21 \pm 2$ \\
04158+2805\textsuperscript{a} & 287.82 & 6.36 & 0.142 & 6.72 $\times$ 2.29 & 93.2 ± 1.3 & 70.1 ± 2.7 & $32 \pm 7$ & $736 \pm 80$ \\
04166+2706 & 143.44 ± 1.64 & 107.66 ± 0.77 & 0.123 & 0.14 $\times$ 0.11 & 121.1 ± 9.4 & 39.1 ± 6.2 & $11 \pm 2$ & $16 \pm 2$ \\
04169+2702 & 207.41 ± 0.88 & 129.11 ± 0.37 & 0.113 & 0.20 $\times$ 0.15 & 140.2 ± 1.8 & 42.1 ± 1.5 & $11 \pm 2$ & $22 \pm 2$ \\
04181+2654A & 16.08 ± 0.26 & 12.33 ± 0.12 & 0.103 & 0.17 $\times$ 0.05 & 75.6 ± 3.8 & 74.0 ± 7.6 & $1.1 \pm 0.2$ & $19 \pm 2$ \\
04191+1523A & 176.17 ± 2.88 & 52.49 ± 0.68 & 0.114 & 0.61 $\times$ 0.31 & 123.1 ± 1.1 & 59.3 ± 1.5 & $20 \pm 4$ & $82 \pm 8$ \\
04191+1523B & 107.03 ± 0.29 & 84.11 ± 0.14 & 0.114 & 0.19 $\times$ 0.11 & 176.2 ± 0.8 & 54.5 ± 0.8 & $12 \pm 3$ & $26 \pm 3$ \\
04239+2436A & 48.59 ± 0.31 & 44.95 ± 0.17 & 0.103 & 0.08 $\times$ 0.05 & 160.4 ± 9.8 & 47.8 ± 8.9 & $3.6 \pm 0.8$ & $10 \pm 1$ \\
04239+2436B & 46.04 ± 0.31 & 43.10 ± 0.17 & 0.103 & 0.07 $\times$ 0.04 & 157.6 ± 10.1 & 52.6 ± 10.8 & $3.4 \pm 0.8$ & $9 \pm 1$ \\
04248+2612A & 6.13 ± 0.39 & 4.93 ± 0.19 & 0.100 & 0.15 $\times$ 0.09 & 23.4 ± 32.7 & 54.0 ± 21.9 & $0.5 \pm 0.1$ & $16 \pm 4$ \\
04248+2612B & 4.92 ± 0.36 & 4.33 ± 0.19 & 0.100 & 0.14 $\times$ 0.02 & 35.4 ± 11.1 & 82.8 ± 26.6 & $0.4 \pm 0.1$ & $15 \pm 5$ \\
04248+2612C & 2.53 ± 0.35 & 1.25 ± 0.12 & 0.100 & 0.31 $\times$ 0.16 & 48.8 ± 22.3 & 59.5 ± 21.9 & $0.22 \pm 0.06$ & $35 \pm 9$ \\
04260+2642 & 183.38 ± 3.40 & 25.41 ± 0.42 & 0.110 & 1.11 $\times$ 0.23 & 137.3 ± 0.4 & 78.3 ± 1.3 & $28 \pm 6$ & $129 \pm 13$ \\
04263+2426A & 93.94 ± 0.49 & 87.58 ± 0.26 & 0.111 & 0.06 $\times$ 0.06 & 147.6 ± 34.1 & 26.7 ± 14.8 & $4.7 \pm 1.1$ & $8 \pm 1$ \\
04263+2426B & 72.91 ± 0.39 & 66.81 ± 0.21 & 0.111 & 0.07 $\times$ 0.06 & 112.4 ± 64.4 & 20.9 ± 13.6 & $3.7 \pm 0.8$ & $9 \pm 1$ \\
04264+2433A & 20.01 ± 0.59 & 16.75 ± 0.29 & 0.108 & 0.16 $\times$ 0.05 & 17.2 ± 3.8 & 71.8 ± 7.4 & $2.4 \pm 0.6$ & $21 \pm 3$ \\
04264+2433B & 5.94 ± 0.30 & 4.78 ± 0.14 & 0.108 & 0.19 $\times$ 0.04 & 11.4 ± 5.7 & 77.9 ± 9.7 & $0.7 \pm 0.2$ & $25 \pm 4$ \\
04287+1801\textsuperscript{a} & 2368.03 & 491.26 & 0.466 & 1.00 $\times$ 0.41 & 165.8 ± 1.9 & 65.8 ± 3.6 & $84 \pm 19$ & $136 \pm 15$ \\
04288+1802\textsuperscript{a} & 1505.28 & 257.81 & 0.216 & 0.39 $\times$ 0.20 & 128.8 ± 5.6 & 59.6 ± 6.3 & $97 \pm 22$ & $53 \pm 6$ \\
04295+2251\textsuperscript{a} & 234.995 & 23.21 & 0.115 & 1.30 $\times$ 0.53 & 71.7 ± 3.7 & 66.0 ± 6.3 & $24 \pm 5$ & $176 \pm 24$ \\
04302+2247 & 378.82 ± 3.95 & 23.78 ± 0.23 & 0.139 & 2.33 $\times$ 0.27 & 175.0 ± 0.1 & 83.2 ± 0.7 & $45 \pm 10$ & $314 \pm 32$ \\
04325+2402A & 57.47 ± 0.90 & 29.57 ± 0.33 & 0.104 & 0.31 $\times$ 0.06 & 106.0 ± 1.3 & 78.1 ± 2.8 & $5.5 \pm 1.2$ & $43 \pm 4$ \\
04325+2402B & 31.97 ± 0.49 & 20.39 ± 0.20 & 0.104 & 0.29 $\times$ 0.07 & 29.4 ± 0.8 & 76.4 ± 1.8 & $2.5 \pm 0.6$ & $39 \pm 4$ \\
04361+2547 & 126.34 ± 1.02 & 75.70 ± 0.41 & 0.104 & 0.26 $\times$ 0.07 & 89.9 ± 1.1 & 73.0 ± 2.2 & $6.9 \pm 1.5$ & $32 \pm 3$ \\
04365+2535 & 429.15 ± 5.72 & 175.93 ± 1.73 & 0.130 & 0.32 $\times$ 0.21 & 73.2 ± 2.8 & 49.3 ± 2.3 & $26 \pm 6$ & $41 \pm 4$ \\
04368+2557 & 358.26 ± 5.12 & 115.02 ± 1.25 & 0.104 & 0.60 $\times$ 0.16 & 2.1 ± 0.4 & 74.3 ± 1.1 & $25 \pm 6$ & $75 \pm 8$ \\
04381+2540A & 32.36 ± 1.65 & 28.45 ± 0.86 & 0.107 & 0.09 $\times$ 0.08 & 49.9 ± 77.5 & 32.1 ± 71.9 & $2.9 \pm 0.7$ & $11 \pm 4$ \\
04381+2540B & 18.21 ± 1.15 & 15.15 ± 0.58 & 0.107 & 0.12 $\times$ 0.09 & 58.2 ± 50.8 & 41.9 ± 46.1 & $1.6 \pm 0.4$ & $14 \pm 5$ \\
04385+2550 & 54.00 ± 0.30 & 42.66 ± 0.14 & 0.105 & 0.16 $\times$ 0.07 & 159.9 ± 1.7 & 63.5 ± 2.1 & $5.3 \pm 1.2$ & $20 \pm 2$ \\
04489+3042A & 28.32 ± 0.26 & 23.68 ± 0.13 & 0.091 & 0.18 $\times$ 0.15 & 11.5 ± 13.5 & 35.8 ± 8.1 & $3.9 \pm 0.9$ & $26 \pm 3$ \\
04489+3042B & 1.21 ± 0.27 & 0.51 ± 0.08 & 0.091 & 0.64 $\times$ 0.31 & 170.3 ± 32.2 & 61.6 ± 25.0 & $0.17 \pm 0.05$ & $93 \pm 33$ \\
DG TauB & 765.98 ± 21.45 & 124.69 ± 3.03 & 0.172 & 0.75 $\times$ 0.36 & 24.5 ± 1.4 & 61.5 ± 2.2 & $48 \pm 11$ & $83 \pm 9$ \\
HH 30 & 58.21 ± 2.05 & 9.09 ± 0.28 & 0.099 & 1.36 $\times$ 0.22 & 121.5 ± 0.6 & 80.8 ± 2.4 & $19 \pm 4$ & $184 \pm 20$ \\
IRAM 04191 & 17.18 ± 0.68 & 11.74 ± 0.30 & 0.105 & 0.20 $\times$ 0.15 & 55.9 ± 17.8 & 40.9 ± 16.7 & $3.3 \pm 0.7$ & $28 \pm 4$ \\
\bottomrule
\end{tabular}%
}

\vspace{2pt}
\begin{minipage}{\textwidth}
\footnotesize
\textsuperscript{a} These sources are binary systems with circumbinary disks, confirmed by the ALMA continuum images. The total flux density ($F_\nu$) and peak intensity are measured using \texttt{imstat} rather than \texttt{imfit}, because the emission is not well described by a Gaussian profile. The size and geometric parameters (deconvolved size, position angle, and inclination) are derived from \texttt{imfit}.

\textsuperscript{b} Uncertainties on the fitted continuum properties are the 2D Gaussian fitting uncertainties returned by CASA \textit{imfit}. Radius and inclination uncertainties are propagated from the deconvolved size uncertainty.

\textsuperscript{c} Uncertainties on $M_{\mathrm{dust}}$ are statistical only; see text for details.
\end{minipage}
\end{table*}
\begin{table*}[t]
\caption{VLA Band Ka (33 GHz) dust continuum properties of Taurus Class 0/I disks.}
\label{tab:vla_continuum}
\small
\centering

\resizebox{\textwidth}{!}{%
\begin{tabular}{lcccccccc}
\toprule
Source &
\shortstack{$F_\nu^{\mathrm{b}}$ \\ (mJy)} &
\shortstack{Peak $I_\nu^{\mathrm{b}}$ \\ (mJy/beam)} &
\shortstack{rms \\ (mJy/beam)} &
\shortstack{Deconv. Size \\ ($''$)} &
\shortstack{PA$^{\mathrm{b}}$ \\ (deg)} &
\shortstack{Incl.$^{\mathrm{b}}$ \\ (deg)} &
\shortstack{$M_{\mathrm{dust}}^{\mathrm{c}}$ \\ ($M_\oplus$)} &
\shortstack{Radius$^{\mathrm{b}}$ \\ (AU)} \\
\midrule
04016+2610 & 0.19 ± 0.03 & 0.17 ± 0.02 & 0.019 & 0.18 $\times$ 0.04 & 100.4 ± 19.5 & 78.8 ± 53.0 & $8.7 \pm 2.5$ & $\leq 20$ \\
04108+2803B & 0.53 ± 0.05 & 0.36 ± 0.02 & 0.019 & 0.28 $\times$ 0.18 & 121.9 ± 71.3 & 50.0 ± 29.6 & $41 \pm 10$ & $33 \pm 12$ \\
04158+2805\textsuperscript{a} & 0.85 & 0.09 & 0.019 & 3.39 $\times$ 1.00 & 85.7 ± 7.0 & 72.9 ± 16.2 & $79 \pm 18$ & $372 \pm 103$ \\
04166+2706 & 1.27 ± 0.05 & 1.05 ± 0.03 & 0.019 & 0.16 $\times$ 0.08 & 143.9 ± 16.6 & 59.2 ± 17.5 & $90 \pm 21$ & $17 \pm 4$ \\
04169+2702 & 1.03 ± 0.08 & 0.51 ± 0.03 & 0.019 & 0.29 $\times$ 0.23 & 134.1 ± 64.9 & 38.7 ± 17.5 & $52 \pm 12$ & $32 \pm 6$ \\
04181+2654A & 0.36 ± 0.06 & 0.16 ± 0.02 & 0.019 & 0.40 $\times$ 0.16 & 96.0 ± 15.0 & 66.9 ± 18.9 & $23 \pm 6$ & $45 \pm 11$ \\
04191+1523A & $\leq$ 0.02 & - & 0.019 & - & - & - & $\leq 2.0$ & - \\
04191+1523B & 0.22 ± 0.04 & 0.16 ± 0.02 & 0.019 & - & - & - & $22 \pm 6$ & $\leq 23$ \\
04239+2436AB & 1.32 ± 0.06 & 0.80 ± 0.02 & 0.021 & 0.40 $\times$ 0.09 & 98.4 ± 3.1 & 77.4 ± 6.5 & $95 \pm 22$ & $50 \pm 6$ \\
04248+2612A & $\leq$ 0.02 & - & 0.021 & - & - & - & $\leq 1.7$ & - \\
04248+2612B & $\leq$ 0.02 & - & 0.021 & - & - & - & $\leq 1.7$ & - \\
04248+2612C & $\leq$ 0.02 & - & 0.021 & - & - & - & $\leq 1.7$ & - \\
04260+2642 & 0.53 ± 0.09 & 0.15 ± 0.02 & 0.017 & 0.73 $\times$ 0.32 & 128.8 ± 12.2 & 63.8 ± 16.0 & $65 \pm 18$ & $85 \pm 22$ \\
04263+2426A & 0.85 ± 0.03 & 0.83 ± 0.02 & 0.016 & - & - & - & $44 \pm 10$ & $\leq 5$ \\
04263+2426B & 1.70 ± 0.04 & 1.54 ± 0.02 & 0.016 & 0.10 $\times$ 0.04 & 155.0 ± 15.2 & 62.5 ± 18.8 & $88 \pm 20$ & $13 \pm 2$ \\
04264+2433A & 0.41 ± 0.04 & 0.24 ± 0.02 & 0.016 & 0.24 $\times$ 0.10 & 97.7 ± 14.1 & 63.8 ± 18.1 & $45 \pm 11$ & $32 \pm 6$ \\
04264+2433B & 0.11 ± 0.02 & 0.08 ± 0.01 & 0.016 & 0.22 $\times$ 0.08 & 135.5 ± 24.6 & 69.2 ± 22.2 & $13 \pm 4$ & $30 \pm 10$ \\
04287+1801\textsuperscript{a} & 8.15 & 0.02 & 0.022 & 0.58 $\times$ 0.18 & 172.7 ± 7.9 & 71.8 ± 13.0 & $307 \pm 69$ & $79 \pm 14$ \\
04288+1802\textsuperscript{a} & 4.66 & 2.49 & 0.021 & 0.40 $\times$ 0.14 & 122.0 ± 3.9 & 70.4 ± 8.7 & $300 \pm 67$ & $54 \pm 9$ \\
04295+2251\textsuperscript{a} & 0.8 & 0.56 & 0.016 & 0.19 $\times$ 0.08 & 143.0 ± 9.5 & 64.5 ± 13.0 & $76 \pm 17$ & $26 \pm 5$ \\
04302+2247 & 0.91 ± 0.16 & 0.11 ± 0.02 & 0.016 & 0.75 $\times$ 0.42 & 166.7 ± 12.9 & 55.8 ± 14.5 & $99 \pm 28$ & $102 \pm 21$ \\
04325+2402A & 0.43 ± 0.07 & 0.18 ± 0.02 & 0.017 & 0.46 $\times$ 0.21 & 145.9 ± 20.0 & 62.4 ± 20.6 & $39 \pm 11$ & $62 \pm 16$ \\
04325+2402B & 0.34 ± 0.06 & 0.11 ± 0.01 & 0.017 & 0.49 $\times$ 0.35 & 106.1 ± 86.3 & 44.2 ± 26.2 & $26 \pm 7$ & $66 \pm 18$ \\
04361+2547 & 1.30 ± 0.04 & 1.11 ± 0.02 & 0.017 & 0.23 $\times$ 0.06 & 109.0 ± 4.9 & 73.3 ± 8.5 & $71 \pm 16$ & $28 \pm 4$ \\
04365+2535 & 1.50 ± 0.05 & 1.25 ± 0.03 & 0.019 & 0.18 $\times$ 0.09 & 88.8 ± 17.2 & 59.7 ± 21.4 & $89 \pm 20$ & $23 \pm 5$ \\
04368+2557 & 2.03 ± 0.10 & 1.20 ± 0.04 & 0.017 & 0.32 $\times$ 0.13 & 174.4 ± 9.9 & 64.9 ± 20.1 & $135 \pm 31$ & $39 \pm 6$ \\
04381+2540A & 0.44 ± 0.08 & 0.32 ± 0.04 & 0.017 & 0.19 $\times$ 0.15 & 160.4 ± 50.0 & 41.9 ± 69.5 & $36 \pm 10$ & $\leq 15$ \\
04381+2540B & 0.41 ± 0.07 & 0.26 ± 0.03 & 0.017 & 0.26 $\times$ 0.14 & 0.8 ± 63.9 & 56.8 ± 46.4 & $34 \pm 10$ & $32 \pm 15$ \\
04385+2550 & 0.23 ± 0.04 & 0.16 ± 0.02 & 0.017 & 0.24 $\times$ 0.09 & 159.2 ± 32.2 & 66.8 ± 38.5 & $21 \pm 6$ & $29 \pm 14$ \\
04489+3042A & 0.07 ± 0.02 & 0.08 ± 0.01 & 0.017 & - & - & - & $8.3 \pm 2.9$ & $\leq 20$ \\
04489+3042B & $\leq$ 0.02 & - & 0.017 & - & - & - & $\leq 2.2$ & - \\
DG TauB & 2.15 ± 0.11 & 1.33 ± 0.04 & 0.017 & 0.23 $\times$ 0.21 & 163.5 ± 50.4 & 24.8 ± 39.5 & $127 \pm 29$ & $26 \pm 5$ \\
HH 30 & $\leq$ 0.02 & - & 0.020 & - & - & - & $\leq 4.7$ & - \\
IRAM 04191 & $\leq$ 0.02 & - & 0.020 & - & - & - & $\leq 3.1$ & - \\
\bottomrule
\end{tabular}%
}
\vspace{2pt}
\begin{minipage}{\textwidth}
\footnotesize
\textsuperscript{a} These sources are binary systems with circumbinary disks, confirmed by the ALMA continuum images. Although the circumbinary emission is not clear in the VLA data, the total flux density ($F_\nu$) and peak intensity were measured using \texttt{imstat} to maintain consistency with the ALMA continuum flux measurements. The size and geometric parameters (deconvolved size, position angle, and inclination) are derived from \texttt{imfit}.

\textsuperscript{b} Uncertainties on the fitted continuum properties are the 2D Gaussian fitting uncertainties returned by CASA \textit{imfit}. Radius and inclination uncertainties are propagated from the deconvolved size uncertainty.

\textsuperscript{c} Uncertainties on $M_{\mathrm{dust}}$ are statistical only; see text for details.
\end{minipage}
\end{table*}
We do not see substantial extended envelope-scale continuum emission toward most sources in our high-angular-resolution ALMA and VLA continuum images ($\sim$0.2$-$0.3\arcsec); instead, the detected emission is dominated by a compact component centered on the protostar. We therefore assume that this compact emission is primarily associated with the circumstellar disk and use it to measure disk fluxes and sizes. This assumption is supported by the detailed radiative-transfer modeling of embedded systems by \citet{VANDAMOrionII_Sheehan22}, who found that the flux of the compact dust emission in their models is very strongly correlated with the disk emission. We used the CASA \texttt{imfit} task to fit a two-dimensional Gaussian model to each continuum image and measure the integrated and peak flux density, deconvolved major and minor axes, and position angle. We then derived the disk inclination from the ratio of the deconvolved minor and major axes. For systems with circumbinary disks (04287+1801, 04158+2805, 04288+1802, and 04295+2251), the continuum emission is not well described by a Gaussian profile; we therefore measure the integrated flux density using \texttt{imstat}. For these sources, we nevertheless use the \texttt{imfit} results for the deconvolved size and geometric parameters (position angle and inclination), as the Gaussian fits still provide a useful measure of the overall disk extent and orientation. All of these quantities are listed in Tables~\ref{tab:alma_continuum} and \ref{tab:vla_continuum}, and the reported uncertainties are the fitting uncertainties returned by \texttt{imfit}. We also calculate the 33--345~GHz spectral index for each source from the integrated flux densities measured in the ALMA and VLA images; the reported uncertainties are estimated using Monte Carlo propagation of the flux uncertainties, including an additional 10$\%$ calibration uncertainty added in quadrature, and the resulting values are listed in Table~\ref{tab:source_properties}, with lower limits reported for sources not detected at 33~GHz.

\par We estimated dust masses using the standard optically thin, isothermal flux-to-mass relation \citep{ftom_Hildebrand83}:
\begin{equation}
    M_\mathrm{dust} = \frac{F_\nu   d^2}{\kappa_\nu   B_\nu(T_\mathrm{dust})},
\end{equation}
where $F_\nu$ is the integrated continuum flux from the \texttt{imfit} measurements, $d$ is the source distance (from Table~\ref{tab:source_properties}), $\kappa_\nu$ is the dust opacity, and $B_\nu(T_\mathrm{dust})$ is the Planck function at the dust temperature. For the dust opacity, we adopt $\kappa_\nu = 10\,(\nu/1000~\mathrm{GHz})~\mathrm{cm}^2~\mathrm{g}^{-1}$ from \citet{dust-opacity_Beckwith&Sargent91} for both our ALMA and VLA measurements, corresponding to $\kappa_{345} = 3.45~\mathrm{cm}^2~\mathrm{g}^{-1}$ at 345 GHz and $\kappa_{33} = 0.33~\mathrm{cm}^2~\mathrm{g}^{-1}$ at 33 GHz, to remain consistent with values widely used in the literature \citep[e.g.,][]{mstar-mdisk_Andrews13, gasmodelgrid_Williams14, CAMPOSII_Hsieh25, AGE-PRO_Zhang25}. We note, however, that this assumes a common dust opacity across all sources and evolutionary stages, whereas in reality $\kappa_\nu$ may vary with dust properties across star-forming regions and across ages.
We scaled the dust temperature with bolometric luminosity using:
\begin{equation}\label{eq:dust_mass}
    T_\mathrm{dust} = T_0 \left( \frac{L_\mathrm{bol}}{1 L_\odot} \right)^{0.25},
\end{equation}
where $T_0 = 43\ \mathrm{K}$. For a detailed discussion of this scaling and its assumptions, see \citet[][Appendix~B]{VANDAMOrionI_2020}. To estimate the uncertainty on $M_{\rm dust}$, we propagated the errors in flux and distance while treating $\kappa_\nu$ and $T_{\rm dust}$ as fixed. Because $M_{\rm dust} \propto F_\nu d^{2}$, standard error propagation gives
\begin{equation}
\left(\frac{\sigma_M}{M_{\rm dust}}\right)^2
= \left(\frac{\sigma_{F_\nu}}{F_\nu}\right)^2
+ \left(2 \frac{\sigma_d}{d}\right)^2 .
\end{equation}
We estimated the flux uncertainty as the quadrature sum of the \texttt{imfit} integrated flux error and a 10\% absolute calibration term, $\sigma_{F_\nu}^2 = \sigma_{\rm imfit}^2 + (0.1 F_\nu)^2$, where the 10\% reflects the typical ALMA/VLA absolute flux-scale accuracy. We also adopted a 10\% distance uncertainty ($\sigma_d = 0.1 d$), consistent with the dispersion among Taurus subregions reported by \citet{taurus-distance-zucker20}. Uncertainty in the bolometric luminosity, $\sigma_{L_{\rm bol}}$, would additionally propagate through the temperature scaling in Equation~\ref{eq:dust_mass}; in the Rayleigh--Jeans limit this contribution is $\frac{\sigma_M}{M_{\rm dust}} \approx \frac{1}{4}\frac{\sigma_{L_{\rm bol}}}{L_{\rm bol}}$. Because $\sigma_{L_{\rm bol}}$ is not uniformly available across the comparison samples and this term enters with only a factor of $\frac{1}{4}$ and is therefore quite small, we do not include it in our quoted statistical uncertainty for $M_{\rm dust}$. 
\par In addition to the above statistical uncertainties, there are several systematic effects that can bias continuum-based dust-mass estimates. The largest uncertainty is likely the dust opacity: opacity values have substantial uncertainties and can vary depending on grain composition, size distribution, and structure, and different prescriptions in the literature can vary by up to an order of magnitude (e.g., \citealt{opacity-Ossenkopf94, dust_Woitke16}). A second important systematic is the dust temperature. While we adopt a luminosity-based temperature scaling, more detailed radiative transfer treatments for embedded disks suggest that the characteristic dust temperature can also depend on the dust disk radius \citet{VANDAMOrionII_Sheehan22}. Because dust radii are not available for all comparison samples in this work, we do not include an explicit radius dependence and instead apply the same luminosity-based prescription uniformly for consistency. For Class 0/I sources, we also assume that bolometric luminosity ($L_\mathrm{bol}$) is a proxy for the stellar luminosity ($L_\ast$) because the stellar emission is reprocessed by dust in embedded systems. Finally, we assume the millimeter continuum is optically thin; however, this assumption may not always hold (e.g., \citealt{rt_Ballering19, polarization_Ko20, OMC3_Liu21, IRAS16293-2422B_Zamponi21, FAUST:XVIII_Maureira24}), and any unaccounted-for optically thick emission would bias the inferred $M_\mathrm{dust}$ towards lower values. Applying the same opacity prescription and temperature assumptions across all samples improves the internal consistency of the comparison, but it does not eliminate source-to-source or sample-to-sample biases. In particular, variations in optical depth between individual disks, or systematic differences in optical depth between samples, may affect not only the absolute mass scale but also the inferred mass distributions, and therefore the relative comparisons between samples. As a rough estimate, the finite-optical-depth relation $F_\nu = B_\nu(T_d)[1-\exp(-\tau_\nu)]\Omega$, where $\tau_\nu$ is the dust optical depth and $\Omega$ is the emitting solid angle, implies an optically thin dust-mass correction of $\tau_\nu/[1-\exp(-\tau_\nu)]$. Thus, sources with $\tau_\nu \sim 1$ and 2 would have their dust masses underestimated by factors of $\sim$1.6 and $\sim$2.3, respectively, while more optically thick sources could be biased low by factors of a few. A fully self-consistent treatment of these effects would require detailed radiative transfer modeling for individual disks, which is beyond the scope of this work. We therefore interpret relative dust-mass trends as suggestive rather than exact.

\par To estimate dust disk sizes, we followed the approach used in other embedded Class~0/I disk studies, which derives the radius from the deconvolved major axis of a two-dimensional Gaussian fit \citep{CAMPOSI_Hseih24, VANDAMOrionI_2020, VANDAMPerseus_Tobin16}.  We determined whether a disk is spatially resolved using the criterion $\theta_{\mathrm{maj,decon}} > 2\sigma_{\theta_{\mathrm{maj,decon}}}$, where $\theta_{\rm maj,decon}$ is the deconvolved major-axis and $\sigma_{\theta_{\rm maj,decon}}$ is its fitting uncertainty returned by \texttt{imfit}. All disks in the ALMA data are resolved, while in the VLA data 5 of the 27 detected disks are unresolved. For resolved sources, we defined the disk dust radius as $R_{\mathrm{dust}} = (2/2.355) \theta_{\mathrm{maj,decon}}  d$, where $d$ is the source distance, to approximate the $2\sigma$ extent of the fitted Gaussian, which encloses 95\% of the total flux. The fractional uncertainty on the radius combines in quadrature the relative uncertainty of the deconvolved major axis and the 10\% distance uncertainty. For unresolved sources, we report an upper limit on the dust radius using the same Gaussian definition, $R_{\mathrm{dust}} < (2/2.355) \frac{\theta_{\mathrm{maj,conv}}}{\sqrt{\mathrm{SNR}}} d$,
where $\theta_{\mathrm{maj,conv}}$ is the convolved major axis of the restoring beam and SNR is the peak signal-to-noise ratio. Here, the SNR is the ratio of the \texttt{imfit} peak flux density to the root-mean-square (rms) noise measured from the \texttt{tclean} residual image, which is effectively source-free and represents the noise level of the primary-beam–corrected continuum map. Tables~\ref{tab:alma_continuum} (ALMA) and~\ref{tab:vla_continuum} (VLA) list the measured continuum properties, together with the derived dust masses and radii for each disk.

\begin{figure*}[ht!]
\centering

\begin{minipage}{0.48\textwidth}
    \centering
    \includegraphics[width=\linewidth]{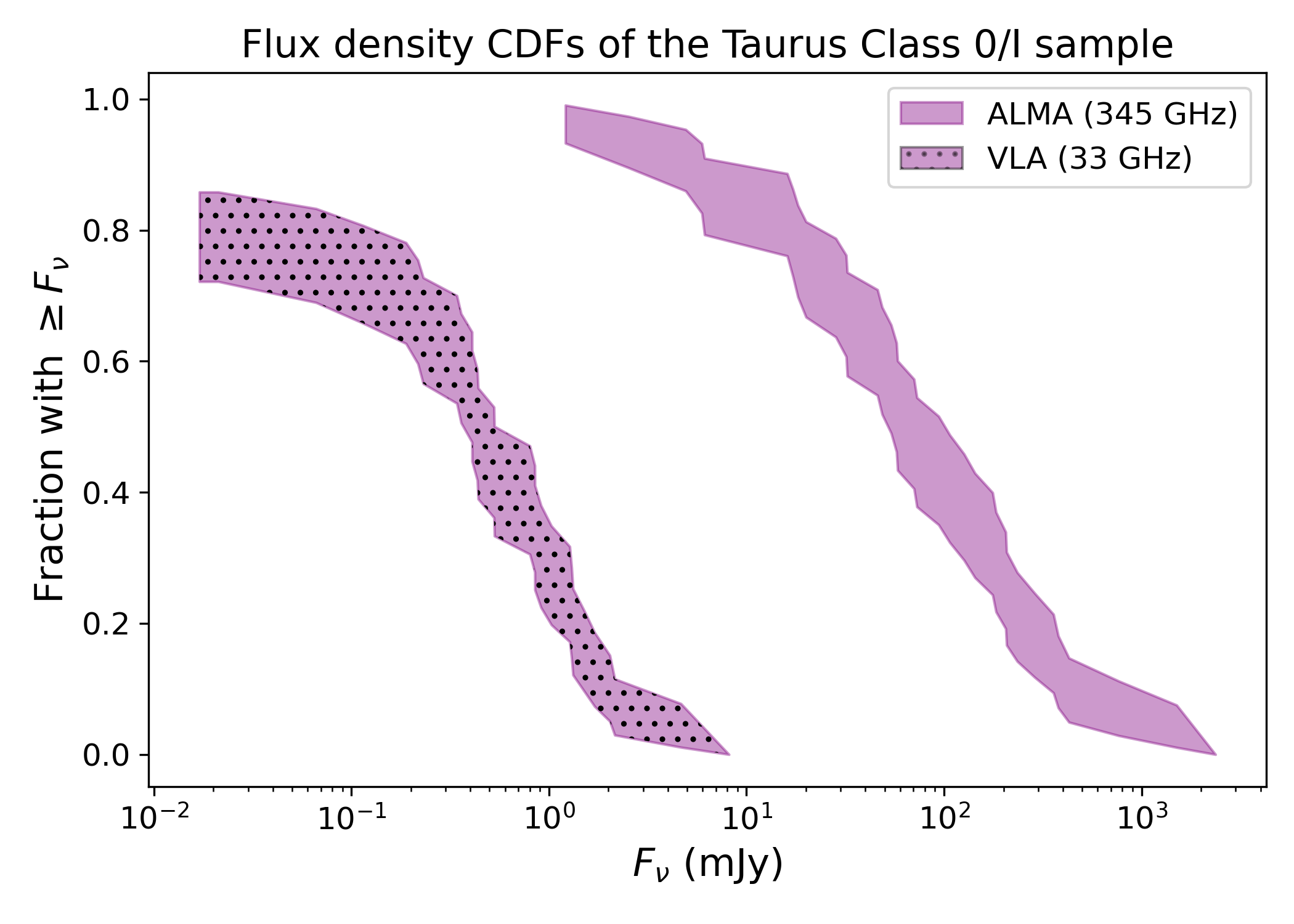}
    \\[-0.5ex]
    {\small (a) Flux Desnity CDF}
\end{minipage}
\hfill
\begin{minipage}{0.48\textwidth}
    \centering
    \includegraphics[width=\linewidth]{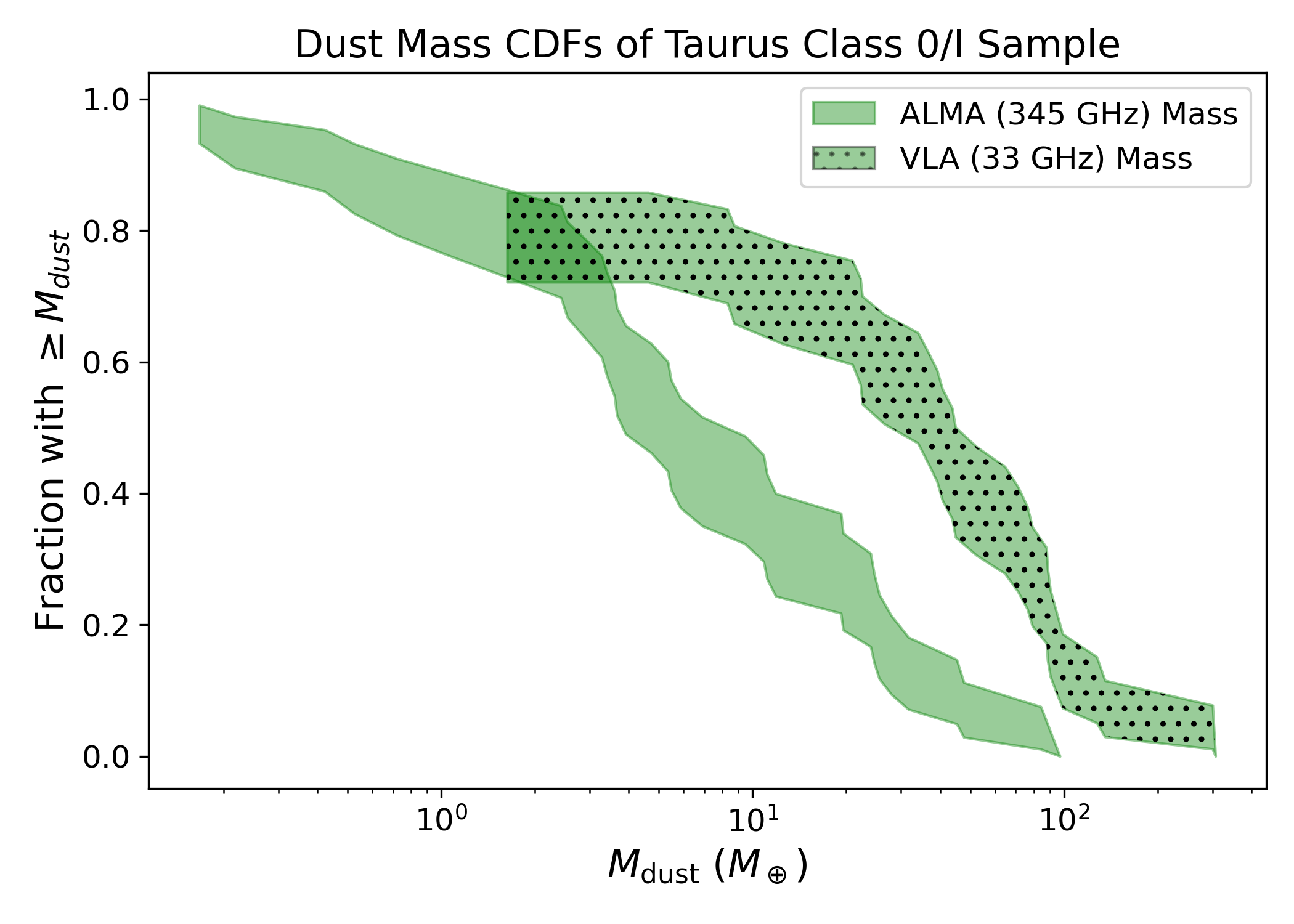}
    \\[-0.5ex]
    {\small (b) Dust mass CDF}
\end{minipage}

\vspace{1em}

\begin{minipage}{0.5\textwidth}
    \centering
    \includegraphics[width=\linewidth]{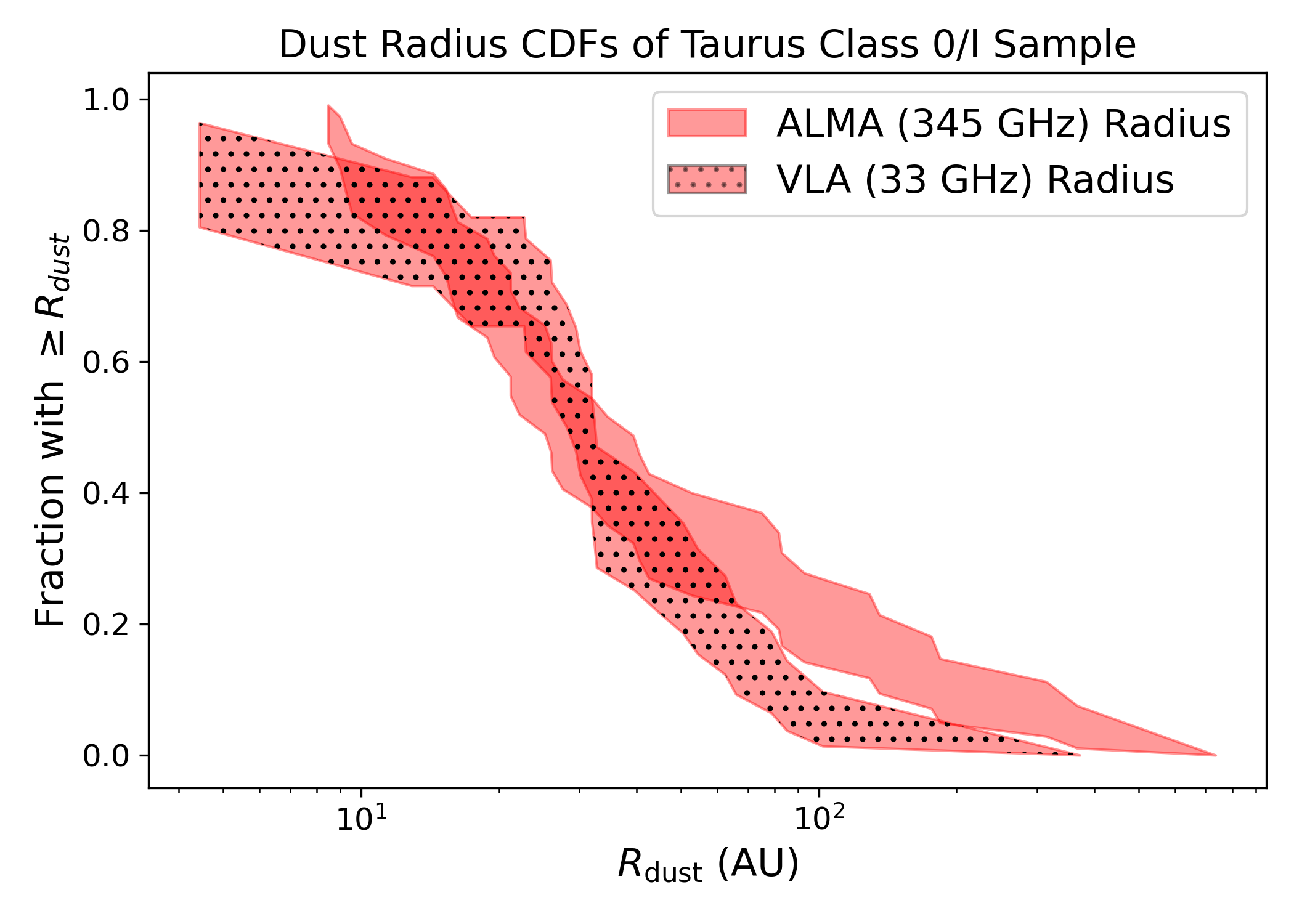}
    \\[-0.5ex]
    {\small (c) Radius CDF}
\end{minipage}

\caption{Cumulative distribution functions (CDFs) of ALMA and VLA continuum properties for the Taurus Class~0/I sample: (a) flux densities, (b) dust masses, and (c) dust radii. Shaded regions show 68\% confidence intervals from Kaplan--Meier survival analysis (i.e., statistical uncertainty from finite sample size and censoring), and do not include measurement uncertainties in the individual fluxes/masses/radii.}
\label{fig:our-cdfs}
\end{figure*}

\par We computed cumulative distribution functions (CDFs) of the continuum fluxes, dust masses, and disk radii using the \texttt{lifelines} Python package \citep{lifelines_Davidson-Pilon2019}, which implements the Kaplan–Meier (KM) estimator\footnote{https://lifelines.readthedocs.io/en/latest/fitters/univariate/KaplanMeierFitter.html} to account for upper limits. In our sample, non-detections enter as upper limits and were therefore treated as left-censored measurements, meaning that their exact values are not known but are constrained to lie below the reported limits. The KM estimator incorporates both detections and upper limits to construct a non-parametric, stepwise estimate of the survival function, from which the CDFs shown in Figure \ref{fig:our-cdfs} were derived. For each distribution, we report the median value $(m)$, defined as the value where the Kaplan--Meier survival function crosses 0.5. We estimated the 68\% confidence interval on the median from the values at which the lower and upper Kaplan–Meier confidence bounds intersect the 0.5 survival-probability level.

\par The ALMA flux densities span a wide dynamic range, from 1.2~mJy to 2.4~Jy, with a median flux density of 71~mJy (68\% confidence interval: 54--107~mJy). The VLA flux densities are systematically lower, with a median of 0.43~mJy (68\% confidence interval: 0.41--0.80~mJy) and values extending up to 8.15~mJy, with the lower end of the distribution set by upper limits for the non-detections. Several disks---04191+1523A, 04248+2612A, 04248+2612B, 04248+2612C, 04489+3042B, HH~30, and IRAM~04191---are not detected in the VLA images; for these sources, we adopt the image rms as the upper limit on the flux density. The difference in flux density levels between the ALMA and VLA measurements is expected because dust continuum emission scales as $F_\nu \propto \nu^{2+\beta}$ (with $0 \lesssim \beta \lesssim 2$), which causes the continuum to be much fainter at 33~GHz than at 345~GHz. The ALMA continuum measurements correspond to dust masses between 0.17 and 97~$M_\oplus$, with a median of 5.5~$M_\oplus$ (68\% confidence interval: 3.9--9.4~$M_\oplus$), while the VLA dust masses reach up to 307~$M_\oplus$ and have a median of 39~$M_\oplus$ (68\% confidence interval: 34--52~$M_\oplus$). The VLA-derived masses are generally offset from the ALMA values for individual sources, and over the portion of the KM-estimated dust-mass CDFs below a fraction of 0.6, the ALMA-based dust masses would need to be multiplied by $\sim$5.7 to approximately match the VLA-based distribution. This systematic difference may reflect uncertainties in the frequency dependence of the dust opacity \citep[e.g.,][]{VANDAMOrionI_2020, VANDAMOrionII_Sheehan22}, the presence of optically thick dust emission, especially at 345~GHz, which would cause the millimeter-wavelength fluxes to underestimate the true dust mass \citep[e.g.,][]{FAUST:XVIII_Maureira24, IRAS16293-2422B_Zamponi21}, or free-free contamination at 33~GHz \citep[e.g.,][]{VANDAMPerseus_Tychoniec18, ORANGES_Bouvier21, VLA-taurus-classII_Garufi25} that causes the VLA-based dust masses to be overestimated.

\par We do not correct the VLA flux densities for free--free emission because we do not have longer-wavelength data for individual sources that would allow us to separate free--free emission from thermal dust emission. We can therefore only make approximate estimates of its possible contribution. In this context, the 33--345~GHz spectral indices listed in Table~\ref{tab:source_properties} are below 2 for several sources, which may suggest a non-negligible free--free contribution at 33~GHz. However, such low spectral indices are not unique evidence for free--free emission, as they can also arise from optically thick dust self-scattering or self-absorption at 345~GHz \citep[e.g.,][]{beta_Li17, alpha_Liu19}. As an approximate estimate of the possible free--free contribution, we use the empirical 4.1 and 6.4 cm radio luminosity--bolometric luminosity relations from \cite{VANDAMPerseus_Tychoniec18}, implicitly assuming that the centimeter emission traced by these relations is dominated by free--free emission. These relations suggest median contributions of $\sim$10--15$\%$ to the observed 33 GHz fluxes in our Taurus sample, although in some sources the inferred contribution could be as high as $\sim$100$\%$. Because these relations were derived for Class~0/I disks in Perseus rather than Taurus, and because the inferred free--free contribution can differ substantially from source to source, they provide only an approximate guide to the average free--free contribution across our sample rather than a robust estimate for any individual source. This source-to-source variation is also seen in \cite{AMI-taurus0/I_Scaife12}, who found that in a sample of seven Taurus protostellar disks, the Class~I disks are often dust-dominated even at 1.8~cm, whereas at least one Class~0 source (IRAS~04368+2557) shows a much larger non-dust contribution. Since our VLA data are at 33~GHz ($\lambda\approx9$~mm), shorter than 1.8~cm, we would therefore expect the non-dust contribution to be smaller for many sources (as in the Class~I systems), though it could remain significant for the three Class~0 systems in our sample. \citet{ORANGES_Bouvier21} found evidence for source-associated free-free emission in 5 of 16 Class~0/I protostars in Orion, with inferred 32.9~GHz free-free fractions of 47--100$\%$ in those sources and up to 24$\%$ in the remainder. For Taurus-specific centimeter studies, \citet{VLA-taurus-classII_Garufi25} found that the free-free contribution spans $\sim$10--75$\%$ of the total 1~cm flux, with an average of $\sim$35$\%$, although their sample consists of the more evolved Class II population. Taken together, these results suggest that free-free emission may be significant at 33 GHz for at least some sources in our sample, and centimeter-wavelength measurements would be needed to quantify its contribution robustly.

\par The ALMA radii span from 8~AU up to 736~AU, with a median radius of 28~AU (68\% confidence interval: 25--39~AU). The VLA radii extend up to 372~AU, with a median radius of 32~AU (68\% confidence interval: 29--33~AU). The ALMA radii are expected to be more reliable tracers of the full extent of the disk because the 345~GHz emission is brighter and traces extended disk structures, providing higher SNR per beam. At 33~GHz, the fainter continuum results in lower SNR, so size measurements are more uncertain and more sources yield upper limits. The discrepancy between the ALMA- and VLA-derived radii becomes more pronounced at larger disk sizes (see Figure~\ref{fig:our-cdfs}c), where the low surface brightness of the outer disk at 33~GHz causes the VLA to miss faint emission that remains detectable at 345~GHz, leading to an underestimation of disk extents for the most extended sources.

\subsection{CO Line Analysis and Keplerian Masking}\label{subsec:line-analysis}
\begin{table*}[t]
\caption{Estimated \textsuperscript{13}CO and C\textsuperscript{18}O (J=3-2) line fluxes, stellar masses, and derived gas masses.}
\label{tab:mask_summary}
\centering

\resizebox{\textwidth}{!}{%
\begin{tabular}{lccccccccc}
\hline\hline
Source &
$v_{\rm sys}$ &
$M_{\ast,\,^{13}\mathrm{CO}}$ &
$M_{\ast,\,\mathrm{C^{18}O}}$ &
Adopted $M_\ast$ &
$F_{^{13}\mathrm{CO}}$ &
$\sigma_{^{13}\mathrm{CO}}$ &
$F_{\mathrm{C^{18}O}}$ &
$\sigma_{\mathrm{C^{18}O}}$ &
$M_{\rm gas}$\textsuperscript{g} \\
& (km s$^{-1}$) &
($M_\odot$) &
($M_\odot$) &
($M_\odot$) &
(Jy km s$^{-1}$) &
(Jy km s$^{-1}$) &
(Jy km s$^{-1}$) &
(Jy km s$^{-1}$) &
($10^{-4}\,M_\odot$) \\
\hline
04016+2610 & 7.0 & 1.45 & 1.40 & 1.45 & 89.89 & 31.24 & 35.46 & 9.373 & $260^{+380}_{-140}$ \\
04108+2803B\textsuperscript{a} & 7.2 & 0.7 & 0.85 & 0.7 & 2.30 & 0.51 & 0.78 & 0.140 & $4.7^{+42}_{-3.1}$ \\
04158+2805\textsuperscript{b,c} & 7.0\textsuperscript{d} & 0.15 & 0.6 & 0.35 & 29.99 & 7.362 & 9.102 & 1.512 & $15^{+25}_{-13}$  \\
04166+2706 & 6.6 & 0.3 & 0.15 & 0.3 & 2.95 & 0.797 & 1.15 & 0.328 & $13^{+16}_{-8}$ \\
04169+2702\textsuperscript{a} & 6.2 & 0.35 & 0.2 & 0.35 & 4.77 & 1.17 & 1.61 & 0.426 & $6.3^{+11}_{-2.6}$\\
04181+2654A & 6.9 & 0.15 & 0.3\textsuperscript{e} & 0.15 & 1.04 & 0.338 & 0.129 & 0.0444 & $0.38^{+0.35}_{-0.26}$\\
04191+1523A & 7.1 & 0.25 & 0.20 & 0.25 & 4.97 & 1.29 & 1.77 & 0.436 & $9.1^{+53}_{-5.5}$\\
04191+1523B & 7.4 & 0.3 & 0.05 & 0.3 & 2.35 & 0.687 & 0.959 & 0.261 & $8.2^{+30}_{-4}$\\
04239+2436AB & 6.2 & 0.35 & 0.15 & 0.35 & 18.3 & 5.46 & 4.73 & 1.58 & $17^{+7.8}_{-5.4}$\\
04248+2612AB & 6.7 & - & 0.2\textsuperscript{e} & 0.1 & 3.11 & 1.21 & 0.687 & 0.322 & $1.8^{+12}_{-1.4}$\\
04260+2642\textsuperscript{c} & 6.1 & 0.35 & 1.0 & 0.65 & 4.73 & 1.13 & 1.47 & 0.235 & $6.7^{+42}_{-6}$\\
04263+2426A\textsuperscript{b} & 6.5\textsuperscript{d} & - & 0.95 & 1.8 & 4.79 & 1.29 & 0.674 & 0.198 & $4.4^{+3.3}_{-2.5}$\\
04263+2426B\textsuperscript{b} & 6.5 & - & 1.15 & 1.8 & 2.92 & 0.928 & 0.377 & 0.114 & $0.48^{+1.2}_{-0.38}$\\
04264+2433AB\textsuperscript{b} & 6.8, 5.0 & 0.35, 0.2 & 0.2, 0.4\textsuperscript{e} & 0.35, 0.2 & 2.37 & 0.438 & 0.315 & 0.0607 & $0.29^{+0.54}_{-0.19}$\\
04287+1801\textsuperscript{f} & 6.5, 6.3 & 0.85, 0.95 & 0.25, 0.3 & 0.85, 0.4 & 22.53 & 6.274 & 18.14 & 4.97 & $340^{+370}_{-190}$\\
04288+1802\textsuperscript{a,c,f} & 7.0 & 0.55 & 0.65 & 0.65 & 20.7 & 2.56 & 10.9 & 2.07 & $200^{+240}_{-81}$\\
04295+2251 & 6.0 & 0.55 & 0.8 & 0.55 & 13.1 & 3.34 & 3.72 & 0.739 & $18^{+77}_{-13}$\\
04302+2247\textsuperscript{a} & 5.6 & 1.25 & 1.3 & 1.3 & 28.3 & 9.16 & 10 & 1.98 & $63^{+230}_{-28}$\\
04325+2402A\textsuperscript{a,c} & 4.7 & 0.95 & 0.9 & 0.95 & 2.7 & 0.929 & 0.595 & 0.177 & $1.5^{+1.3}_{-1.2}$\\
04325+2402B\textsuperscript{a} & 6.4 & 0.05 & 0.05\textsuperscript{e} & 0.1 & 1.3 & 0.331 & 0.332 & 0.0709 & $1.4^{+1}_{-1.1}$\\
04361+2547\textsuperscript{b,f} & 5.5\textsuperscript{d} & 1.5 & 1.25 & 1.5 & 8.64 & 2.35 & 1.57 & 0.452 & $0.91^{+4.8}_{-0.81}$\\
04365+2535\textsuperscript{c} & 6.0\textsuperscript{d} & 0.75 & 0.6 & 0.6 & 9.74 & 3.44 & 5.46 & 1.82 & $160^{+380}_{-110}$\\
04368+2557\textsuperscript{a} & 5.9 & 0.5 & 0.15 & 0.5 & 6.26 & 1.98 & 3.99 & 1.19 & $120^{+370}_{-88}$\\
04381+2540AB\textsuperscript{f} & 5.2\textsuperscript{d} & 1.0 & 0.3 & 1.35 & 7.85 & 2.97 & 4.13 & 1.79 & $140^{+390}_{-100}$\\
04385+2550 & 5.2 & 0.35 & 0.25\textsuperscript{e} & 0.35 & 1.9 & 0.439 & 0.281 & 0.045 & $0.28^{+1.3}_{-0.18}$\\
04489+3042A & 6.1 & 0.25 & 0.3 & 0.3 & 1.36 & 0.332 & 0.34 & 0.0736 & $1.5^{+0.94}_{-1}$\\
DG TauB\textsuperscript{a} & 6.5 & 0.45 & 0.45 & 0.45 & 15.5 & 6.18 & 7.59 & 2.82 & $120^{+370}_{-84}$\\
HH 30\textsuperscript{c} & 6.7 & 0.45 & 0.55 & 0.45 & 5.13 & 1.06 & 1.34 & 0.137 & $4.4^{+5.1}_{-3.9}$\\
IRAM 04191 & 6.5 & 0.25 & 0.25 & 0.25 & 1.32 & 0.258 & 0.496 & 0.129 & $140^{+390}_{-100}$\\
\hline
\end{tabular}%
}

\vspace{2pt}
\begin{minipage}{\textwidth}
\footnotesize
\textsuperscript{a} For these sources, no $180^\circ$ offset to the position angle (PA) was required. For all other sources, the position angle (PA) used for the Keplerian mask was obtained by adding $180^\circ$ to the original PA, since the mask convention defines the PA as that of the redshifted side of the disk.

\textsuperscript{b} For these sources, $r_{\rm out}$ was not set to $R_{\rm dust}$ when constructing the Keplerian masks for the stellar-mass search. For 04158+2805, we used $r_{\rm out}=0.5R_{\rm dust}$ so that the mask more closely traces the central binary system, because using the full circumbinary dust radius produced a mask that was too large and extended into substantial regions not associated with any emission. For 04263+2426A we adopted $r_{\rm out}=3R_{\rm dust}$, and for 04263+2426B, 04264+2433AB, and 04361+2547 we adopted $r_{\rm out}=2R_{\rm dust}$ because their small dust radii caused the Keplerian mask to miss emission in several velocity channels.

\textsuperscript{c} For all sources we adopted $z/r=0.3$ except for these noted disks. The following disks required different values based on emission: 04158+2805 ($z/r=0.5$), 04260+2642 ($z/r=0.1$), 04288+1802 ($z/r=0.4$), 04325+2402A ($z/r=0.2$), 04365+2535 ($z/r=0.5$), and HH~30 ($z/r=0.1$).

\textsuperscript{d} For these sources, the systemic velocity was determined manually from the channel maps because the moment~1 median was unreliable (see Section~\ref{subsec:line-analysis}).

\textsuperscript{e} For these sources, C$^{18}$O was not used in the $\chi^2$-like calculation because the C$^{18}$O emission is too noisy to reliably trace the disk kinematics; $^{13}$CO was used instead.

\textsuperscript{f} For the following sources, slightly different source centers than those listed in Table~\ref{tab:source_properties} were adopted when constructing the Keplerian masks in order to ensure that the masks fully captured the relevant emission regions. 04287+1801: This highly luminous source ($L_{\rm bol}\sim27\,L_\odot$) exhibits emission that is not consistent with the $^{13}$CO or C$^{18}$O emission, likely including contributions from other molecular species, which complicates the mask determination. The system is a binary surrounded by a circumbinary disk; for mask construction, we adopted the positions of the brighter component A (RA = 04:31:34.163, Dec = +18:08:04.8) and the fainter component B (RA = 04:31:34.167, Dec = +18:08:04.2). 04288+1802: This is a binary system surrounded by a circumbinary disk. To construct the Keplerian mask, we combined two masks: one centered on the binary center, and a second centered at RA = 04:31:44.51297, Dec = +18:08:31.32937, to ensure that all disk emission was fully captured. 04361+2547: The Keplerian mask was centered at RA = 04:39:13.926, Dec = +25:53:20.4 to capture the emission. 04381+2540AB: The declination was updated to Dec = +25:46:34.66. The two disks are not resolved in the line data, so we adopted the full radius encompassing both sources.

\textsuperscript{g} The reported gas masses are the medians of the match-weighted gas-mass distributions, with superscript and subscript giving the differences between the median and the 84th and 16th percentiles, respectively. See Section~\ref{subsec:gas-mass} for details.
\end{minipage}
\end{table*}

\begin{figure*}[ht!]
    \centering
    \includegraphics[width=0.80\textwidth]{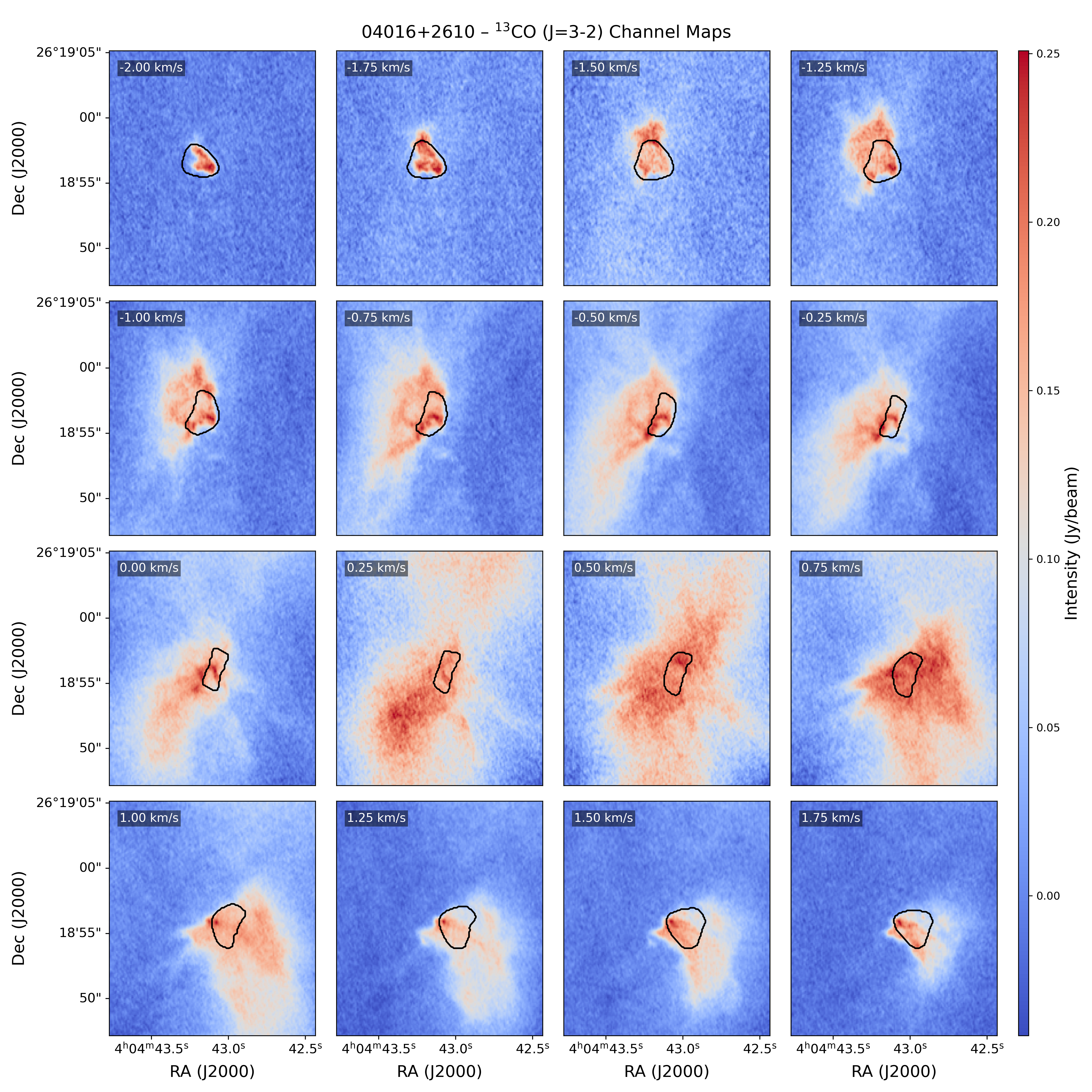}
    \caption{Example channel maps illustrating the Keplerian mask used in our analysis. Shown here are $^{13}$CO (J=3-2) channel maps for IRAS 04016+2610 at velocities reported relative to the systemic velocity ($v_{\rm sys} \approx 7$~km s$^{-1}$). The black contour shows the Keplerian mask computed for the best-fit stellar mass ($1.45 M_\odot$) and an outer radius $r_{\rm out} = 1.0 \times R_{\rm dust}$. At each velocity channel, the mask isolates emission consistent with Keplerian rotation, enabling extraction of disk emission while excluding envelope contamination. The full set of channel maps with the adopted Keplerian masks for all sources is presented as an image set in Appendix~\ref{app:channel_maps}.} 
    \label{fig:keplerian_mask}
\end{figure*}

We use the $^{13}$CO and C$^{18}$O line observations to trace the gas disk component because they provide complementary leverage on disk gas column densities. CO is the most abundant observable molecule in disks after H$_2$, which remains undetectable under the cold conditions typical of protoplanetary disks ($T \lesssim 100$ K). However, the most abundant CO isotopologue, $^{12}$CO, is generally optically thick and therefore not a reliable tracer of total gas mass. While $^{13}$CO can also reach moderate optical depths \citep[e.g.,][]{13co-dutrey96,13co-Zadelhoff01}, it is typically less saturated than $^{12}$CO, and in combination with the rarer C$^{18}$O line, which is often optically thinner, provides a more accurate probe of the overall molecular gas reservoir \citep[e.g.,][]{13co-c18o-Goldsmith97, gasmodelgrid_Williams14, NOISO_Miotello16}. However, both $^{13}$CO and C$^{18}$O can become optically thick in dense regions of embedded disks \citep[e.g.,][]{L1527_vantHoff18}, making gas mass estimates more challenging; we discuss these optical depth effects in Section \ref{subsec: gas-dust-ratio}.
\par Our targets are deeply embedded Class 0/I systems in which the low-velocity channels are often contaminated by extended envelope emission. To isolate the disk emission, we applied a physically motivated Keplerian mask under the assumption that the disk follows Keplerian rotation about the central protostar, while the envelope experiences a more complicated structure of infall and rotation \citep[e.g.,][]{envelope_model_Ulrich76, core-collapse_Terebey84, envelope-model_Oya22}. In reality, the innermost envelope can exhibit rotational signatures that overlap with Keplerian disk kinematics, so a clean separation of disk and envelope emission is not always possible with masking alone; our masks are therefore designed to minimize envelope contamination while retaining the disk-dominated signal. We generated 3D ($x$–$y$–$v$) Keplerian masks for each source using \texttt{keplerian\_mask.py} \citep{keplerian-masks_Teague20} \footnote{\url{https://github.com/richteague/keplerian_mask}; written by Richard Teague}. The key mask input parameters adopted for each source are listed in Table~\ref{tab:mask_summary}. To construct each mask, the code requires the disk inclination ($i$) and position angle (PA), the stellar mass ($M_\star$), distance, and systemic velocity ($v_{\rm sys}$), which together define the projected Keplerian velocity field. We also specified radial limits ($r_{\rm min}$–$r_{\rm max}$) for the masked region, a target spatial resolution of 0.3$''$ to match our observations, and an assumed $z/r$ aspect ratio to account for the disk’s vertical structure. 
\par We defined the disk geometry using the deconvolved position angle and inclination (PA and $i$) from \texttt{imfit} on the continuum image (Table \ref{tab:alma_continuum}). Because the code defines the position angle along the redshifted major axis, we added $180^\circ$ when the continuum PA referred to the blueshifted side. This adjustment is necessary because the continuum PA is intrinsically $180^\circ$ degenerate, in contrast to the line emission, which uniquely identifies the redshifted and blueshifted sides. We adopted a uniform distance of 140 pc (the mean Taurus distance) for all sources, since the typical mask uncertainties are large enough that individual distance variations within the region are negligible. We set the systemic velocity ($v_{\rm sys}$) to the median of the moment~1 map within the continuum disk region. However, this approach is not always reliable in embedded systems, where envelope emission can distort the moment~1 structure. For sources where the initial estimate fell outside the expected Taurus range (5.5–7.0~km~s$^{-1}$), we refined $v_{\rm sys}$ by visually identifying, in the channel maps, the velocity at which the emission transitions between blueshifted and redshifted. We set the emitting layer height to $z/r = 0.3$ to represent emission from both disk surfaces and adjusted it slightly for sources whose channel maps showed systematically broader or narrower emission. We fixed the inner radius at $r_{\rm in}=0$, which does not affect the masked emission at our angular resolution. For the initial stellar-mass estimate, we set the gas outer radius to $r_{\rm out} = 1.0 \times R_{\rm dust}$ (Table \ref{tab:alma_continuum}), motivated by the assumption that gas disks are observationally at least comparable in size to, and often larger than, their dust disks \citep[e.g.,][]{lupusI_Andsell16, gas-size_Trapman19}. Later, when measuring line fluxes, we varied larger outer radii up to $3.0 \times R_{\rm dust}$. In a few cases, this prescription required adjustment: (i) when $R_{\rm dust}$ was so small that $r_{\rm out}$ spanned only a few image pixels and the resulting mask was not even visible in some channels, and (ii) when circumbinary structure inflated $R_{\rm dust}$. In the first case, we increased $r_{\rm out}$ to ensure that disk emission was captured across all velocity channels; specifically, we adopted $r_{\rm out} = 3R_{\rm dust}$ for 04263+2426A and $r_{\rm out} = 2R_{\rm dust}$ for 04263+2426B, 04264+2433B, and 04361+2547. Because the dust radii of these disks are very small, increasing $r_{\rm out}$ by a modest factor does not significantly affect the measured line flux, but prevents emission from being excluded by an undersized Keplerian mask. In the second case, for 04158+2805, which hosts a circumbinary disk, we adopted a smaller value of $r_{\rm out}=0.5R_{\rm dust}$ so that the mask more closely traces the central binary system, because using the full circumbinary dust radius produced a mask that was too large and included substantial regions not associated with any emission. All adopted values of $r_{\rm out}$ are listed in Table~\ref{tab:mask_summary}. After fixing $r_{\rm out}$, we searched a grid of stellar masses ($M_\star \in [0.05,2] M_\odot$ in 0.05 M$_\odot$ steps), generating a Keplerian mask for each trial mass and evaluating two complementary criteria. 
\begin{enumerate}

\item {$^{13}$CO channel-extent method (S/N-driven):} For each trial $M_\star$, we evaluated whether the observed $^{13}$CO emission was consistent with the velocity range predicted by the corresponding Keplerian mask. We required the first and last masked velocity channels for a given $M_\star$ to contain $\ge 3\sigma$ emission. If either edge failed this criterion, the trial mass was deemed too large. Conversely, if $\ge 3\sigma$ emission extended beyond the masked region on both the blue and red sides, the trial mass was considered too small. As the trial mass increased, the mask’s velocity extent broadened, producing a characteristic transition from under-predicting the observed velocity range (too small) to over-predicting it (too large). We adopted the stellar mass ($M_\star$) at the transition that satisfies the edge-emission criterion, corresponding to the point where the mask transitions from under- to over-predicting the observed velocity extent. We used $^{13}$CO for this step because its higher signal-to-noise (S/N), particularly at large velocity offsets from the systemic velocity, provided a more reliable measure of the full velocity extent. However, since a $3\sigma$ threshold can be arbitrary in some channels, we also evaluate the C$^{18}$O $\chi^2$-like method.
\item {C$^{18}$O $\chi^2$-like method:} To complement the S/N-based $^{13}$CO test, we used a C$^{18}$O-based metric to evaluate how well each trial stellar mass reproduced the observed velocity structure. For each trial mass, we compared the C$^{18}$O cube to the corresponding Keplerian mask. In the remaining channels, we computed a $\chi^{2}$-like statistic that measures the amount of residual emission outside the mask, where disk emission is not expected. We also added a small penalty term proportional to the masked area to avoid favoring unrealistically large masks. The overall cost for a given stellar mass was taken as the mean of this quantity across all included channels, and we adopted the mass that minimized this cost. We used C$^{18}$O because it is less optically thick and therefore less affected by extended-envelope contamination than $^{13}$CO. We also excluded channels near the systemic velocity to down-weight envelope emission, since these central channels are heavily contaminated by large-scale envelope emission.
\end{enumerate}
\par The $^{13}$CO and C$^{18}$O-based stellar mass estimates provided approximate initial values for $M_\star$ to use in constructing the Keplerian masks. We then inspected the resulting masks in the channel maps by eye and adopted the stellar-mass value whose mask best traced the observed disk emission. The $^{13}$CO and C$^{18}$O-based estimates, as well as the final adopted $M_\star$ values, are listed in Table~\ref{tab:mask_summary}. A more rigorous determination of stellar mass from the molecular-line kinematics would require modeling of the full disk+envelope velocity structure (and, ideally, the line emission) in a self-consistent framework, which is beyond the scope of this work. A comprehensive radiative-transfer analysis of the line emission and kinematics for this sample is currently in progress (C.~Plante et al., in prep.) and will provide more physically motivated stellar-mass constraints for sources in this sample. At the time this work was carried out, those radiative-transfer-based stellar masses were not yet available; however, they are now available for a subset of the sample, and we find that gas-mass estimates inferred using those radiative-transfer stellar masses are consistent with our fiducial gas masses within the quoted uncertainties. The adopted $M_\star$ values should therefore be interpreted as practical mask-construction choices, rather than precise measurements of the true stellar masses. An example of a Keplerian mask overlaid on the channel maps is shown in Figure~\ref{fig:keplerian_mask}, while the corresponding adopted mask overlays for all sources in both the $^{13}$CO and C$^{18}$O lines are presented in Appendix~\ref{app:channel_maps}. 

\par With $M_\star$ fixed to the adopted value, we refined the outer radius by scanning $r_{\rm out} = f \times R_{\rm dust}$ for $f = 1.0$-$3.0$, motivated by the fact that gas disks generally appear larger than dust disks \citep{gas-size_Trapman19}. For each $r_{\rm out}$, we regenerated the Keplerian mask and applied it to the observed line cubes, created masked moment~0 maps, and measured the integrated line flux using \texttt{imstat}. We adopted the mean flux across the $r_{\rm out}$ grid as the measured value and the standard deviation as the associated uncertainty. The adopted stellar masses ($M_\star$) and final fluxes with uncertainties are reported in Table~\ref{tab:mask_summary}.

\subsection{Estimating Gas Mass from Radiative--Transfer Model Grids}\label{subsec:gas-mass}
\begin{table}[ht!]\label{tab: model-param}
\centering
\caption{Parameter Values of the Model Grid}
\begin{tabular}{ll}
\hline
\hline
Parameter & Range \\
\hline
Stellar Mass $(M_{\star})$ & 0.5, 1.0, 1.5~$M_{\odot}$ \\
Stellar Luminosity $(L_{\star})$ & 0.1, 1.0, 5.0, 25.0~$L_{\odot}$ \\
Disk Dust Mass ($M_{\mathrm{dust}}$) & $1\times10^{-7}$, $3\times10^{-7}$, $1\times10^{-6}$, $3\times10^{-6}$, $1\times10^{-5}$, $3\times10^{-5}$, $1\times10^{-4}$, $3\times10^{-4}$, $1\times10^{-3}$~$M_{\odot}$\\
Disk Critical Radius $(r_c)$ & 10, 30, 60, 100, 200~AU \\
Surface Density Index ($\gamma$) & 0.0, 0.75, 1.5 \\
Flaring Index ($\psi)$ & 1.0, 1.2 \\
Envelope Mass $(M_{\mathrm{env}})$ & $1\times10^{-5}$, $1\times10^{-4}$, $1\times10^{-3}$~$M_{\odot}$ \\
Inclination $(i)$ & 0$^{\circ}$, 45$^{\circ}$, 75$^{\circ}$ \\
$[^{13}\mathrm{CO}]$ abundance ratio & 70 \\
$[{\mathrm{C}}^{18}\mathrm{O}]$ abundance ratio & 550 \\
\hline
\end{tabular}
\end{table}

\begin{figure*}[t]
    \centering
    \includegraphics[width=0.85\textwidth]{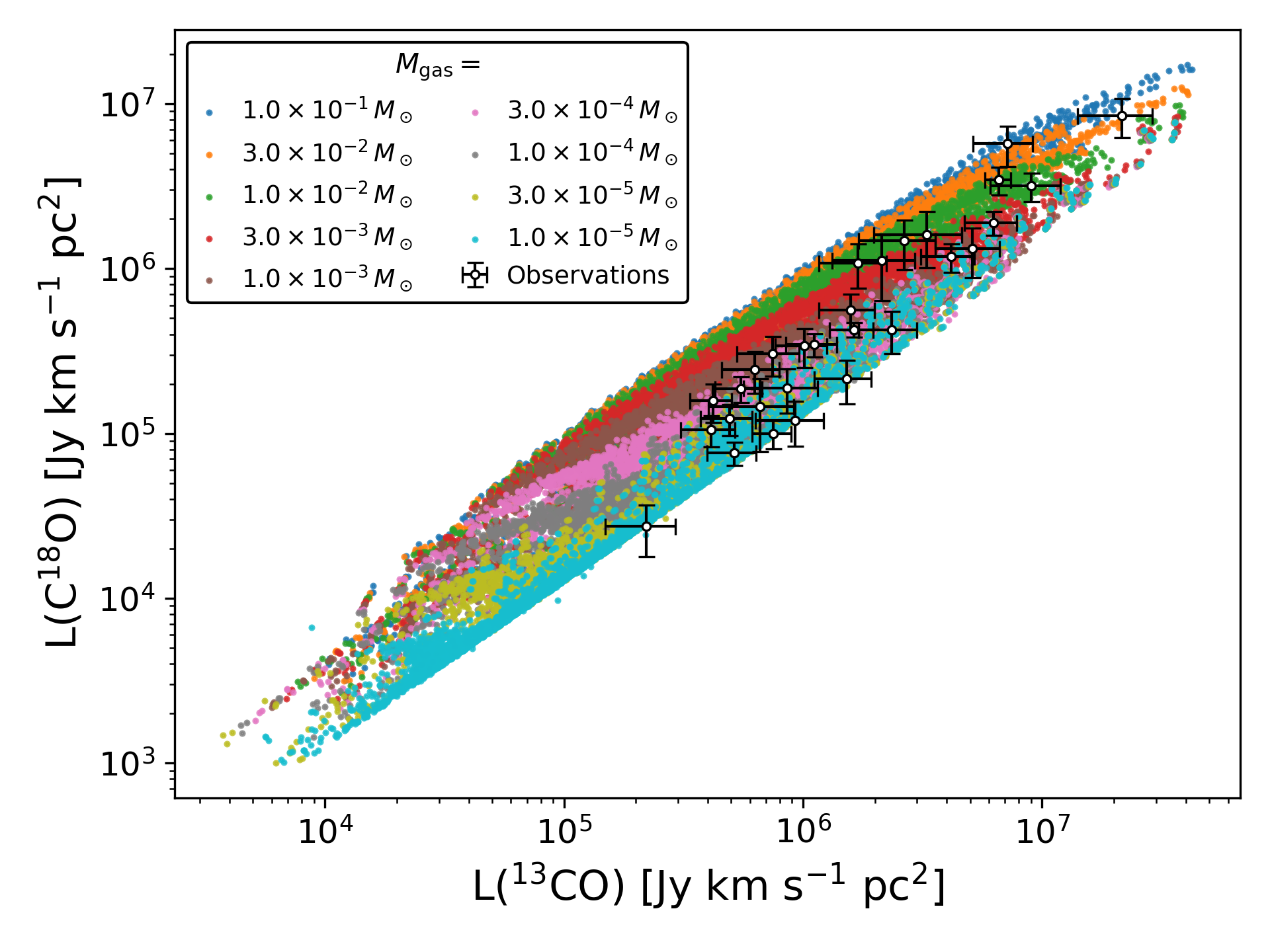}
    \caption{
    Comparison of model grid luminosities and observed CO isotopologue luminosities.
    The colored points show the modeled $L(\mathrm{C^{18}O})$ versus $L(\mathrm{^{13}CO})$ values for a range of gas masses, based on a grid of 29,160 paired $^{13}$CO and C$^{18}$O models. The black points with error bars represent the observed luminosities. For the lowest gas-mass models, a small number of outliers fall below $10^{3}$ in luminosity and are not shown: 12 models for $M_{\rm gas}=3\times10^{-4} M_\odot$, 60 models for $M_{\rm gas}=10^{-4} M_\odot$, 75 models for $M_{\rm gas}=3\times10^{-5} M_\odot$, and 244 models for $M_{\rm gas}=10^{-5} M_\odot$. These low-luminosity outliers do not affect the inferred gas mass estimates.}
    \label{fig:model-grid-co-luminosities}
\end{figure*}

We used the \texttt{pdspy} package \citep{pdspycode_Sheehan18}, which follows the radiative transfer modeling framework outlined by \citet{ClassI_Taurus_Sheehan17}, to construct a grid of disk + envelope models. In practice, \texttt{pdspy} provides a high-level interface to RADMC-3D \citep{radmc3d_Dullemond12}: the radiative transfer calculations are performed with RADMC-3D, while the grid setup, parameter control, and output handling are managed within pdspy. Although the code is capable of full MCMC fitting to multiwavelength data sets, here we used it to construct a model grid and compare the resulting $^{13}$CO and C$^{18}$O ($J{=}3$--$2$) fluxes to our observations.
\par Each model was calculated on a spherical--polar grid extending from 0.1~AU to 4000~AU in radius, consisting of 99 logarithmically spaced radial cells, 99 polar cells, and a single azimuthal cell. The wavelength grid spanned 0.1–10$^5$$\mu$m with 500 logarithmically spaced bins. Our models consisted of three main components: a central protostar, a circumstellar disk, and an infalling envelope. Because the stellar properties are poorly constrained for most targets, we adopted a simplified protostar model. We fix the effective temperature to $T_{\rm eff}=4000$~K (typical for young, low-mass stars) and treat both the stellar luminosity $L_\star$ and stellar mass $M_\star$ as free parameters in the model grid. We set the $M_\star$ range based on the distribution of estimates from our Keplerian-mask analysis (see Table \ref{tab:mask_summary}), which is also consistent with expectations for low-mass protostars. We set the range for $L_\star$ from the observed bolometric luminosity $L_{bol}$ (Table \ref{tab:source_properties}). Since the central protostar’s radiation is reprocessed by the surrounding envelope, the observed $L_{bol}$ provides a proxy for the $L_\star$ range we explore for our model grid. The disk component followed the standard prescription for a viscously evolving accretion disk \citep{disk_model_Lyden-Bell74}, where the surface density decreases as a power law in radius ($\gamma$ controls the power-law) and tapers exponentially beyond a characteristic radius ($r_c$). The disk’s vertical density structure is Gaussian, as expected for a vertically isothermal disk in approximate hydrostatic equilibrium \citep{disk-evolution_Williams&Cieza11}, and the scale height increases with radius as a power law set by the flaring index, $\psi$, in our model setup. We fixed the inner radius to 0.1~AU (dust sublimation boundary). Because our sources are deeply embedded Class~0/I systems, we included an infalling envelope in our model setup based on the rotating–collapse solution of \citet{envelope_model_Ulrich76}. We fixed the envelope outer radius to 3000~AU in all models, while treating the envelope mass $M_{\rm env}$ as a free parameter in the model grid. \cite{VANDAMOrionII_Sheehan22} provides further details about the model setup and underlying physics. 

\par We assumed fixed molecular abundances for the model grid. Specifically, we adopted a CO abundance relative to H$_2$ of $x_{\rm CO} \equiv [{\rm CO}]/[{\rm H}_2] = 10^{-4}$, a canonical ISM-like gas-phase CO abundance \citep{Frerking1982}, and we set the $^{13}$CO and C$^{18}$O abundances by scaling CO using solar-neighborhood ISM isotopologue ratios, $[{\rm CO}]/[^{13}{\rm CO}] = 70$ and $[{\rm CO}]/[{\rm C}^{18}{\rm O}] = 550$ \citep{ISM-ratio_Wilson94}. We also include CO freeze-out at 20~K for both $^{13}$CO and C$^{18}$O, since the freeze-out threshold is the same for both isotopologues, consistent with observational studies of CO depletion and the CO snowline in disks \citep[e.g.,][]{Qi2008,Qi2011,Rosenfeld2013}. Below this temperature, we do not set the gas-phase CO abundance strictly to zero; instead, we multiply the unfrozen value by a factor of $10^{-8}$, so that frozen-out CO contributes negligibly to the modeled isotopologue emission. We also note that young Class~0/I disks are expected to be warmer on average (e.g., \citealt{L1527_vantHoff18, VANDAMOrionI_2020}), suggesting that freeze-out should be less efficient. We do not, however, include isotope-selective photodissociation, chemical conversion of CO, or redistribution of icy CO-bearing material within the disk. This simplified treatment of changes in the CO abundance may be reasonable for young Class~0/I disks, because these mechanisms may not yet have had sufficient time to substantially deplete the gas-phase CO reservoir \citep{co-evolution_Zhang2020}. In addition, we find that model $^{13}$CO and C$^{18}$O luminosities computed using ISM-level abundances for both isotopologues reproduce the observed line luminosities in our Class~0/I sample (Figure \ref{fig:model-grid-co-luminosities}), suggesting that selective photodissociation may not strongly affect the CO isotopologue emission in these disks.

\par To compare models and observed line cube on equal footing, we generated an ensemble of model-specific Keplerian masks using the same procedure as for the observations (Section \ref{subsec:line-analysis}), but with the model input parameters ($M_\star$, $i$, $r_c$, $z/r$ = 0.3, and position angle $(PA$ = 30 degrees). We then applied this model–specific mask to the corresponding synthetic $^{13}$CO and C$^{18}$O $J{=}$3--2 cubes to isolate disk emission and exclude the envelope in a consistent way. We then spatially and spectrally integrated the masked cubes to obtain the model line luminosities,
\begin{equation}
    L_{\rm model}=4\pi d^2 \int F(x,y,v) dx dy dv,
\end{equation}
where $d$ is the source distance and $F(x,y,v)$ is the model cube flux density as a function of sky position and velocity. In practice, for each model in the grid, we applied a Keplerian mask to the synthetic line cube, constructed a masked moment 0 map by multiplying by the channel width of 0.25~km~s$^{-1}$ and summing along the velocity axis. We then measured the integrated line flux from the masked moment~0 map using \texttt{imstat} and converted it to a line luminosity using the source distance, yielding a pair of model luminosities for $^{13}$CO and C$^{18}$O at each grid point. To estimate the disk gas mass, we compared the observed line luminosities directly to the model grid (Figure \ref{fig:model-grid-co-luminosities}). For a given gas mass, we restricted the comparison to models whose emitting radii were consistent with the observed dust continuum radius (Table \ref{tab:alma_continuum}), selecting models within $\pm50\%$ (i.e., within half of the measured radius). Within this radius-consistent subset, we used two observational conditions to constrain the gas mass: consistency with the observed $^{13}$CO/C$^{18}$O luminosity ratio and with the observed C$^{18}$O luminosity. The tolerance on the luminosity ratio is set by the quadrature sum of the uncertainties on the observed $^{13}$CO and C$^{18}$O luminosities, while the tolerance on the C$^{18}$O luminosity is given by its observational uncertainty (for 04239+2436AB, we adopt a tolerance of $2\times\sigma_{\rm C^{18}O}$, as no models satisfy the nominal constraint). For each discrete gas mass $M_i$ in the grid, we defined a relative weight $w_i \equiv N_{\rm match}(M_i)/N_{\rm rad}(M_i)$, where $N_{\rm rad}$ is the number of radius-consistent models and $N_{\rm match}$ is the subset that satisfies both luminosity constraints. After normalizing the weights such that $\tilde w_i = w_i / \sum_j w_j$, we obtained a discrete gas-mass distribution $p(M_i)=\tilde w_i$ and adopted its median and 16th--84th percentiles as the inferred gas mass and associated uncertainty. Table~\ref{tab:mask_summary} presents the inferred gas masses for the full sample, with superscript and subscript indicating the offsets from the median to the 84th and 16th percentiles, respectively.

\par Using the CO-based ($^{13}$CO and C$^{18}$O) gas mass estimates derived with this method, we find a median gas mass of $M_{\rm gas,med}=6.7\times10^{-4}$~$M_\odot$ (16th--84th percentile range: $1.1\times10^{-4}$--$1.3\times10^{-2}$~$M_\odot$). Combined with the dust masses derived from the ALMA Band~7 continuum emission (Section~\ref{subsec:continuum}), this yields an average gas-to-dust mass ratio of $147 \pm 75$ a median of 26, and a 16th–84th percentile range of 8–147 for the Taurus Class 0/I sample. 

\par Because the CO-based gas masses depend directly on the adopted CO abundance and freeze-out temperature, we performed sensitivity tests for these two assumptions to quantify how they affect the inferred gas masses. Since recomputing the full multidimensional model grid for each alternative chemical assumption would be computationally expensive, we instead considered a representative grid sequence with fixed parameters $M_\star = 1.0 M_\odot$, $L_\star = 1.0 L_\odot$, $r_c = 60$ au, $\gamma = 0.75$, $\psi = 1.2$, $M_{\rm env} = 10^{-4} M_\odot$, and $i = 45^\circ$, and varied only the gas mass. For the CO abundance test, we adopted $x_{\rm CO} = 10^{-5}$ and $10^{-3}$ around the fiducial value of $10^{-4}$. Across this representative sequence, changing $x_{\rm CO}$ from $10^{-4}$ to $10^{-5}$ reduces the model $^{13}$CO and C$^{18}$O luminosities to $\sim$0.11--0.70 and $\sim$0.10--0.65 times the fiducial values, respectively, while changing to $10^{-3}$ increases them to $\sim$1.4--6.1 and $\sim$1.4--9.2 times the fiducial values. Interpreting these luminosity shifts approximately inversely in terms of gas mass, as expected in the optically thin limit, implies that adopting $x_{\rm CO}=10^{-5}$ would increase the inferred gas mass by roughly factors of $\sim$1.4--10, whereas adopting $x_{\rm CO}=10^{-3}$ would decrease it to $\sim$0.11--0.74 times the fiducial value. The largest shifts occur at the low-mass end, where the line emission is closer to optically thin, while the effect is more modest at the high-mass end due to optical depth.
\par For the freeze-out test, lowering the adopted CO freeze-out temperature from 20~K to 15~K increases the model $^{13}$CO and C$^{18}$O luminosities by about 1--12\% and 1--24\%, respectively, while raising it from 20~K to 25~K decreases them by about 6--26\% and 6--31\%, respectively. To gauge the corresponding effect on the inferred gas mass, we also compared these results with additional $\pm 25\%$ and $\pm 50\%$ gas-mass perturbation tests for two representative cases, with fiducial gas masses of $3\times10^{-5}\,M_\odot$ and $10^{-3}\,M_\odot$. In the lower-mass case, the luminosity changes produced by adopting 15~K or 25~K are both very close to those produced by a $\pm 25\%$ gas-mass perturbation and remain smaller than the $\pm 50\%$ case. In the higher-mass case, adopting 15~K produces only a minimal change, smaller than the $+25\%$ mass perturbation, whereas adopting 25~K produces a larger decrease, lying between the $-25\%$ and $-50\%$ mass-perturbation cases, depending on the CO tracer. This asymmetric behavior indicates that the response of the line luminosities, and therefore the inferred gas masses, to changes in the freeze-out temperature is nonlinear, particularly at higher gas masses. Overall, a 5~K change in the adopted freeze-out temperature can correspond to $\sim$25--50\% although the effect can be smaller in some cases.
\par Overall, these tests show that the assumed CO abundance and freeze-out temperature introduce systematic uncertainties in the inferred gas masses beyond the statistical uncertainties reported in Table~\ref{tab:mask_summary} . The CO abundance is the dominant chemical systematic, while the freeze-out-temperature effect is generally smaller.
\section{Discussion}\label{sec:discussions}
\subsection{Continuum Emission and Dust Properties Across Star-Forming Regions} \label{subsec: cont_comparison}
\begin{figure*}[ht!]
    \centering

    \begin{minipage}{0.48\textwidth}
        \centering
        \includegraphics[width=\linewidth]{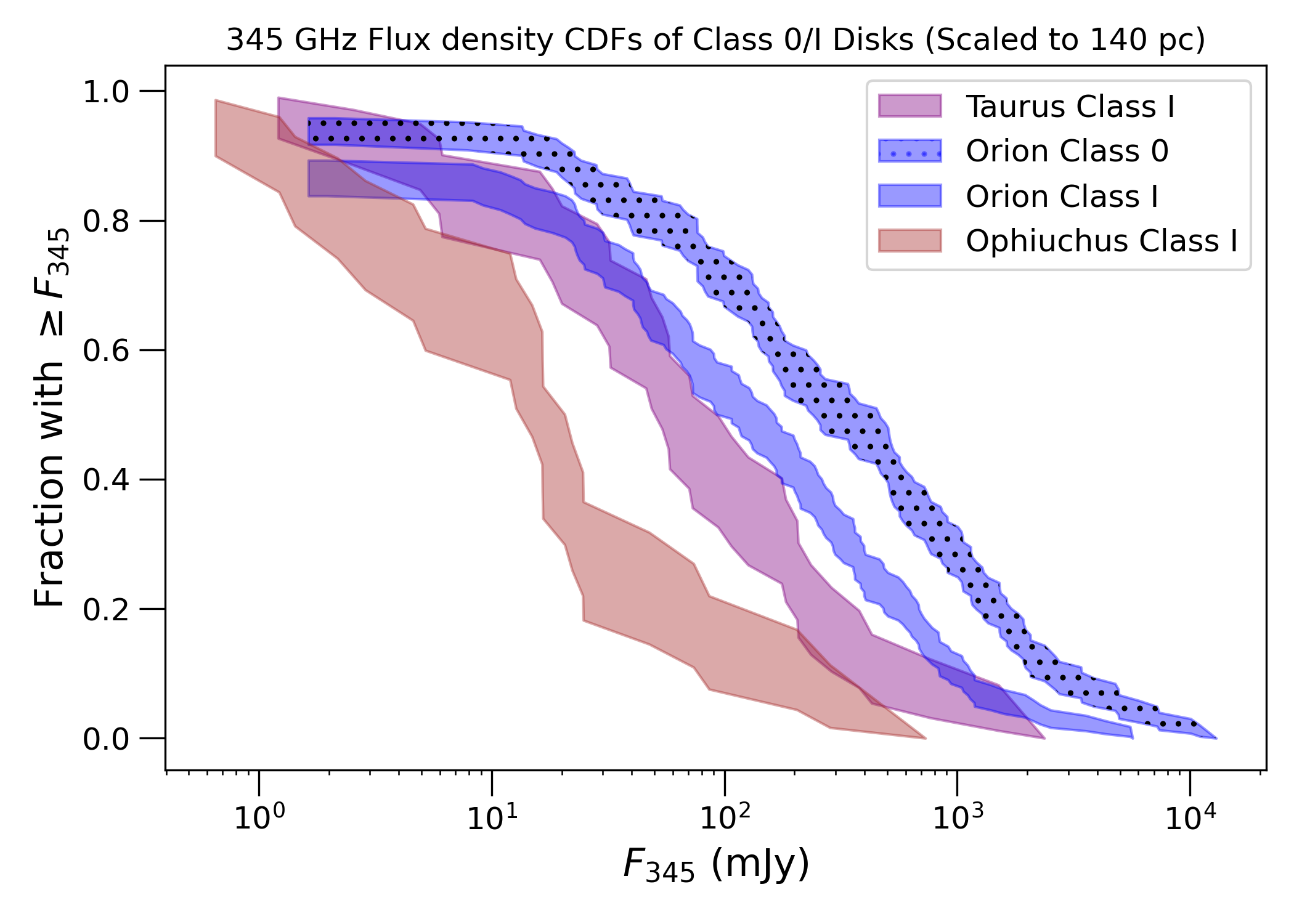}
        \\[-0.5ex]
        {\small (a) 345~GHz flux density CDFs}
    \end{minipage}
    \hfill
    \begin{minipage}{0.48\textwidth}
        \centering
        \includegraphics[width=\linewidth]{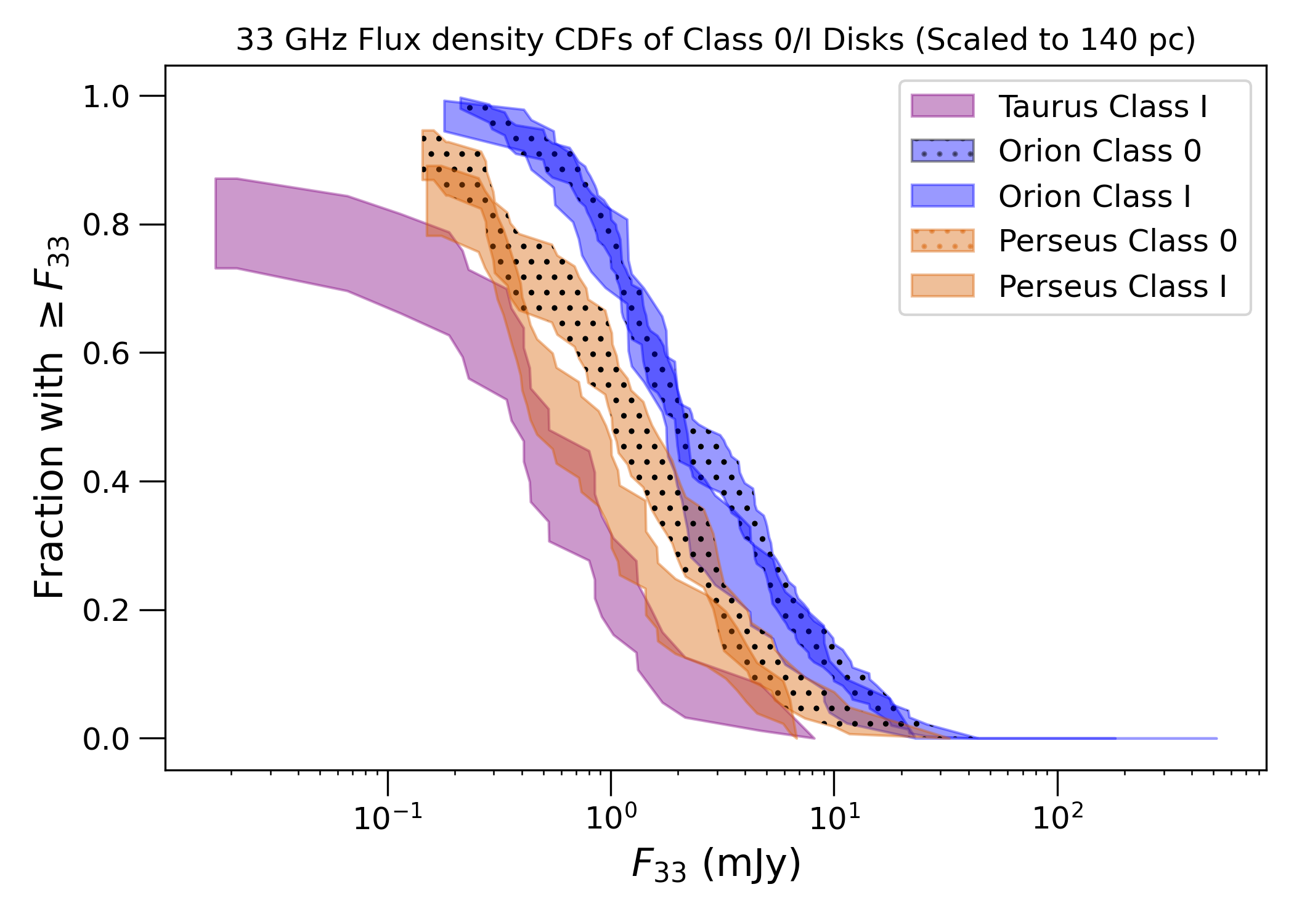}
        \\[-0.5ex]
        {\small (b) 33~GHz flux density CDFs}
    \end{minipage}

    \caption{Cumulative distribution functions (CDFs) of continuum flux densities for Class~0/I disks across different star-forming regions. Panels (a) and (b) show 345~GHz and 33~GHz flux densities, respectively. All flux densities are scaled to a common distance of 140~pc for direct comparison. Shaded regions show 68\% confidence intervals from Kaplan--Meier survival analysis, which incorporates upper limits. For the Taurus sample, only the Class~I disks are included in these distributions, since the number of Class~0 disks in our sample is too small ($N=3$) for a meaningful statistical comparison.}
    \label{fig:flux-cdfs}
\end{figure*}

\begin{figure*}[ht!]
    \centering

    \begin{minipage}{0.48\textwidth}
        \centering
        \includegraphics[width=\linewidth]{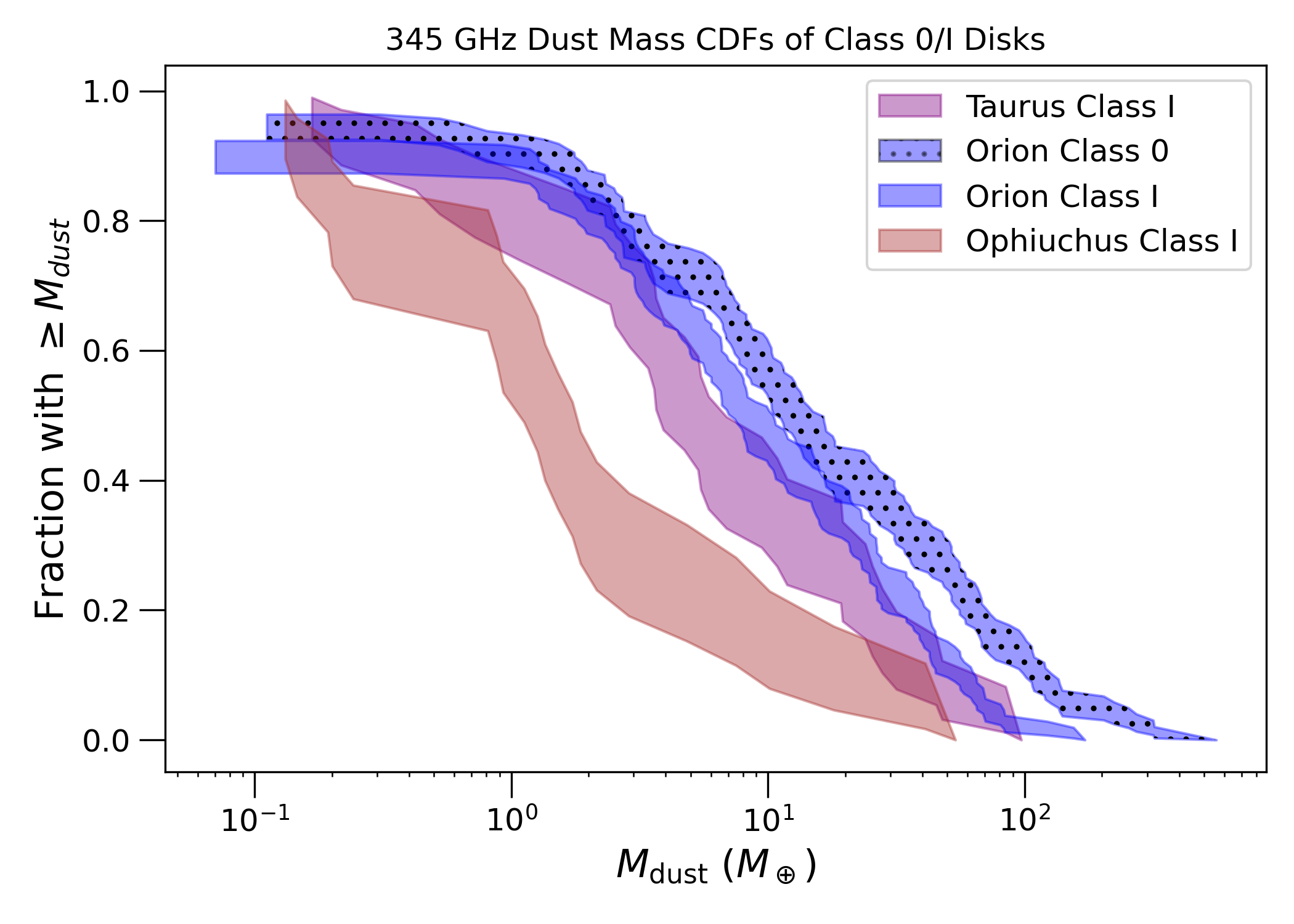}
        \\[-0.5ex]
        {\small (a) 345~GHz dust mass CDFs}
    \end{minipage}
    \hfill
    \begin{minipage}{0.48\textwidth}
        \centering
        \includegraphics[width=\linewidth]{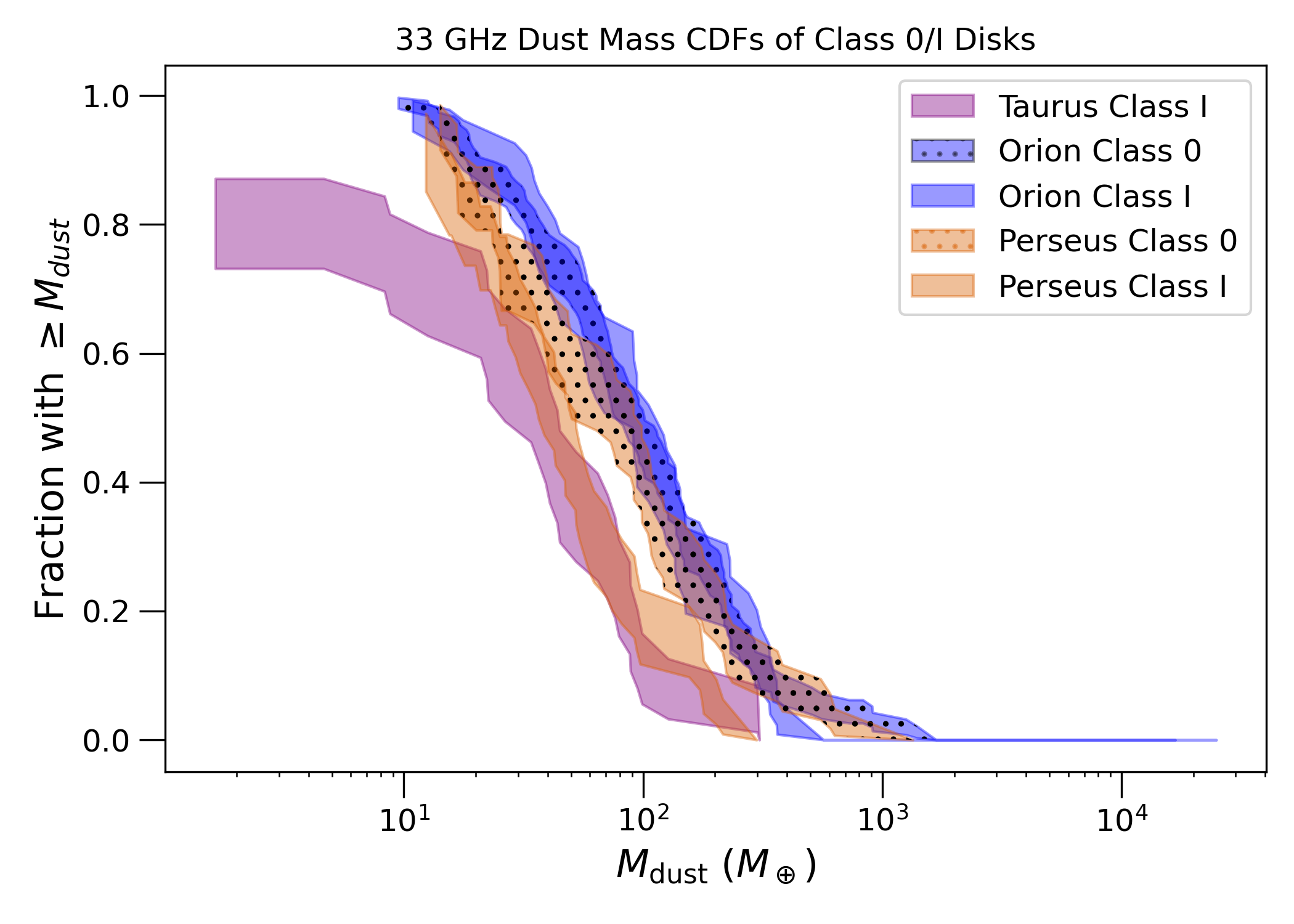}
        {\small (b) 33~GHz dust mass CDFs}
    \end{minipage}

    \caption{Cumulative distribution functions (CDFs) of dust masses inferred from continuum emission for Class~0/I disks across different star-forming regions. Panels (a) and (b) show dust masses derived from 345~GHz and 33~GHz data, respectively. Shaded regions show 68\% confidence intervals from Kaplan--Meier survival analysis, which incorporates upper limits. For the Taurus sample, only the Class~I disks are included in these distributions, since the number of Class~0 disks in our sample is too small ($N=3$) for a meaningful statistical comparison.}
    \label{fig:mass-cdfs}
\end{figure*}

\begin{figure}[ht!]
    \centering
    \includegraphics[width=0.6\textwidth]{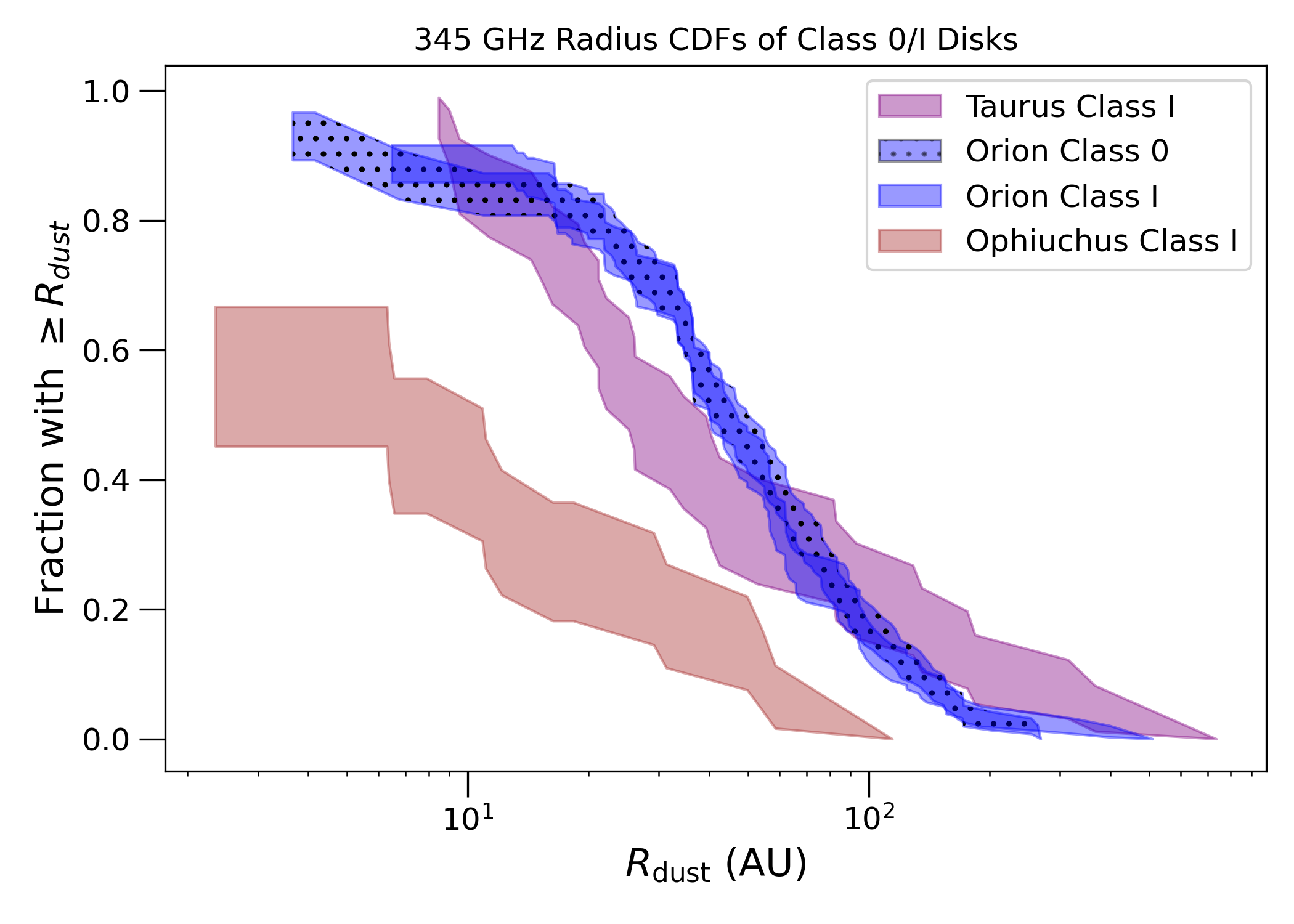}
    \caption{Cumulative distribution functions (CDFs) of dust disk radii at 345 GHz for Class 0/I disks in Taurus, Orion, and Ophiuchus. Shaded regions show 68\% confidence intervals from Kaplan–Meier survival analysis, which incorporates upper limits. For the Taurus sample, only the Class~I disks are included in these distributions, since the number of Class~0 disks in our sample is too small ($N=3$) for a meaningful statistical comparison.}
    \label{fig: radii-cdf}
\end{figure}

In this section, we compare the dust properties of our Taurus Class 0/I sample with those from well-studied protostellar disk surveys in Orion, Ophiuchus, and Perseus. These surveys, along with our Taurus sample, represent broadly unbiased selections of Class 0/I sources within each region, allowing for a meaningful comparison of dust properties across different environments. To ensure consistency, we perform the comparisons separately at each observing frequency. Combining measurements taken at different frequencies can introduce additional uncertainties, as the dust opacity and emission characteristics vary with frequency in ways that are not yet well constrained. Restricting the analysis to frequency-matched datasets minimizes these systematic effects. We include the Orion Class 0/I protostar sample from the VANDAM:Orion survey \citep{VANDAMOrionI_2020}, which contains 328 systems observed with ALMA at 345 GHz and the VLA at 33 GHz. The Ophiuchus Class 0/I sample from \citet{Ophiuchus_Encalada21} targeted 25 systems with ALMA at 345 GHz, yielding 31 detections including binaries and candidates; we note that we do not include the ODISEA Ophiuchus sample \citep{ODISEA_Williams19} because it was observed in ALMA Band~6 (225~GHz). For Perseus, we use the VANDAM survey \citep{VANDAMPerseus_Tobin16}, a VLA census of 94 protostars observed in Ka band (8 mm and 1 cm) at $\sim$15 AU resolution, comprising 37 Class 0, 8 Class 0/I, 37 Class I (including flat-spectrum), and 12 Class II sources. Since the VANDAM–Perseus survey provides continuum measurements at 8 and 10 mm, we estimate 9 mm fluxes by averaging the two bands and use these values to derive dust masses. Although \citet{VANDAMPerseus_Tychoniec18} applied free–free emission corrections to the VANDAM:Perseus sample, we adopt the uncorrected continuum flux densities reported by \citet{VANDAMPerseus_Tobin16} to ensure consistency with our Taurus Class 0/I sample analysis, for which free–free emission corrections are not applied. Since our analysis focuses on the earliest phases of disk evolution, the Class II objects are excluded from further comparison.

\par All fluxes were scaled to a common distance of 140~pc for direct comparison across regions (Figure \ref{fig:flux-cdfs}). At 345~GHz, the median flux densities are $58$~mJy (68\% range: $54$--$94$~mJy) for Taurus~Class~I (excluding Taurus~Class~0 sources, as there are only three Class 0 disks in our sample), $343$~mJy (68\% range: $240$--$468$~mJy) for Orion Class 0, $128$~mJy (68\% range: $107$--$167$~mJy) for Orion Class I, $17$~mJy (68\% range: $15$--$22$~mJy) for Ophiuchus~Class~I. Because some comparison samples include nondetections, we used the Peto--Peto test \citep{Peto_Peto1972} to compare the censored flux distributions while accounting for upper limits. The test evaluates whether censored samples are consistent with being drawn from the same underlying distribution, and the resulting $p$-value provides a guideline for how incompatible the distributions are with that assumption, with smaller values indicating greater incompatibility. At 345~GHz, an overall Peto--Peto test comparing the Taurus~Class~I, Orion~Class~0, Orion~Class~I, and Ophiuchus~Class~I samples yields $p=1.31\times10^{-8}$, providing strong evidence that the four censored flux distributions are not all drawn from the same underlying distribution. Pairwise comparisons indicate that the Taurus~Class~I flux distribution is unlikely to be drawn from the same underlying distribution as the Orion~Class~0 ($p=0.0012$) and Ophiuchus~Class~I ($p=0.0056$) samples, but is more consistent with being drawn from the same underlying distribution as the Orion~Class~I sample ($p=0.2201$). At 33~GHz, the corresponding medians are $0.43$~mJy (68\% range: $0.36$--$0.52$~mJy) for Taurus~Class~I, $2.0$~mJy (68\% range: $1.9$--$2.3$~mJy) for Orion~Class~0, $1.9$~mJy (68\% range: $1.8$--$2.1$~mJy) for Orion~Class~I, $1.2$~mJy (68\% range: $1.0$--$1.5$~mJy) for Perseus~Class~0, $0.57$~mJy (68\% range: $0.44$--$0.95$~mJy) for Perseus~Class~I. At 33~GHz, pairwise Peto--Peto comparisons indicate that the Taurus~Class~I flux distribution is incompatible with being drawn from the same underlying distribution as the Orion~Class~0 ($p=1.0\times10^{-6}$), Orion~Class~I ($p=2.9\times10^{-6}$), and Perseus~Class~0 ($p=0.0062$) samples, but remains compatible with the Perseus~Class~I sample ($p=0.2096$). At both frequencies, Taurus~Class~I disks have lower median flux densities than the Orion~Class~0 and Class~I samples. At 345 GHz, Taurus~Class~I disks are brighter than those in Ophiuchus, whereas at 33~GHz they are fainter than Perseus~Class~0 disks but more comparable to the Perseus~Class~I population.

\par We converted the measured flux densities into dust masses following the procedure described in Section~\ref{subsec:continuum}, which requires flux densities, source distances, and dust temperatures. For Orion, we adopted distances to individual protostellar systems from \citet{VANDAMOrionI_2020}, based on Gaia measurements for more evolved members in the region. For Ophiuchus, we used distances from the CAMPOS~II survey \citep{CAMPOSII_Hsieh25}, while for Perseus, where individual estimates were unavailable, we assumed a uniform distance of 294~pc \citep{Perseus-distance-Zucker18}. We calculated dust temperatures from bolometric luminosities when available and excluded sources lacking $L_{\rm bol}$ (one in Ophiuchus and two in Perseus) from the mass analysis. At 345~GHz, the median dust masses are $5.3$~M$\oplus$ (68\% range: $3.9$--$6.9$~M$\oplus$) for Taurus~Class~I, $13$~M$\oplus$ (68\% range: $11$--$16$~M$\oplus$) for Orion~Class~0, $8.2$~M$\oplus$ (68\% range: $7.4$--$11$~M$\oplus$) for Orion~Class~I, and $1.4$~M$\oplus$ (68\% range: $1.1$--$1.7$~M$\oplus$) for Ophiuchus~Class~I. We again use the Peto--Peto test to compare the censored dust-mass distributions. At 345~GHz, pairwise comparisons indicate that the Taurus~Class~I dust-mass distribution is unlikely to be drawn from the same underlying distribution as the Orion~Class~0 ($p=0.0317$) and Ophiuchus~Class~I ($p=0.0231$) samples, but is more consistent with being drawn from the same underlying distribution as the Orion~Class~I sample ($p=0.3233$). At 33~GHz, the corresponding median dust masses are $39$~M$\oplus$ (68\% range: $26$--$45$~M$\oplus$) for Taurus~Class~I, $92$~M$\oplus$ (68\% range: $78$--$100$~M$\oplus$) for Orion~Class~0, $94$~M$\oplus$ (68\% range: $90$--$110$~M$\oplus$) for Orion~Class~I, $77$~M$\oplus$ (68\% range: $50$--$99$~M$\oplus$) for Perseus~Class~0, and $43$~M$\oplus$ (68\% range: $37$--$53$~M$\oplus$) for Perseus~Class~I. At 33~GHz, pairwise Peto--Peto comparisons indicate that the Taurus~Class~I dust-mass distribution is unlikely to be drawn from the same underlying distribution as the Orion~Class~0 ($p=0.0015$), Orion~Class~I ($p=0.0015$), and Perseus~Class~0 ($p=0.0177$) samples, but is more consistent with being drawn from the same underlying distribution as the Perseus~Class~I sample ($p=0.4497$). Despite scaling the mass estimates based on luminosity and distance, the overall comparison follows the same trend observed in the flux distributions. Taurus disks are on average less massive than the Orion Class 0/I population at both frequencies and more massive than the Ophiuchus Class I disks at 345 GHz. At 33 GHz, Taurus Class I disks are less massive than Perseus Class 0 disks but broadly comparable to Perseus Class I disks.

\par The VANDAM:Orion survey is an unbiased census that uniformly targeted most protostars identified by the Herschel Orion Protostar Survey (HOPS; e.g., \citealt{HOPS_Furlan16}), and some other bona fide protostars, minimizing selection effects. Since we also compare the fluxes at the same frequencies and derive masses using the same method, the differences we observe are unlikely to arise from analysis biases, but rather from the distinct environments in which these systems form. Orion spans a wide range of environments \citep[e.g.,][]{OrionAB_Megeath12, OrionA_Hsu13, VIENNAII_Meingast18, SODAI_Terwisga22}, from relatively dense, clustered regions (e.g., Orion Molecular Cloud 2/3; OMC-2/3) to more distributed, lower-density populations that are not unlike Taurus. Nevertheless, Orion forms a much larger number of protostars overall than Taurus, with the Class~0/I samples containing $\sim$300 sources in Orion compared to $\sim$30 in Taurus. If the protostellar populations in Taurus and Orion are assumed to be drawn from the same underlying stellar initial mass function (IMF), then a region that forms many more protostars overall, such as Orion, will also naturally contain more massive protostars than a lower-mass region such as Taurus. This is also consistent with the bolometric luminosity distributions of the embedded samples considered here, if bolometric luminosity is assumed to trace protostellar mass \citep{mass-luminosity_Hartmann25}: the Orion sample has a median bolometric luminosity approximately three times higher than the Taurus sample ($L_{\rm bol,med}\approx1.9 L_\odot$ versus $0.63 L_\odot$), which is broadly consistent with Orion hosting more massive protostars. Empirical studies of more evolved Class II disks show a strong and nearly linear correlation between stellar mass and disk mass \citep[e.g.,][]{mstar-mdisk_Andrews13}, which means that more massive young stars consistently host more massive disks. If the same scaling holds at earlier evolutionary stages, then a region like Orion, with a higher number of massive protostars, should naturally host more massive and intrinsically brighter Class 0/I disks than Taurus. This is consistent with our results because we find that the Orion Class~0/I disks are, on average, brighter and more massive than their Taurus counterparts. 

\par In addition to these differences in the stellar mass distribution, it is important to note that the Orion complex includes subregions with physical conditions that are distinct from those in Taurus, particularly in the dense, clustered environments in the OMC2/3. In these regions, the high stellar densities and the presence of numerous intermediate- and high-mass stars can produce strong external ultraviolet (UV) radiation fields that may drive external photoevaporation \citep{ONC_Scally01, ONC_Eisner18, external-photoevaporation_Winter22}, dynamical interactions, and enhanced feedback from winds and outflows. These environmental processes are expected to truncate disks and reduce their dust masses through external photoevaporation \citep{SODAII_Terwisga23, ONC_Eisner18} and dynamical interactions \citep{dynamical-interaction-ONC_Zwart16}. If such effects dominated disk mass in these regions, disks in Orion would be expected to be systematically less massive than those in Taurus. To verify this would require an environment-based comparative study (e.g., separating sources by high and low local stellar density) to compare more fairly with Taurus. However, there are also many low-density regions in Orion that are distributed and not unlike Taurus, and \citet{VANDAMOrionI_2020} did not find strong differences in protostellar disk properties as a function of local environment within Orion. In addition, Class~0/I disks are deeply embedded, and their surrounding envelopes may shield the disks from external irradiation and replenish disk material through continued infall, leaving the extent to which external environmental effects influence disks at these early stages unclear. Based on our analysis, systematically higher brightness and dust masses that we find for Orion Class~0/I disks, compared to our Taurus Class~0/I sample, are driven are more likely by differences in the stellar mass distribution, assuming the protostellar populations in the two regions are drawn from the same underlying IMF, rather than by the presence of dense, clustered subregions in Orion. If this assumption does not hold, however, then this interpretation may not be valid.

In comparison to Taurus, we find that the Ophiuchus Class I disks from the \cite{Ophiuchus_Encalada21} sample are systematically fainter and have correspondingly lower dust masses. Previous studies have noted that extinction toward Ophiuchus can sometimes cause Class II sources to be misclassified as Class I when relying solely on infrared spectral indices \citep[e.g.,][]{Oph-polarization_Sadavoy19, VANDAMOrionI_2020}. However, sample misclassification is unlikely to be a major concern for the \cite{Ophiuchus_Encalada21} Ophiuchus sample, as their stringent selection criteria were specifically designed to minimize contamination from misclassified young stellar objects. Both \cite{VANDAMOrionI_2020} and \cite{Ophiuchus_Encalada21} found that Ophiuchus protostars are significantly less massive than those in Orion, and our results are consistent with this picture of Ophiuchus hosting a population of lower-mass disks. \cite{Ophiuchus_Encalada21} proposed that the low disk masses may partly reflect the region’s evolutionary demographics, since Ophiuchus contains very few Class 0 objects (only one out of their 24 Class~0/I disks). Although Taurus includes more Class 0 disks (three in our 26 Class 0/I systems), the difference is still modest, making it difficult to attribute the lower Ophiuchus disk masses primarily to evolutionary age. This interpretation is also consistent with evidence from other star forming regions: for example, \citet{CrA_Cazzoletti19} found unusually low Class II disk masses in the young Corona Australis region despite its young age, supporting a role for factors beyond age alone. An alternative possibility is that Ophiuchus formed in different environmental conditions from Taurus, for example, a more compact and more highly extinguished cloud environment together with stronger external influence from the nearby Sco OB2 association \citep[e.g.,][]{oph_Vrba77,Oph-region_McClure2010,Oph_Pattle2015}, which may give rise to a population of low-mass protostellar disks. Additionally, \citet{Serpens_Anderson22} showed that the young, clustered Serpens region has disk-mass distributions broadly similar to Taurus, Orion, and Perseus, while Ophiuchus remains the outlier, suggesting that clustering or youth alone is probably not sufficient to explain the unusually low disk masses in Ophiuchus. This may instead point to more specific cloud- or core-scale initial conditions in Ophiuchus, such as a larger role for external pressure in the dense-core environment or other differences in the parent cloud structure and collapse conditions \citep[e.g.,][]{Oph_Pattle2015,Oph_Maruta10,Oph_Johnstone04}.

\par The VANDAM–Perseus survey provides an unbiased census of Class 0/I protostars in Perseus, with a significantly larger Class 0 sample ($\sim$37 sources) than in Taurus. Unlike Taurus—which contains only a small number of Class 0 objects and is dominated by more evolved Class I systems—Perseus has comparable numbers of Class 0 and Class I protostars. Perseus is also a low- to intermediate-mass star-forming region that contains young B stars and two rich clusters, IC 348 \citep{IC348_Muench07} and NGC 1333 \citep{NGC1333_Lada96}. These environmental and evolutionary differences may contribute to variations in disk properties. At 33~GHz, the fluxes and inferred dust masses of Perseus~Class~I sources are broadly consistent with those of Taurus~Class~I sources, whereas the Perseus~Class~0 population is generally brighter and more massive. Additionally, we do not correct the 33~GHz fluxes for free--free emission, and the level of contamination may vary from source to source and across star-forming regions, potentially introducing additional scatter into the comparison.

For the disk radii, we adopt the same procedure described in Section~\ref{subsec:continuum}. Using the Kaplan--Meier estimator (Figure \ref{fig: radii-cdf}) to include upper limits, we find that Taurus Class~I disks have a median characteristic radius of $26$~AU (68\% range: $25$--$39$~AU). Orion Class~0 and Class~I disks are larger, with median radii of $46$~AU (68\% range: $40$--$51$~AU) and $43$~AU (68\% range: $40$--$47$~AU), respectively. In contrast, Ophiuchus Class~I disks are more compact, with a median radius of $7$~AU (68\% range: $2$--$11$~AU). Because unresolved sources provide upper limits on disk radius, we used the Peto--Peto test to compare the censored radius distributions. Pairwise comparisons indicate that the Taurus~Class~I radius distribution is unlikely to be drawn from the same underlying distribution as the Ophiuchus~Class~I sample ($p=2.7\times10^{-4}$), but is more consistent with being drawn from the same underlying distribution as the Orion~Class~0 ($p=0.4430$) and Orion~Class~I ($p=0.4430$) samples. The Orion~Class~0 and Class~I radius distributions are also more consistent with being drawn from the same underlying distribution as each other ($p=0.7586$), whereas each is unlikely to be drawn from the same underlying distribution as the Ophiuchus~Class~I sample ($p=7.1\times10^{-5}$). These results show that Ophiuchus disks are significantly smaller than both Orion and Taurus disks. This environmental trend is consistent with the findings of \citet{CAMPOSI_Hseih24}, who reported that Ophiuchus hosts the smallest disks in their seven-cloud protostellar sample, with a median Class I disk radius of $12$~AU, and demonstrated substantial cloud-to-cloud variation in disk radii even within the same evolutionary class. Such diversity underscores the influence of local environments---such as clustering, stellar density, radiation fields, and turbulence---on disk size evolution. Overall, the Taurus Class~I disks occupy an intermediate regime between the large Orion disks and the very compact Ophiuchus population, reinforcing that disk radii are not universal but shaped by their birth environments.

\subsection{Comparing Taurus Class 0/I and Class II Disks} \label{subsec:classII}
\begin{figure*}[ht!]
    \centering

    \begin{minipage}{0.48\textwidth}
        \centering
        \includegraphics[width=\linewidth]{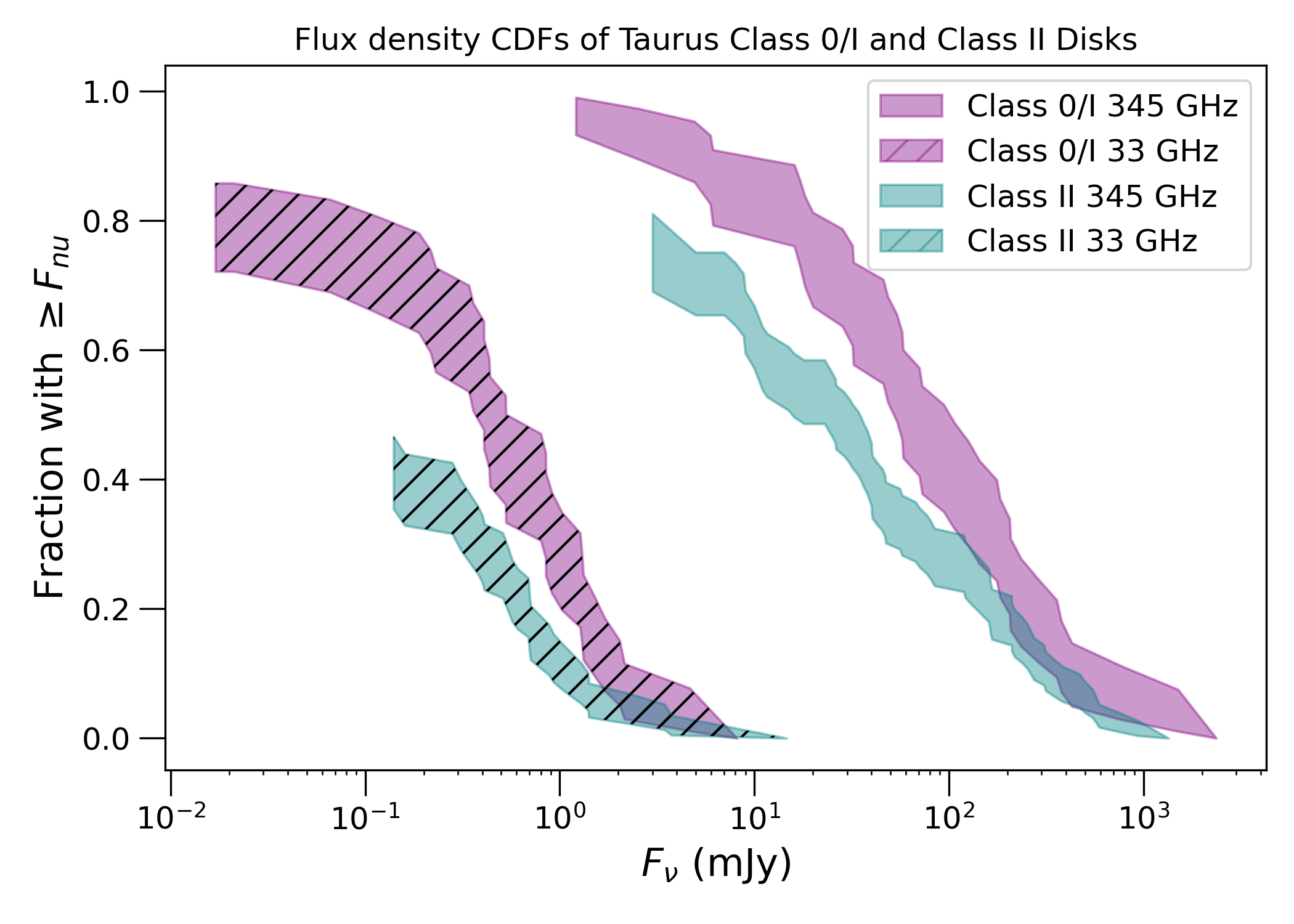}
        \\[-0.5ex]
        {\small (a) Flux density CDFs}
    \end{minipage}
    \hfill
    \begin{minipage}{0.48\textwidth}
        \centering
        \includegraphics[width=\linewidth]{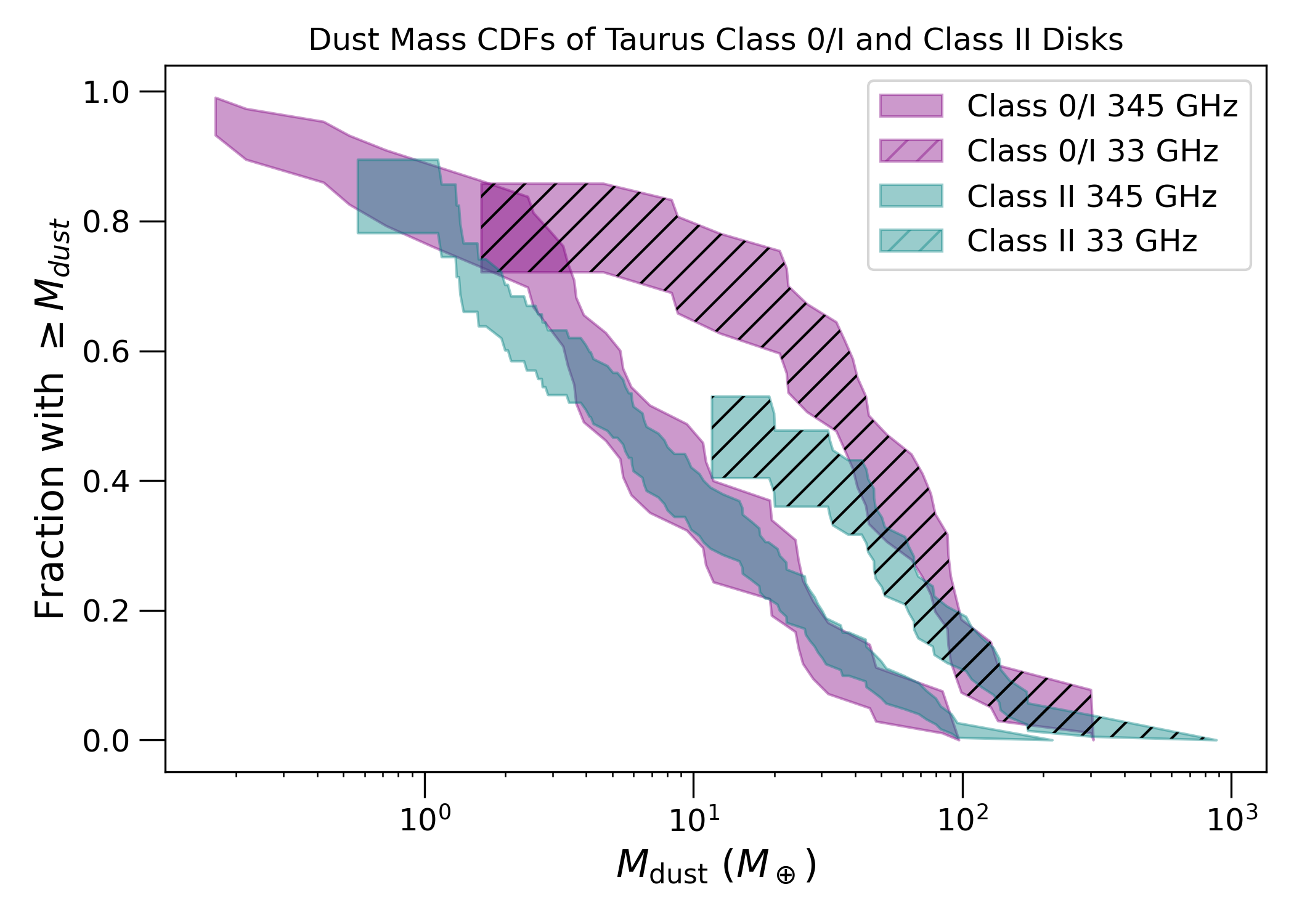}
        \\[-0.5ex]
        {\small (b) Dust mass CDFs}
    \end{minipage}

    \caption{
    Cumulative distribution functions (CDFs) of continuum flux densities and corresponding dust masses for Taurus Class~0/I and Class~II disks at 345 and 33~GHz. The 345~GHz Class~II comparison sample is taken from \cite{mstar-mdisk_Andrews13}, while the 33~GHz Class~II comparison sample is based on the Taurus disks with centimeter-wave measurements from \cite{Chung25_TaurusII-33GHz}, with upper limits adopted for the remaining sources in the \cite{sample_Andrews05} sample as described in the text. The left shows the flux density distributions at 345 and 33~GHz, and the right panel shows the corresponding dust mass distributions derived using consistent assumptions for each comparison. Shaded regions indicate 68\% confidence intervals from Kaplan--Meier survival analysis, which incorporates upper limits.}
    \label{fig:flux-mass-cdfs}
\end{figure*}

\par Because continuum flux and inferred dust mass depend strongly on observing frequency and can vary between star-forming environments, we compare our Class~0/I disks to Class~II disks within Taurus separately at 345 and 33~GHz, using the corresponding Taurus Class~II sample available at each frequency. For the 345~GHz comparison, we use the Taurus Class~II sample from Table~2 of \citet{mstar-mdisk_Andrews13}, while for the 33~GHz comparison, we use the 32 Taurus Class~II disks with centimeter-wave measurements from \citet{Chung25_TaurusII-33GHz}. We use only the direct 345~GHz measurements from \cite{mstar-mdisk_Andrews13} to avoid additional uncertainties from frequency scaling, which depends on the assumed spectral index. The 33~GHz sample represents a subset of the brighter Taurus Class~II disks from \citet{SMA-taurus-classII_Chung24} and, more broadly, from the \citet{sample_Andrews05} sample. For the remaining Taurus Class~II disks in the 74-source sample of \citet{sample_Andrews05} that are not included in \citet{Chung25_TaurusII-33GHz}, we adopt the minimum 33~GHz flux density in the \citet{Chung25_TaurusII-33GHz} sample as an upper limit. For the Class II sources, we adopt individual distances from \citet{SPHERE-Taurus_Garufi24} when available and assume a standard Taurus distance of 140 pc for the remainder. We also adopt the bolometric luminosities from \citet{mstar-mdisk_Andrews13}. Dust masses are then computed using the same methodology (Equation~\ref{eq:dust_mass}; Section~\ref{subsec:continuum}) applied to our Class~0/I disks; for the Class~II disks, we adopt a characteristic dust temperature of $T_0 = 25$~K, following \citet{mstar-mdisk_Andrews13}. We note that we adopted $T_0 = 43$~K for the Class~0/I disks, consistent with the expectation that they are to be warmer on average than more evolved Class~II disks \citep[e.g.,][]{L1527_vantHoff18, VANDAMOrionI_2020}. At 345~GHz, the Taurus Class~II disks show a median flux density of $26$~mJy (68\% range: $16$--$35$~mJy), lower than the median flux density of $71$~mJy (68\% range: $54$--$107$~mJy) for the Taurus Class~0/I disks. For the corresponding dust masses, the Class~II disks have a median of $5.6$~M$\oplus$ (68\% range: $4.1$--$6.5$~M$\oplus$), while the Class~0/I disks have a median of $5.5$~M$\oplus$ (68\% range: $3.9$--$9.4$~M$\oplus$). At 33~GHz, the Taurus Class~II sample is dominated by upper limits, preventing a robust nonzero Kaplan--Meier median from being derived. The corresponding 68\% confidence interval places the median near $0.14$~mJy in flux density and at roughly $12$--$20$~M$\oplus$ in dust mass. By comparison, the Taurus Class~0/I disks have a median flux density of $0.43$~mJy (68\% range: $0.41$--$0.80$~mJy) and a corresponding median dust mass of $39$~M$\oplus$ (68\% range: $34$--$52$~M$_\oplus$). Using the Peto--Peto test to account for upper limits, we find that the Taurus~Class~0/I and Class~II flux distributions are unlikely to be drawn from the same underlying distribution at both 345~GHz ($p=0.0060$) and 33~GHz ($p=0.0011$), with Class~0/I disks being brighter at both frequencies. In contrast, the inferred dust-mass distributions are more consistent with being drawn from the same underlying distribution at 345~GHz ($p=0.6000$) and show only weak evidence against a common underlying distribution at 33~GHz ($p=0.0607$).

\par From Figure \ref{fig:flux-mass-cdfs}, we see that Taurus Class~0/I disks are brighter than Taurus Class~II disks at both 345 and 33~GHz. At 345~GHz, however, the inferred median dust masses of the two populations are comparable, in contrast to many previous studies that report a decrease in dust mass from the embedded Class~0/I to the Class~II phase \citep[e.g.,][]{ClassI_Taurus_Sheehan17, USco_Barenfeld16, VANDAMOrionI_2020, review_Manara23, AGE-PRO_Zhang25}. A key strength of our analysis is that, unlike most previous studies, we compare disks within the same star-forming region and at the same observing frequencies, thereby reducing environmental and frequency-dependent systematics. We also adopt different dust temperatures for the Class~0/I and Class~II samples, consistent with the expectation that Class~0/I disks are warmer on average than Class~II disks. Some systematic uncertainty nevertheless remains in the mass comparison. In particular, our adopted Class~0/I dust temperatures are scaled solely with stellar luminosity, whereas detailed radiative-transfer modeling of a large Orion protostellar sample shows that dust temperature also depends on disk radius \citep{VANDAMOrionII_Sheehan22}. In addition, we assume the same dust opacity for both samples, implicitly adopting similar grain size distributions, even though grain growth and radial drift can alter millimeter opacities and the spatial distribution of continuum emission \citep[e.g.,][]{dust-evolution_Testi14, radial-drift_Takeuchi05}. Optical depth, especially at 345~GHz, is another important consideration, since both Class~0/I and Class~II disks may be at least partially optically thick, and differences in optical depth between the two populations could affect the inferred continuum-based dust masses differently. At 33~GHz, the Class~0/I disks appear to have slightly higher inferred dust masses than the Class~II disks, although this comparison has its own uncertainties because free-free emission may contribute significantly to the measured fluxes, potentially at different levels in the two populations. Overall, our Taurus-only comparison reduces several of the major systematics that affect cross-region studies, but the remaining uncertainties in temperature structure, opacity, optical depth, and free-free contamination highlight the value of future source-by-source radiative-transfer modeling. This is particularly important because \citet{VANDAMOrionII_Sheehan22} found that dust masses derived from radiative-transfer modeling of Class~0/I disks make it less clear whether Class~0/I and flat-spectrum disks are systematically more massive than Class~II disks.

\subsection{Gas-to-Dust Mass Ratios} \label{subsec: gas-dust-ratio}
\begin{figure}[ht!]
    \centering
    \includegraphics[width=\linewidth]{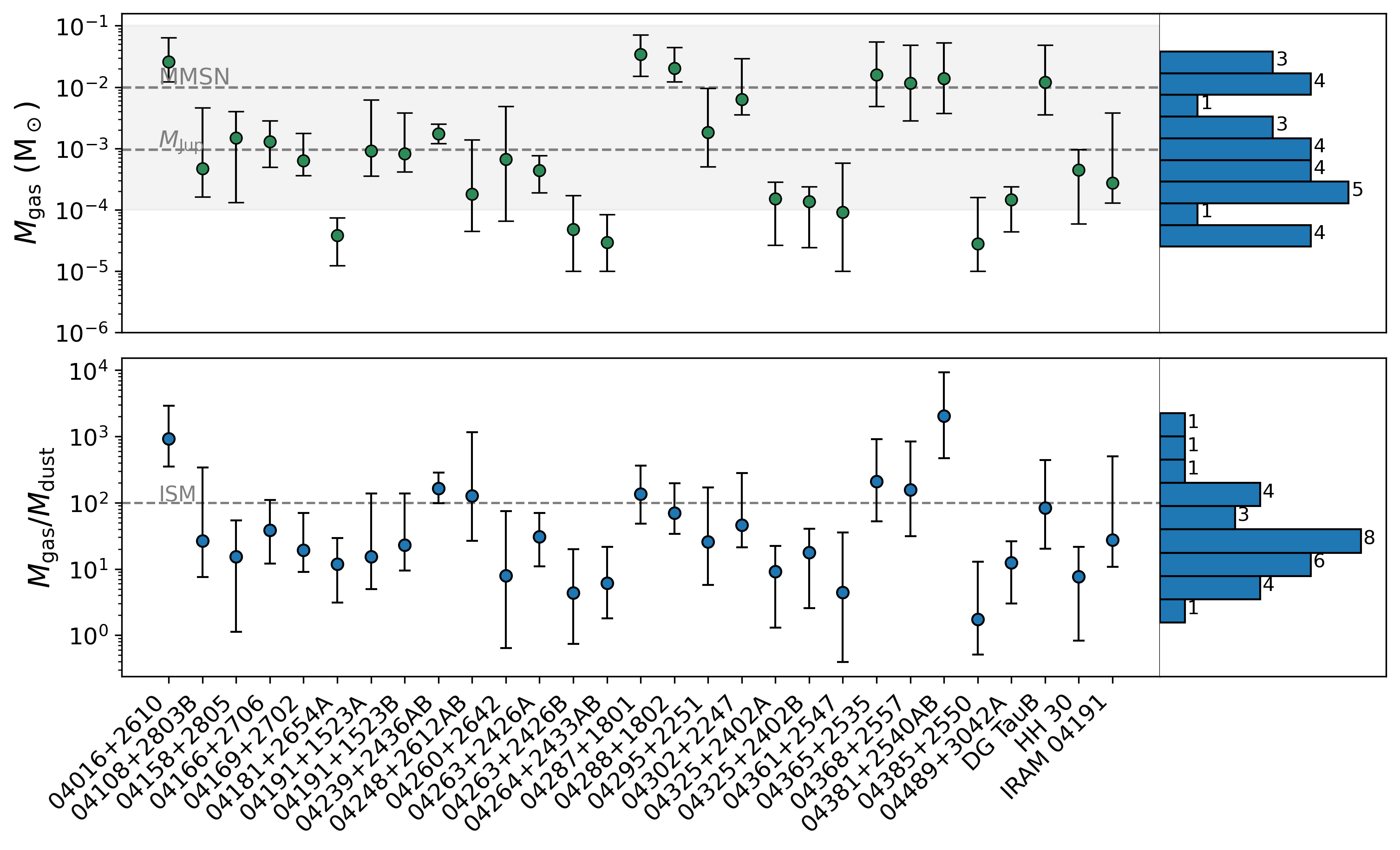}
    \caption{Gas masses (top) and gas-to-dust mass ratios (bottom) for our Taurus Class~0/I sample. The dust masses used to compute these gas-to-dust ratios are derived from the ALMA 345~GHz continuum data (Section \ref{subsec:continuum}). In the top panel, error bars show the asymmetric 16th--84th percentile bounds from the radiative-transfer modeling (Section \ref{subsec:gas-mass}). In the bottom panel, error bars reflect the propagated uncertainties from both the gas and dust masses. Dashed horizontal lines in the top panel mark the minimum-mass solar nebula (MMSN) and Jupiter mass, and the shaded region indicates the range of gas masses spanned by our model grid ($10^{-5}-10^{-1} M_\odot$). In the bottom panel, the dashed gray line marks the canonical ISM gas-to-dust ratio of~100.}
    \label{fig:gas-dust-ratio}
\end{figure}

Using the CO-based ($^{13}$CO and C$^{18}$O) gas mass estimates derived in Section~\ref{subsec:gas-mass}, together with dust masses derived from the ALMA Band~7 continuum emission (Section~\ref{subsec:continuum}), we find a mean gas-to-dust mass ratio of $147 \pm 75$, where the quoted uncertainty is the standard error on the mean across our sample. This sample-average value is broadly consistent with the canonical interstellar medium value of 100, although the ratios vary significantly from source to source (Figure~\ref{fig:gas-dust-ratio}. The gas-to-dust mass ratio distribution in our Taurus Class~0/I sample is broad and skewed, with a median gas-to-dust ratio of 26 and a 16th--84th percentile range of $\sim 8$--147. It is important to note, however, that both the gas and dust mass estimates depend on a number of assumptions, including the adopted dust opacity, dust temperature, optical depth effects, CO abundance, and the details of the gas-mass modeling, as discussed in Sections~\ref{subsec:continuum} and~\ref{subsec:gas-mass}, respectively. We used $^{13}$CO and C$^{18}$O line emission to estimate the gas mass; however, the extent to which CO isotopologue emission reliably traces the total disk gas mass remains uncertain. Using CO as a gas-mass tracer requires adopting a CO abundance relative to H$_2$, typically $x_{\rm CO} \sim 10^{-4}$ in the disk, which is also the abundance assumed in our model setup (Section~\ref{subsec:line-analysis}). But the CO abundance can be reduced by two common processes: photodissociation in the disk surface layers and freeze-out toward the midplane. However, even after accounting for the photodissociation and freeze-out, gas masses inferred from CO in benchmark Class~II disks (TW Hya, DM Tau, and GM Aur) are often lower than those derived from hydrogen deuteride (HD), a direct tracer of the bulk H$_2$ reservoir, by factors of $\sim$5--100 \citep[e.g.,][]{HD_McClure16, HD_Trapman17, TWHya_Calahan21}. This suggests that additional processes can further reduce the observable gas-phase CO abundance, including chemical conversion of CO into other species, CO being locked into solids and/or transported to the optically thick inner disk. As a result, the inferred gas-to-dust ratios should be interpreted with caution rather than as precise absolute values. 

\par Despite the caveats associated with using CO isotopologue emission to trace the total disk gas mass, comparisons with studies that use similar approaches to estimate disk gas masses remain useful. In this context, we compare our results with \citet{gasmodelgrid_Williams14}, who reported CO-inferred gas masses for more evolved Class~II disks in the same star-forming region (Taurus) using a broadly similar modeling approach with radiative transfer calculations. One important difference is that \cite{gasmodelgrid_Williams14} adopted a prescribed parametric temperature structure, whereas in our work we compute the temperature structure with RADMC-3D before generating the synthetic CO isotopologue emission. We also note that the dust masses in both studies were derived following the same procedure (Section~\ref{subsec:continuum}), although at different observing frequencies (230 GHz in \citet{gasmodelgrid_Williams14} and 345 GHz in our work). \citet{gasmodelgrid_Williams14} found that gas-to-dust ratios in Class~II disks are low and showed substantial scatter, reporting a mean of 16 and a standard deviation of 11 for the nine Taurus disks in their survey. Because this sample includes upper limits, and because the distribution is broad, we reanalyzed the ratios using a left-censored Kaplan--Meier estimator. For the original nine-disk sample, we obtain a median gas-to-dust ratio of 9, with 16th and 84th percentiles of 2 and 19, respectively. \citet{gasmodelgrid_Williams14} also considered an extended sample containing six additional disks with previously published CO isotopologue measurements; however, only two of these sources are located in Taurus, yielding an 11-disk Taurus Class~II comparison sample for which the mean gas-to-dust ratio is 21, while the Kaplan--Meier median is 14, with 16th and 84th percentiles of 6 and 43, respectively. In comparison, our Taurus Class~0/I sample has a median gas-to-dust ratio of 26, with corresponding 16th and 84th percentiles of 8 and 147. Thus, Taurus Class 0/I disks have a moderately higher median gas-to-dust ratio than Taurus Class II disks, although the medians of both populations remain below the canonical ISM value, and the Class 0/I distribution extends to substantially higher ratios. \citet{gasmodelgrid_Williams14} found that reduced C$^{18}$O abundances provided a better match to observations of Class~II disks, which they interpreted as possible evidence for isotope-selective photodissociation. However, we find that ISM-level abundances for both $^{13}$CO and C$^{18}$O reproduce the observed line luminosities in our Class~0/I sample, consistent with the young, embedded nature of these systems, for which CO-to-H$_2$ ratios are expected to remain close to canonical values \citep[e.g.,][]{co-evolution_Zhang2020}. Both studies account for CO freeze-out through an assumed freeze-out temperature of $T \sim 20$~K; therefore, differences in freeze-out treatment alone are unlikely to explain observed differences in CO-inferred gas-to-dust ratios between the Class~0/I and Class~II samples. This comparison should nevertheless be interpreted cautiously because the Class II sample is small, and it is therefore unclear how representative it is of the full Taurus Class II population.
\par Possible reasons for low CO-inferred gas-to-dust ratios in Class~II disks include physical and chemical evolution that becomes more important at later stages. Processes such as CO conversion into other species and sequestration into solids that can move within the disk may play a larger role in more evolved Class~II disks, thereby altering the gas-phase CO abundance and complicating the interpretation of CO-based gas mass estimates in Class~II disks. In addition, dust evolution can affect the inferred gas-to-dust ratio in different ways: dust growth and inward drift within the disk can reduce the observed dust mass and thereby increase the inferred gas-to-dust ratio, whereas dust trapping together with gas mass loss can decrease the ratio. For more evolved disks, these effects may therefore make the gas-to-dust ratio more difficult to interpret, and the evolution is not necessarily monotonic. For example, the recent AGE-PRO survey \citep{AGE-PRO_Zhang25} found that the median gas-to-dust ratio changes from 122 in the younger Ophiuchus population to 46 in Lupus and then rises again to 120 in Upper Sco, indicating that gas and dust do not evolve on the same timescale and that different physical processes may dominate at different stages. Viewed in this context, the broad Taurus Class~0/I distribution may indicate that some disks retain ISM-like gas-to-dust ratios, while others already approach the low CO-inferred ratios seen in Class~II disks, if the \citet{gasmodelgrid_Williams14} sample is representative of the broader Taurus Class~II population.

\par We also compare our results with other available gas mass estimates for Class~0/I disks, although such comparisons remain limited because constraining gas masses in embedded systems is observationally challenging. One notable recent effort is the aforementioned AGE-PRO survey \citep{AGE-PRO_Zhang25} of embedded disks in Ophiuchus, for which \cite{gas-mass-AGEPRO_Trapman25} derived gas masses using thermochemical modeling of C$^{17}$O $J{=}2$--1 emission, a more optically thin tracer than the $^{13}$CO and C$^{18}$O lines. Their modeling also explicitly includes selective photodissociation, unlike ours, while still assuming that more complex, long-term CO depletion chemistry, such as conversion of CO into other molecules or ice, is not yet important for these young Class~0/I disks; they therefore consider photodissociation in the surface layers and freeze-out toward the midplane when estimating disk gas masses. Using the dust masses and median gas masses reported in Table~2 of \citet{gas-mass-AGEPRO_Trapman25}, we calculate gas-to-dust mass ratios of $\sim44$--1720, with a median of 136 and a 16th--84th percentile range of 90--1006. As noted by the authors, the highest ratios may be affected by underestimated dust masses due to optical-depth effects and uncertainties in the adopted dust opacity. Their dust masses were derived using a similar standard continuum-based procedure to ours (Section~\ref{subsec:continuum}), although from 230~GHz emission rather than 345~GHz emission. The AGE-PRO median gas-to-dust ratio of 136 is substantially higher than the median of 26 inferred for our Taurus sample. However, at least ten disks in our sample have ratios of order $100$, with several reaching values comparable to the upper end of the AGE-PRO distribution, around $1000$. Given that the AGE-PRO sample contains only ten disks, it may preferentially represent the high-ratio tail of the broader distribution observed in Taurus. Evaluating this possibility would require extending the AGE-PRO analysis to the full Class 0/I disk population in Ophiuchus.

\par \citet{oph-gas-mass_Rodriguez} found that C$^{17}$O can systematically yield slightly higher gas mass estimates than C$^{18}$O when analyzed with simplified slab models, because C$^{17}$O is more optically thin. This may partly contribute to the higher median gas-to-dust ratios inferred for the AGE-PRO sample relative to our Taurus sample; however, because comparable C$^{17}$O data are not available for our full sample, it is difficult to test this directly. In addition, the inferred gas masses are sensitive to the adopted thermochemical modeling framework. For example, \citet{NOISO_Miotello16} compared their results with those of \citet{gasmodelgrid_Williams14} and showed that differences in the treatment of the temperature structure alone---self-consistent in their models versus parameterized in \citet{gasmodelgrid_Williams14}---make the relation between CO isotopologue line luminosity and disk mass more degenerate. They further showed that including isotope-selective effects introduces an additional disk-mass dependence, particularly for C$^{18}$O. As a result, the same line luminosity can correspond to a broader range of disk masses, with differences reaching up to about an order of magnitude at the low disk-mass end. This highlights that part of the difference between the AGE-PRO and Taurus samples may arise from the use of different molecular tracers and thermochemical modeling assumptions, so a more rigorous comparison would require samples analyzed with the same molecular tracer and the same thermochemical modeling framework.


\section{Conclusions}\label{sec:conclusions}
In this work, we carried out a uniform, region-wide analysis of dust and gas in all consistently identified embedded (Class~0/I) systems in the Taurus Molecular Cloud (26 protostellar systems: 3 Class~0 and 23 Class~I), providing a comprehensive view of key disk properties—dust masses, disk sizes, and gas masses—at the earliest stages of protoplanetary disk evolution. By combining ALMA Band~7 and VLA Ka-band data, we measured dust masses and disk sizes for the Taurus Class~0/I disks and compared these properties with other Class~0/I samples to place the Taurus population in the broader context of young disks across different molecular cloud environments. We also estimated disk gas masses using $^{13}$CO and C$^{18}$O molecular line observations and derived gas-to-dust mass ratios, comparing our results with those from an Ophiuchus Class~0/I sample. Finally, we compared the dust and gas masses of the Taurus Class~0/I disks with those of Class~II disks in Taurus to assess differences associated with disk evolutionary stage.
\begin{itemize}
    \item Using uniform ALMA and VLA observations, we find that the ALMA fluxes range from 1.2~mJy to 2.4~Jy with a median flux density of $71$~mJy (68\% range: $54$--$107$~mJy), corresponding to dust masses of $0.17$--$97$~M$\oplus$ with a median dust mass of $5.5$~M$\oplus$ (68\% range: $3.9$--$9.4$~M$\oplus$) and radii from a few au up to $736$~AU with a median radius of $28$~AU (68\% range: $25$--$39$~AU). VLA fluxes are systematically lower, with a median flux density of $0.43$~mJy (68\% range: $0.41$--$0.80$~mJy), with inferred dust masses reaching up to $307$~M$\oplus$ and a median dust mass of $39$~M$\oplus$ (68\% range: $34$--$52$~M$\oplus$) and radii from $3$~AU to $372$~AU with a median radius of $32$~AU (68\% range: $29$--$33$~AU).
    \item Since the continuum measurements are made at two different frequencies, 345~GHz and 33~GHz, we compare the Class~0/I samples separately at each frequency. In all cases, the inferred dust masses depend on the adopted dust opacity and therefore assume broadly similar dust properties across regions. Under this framework, Orion Class~0/I disks are significantly brighter and more massive than those in Taurus at both 345~GHz and 33~GHz, consistent with expectations if the two regions are drawn from the same underlying stellar initial mass function (IMF): a dense region like Orion naturally forms more massive protostars than a low-mass region such as Taurus, and more massive protostars tend to host more massive young disks. At 33~GHz, Perseus~Class~0 disks have higher continuum fluxes and inferred dust masses than Taurus~Class~I disks, whereas the Perseus~Class~I population is broadly comparable to Taurus. At 345~GHz, Ophiuchus Class~I disks are significantly fainter and less massive than those in Taurus, suggesting that Ophiuchus may have formed under different initial physical conditions.
    \item We find that Taurus~Class~0/I disks are brighter than Taurus~Class~II disks at both 345 and 33~GHz. At 345~GHz, however, their inferred dust-mass distributions remain comparable, differing from many previous studies that report a decrease in dust mass from the embedded Class~0/I to the Class~II stage. A key advantage of our analysis is that the two populations are compared within the same star-forming region and at the same observing frequencies, reducing environmental and frequency-dependent systematics. Nevertheless, uncertainties in the adopted dust temperatures and opacities, optical-depth effects, and free--free contamination at 33~GHz limit the interpretation of the inferred dust masses. Source-by-source radiative-transfer modeling will therefore be important for more reliably tracing dust-mass evolution between the Class~0/I and Class~II stages.
    \item We find a mean CO-inferred gas-to-dust mass ratio of $147 \pm 75$, a median of 26, and a 16th--84th percentile range of 8--147 for Taurus Class~0/I disks. The distribution is broad: its lower end overlaps the Taurus Class~II population, which has a median ratio of 14 and a 16th--84th percentile range of 6--43, whereas at least ten Class~0/I disks have ratios of order $100$ or higher, and several reach values of order $1000$, comparable to those found in the AGE-PRO Ophiuchus Class~0/I sample. However, this gas-to-dust ratio is subject to substantial uncertainties in both the gas and dust mass estimates. Dust masses depend sensitively on assumptions about dust opacity, temperature, and grain properties, while gas masses inferred from CO are subject to additional uncertainties related to whether CO reliably traces the total disk gas mass and are inherently model dependent, potentially affected by incomplete treatments of the underlying physical and chemical processes. These uncertainties motivate caution in interpreting absolute gas-to-dust ratios. More detailed modeling that incorporates CO transport within disks, including thermochemical processes such as chemical conversion pathways along with CO freeze-out, will be necessary to place gas-to-dust ratios in embedded disks on a more robust physical footing.

\end{itemize}
\section{Acknowledgments}
\begin{acknowledgments}

The authors would like to thank the anonymous referee for a constructive report that helped to improve the manuscript. P.D.S acknowledges support from NSF AST-2305482. This paper makes use of the following ALMA data: ADS/JAO.ALMA\#2019.1.00857.S. ALMA is a partnership of ESO (representing its member states), NSF (USA) and NINS (Japan), together with NRC (Canada), MOST and ASIAA (Taiwan), and KASI (Republic of Korea), in cooperation with the Republic of Chile. The Joint ALMA Observatory is operated by ESO, AUI/NRAO and NAOJ. The National Radio Astronomy Observatory is a facility of the National Science Foundation operated under cooperative agreement by Associated Universities, Inc.

\end{acknowledgments}

%

\vspace{5mm}
\facilities{ALMA, VLA}


\software{
CASA \citep{CASA:2022}},
pdspy \citep{pdspycode_Sheehan18},
RADMC-3D \citep{radmc3d_Dullemond12},
emcee \citep{mcmc_Foreman-Mackey13},
Astropy \citep{astropy:2013, astropy:2018, astropy:2022},
matplotlib \citep{matplotlib_Hunter07},
lifelines \citep{lifelines_Davidson-Pilon2019}



\appendix
\section{Continuum Images of All Sources}\label{app:disk_images}
This appendix presents a figure set containing continuum images of all sources observed with ALMA at 345~GHz and the VLA at 33~GHz. Each image is centered on the source position; for close binaries, the image is centered on the brighter component, whereas for the wider binary 04191+1523AB, it is centered on the midpoint of the separation between the two components. The plotted field of view is defined using a fixed fractional zoom around the source position, with different zoom fractions adopted for the ALMA and VLA images to appropriately capture the spatial extent of the detected emission; as a result, some sources appear more zoomed in than others. All images are shown in units of Jy~beam$^{-1}$ and displayed in celestial coordinates (RA/Dec; J2000). A synthesized beam is shown in white in the lower left corner of each panel, and a 100~AU scale bar is shown in white in the lower right. Images are displayed with a linear intensity scale, except for the ALMA image of 04489+3042AB and the VLA image of DG~TauB, for which a logarithmic color scale is used to highlight faint emission. In particular, the companion 04489+3042B is very faint and is visible only in the ALMA image with the logarithmic stretch; its position is marked with a small black X. See \citet{NESTI_Plante26} for a more detailed discussion of this source.\\
\figsetstart
\figsetnum{A1}
\figsettitle{ALMA and VLA Continuum Images of All Sources}

\figsetgrpstart
\figsetgrpnum{A1.1}
\figsetgrptitle{04016+2610}
\figsetplot{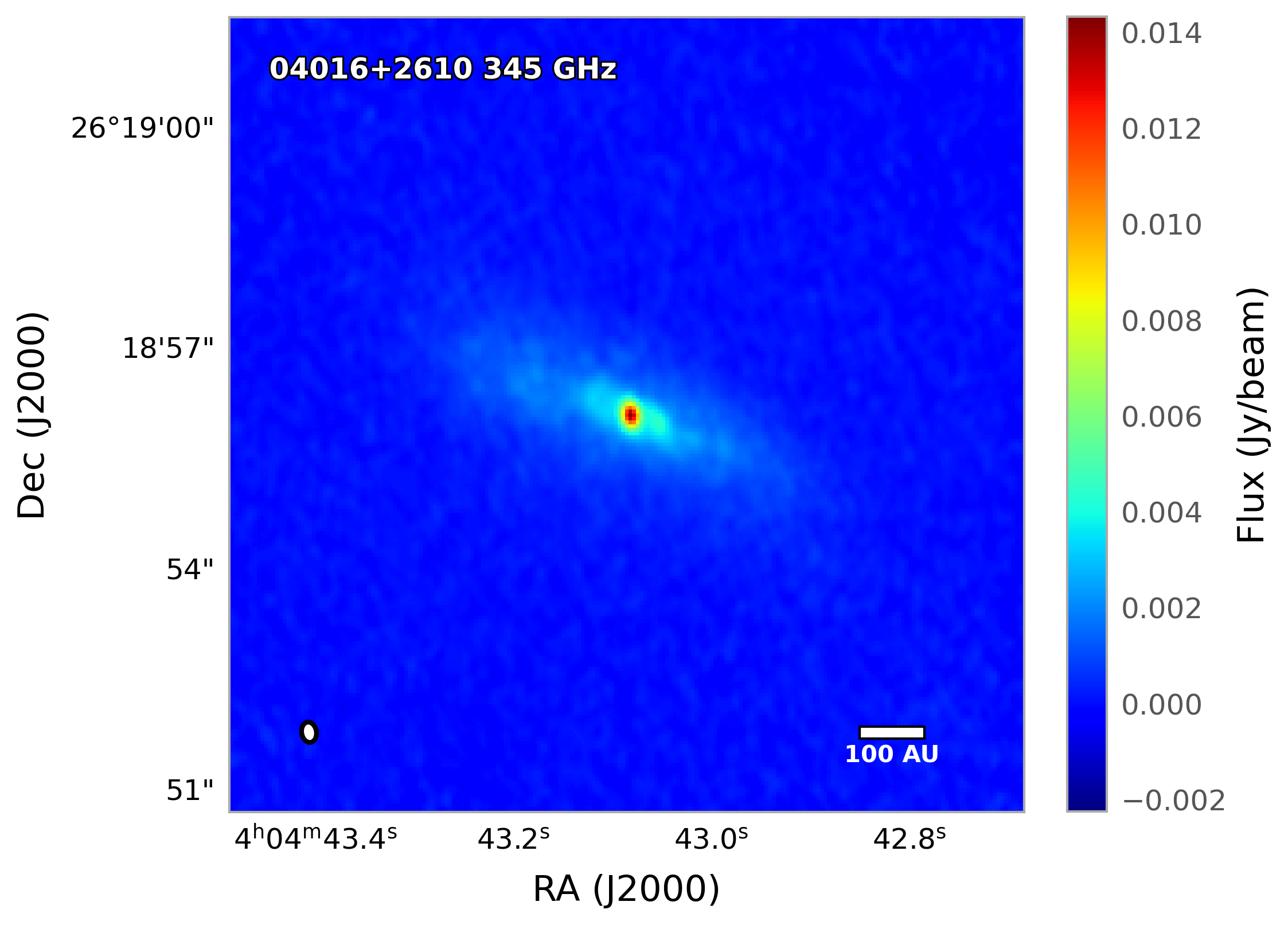}
\figsetplot{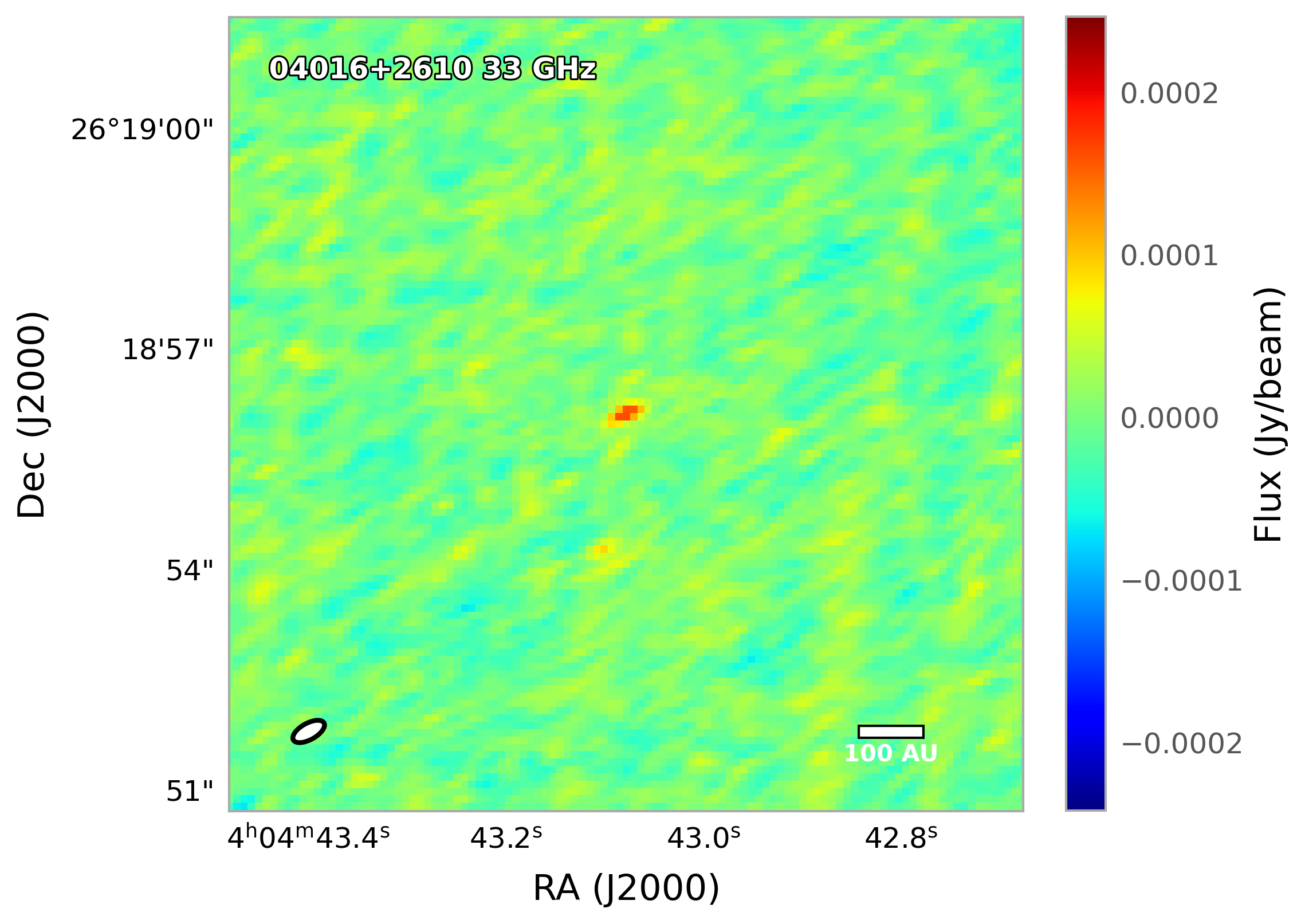}
\figsetgrpnote{ALMA 345~GHz and VLA 33~GHz continuum images of 04016+2610. The images are shown in units of Jy~beam$^{-1}$ in celestial coordinates (R.A./Decl.; J2000). A synthesized beam is shown in white in the lower left of each panel, and a 100~AU scale bar is shown in white in the lower right. Both images use a linear intensity scale.}
\figsetgrpend

\figsetgrpstart
\figsetgrpnum{A1.2}
\figsetgrptitle{04108+2803B}
\figsetplot{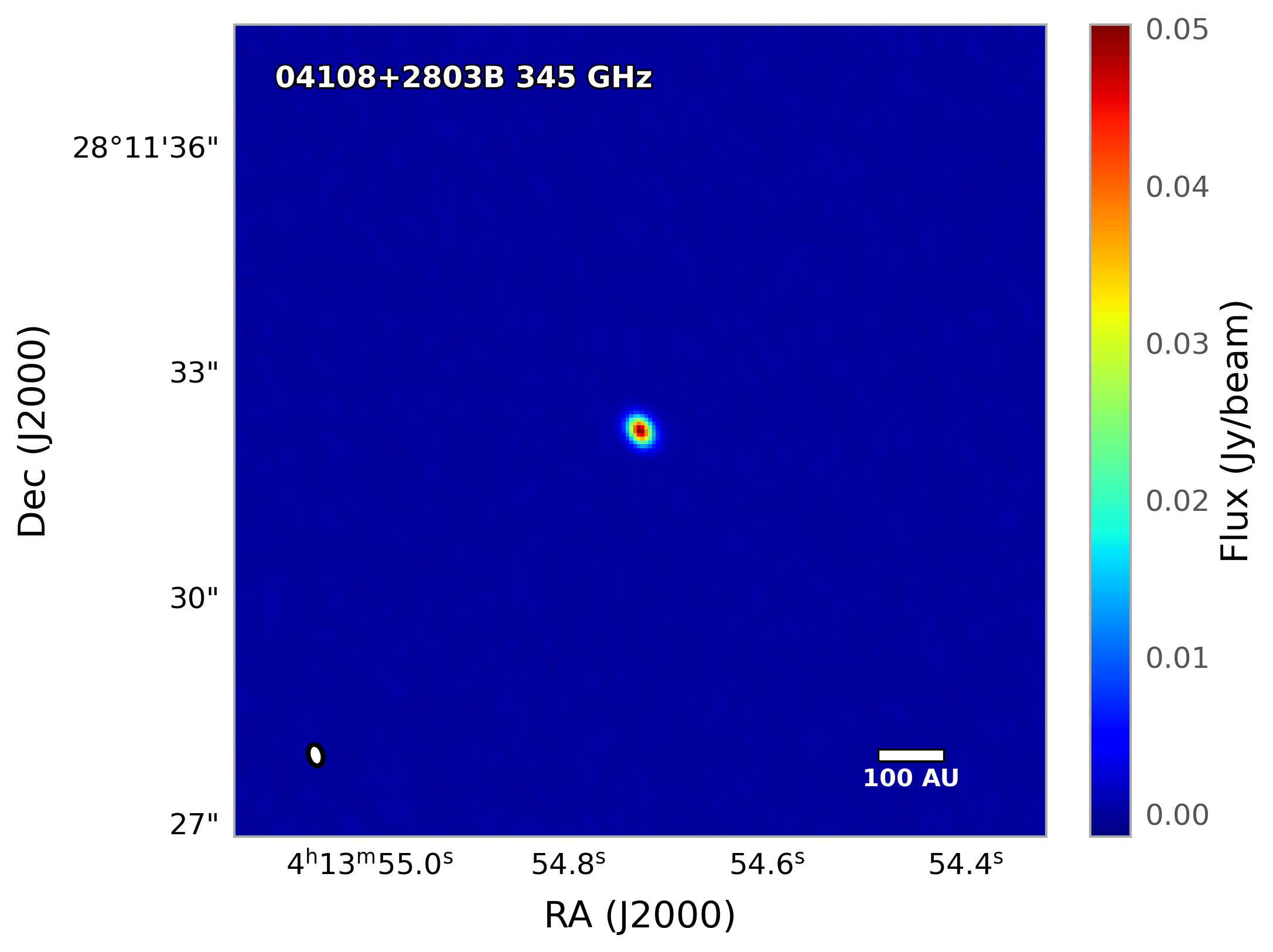}
\figsetplot{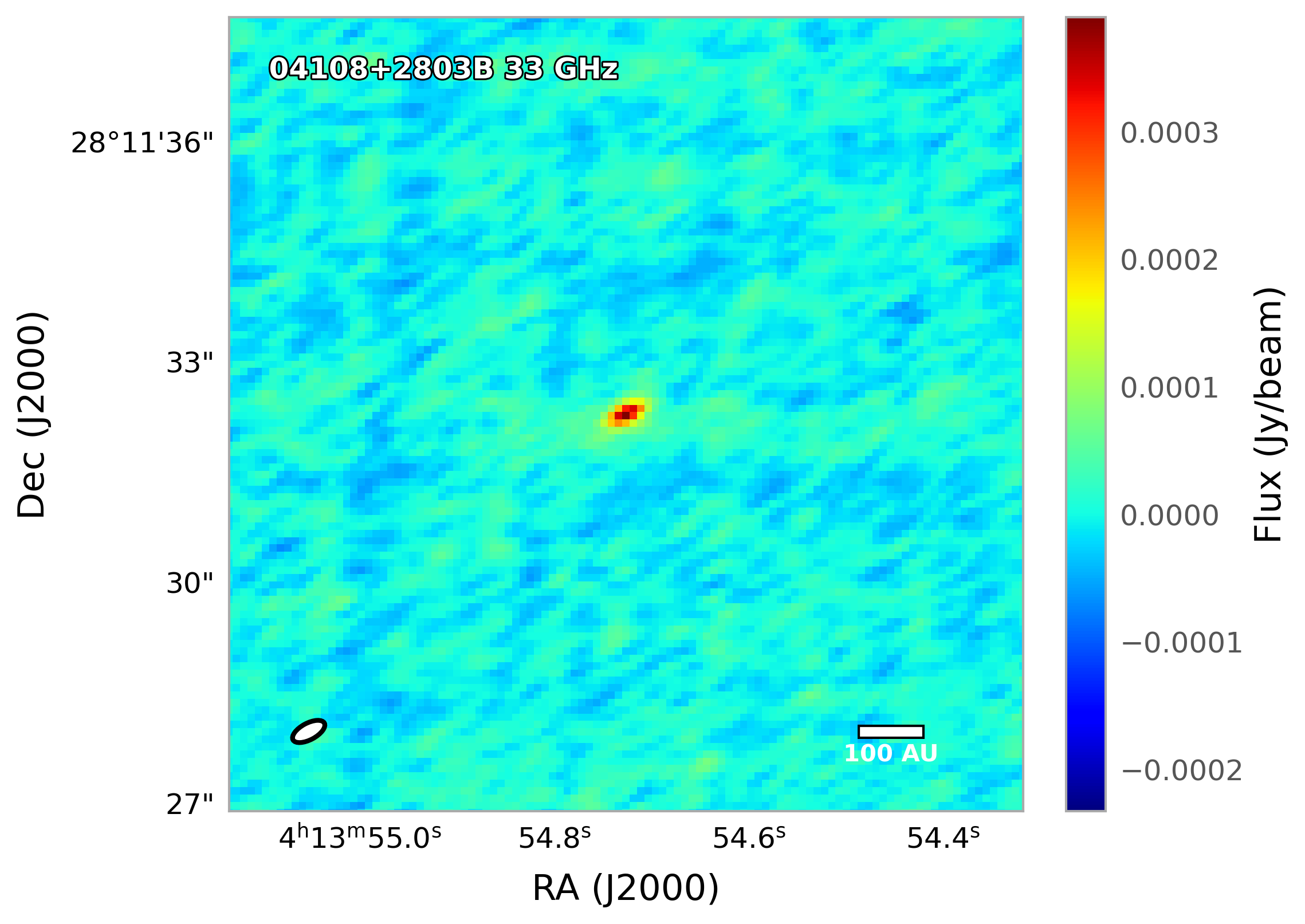}
\figsetgrpnote{ALMA 345~GHz and VLA 33~GHz continuum images of 04108+2803B. The images are shown in units of Jy~beam$^{-1}$ in celestial coordinates (R.A./Decl.; J2000). A synthesized beam is shown in white in the lower left of each panel, and a 100~AU scale bar is shown in white in the lower right. Both images use a linear intensity scale.}
\figsetgrpend

\figsetgrpstart
\figsetgrpnum{A1.3}
\figsetgrptitle{04158+2805}
\figsetplot{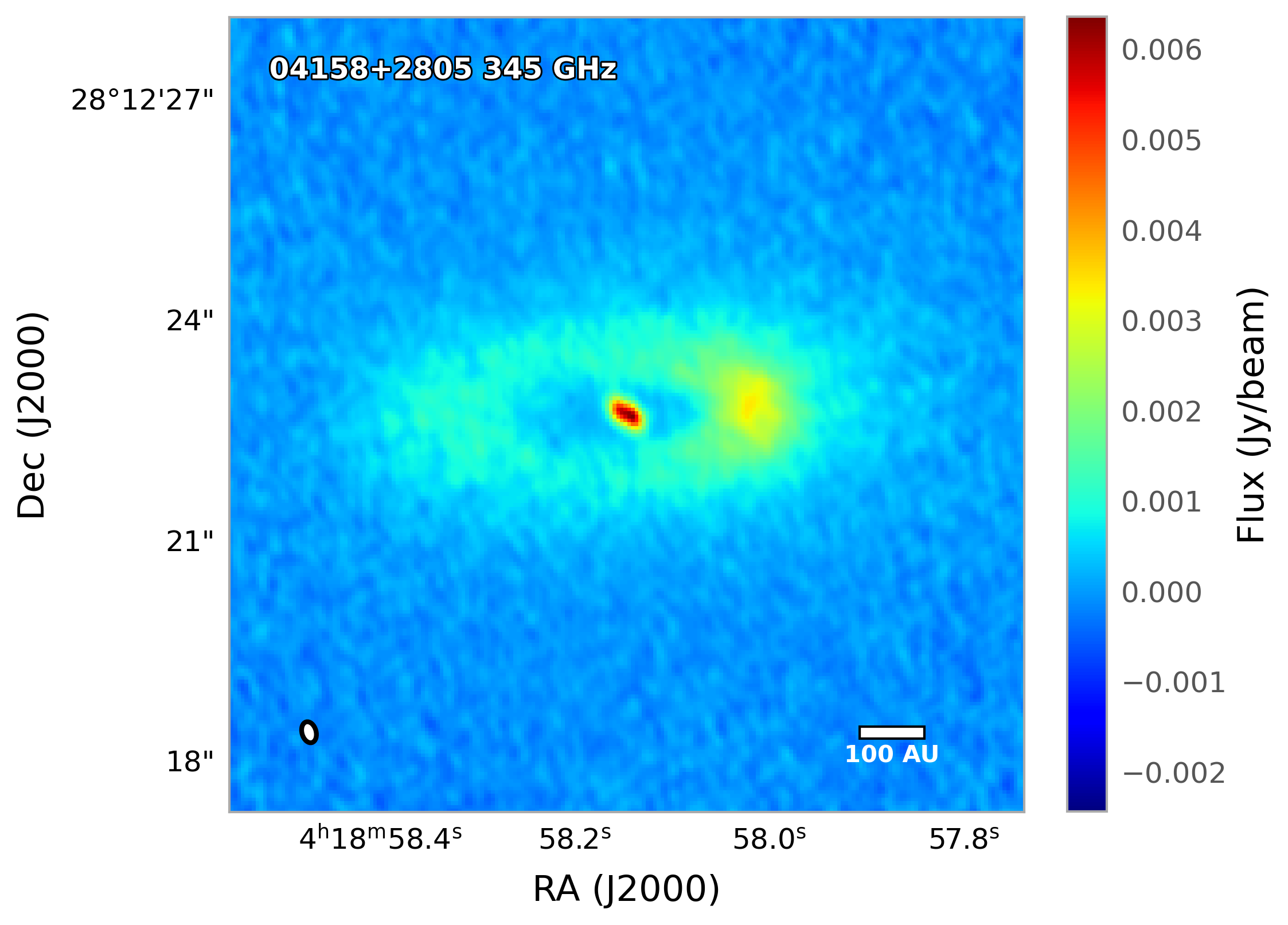}
\figsetplot{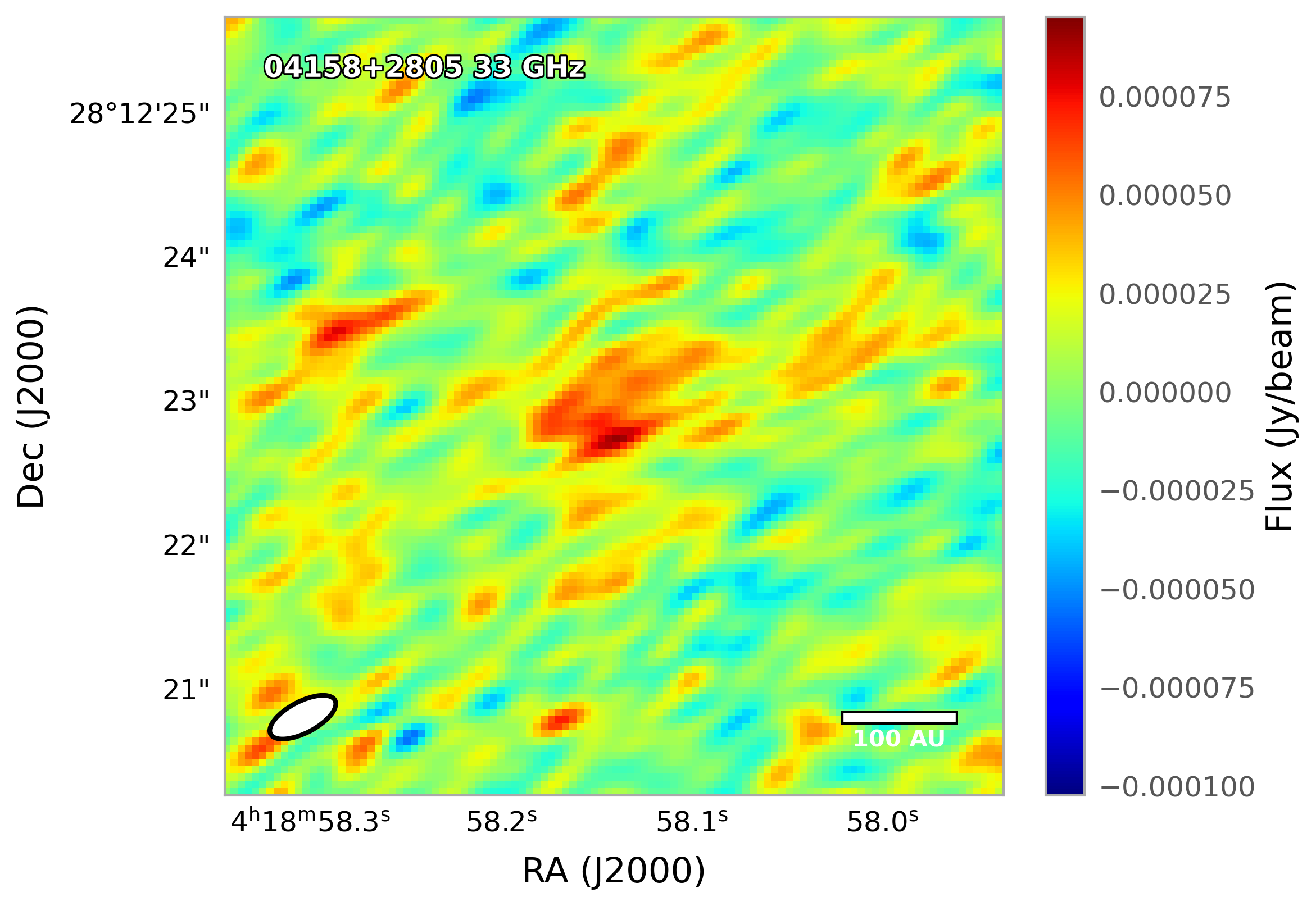}
\figsetgrpnote{ALMA 345~GHz and VLA 33~GHz continuum images of 04158+2805. The images are shown in units of Jy~beam$^{-1}$ in celestial coordinates (R.A./Decl.; J2000). A synthesized beam is shown in white in the lower left of each panel, and a 100~AU scale bar is shown in white in the lower right. Both images use a linear intensity scale.}
\figsetgrpend

\figsetgrpstart
\figsetgrpnum{A1.4}
\figsetgrptitle{04166+2706}
\figsetplot{alma_04166+2706.png}
\figsetplot{vla_04166+2706.png}
\figsetgrpnote{ALMA 345~GHz and VLA 33~GHz continuum images of 04166+2706. The images are shown in units of Jy~beam$^{-1}$ in celestial coordinates (R.A./Decl.; J2000). A synthesized beam is shown in white in the lower left of each panel, and a 100~AU scale bar is shown in white in the lower right. Both images use a linear intensity scale.}
\figsetgrpend

\figsetgrpstart
\figsetgrpnum{A1.5}
\figsetgrptitle{04169+2702}
\figsetplot{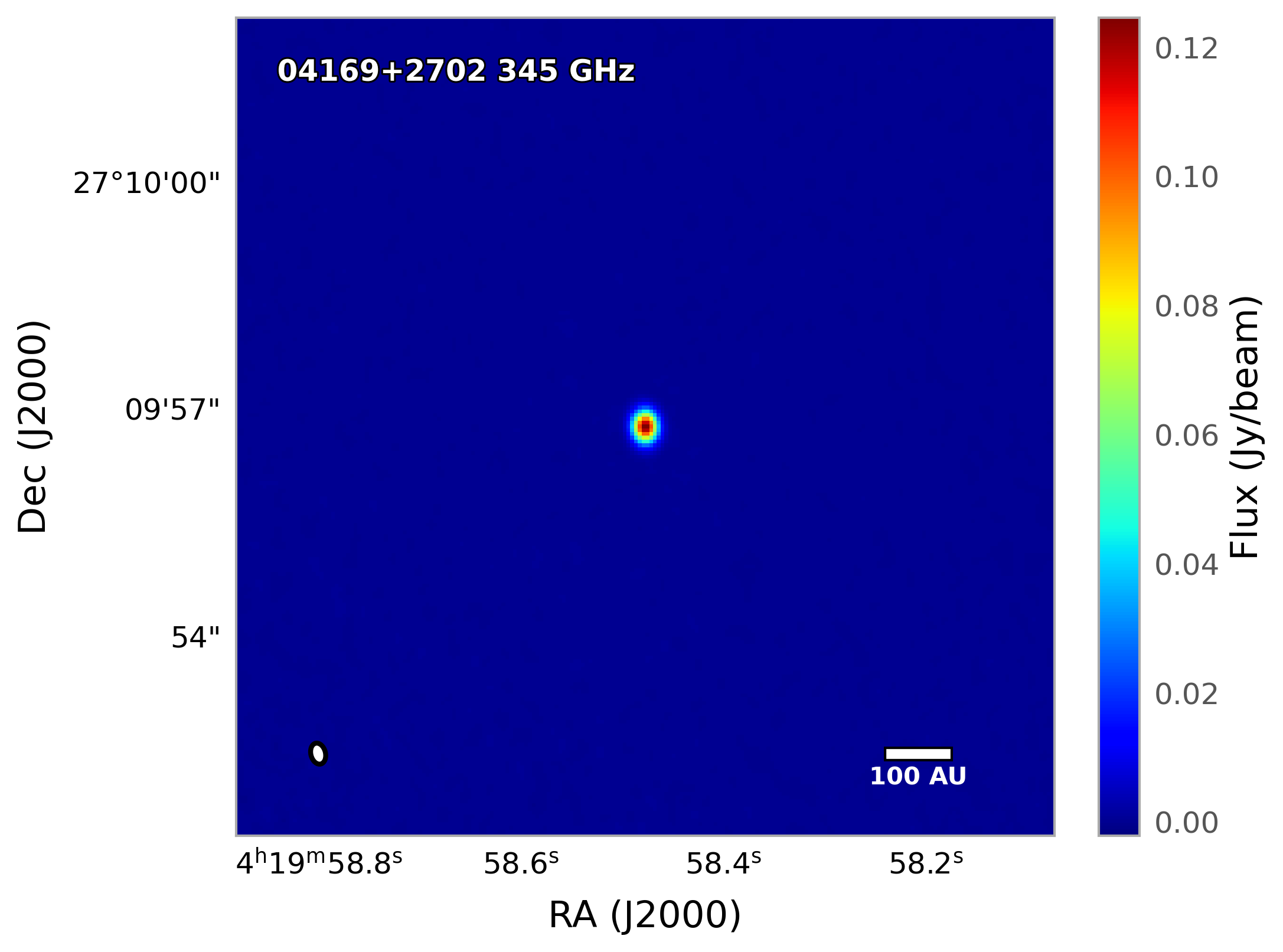}
\figsetplot{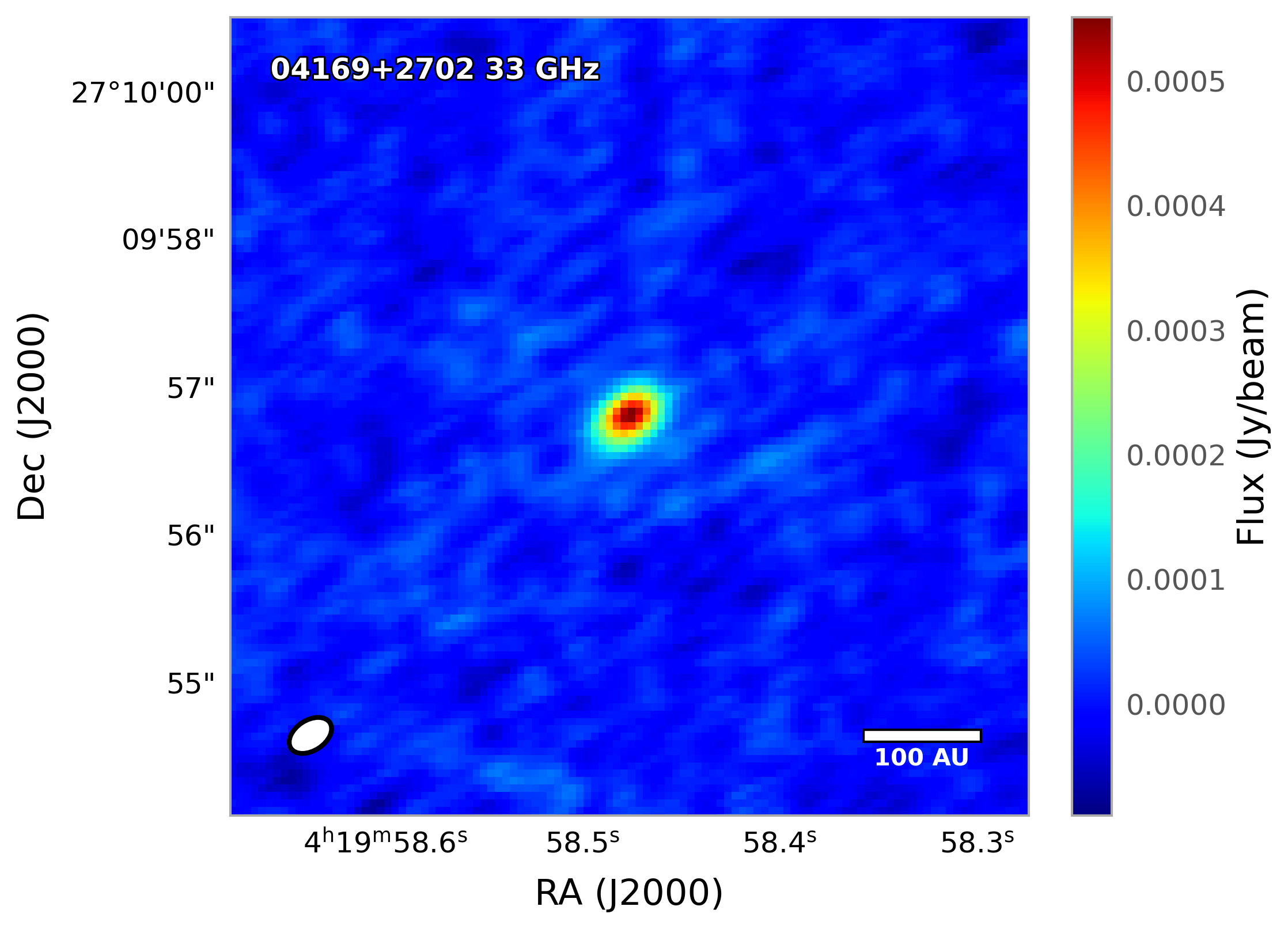}
\figsetgrpnote{ALMA 345~GHz and VLA 33~GHz continuum images of 04169+2702. The images are shown in units of Jy~beam$^{-1}$ in celestial coordinates (R.A./Decl.; J2000). A synthesized beam is shown in white in the lower left of each panel, and a 100~AU scale bar is shown in white in the lower right. Both images use a linear intensity scale.}
\figsetgrpend

\figsetgrpstart
\figsetgrpnum{A1.6}
\figsetgrptitle{04181+2654A}
\figsetplot{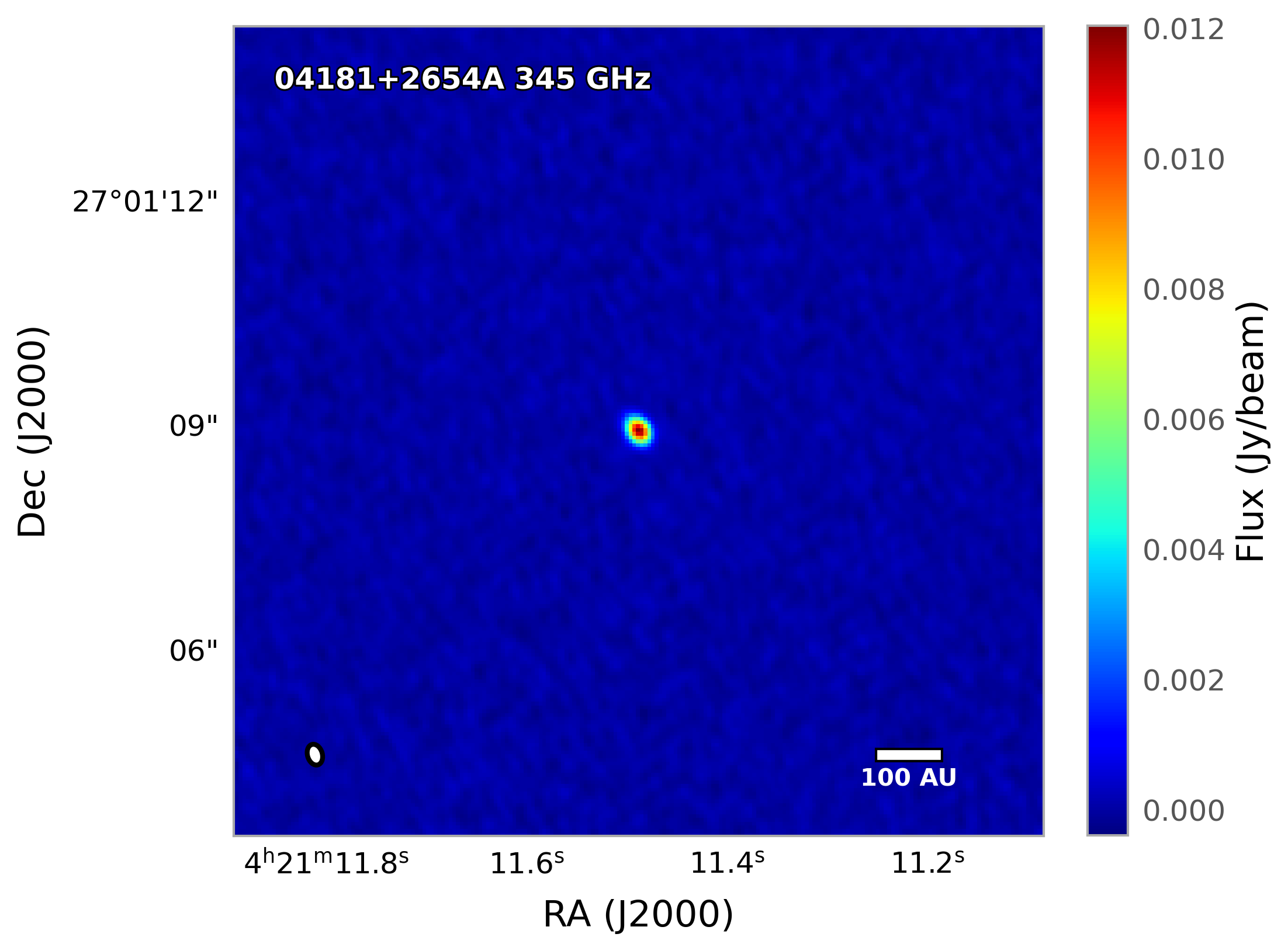}
\figsetplot{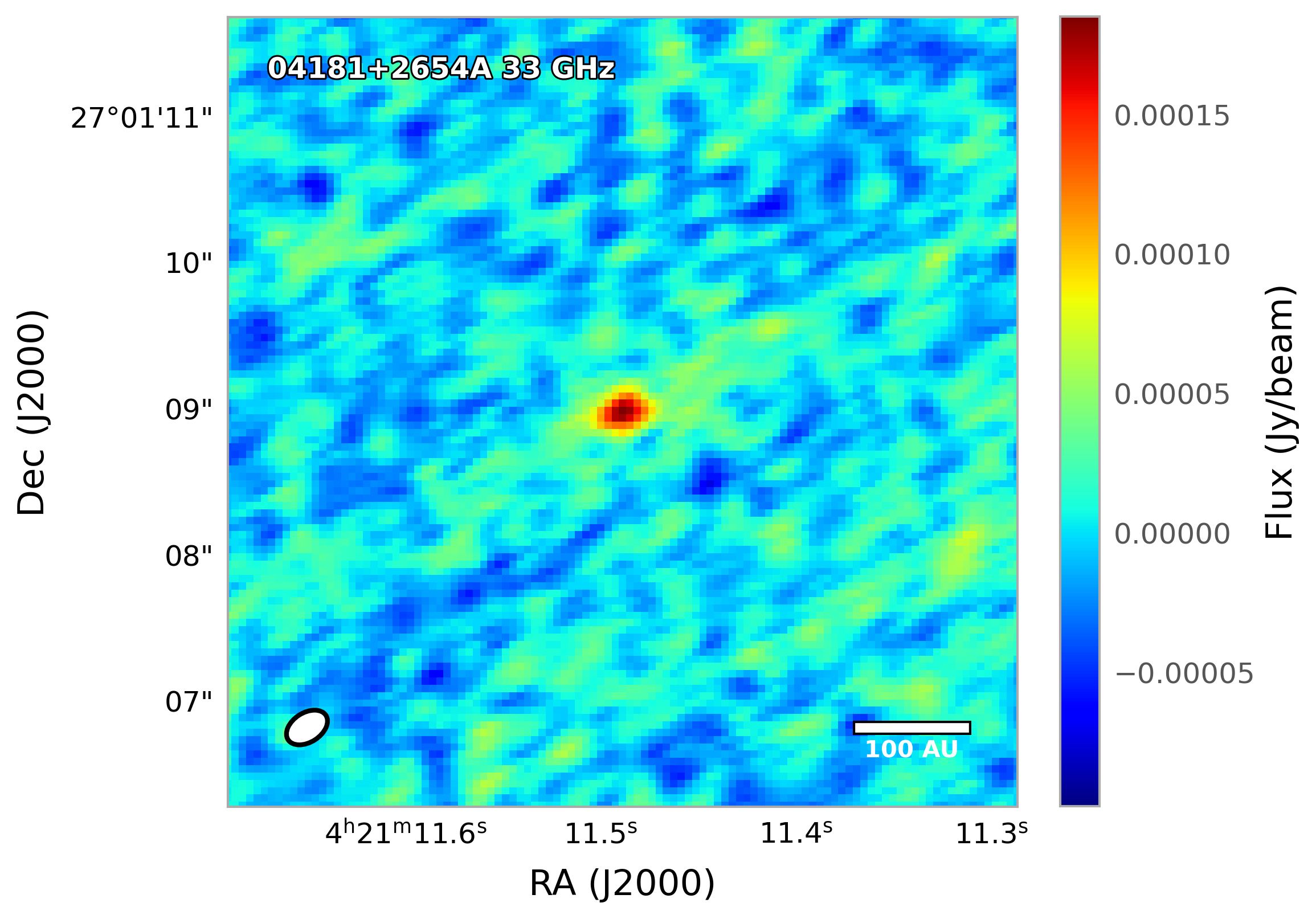}
\figsetgrpnote{ALMA 345~GHz and VLA 33~GHz continuum images of 04181+2654A. The images are shown in units of Jy~beam$^{-1}$ in celestial coordinates (R.A./Decl.; J2000). A synthesized beam is shown in white in the lower left of each panel, and a 100~AU scale bar is shown in white in the lower right. Both images use a linear intensity scale.}
\figsetgrpend

\figsetgrpstart
\figsetgrpnum{A1.7}
\figsetgrptitle{04191+1523AB}
\figsetplot{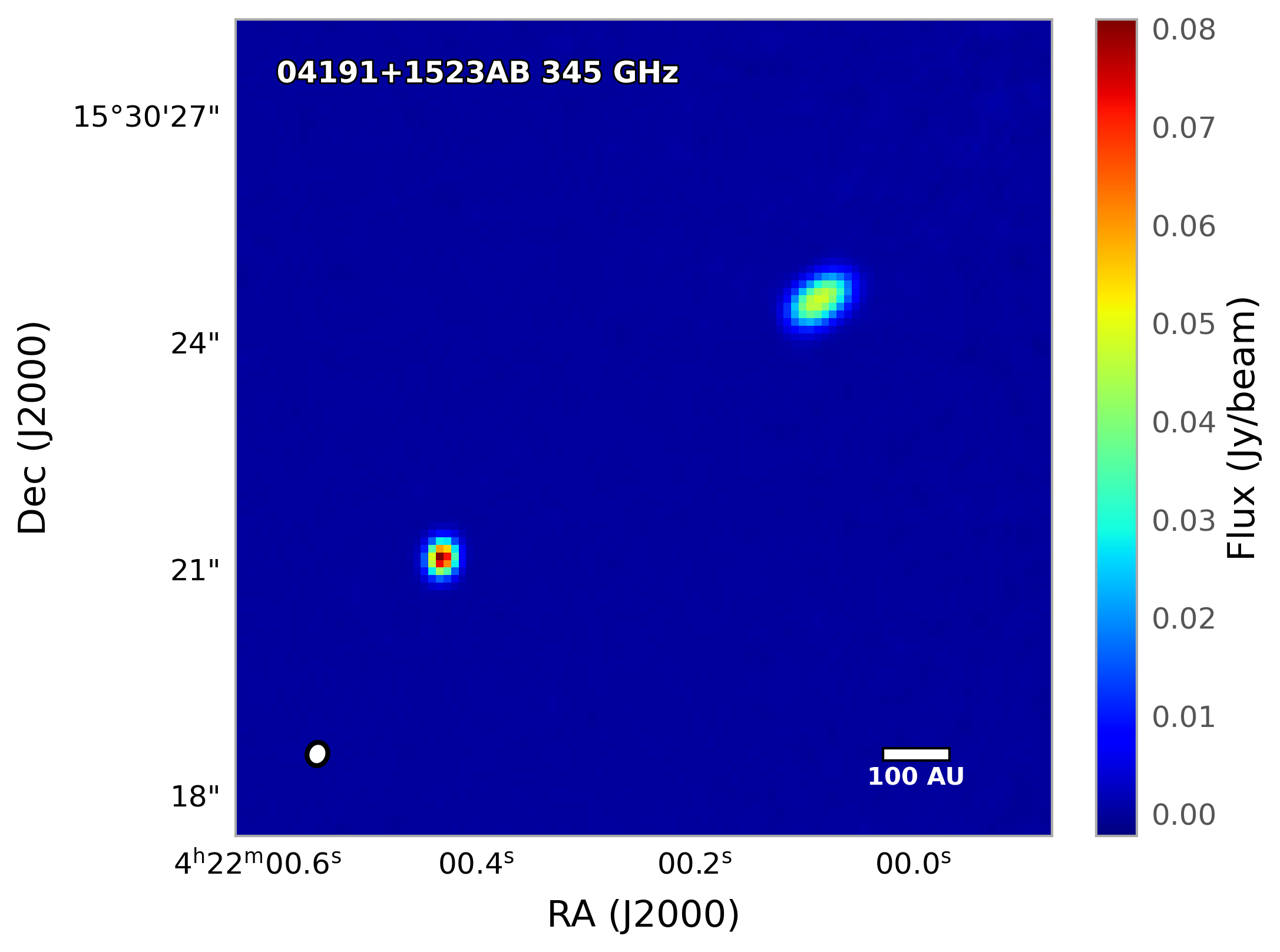}
\figsetplot{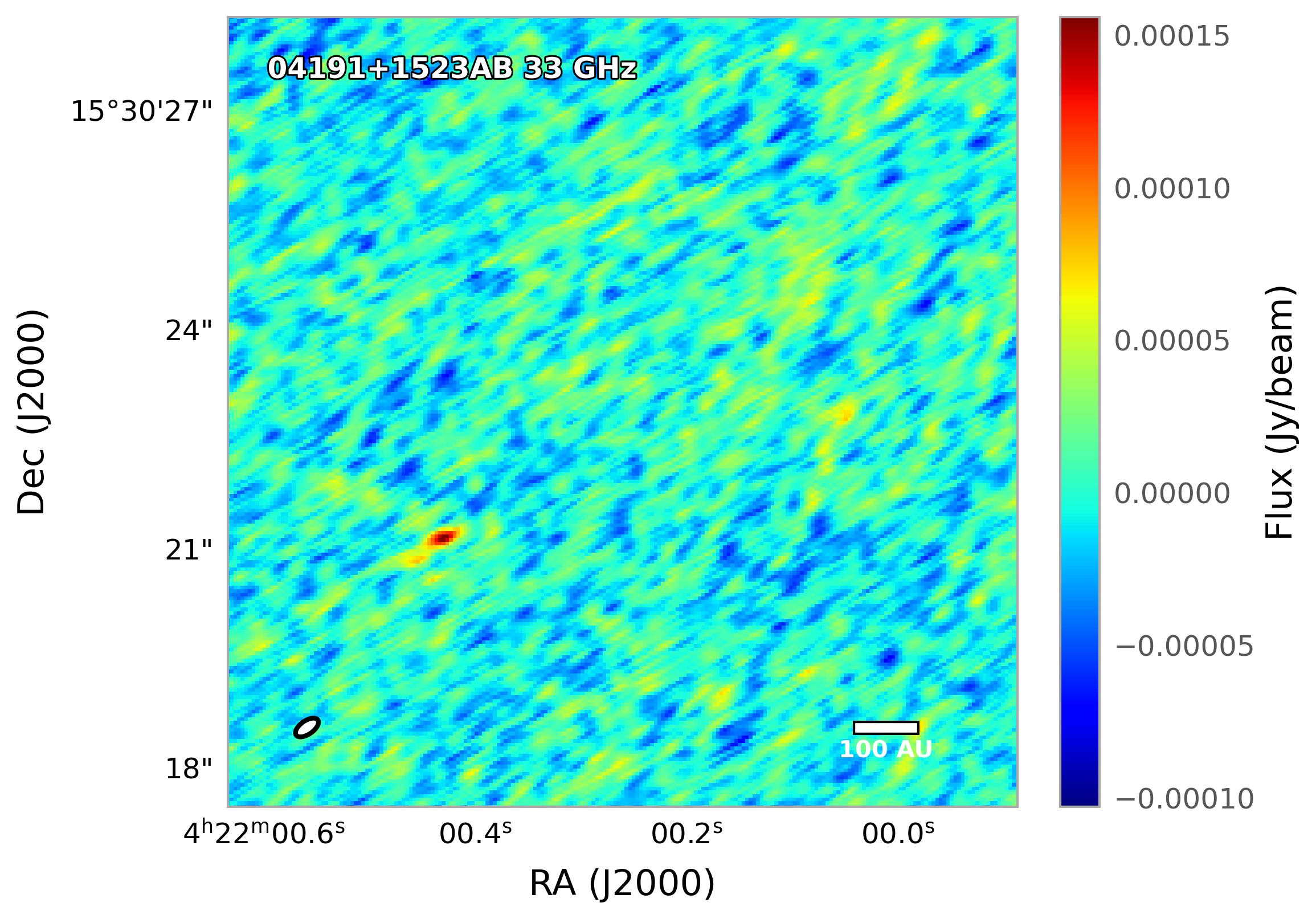}
\figsetgrpnote{ALMA 345~GHz and VLA 33~GHz continuum images of 04191+1523AB. The images are shown in units of Jy~beam$^{-1}$ in celestial coordinates (R.A./Decl.; J2000). A synthesized beam is shown in white in the lower left of each panel, and a 100~AU scale bar is shown in white in the lower right. The images are centered on the midpoint between the two components. Both images use a linear intensity scale.}
\figsetgrpend

\figsetgrpstart
\figsetgrpnum{A1.8}
\figsetgrptitle{04239+2436AB}
\figsetplot{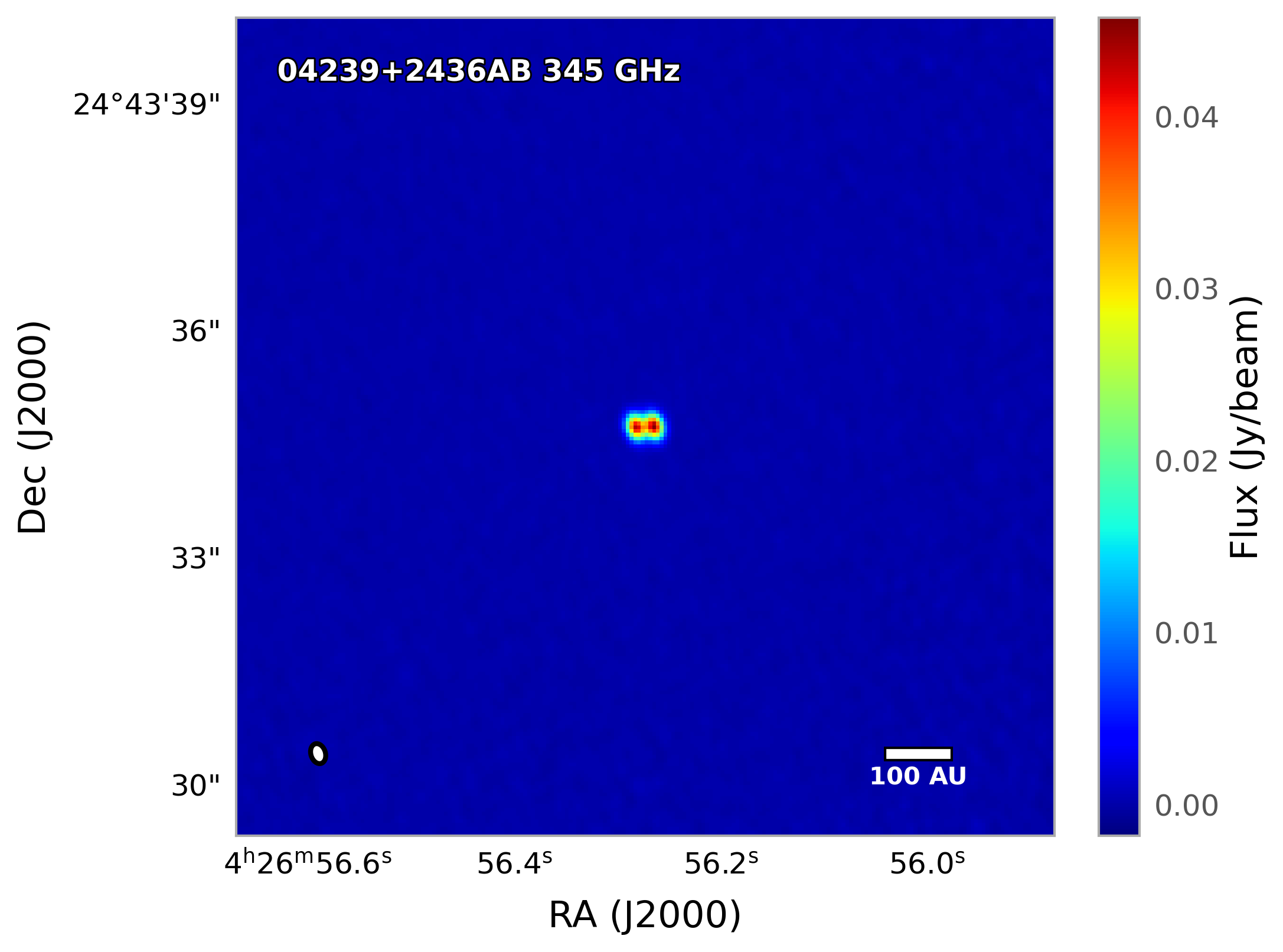}
\figsetplot{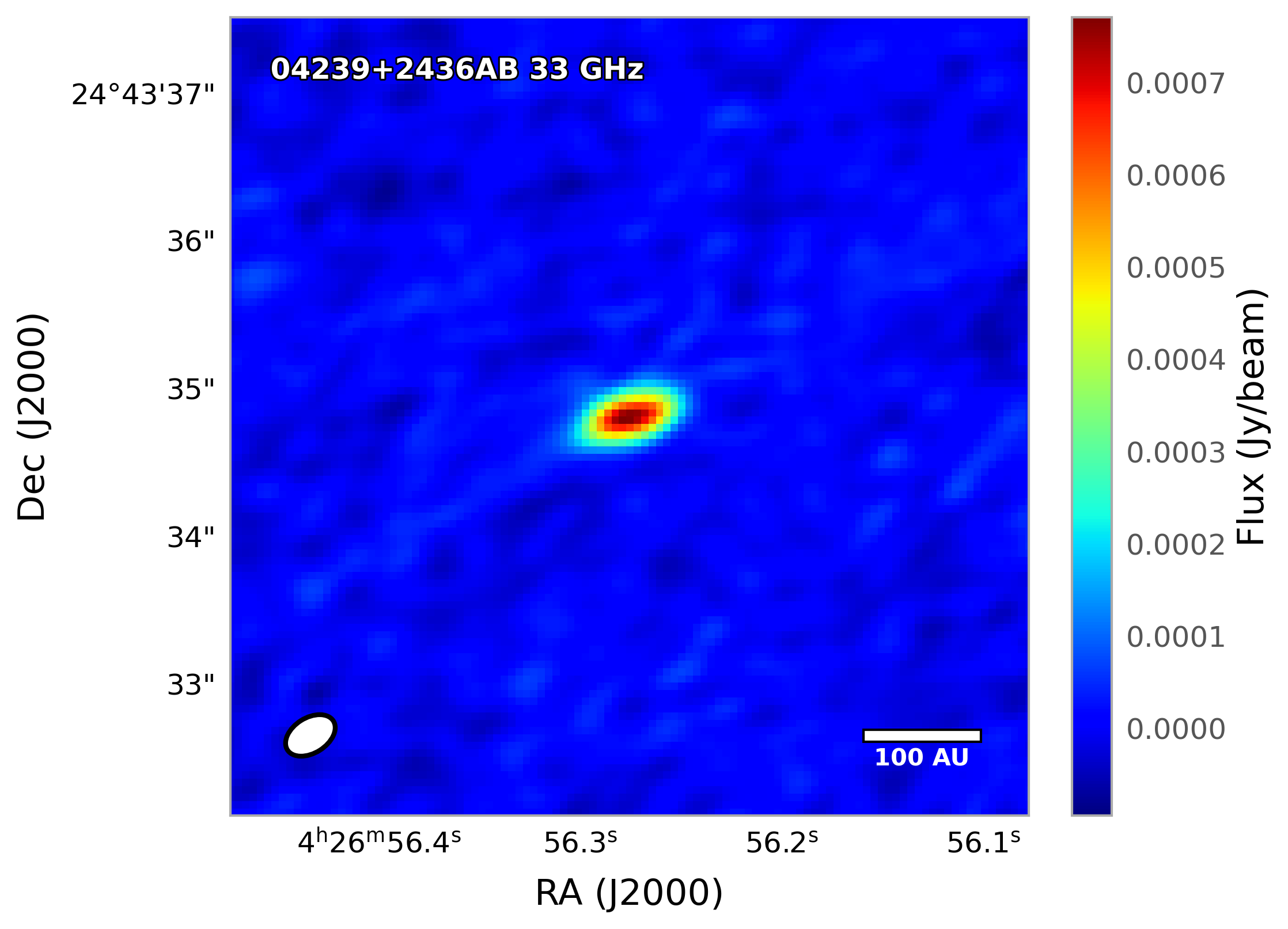}
\figsetgrpnote{ALMA 345~GHz and VLA 33~GHz continuum images of 04239+2436AB. The images are shown in units of Jy~beam$^{-1}$ in celestial coordinates (R.A./Decl.; J2000). A synthesized beam is shown in white in the lower left of each panel, and a 100~AU scale bar is shown in white in the lower right. Both images use a linear intensity scale.}
\figsetgrpend

\figsetgrpstart
\figsetgrpnum{A1.9}
\figsetgrptitle{04248+2612AB}
\figsetplot{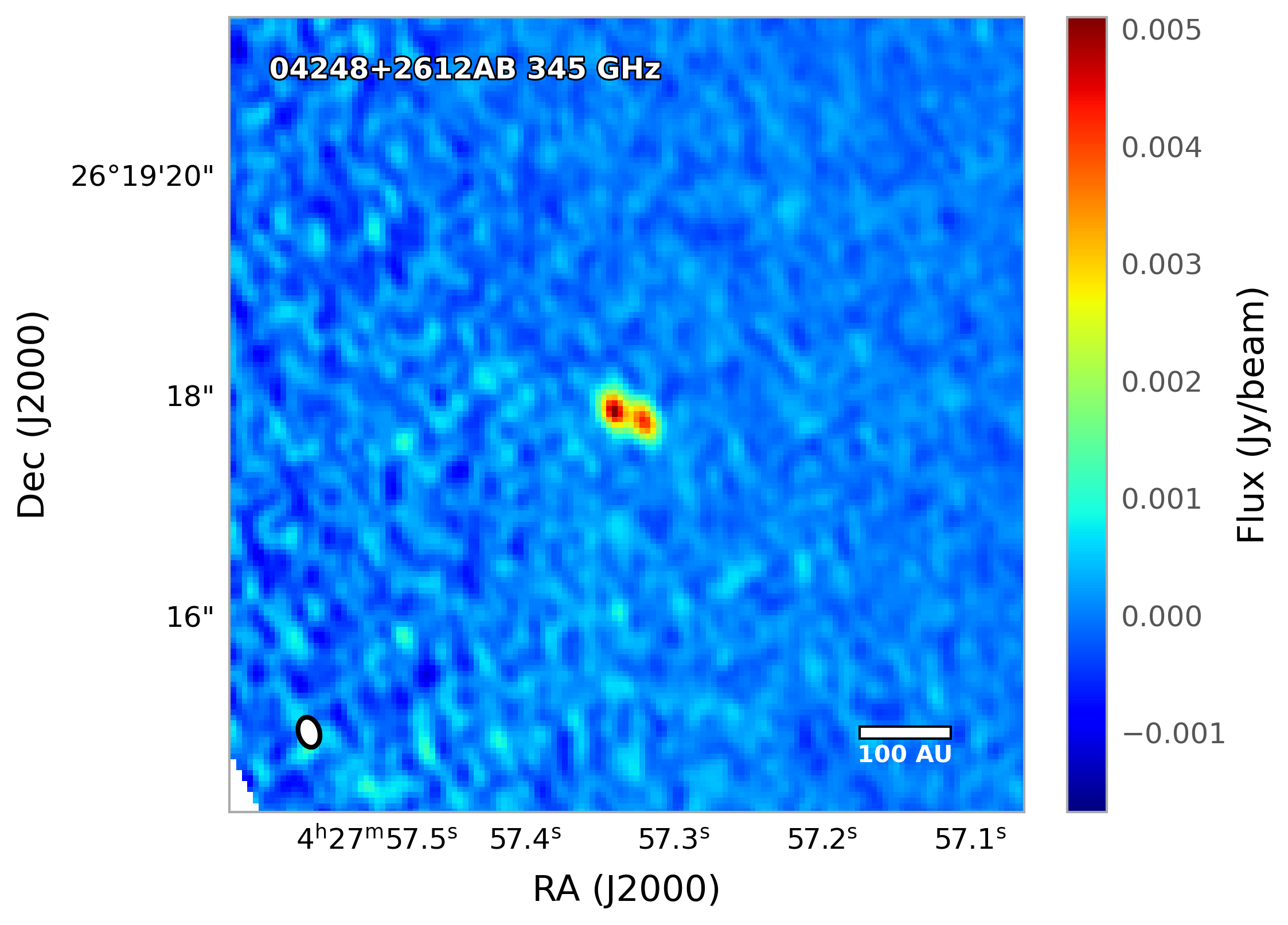}
\figsetplot{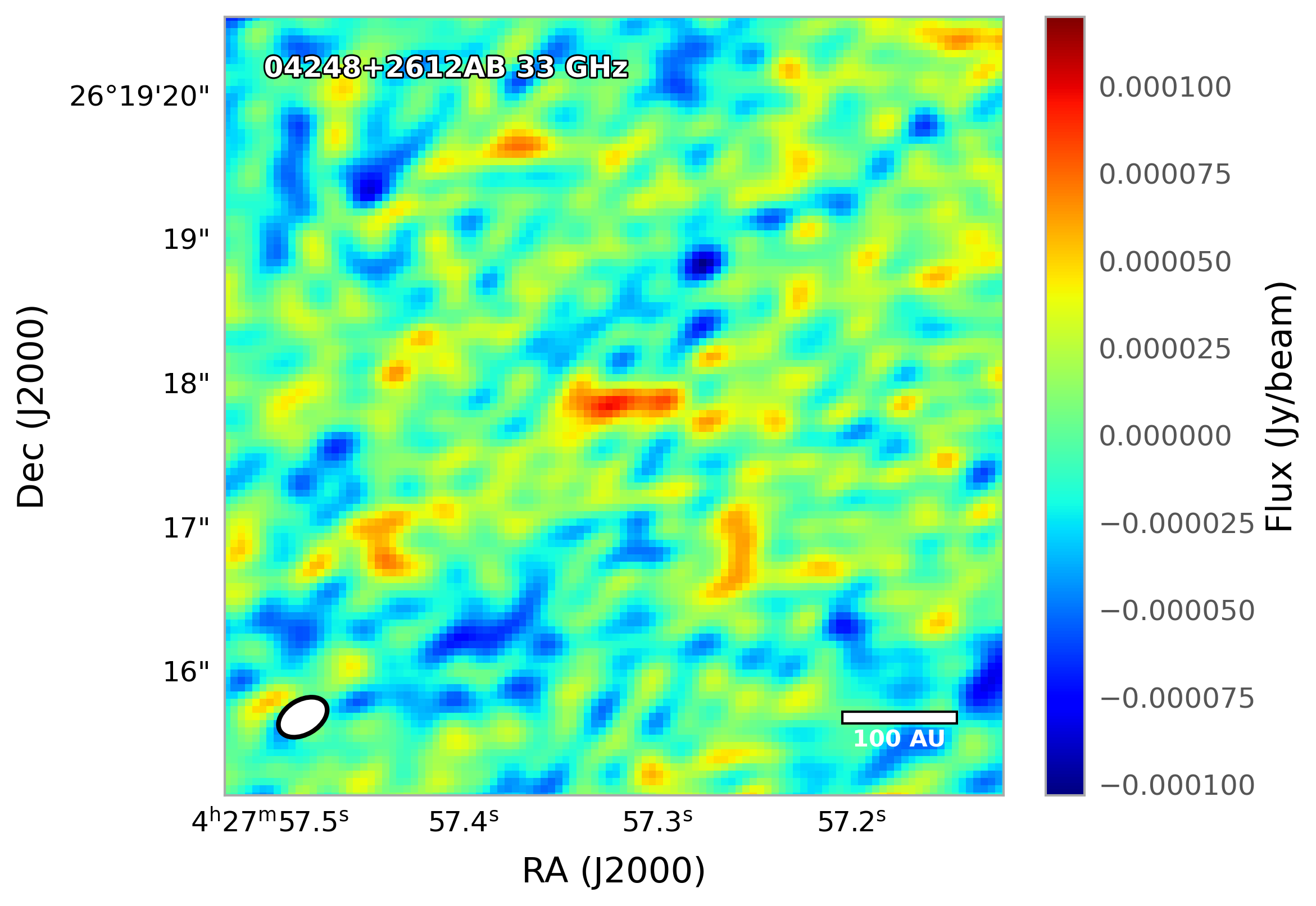}
\figsetgrpnote{ALMA 345~GHz and VLA 33~GHz continuum images of 04248+2612AB. The images are shown in units of Jy~beam$^{-1}$ in celestial coordinates (R.A./Decl.; J2000). A synthesized beam is shown in white in the lower left of each panel, and a 100~AU scale bar is shown in white in the lower right. Both images use a linear intensity scale.}
\figsetgrpend

\figsetgrpstart
\figsetgrpnum{A1.10}
\figsetgrptitle{04248+2612C}
\figsetplot{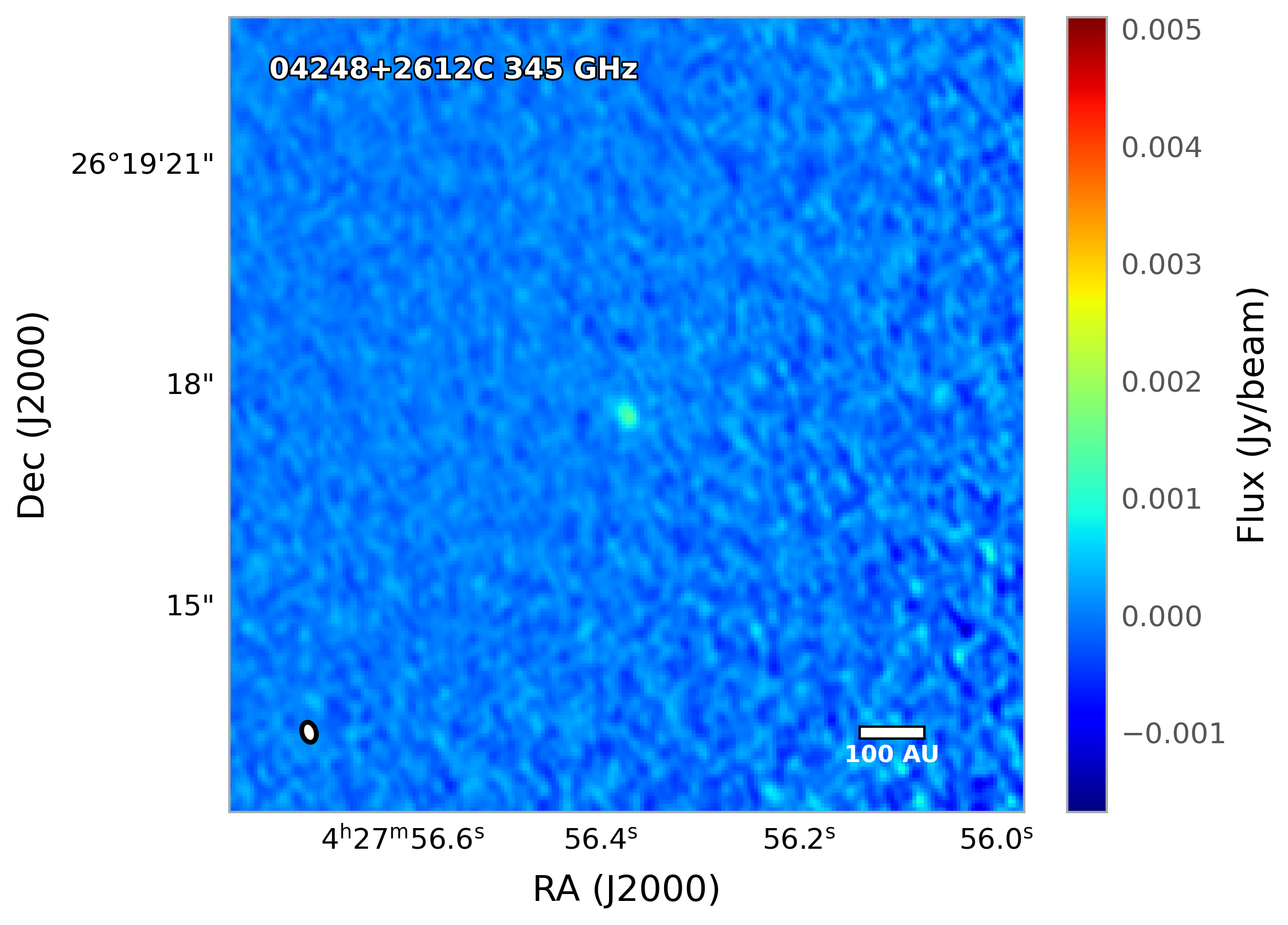}
\figsetplot{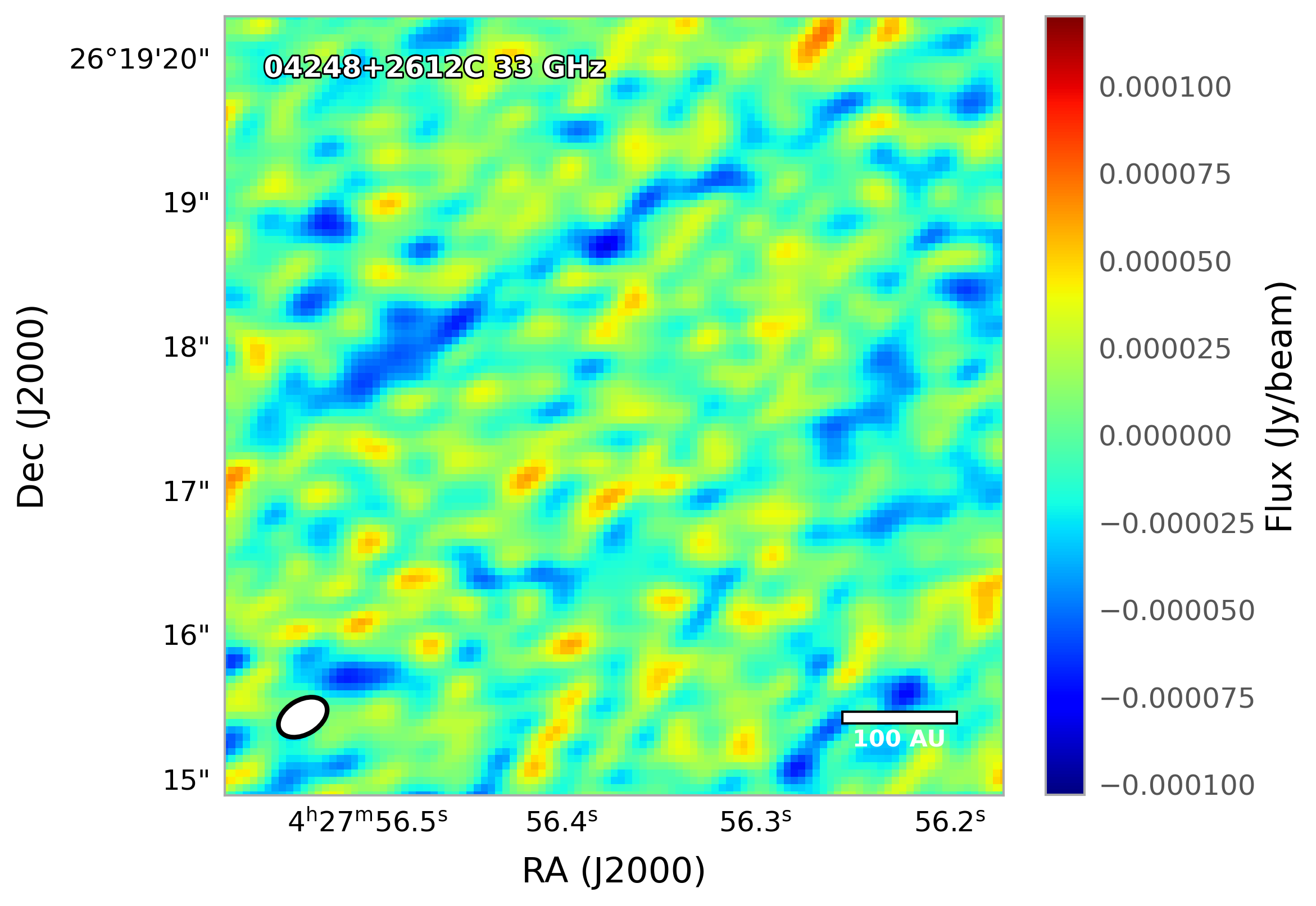}
\figsetgrpnote{ALMA 345~GHz and VLA 33~GHz continuum images of 04248+2612C. The images are shown in units of Jy~beam$^{-1}$ in celestial coordinates (R.A./Decl.; J2000). A synthesized beam is shown in white in the lower left of each panel, and a 100~AU scale bar is shown in white in the lower right. Both images use a linear intensity scale.}
\figsetgrpend

\figsetgrpstart
\figsetgrpnum{A1.11}
\figsetgrptitle{04260+2642}
\figsetplot{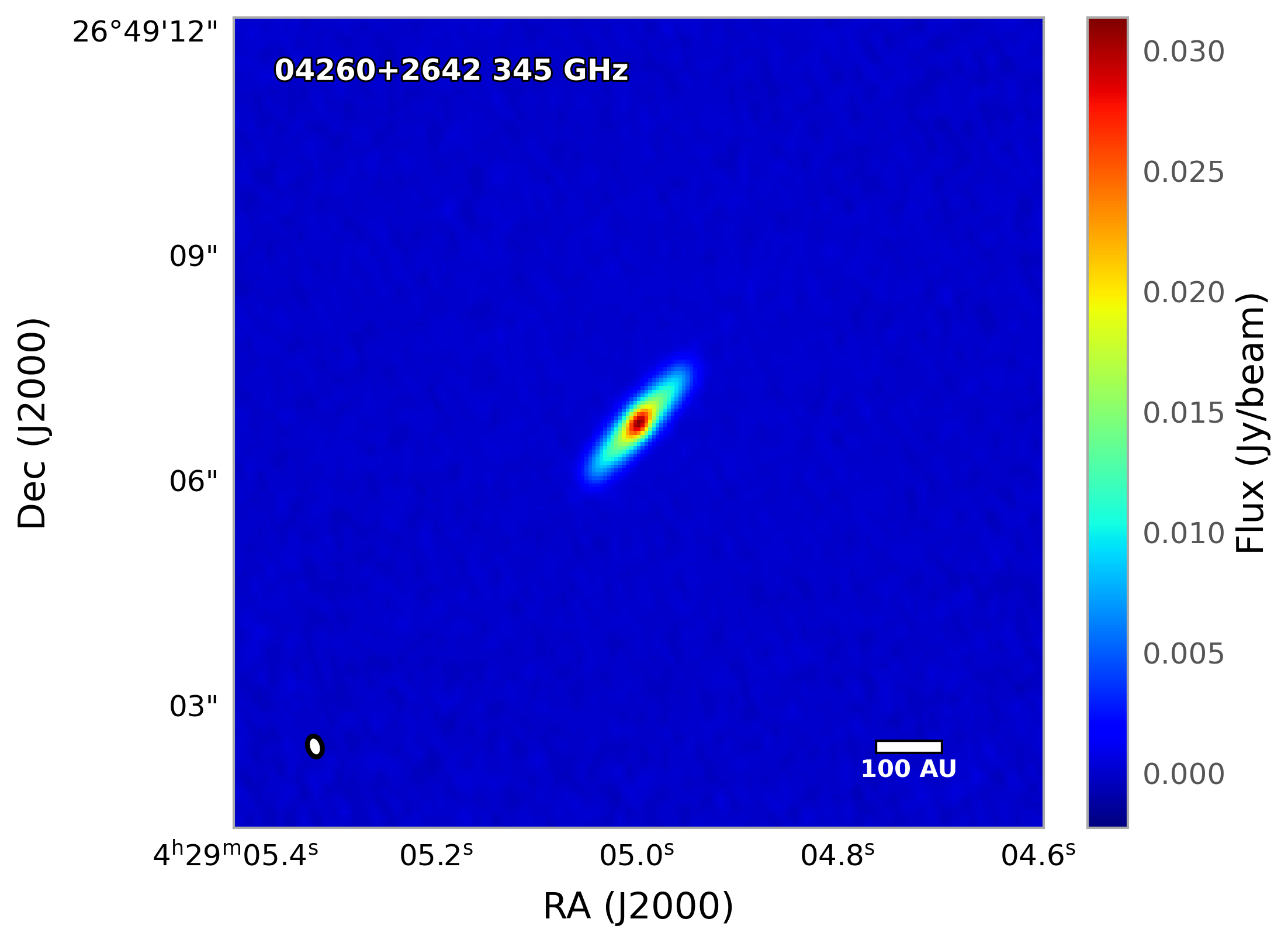}
\figsetplot{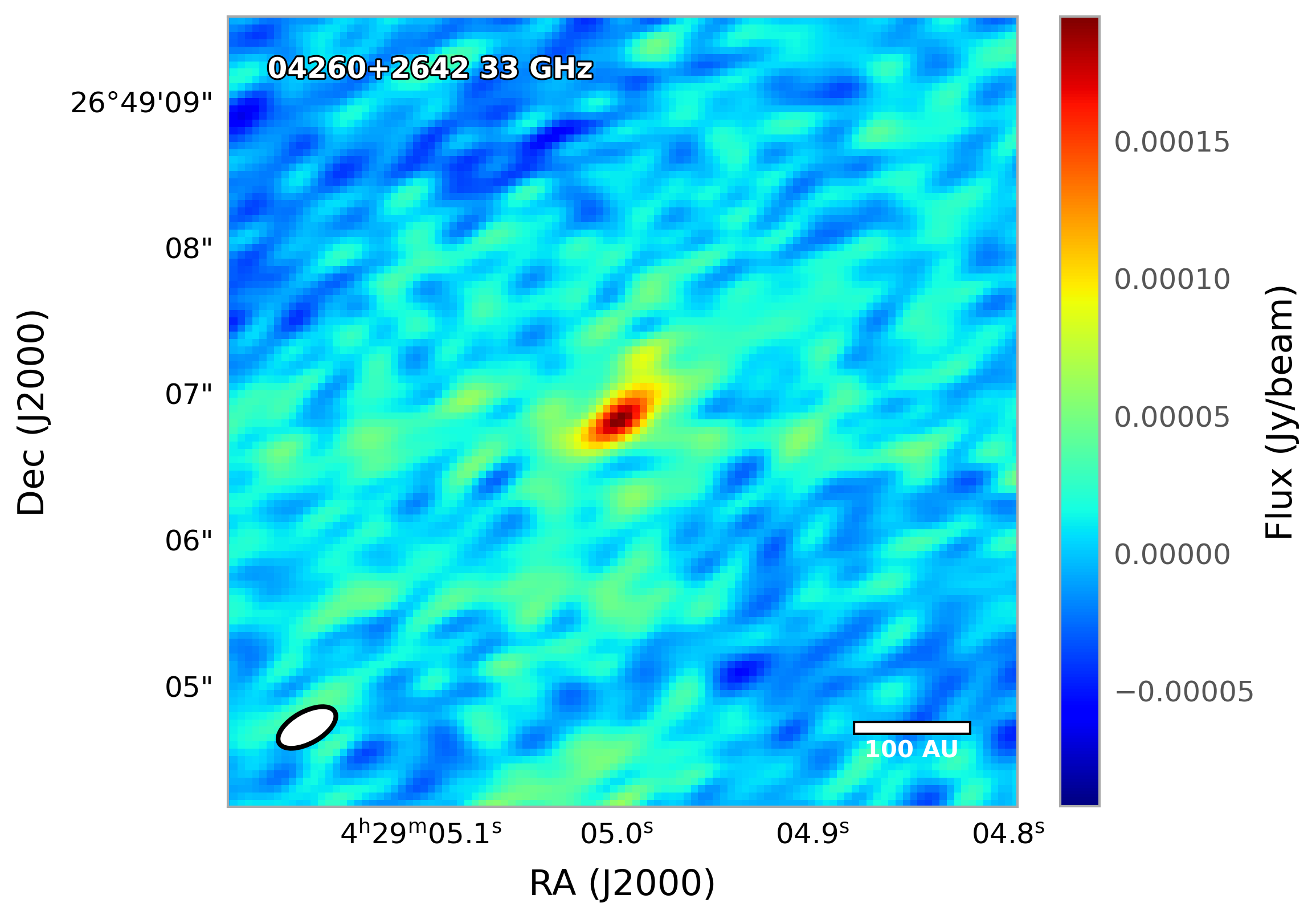}
\figsetgrpnote{ALMA 345~GHz and VLA 33~GHz continuum images of 04260+2642. The images are shown in units of Jy~beam$^{-1}$ in celestial coordinates (R.A./Decl.; J2000). A synthesized beam is shown in white in the lower left of each panel, and a 100~AU scale bar is shown in white in the lower right. Both images use a linear intensity scale.}
\figsetgrpend

\figsetgrpstart
\figsetgrpnum{A1.12}
\figsetgrptitle{04263+2426AB}
\figsetplot{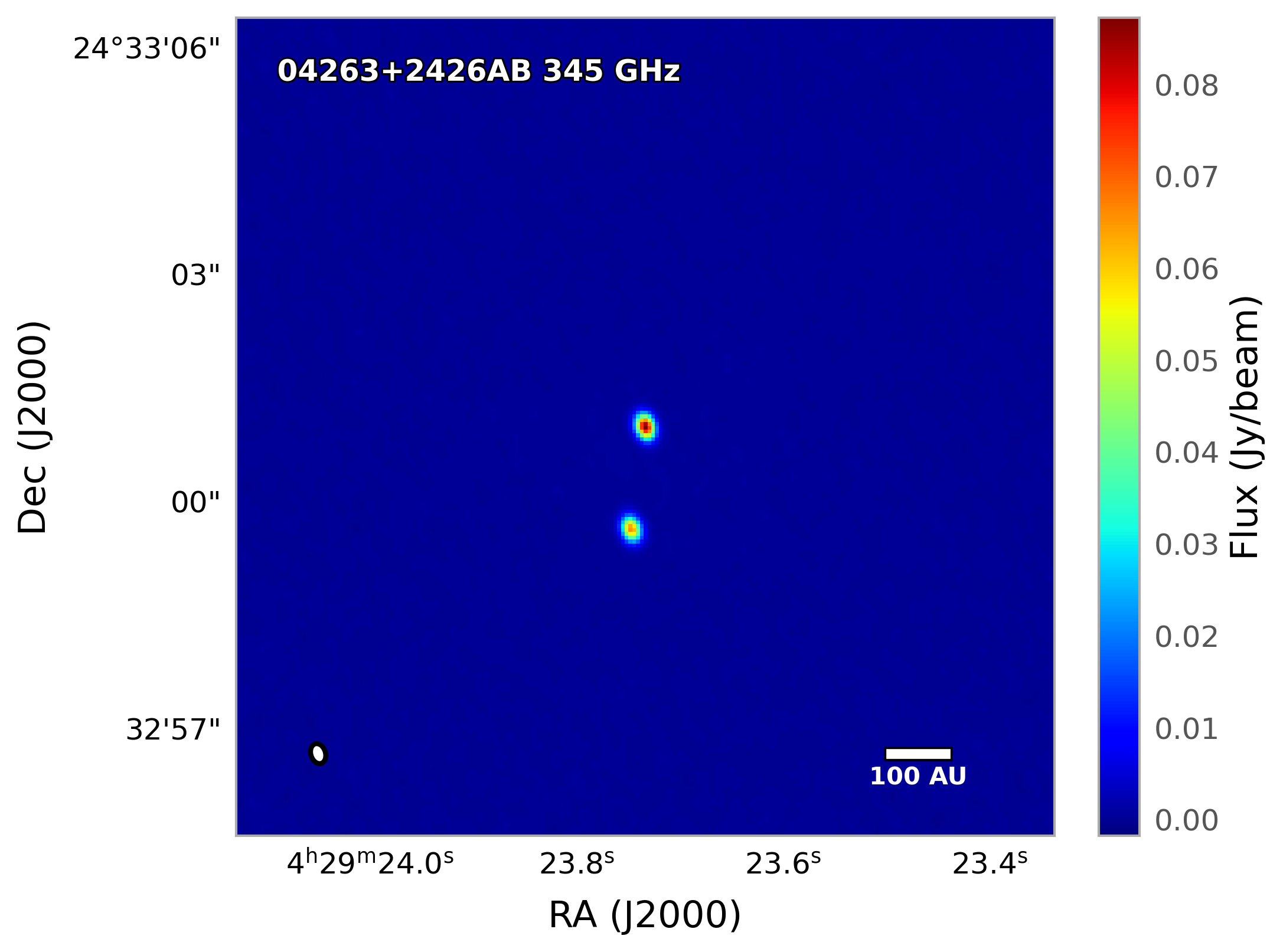}
\figsetplot{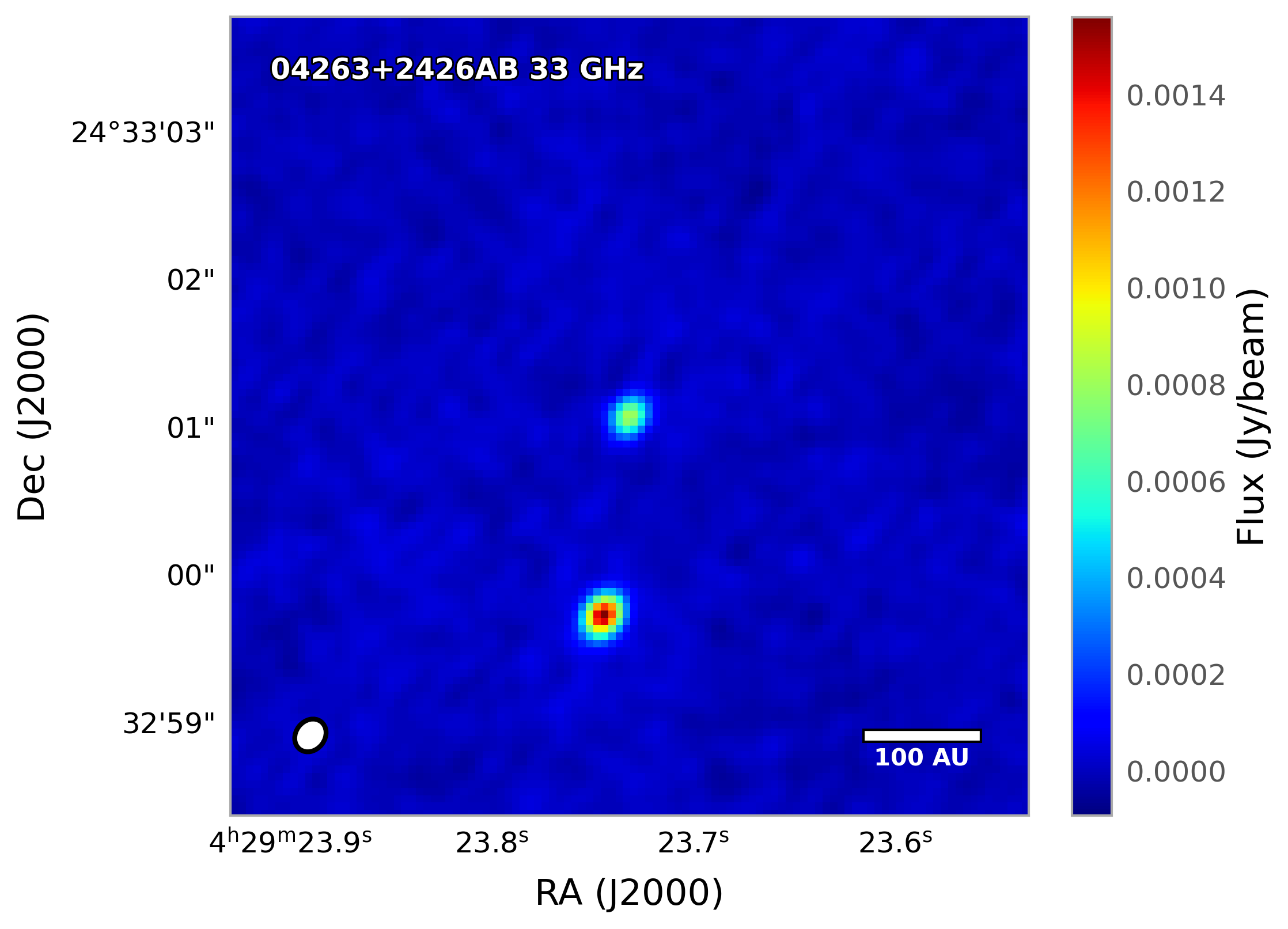}
\figsetgrpnote{ALMA 345~GHz and VLA 33~GHz continuum images of 04263+2426AB. The images are shown in units of Jy~beam$^{-1}$ in celestial coordinates (R.A./Decl.; J2000). A synthesized beam is shown in white in the lower left of each panel, and a 100~AU scale bar is shown in white in the lower right. Both images use a linear intensity scale.}
\figsetgrpend

\figsetgrpstart
\figsetgrpnum{A1.13}
\figsetgrptitle{04264+2433AB}
\figsetplot{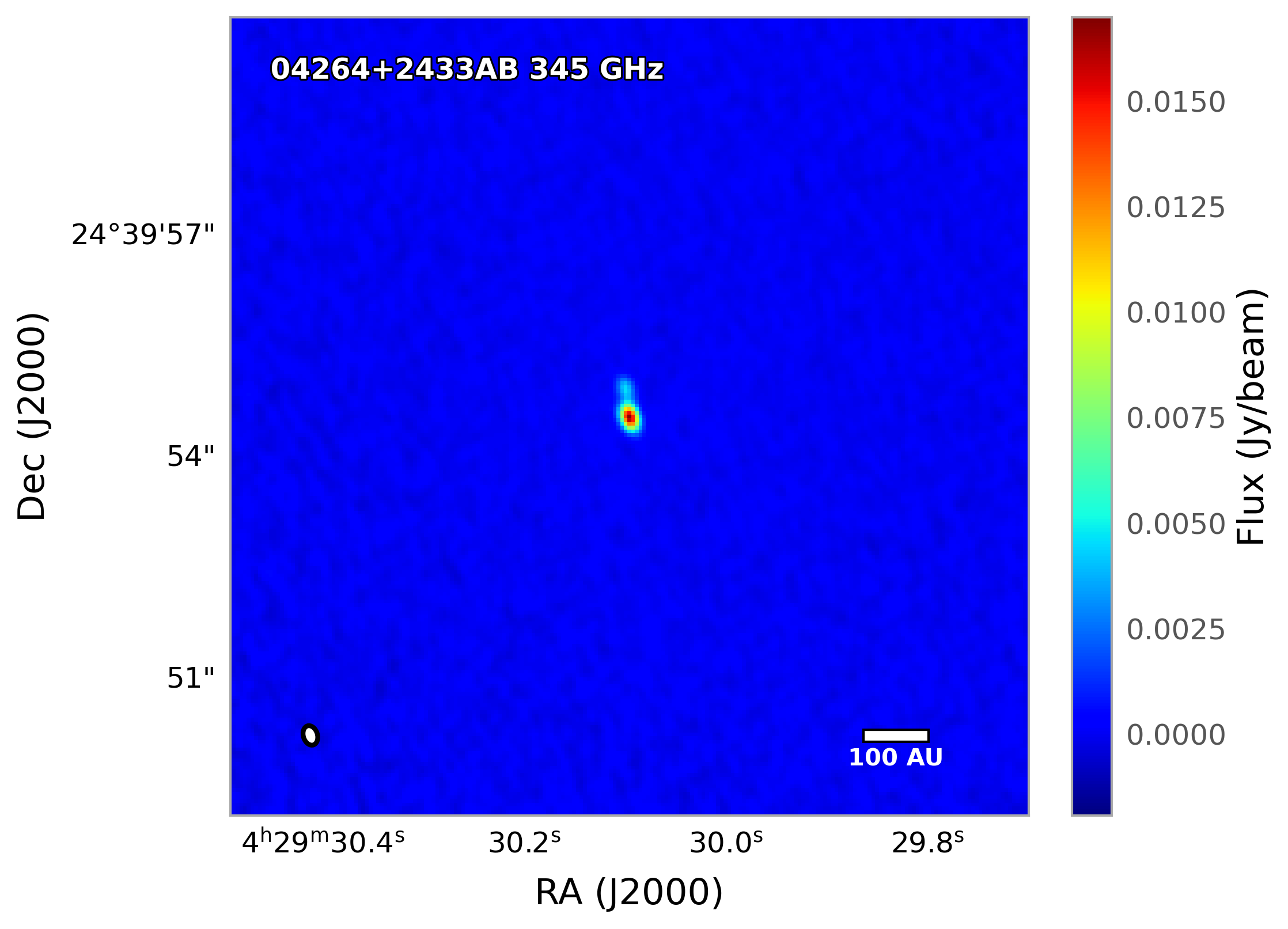}
\figsetplot{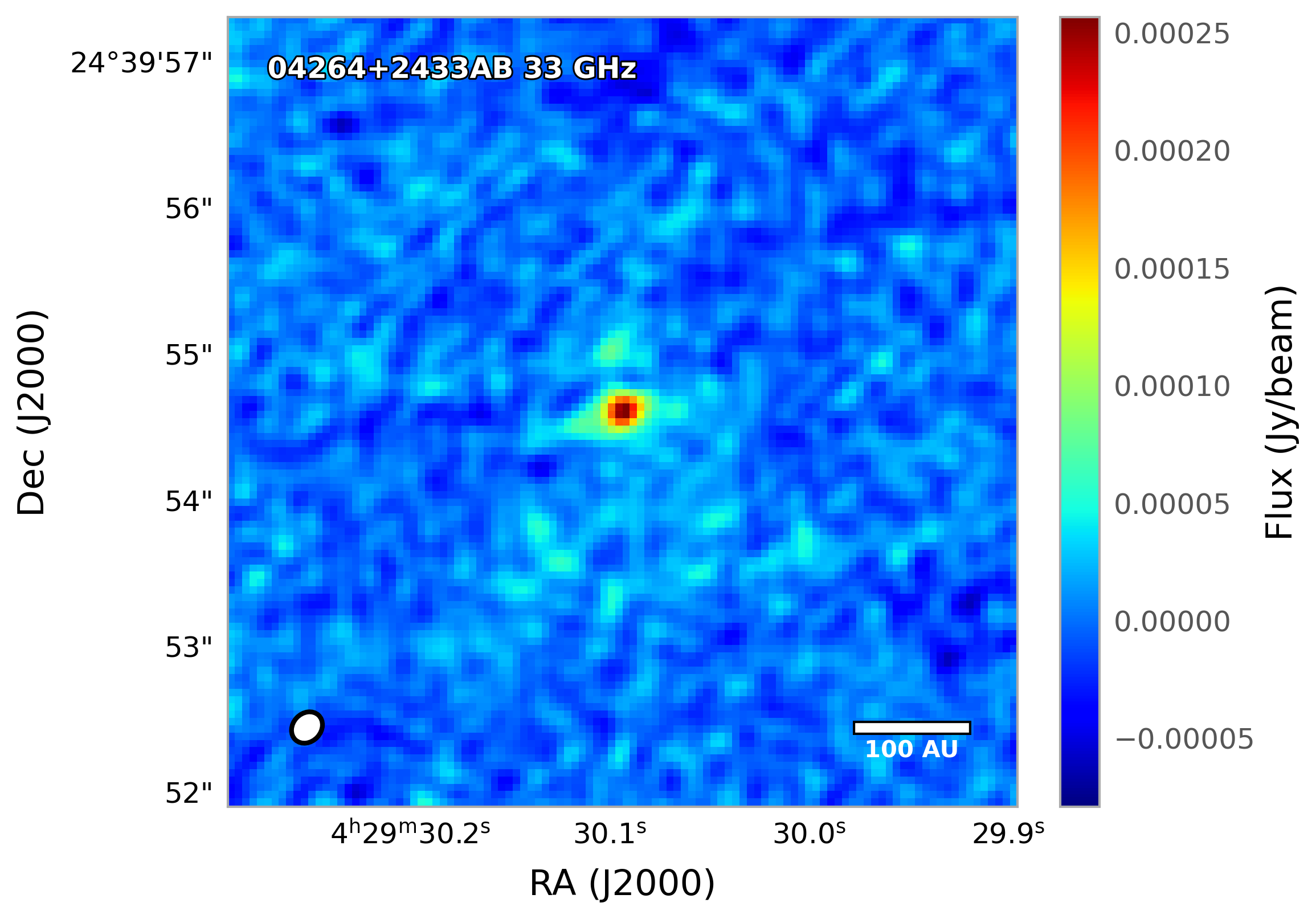}
\figsetgrpnote{ALMA 345~GHz and VLA 33~GHz continuum images of 04264+2433AB. The images are shown in units of Jy~beam$^{-1}$ in celestial coordinates (R.A./Decl.; J2000). A synthesized beam is shown in white in the lower left of each panel, and a 100~AU scale bar is shown in white in the lower right. Both images use a linear intensity scale.}
\figsetgrpend

\figsetgrpstart
\figsetgrpnum{A1.14}
\figsetgrptitle{04287+1801}
\figsetplot{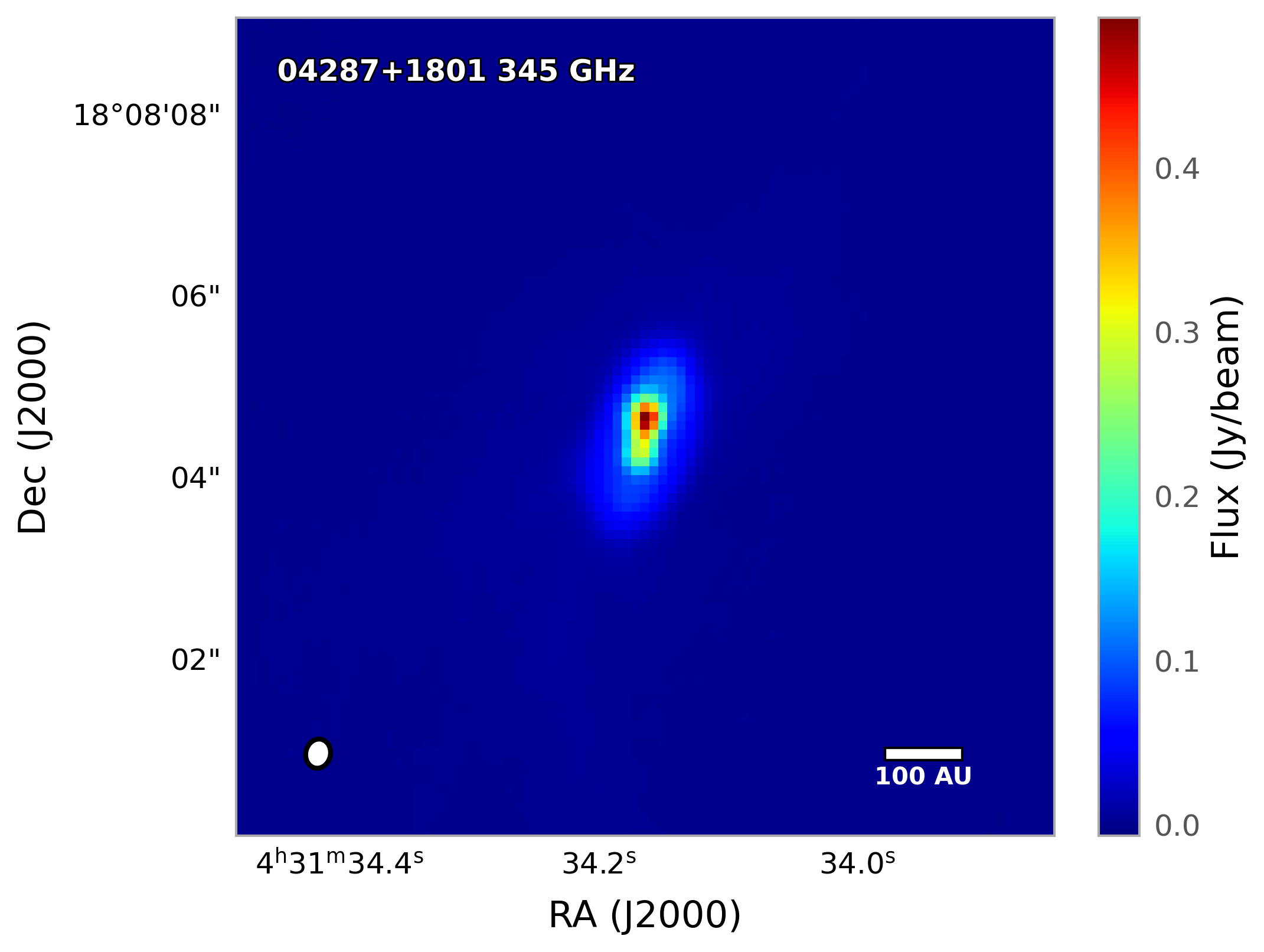}
\figsetplot{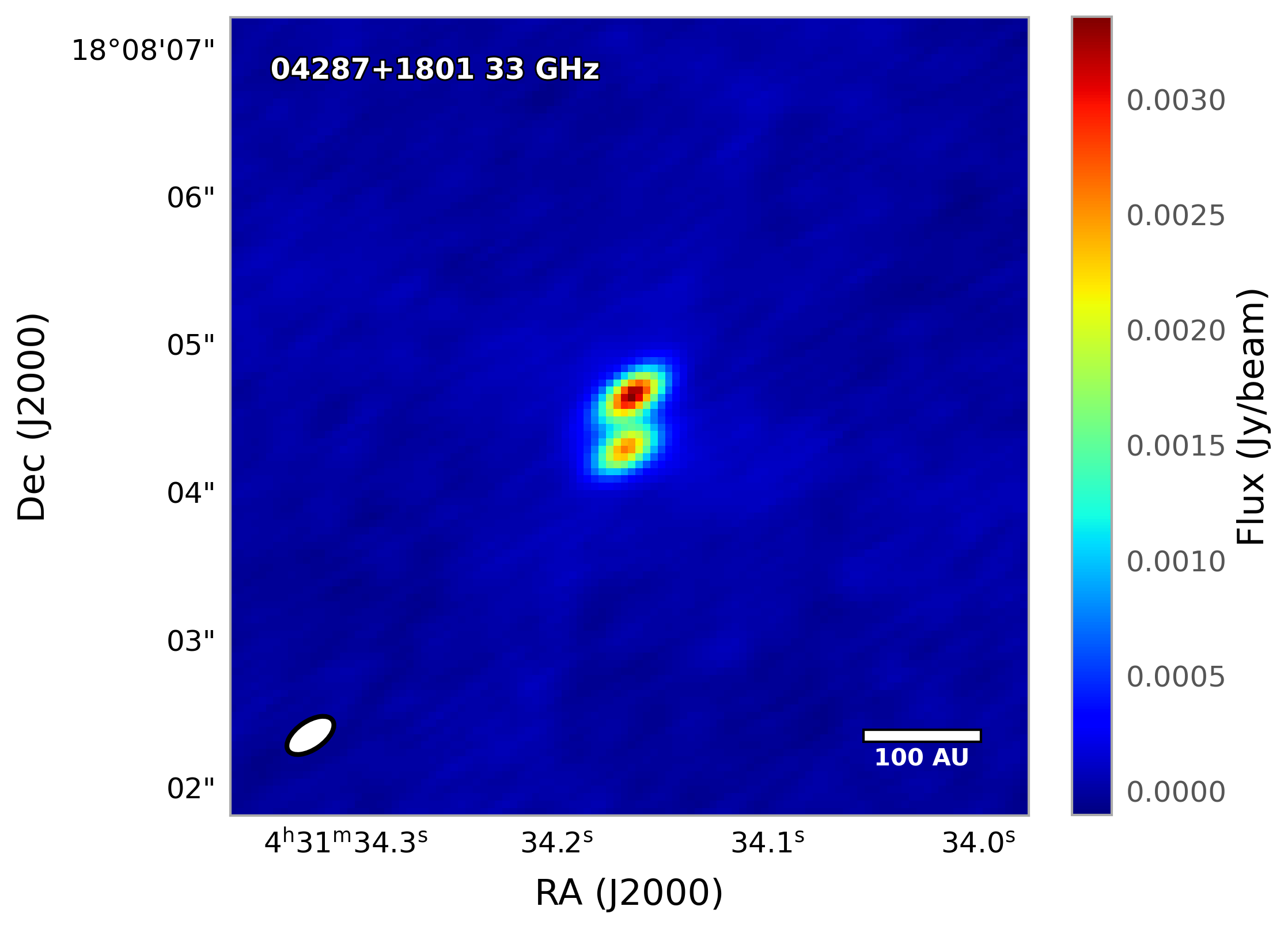}
\figsetgrpnote{ALMA 345~GHz and VLA 33~GHz continuum images of 04287+1801. The images are shown in units of Jy~beam$^{-1}$ in celestial coordinates (R.A./Decl.; J2000). A synthesized beam is shown in white in the lower left of each panel, and a 100~AU scale bar is shown in white in the lower right. Both images use a linear intensity scale.}
\figsetgrpend

\figsetgrpstart
\figsetgrpnum{A1.15}
\figsetgrptitle{04288+1802}
\figsetplot{alma_04288+1802.png}
\figsetplot{vla_04288+1802.png}
\figsetgrpnote{ALMA 345~GHz and VLA 33~GHz continuum images of 04288+1802. The images are shown in units of Jy~beam$^{-1}$ in celestial coordinates (R.A./Decl.; J2000). A synthesized beam is shown in white in the lower left of each panel, and a 100~AU scale bar is shown in white in the lower right. Both images use a linear intensity scale.}
\figsetgrpend

\figsetgrpstart
\figsetgrpnum{A1.16}
\figsetgrptitle{04295+2251}
\figsetplot{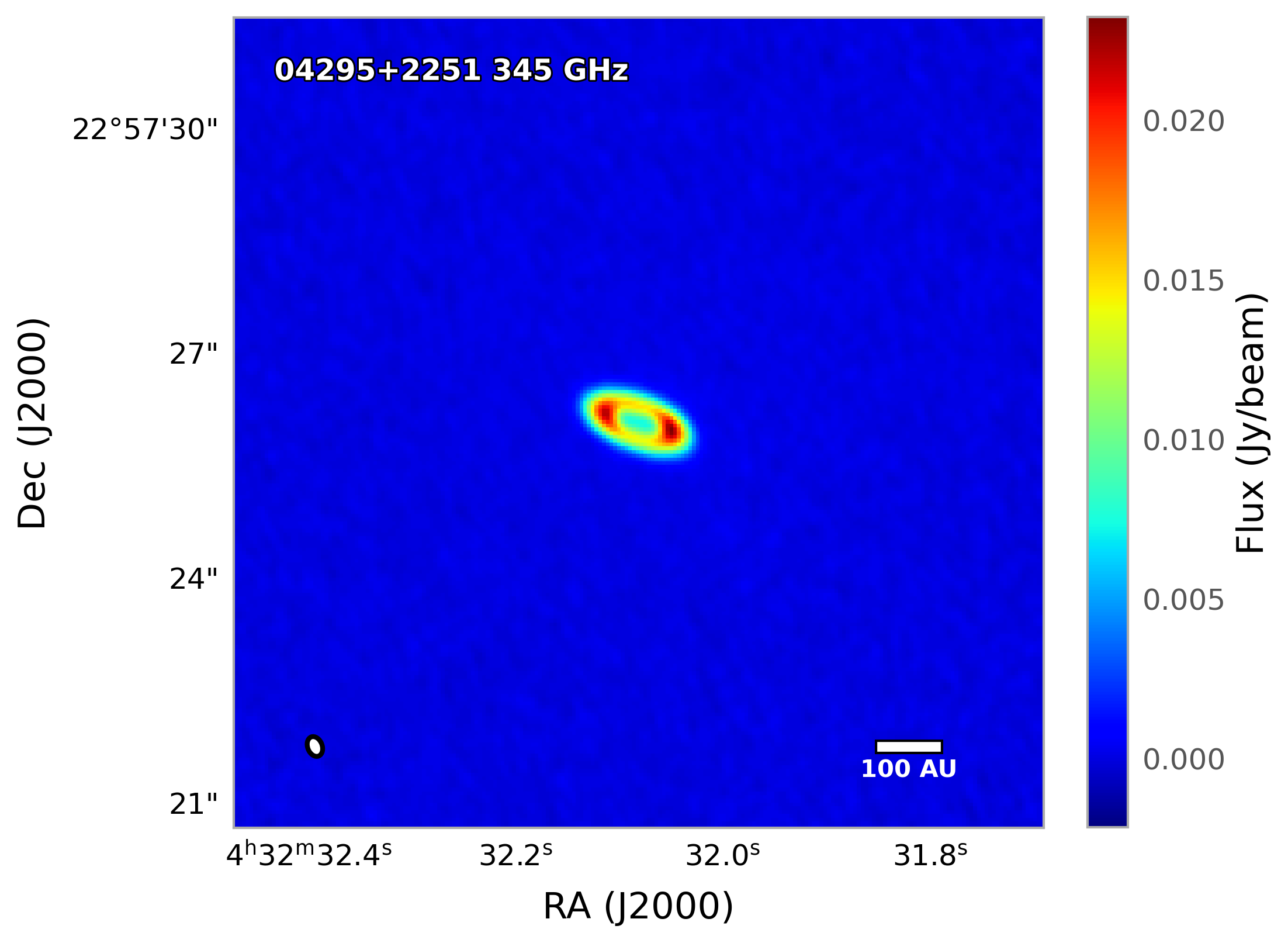}
\figsetplot{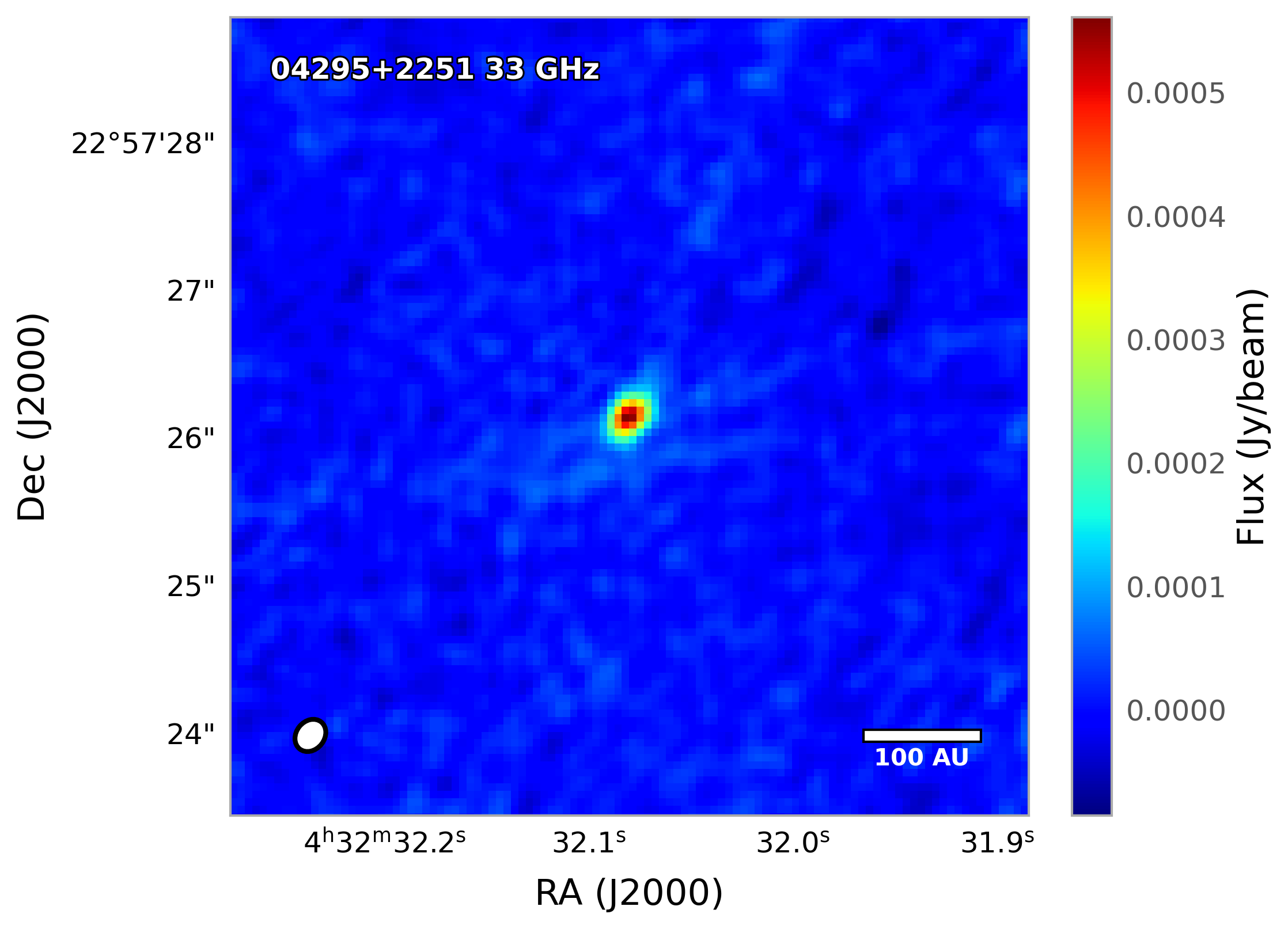}
\figsetgrpnote{ALMA 345~GHz and VLA 33~GHz continuum images of 04295+2251. The images are shown in units of Jy~beam$^{-1}$ in celestial coordinates (R.A./Decl.; J2000). A synthesized beam is shown in white in the lower left of each panel, and a 100~AU scale bar is shown in white in the lower right. Both images use a linear intensity scale.}
\figsetgrpend

\figsetgrpstart
\figsetgrpnum{A1.17}
\figsetgrptitle{04302+2247}
\figsetplot{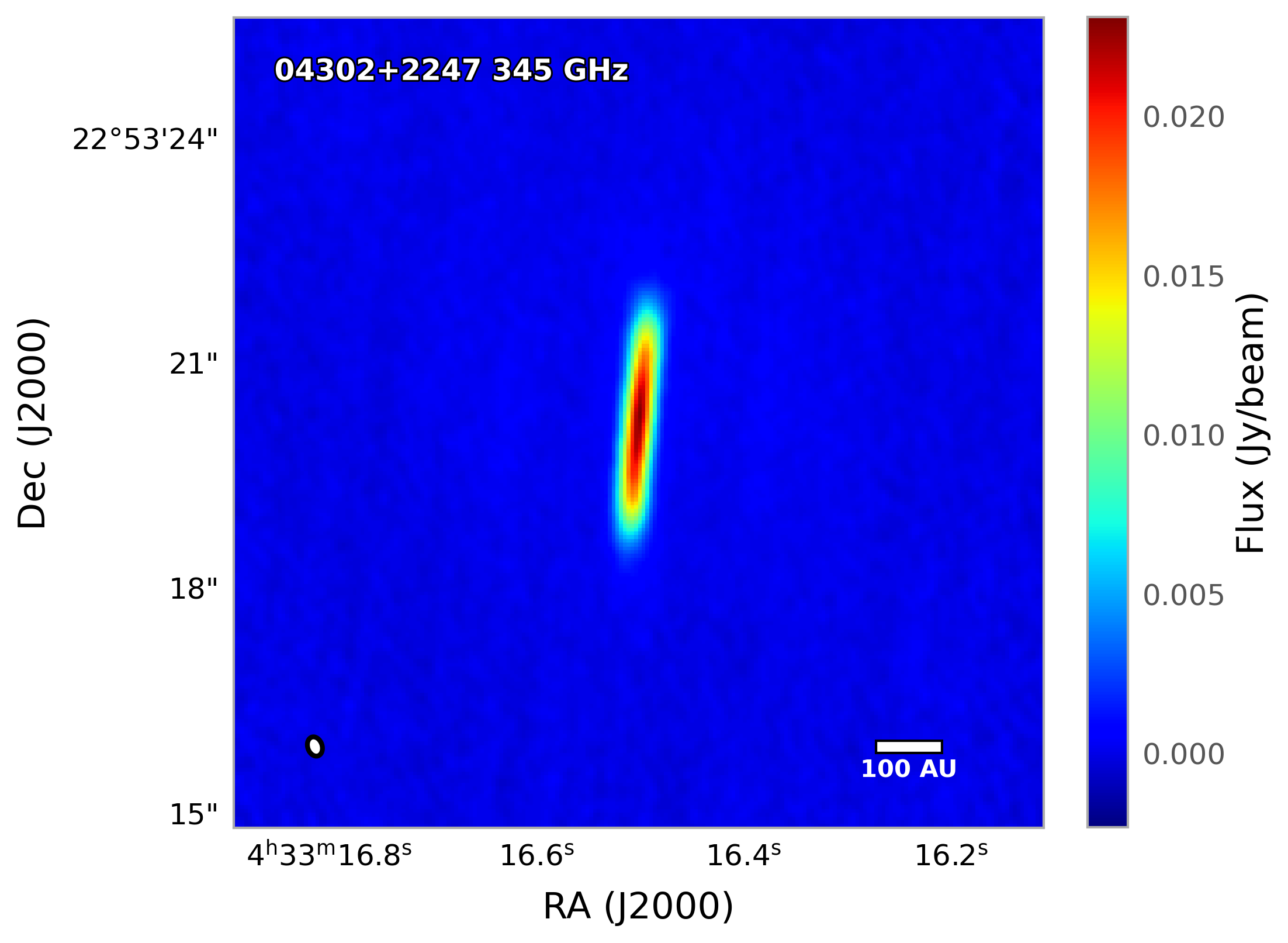}
\figsetplot{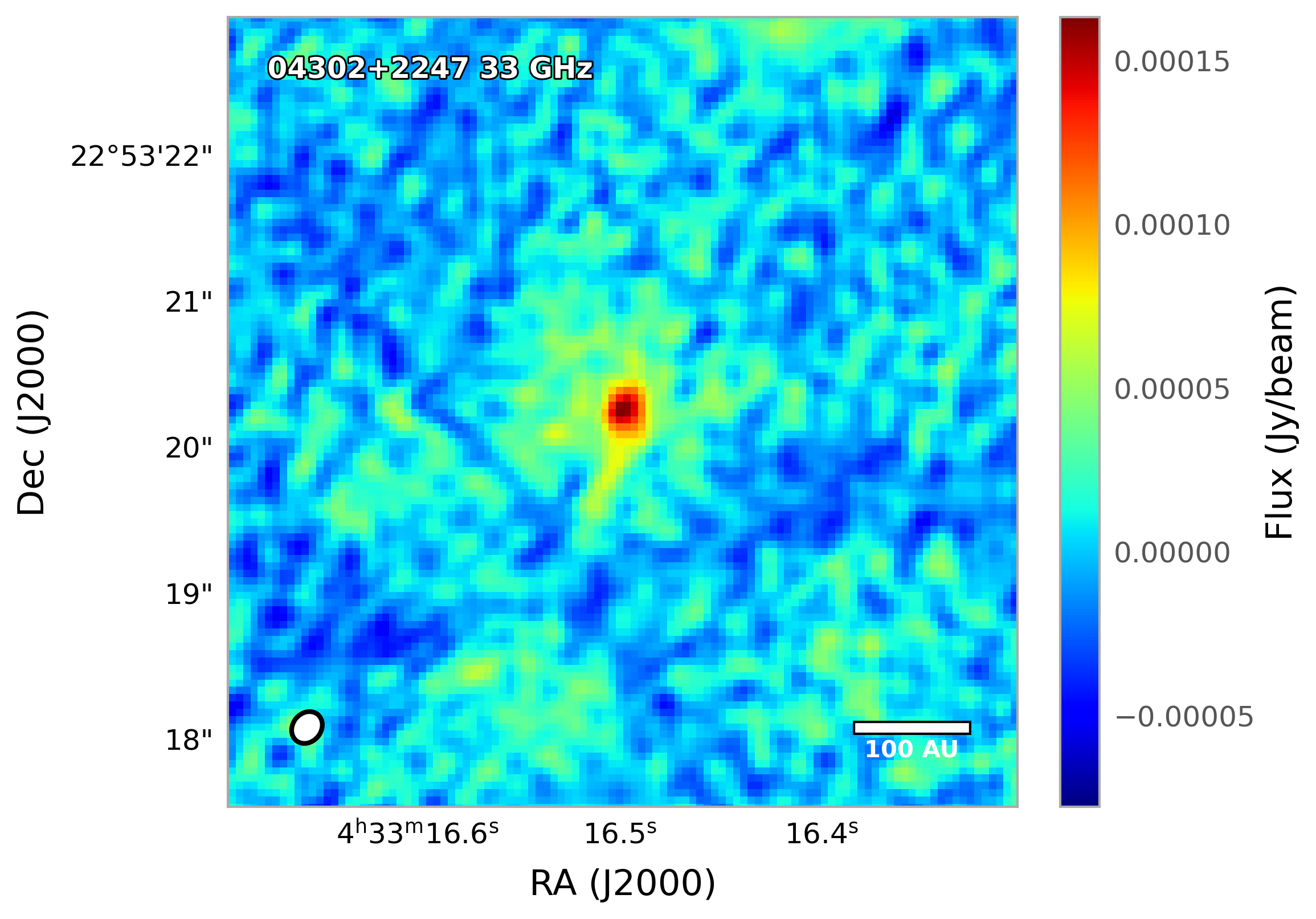}
\figsetgrpnote{ALMA 345~GHz and VLA 33~GHz continuum images of 04302+2247. The images are shown in units of Jy~beam$^{-1}$ in celestial coordinates (R.A./Decl.; J2000). A synthesized beam is shown in white in the lower left of each panel, and a 100~AU scale bar is shown in white in the lower right. Both images use a linear intensity scale.}
\figsetgrpend

\figsetgrpstart
\figsetgrpnum{A1.18}
\figsetgrptitle{04325+2402A}
\figsetplot{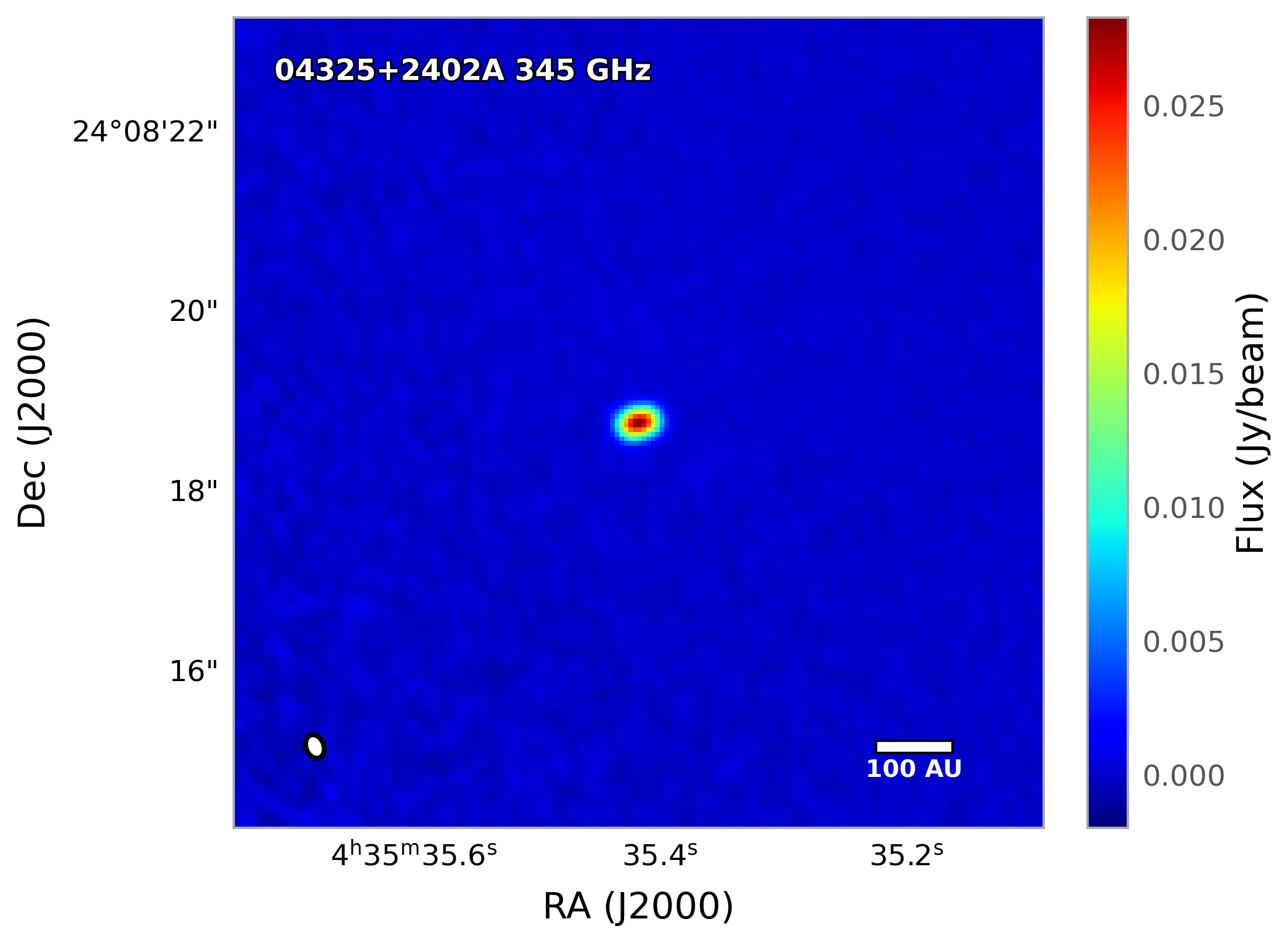}
\figsetplot{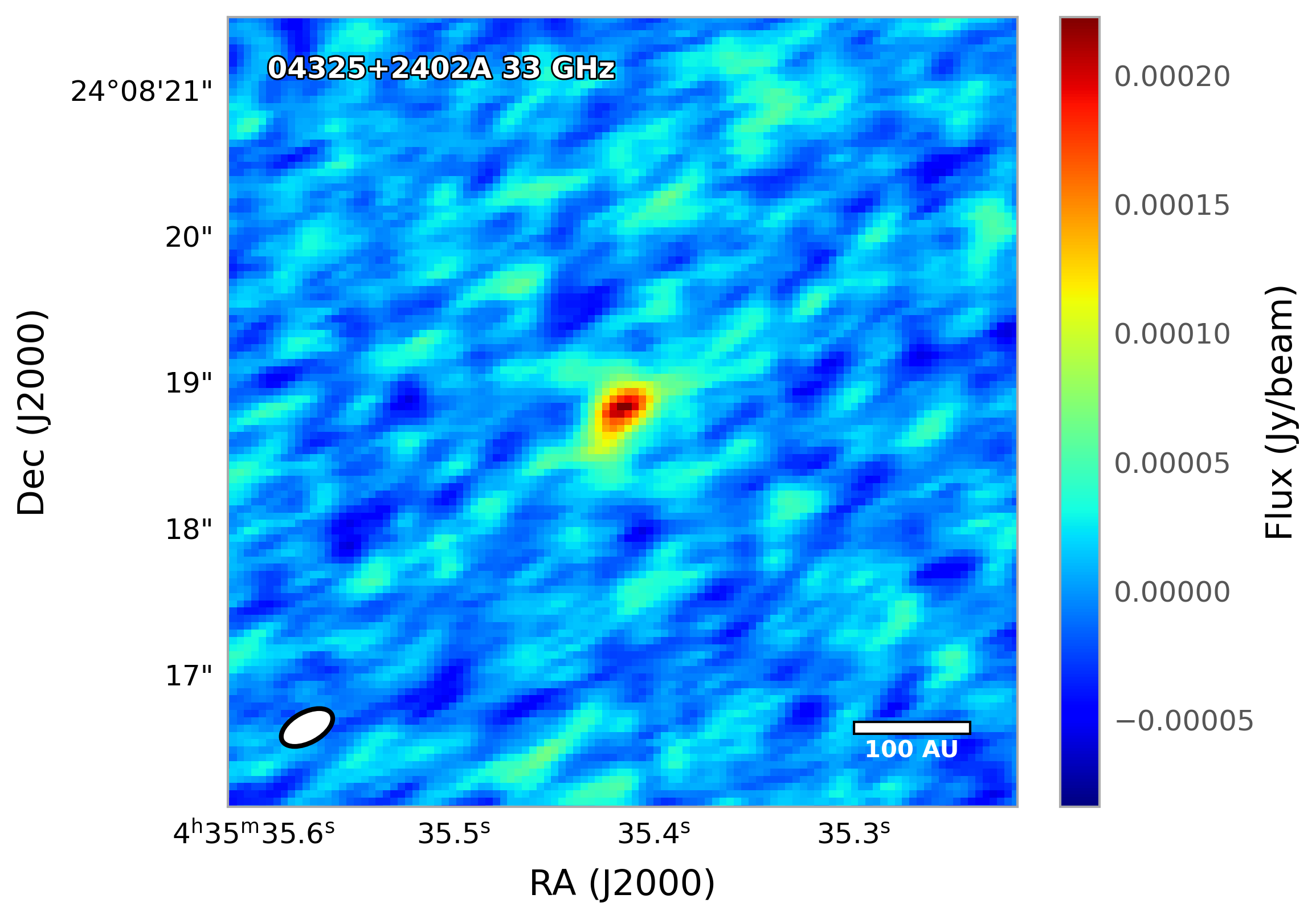}
\figsetgrpnote{ALMA 345~GHz and VLA 33~GHz continuum images of 04325+2402A. The images are shown in units of Jy~beam$^{-1}$ in celestial coordinates (R.A./Decl.; J2000). A synthesized beam is shown in white in the lower left of each panel, and a 100~AU scale bar is shown in white in the lower right. Both images use a linear intensity scale.}
\figsetgrpend

\figsetgrpstart
\figsetgrpnum{A1.19}
\figsetgrptitle{04325+2402B}
\figsetplot{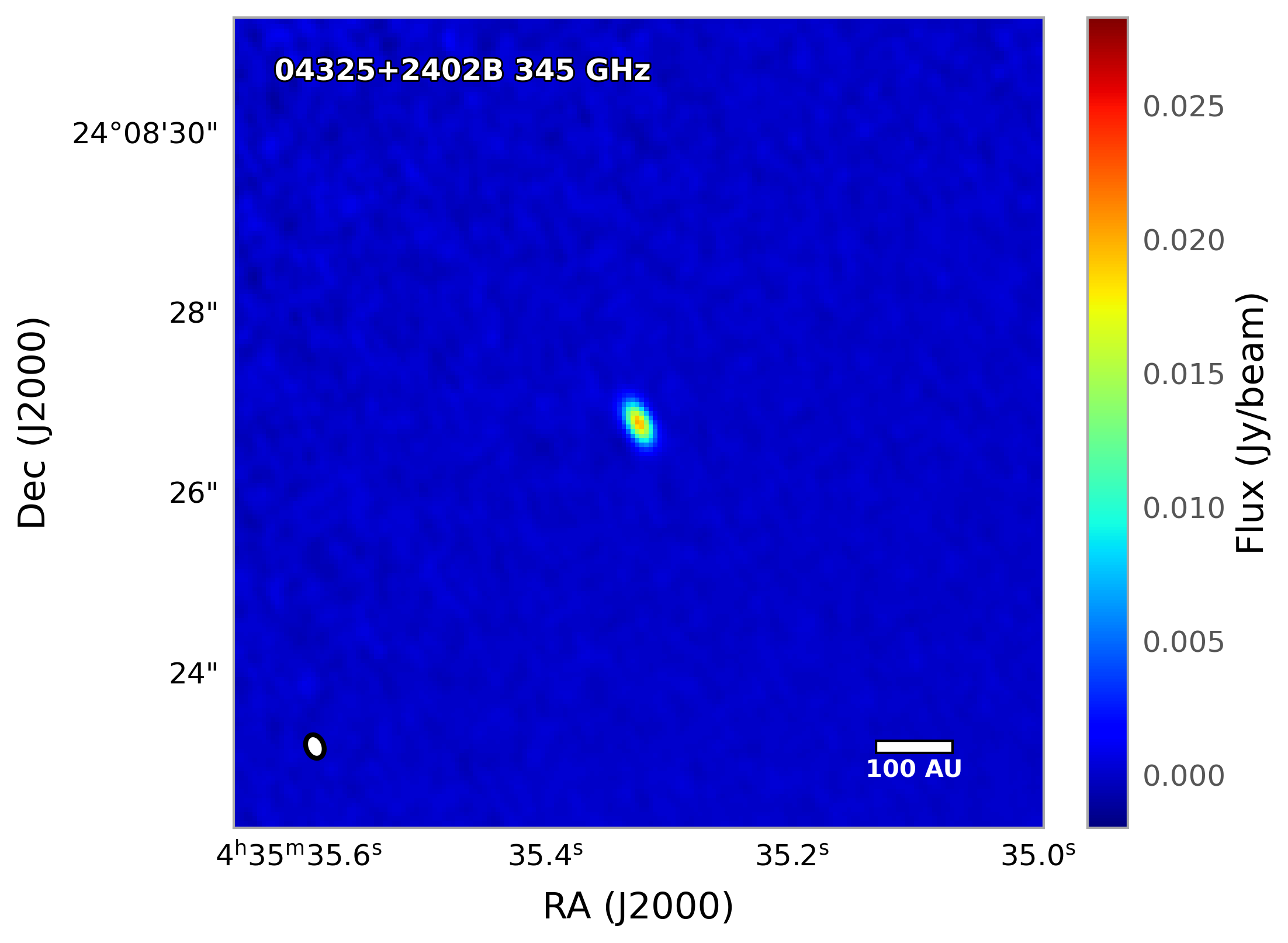}
\figsetplot{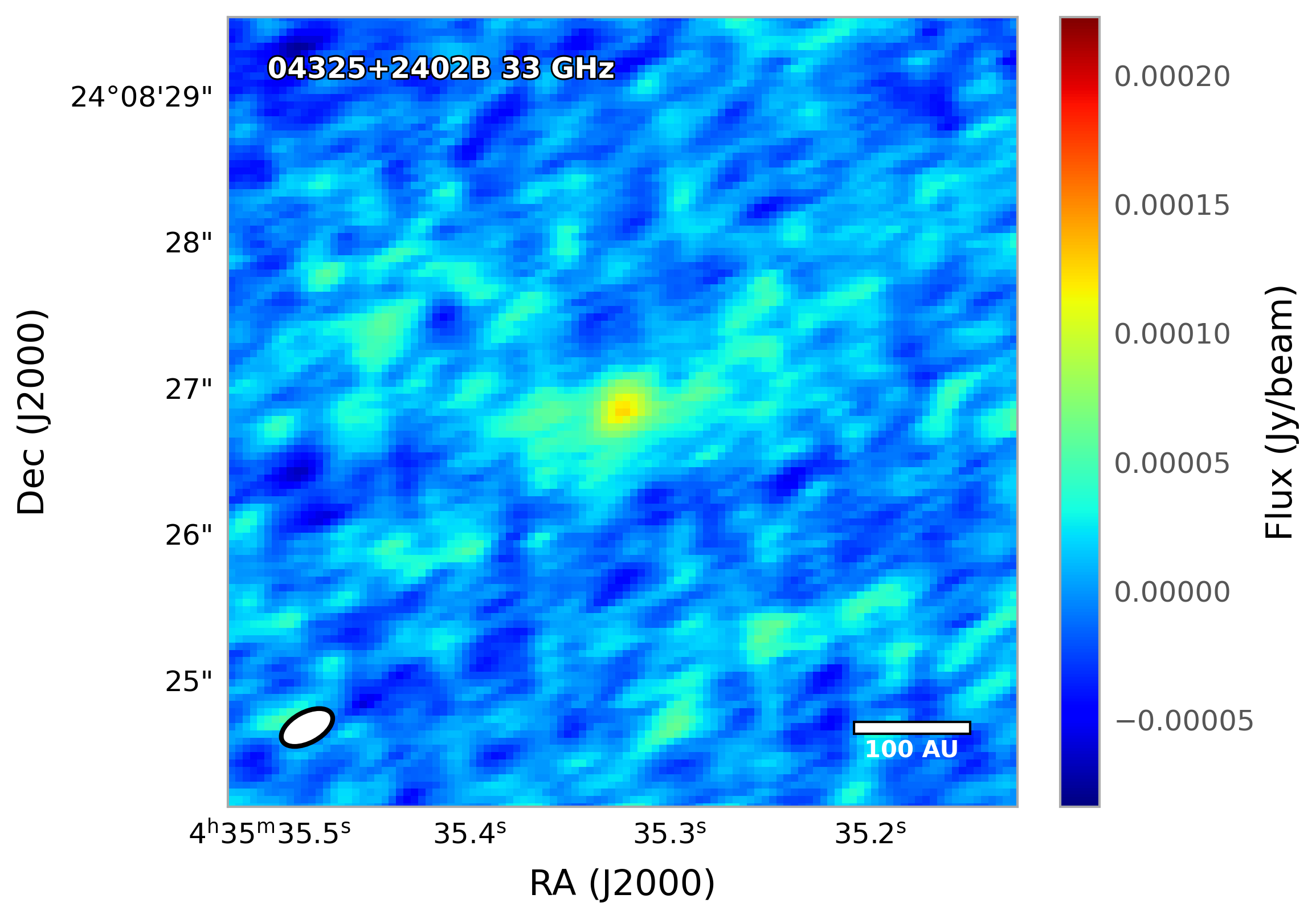}
\figsetgrpnote{ALMA 345~GHz and VLA 33~GHz continuum images of 04325+2402B. The images are shown in units of Jy~beam$^{-1}$ in celestial coordinates (R.A./Decl.; J2000). A synthesized beam is shown in white in the lower left of each panel, and a 100~AU scale bar is shown in white in the lower right. Both images use a linear intensity scale.}
\figsetgrpend

\figsetgrpstart
\figsetgrpnum{A1.20}
\figsetgrptitle{04361+2547}
\figsetplot{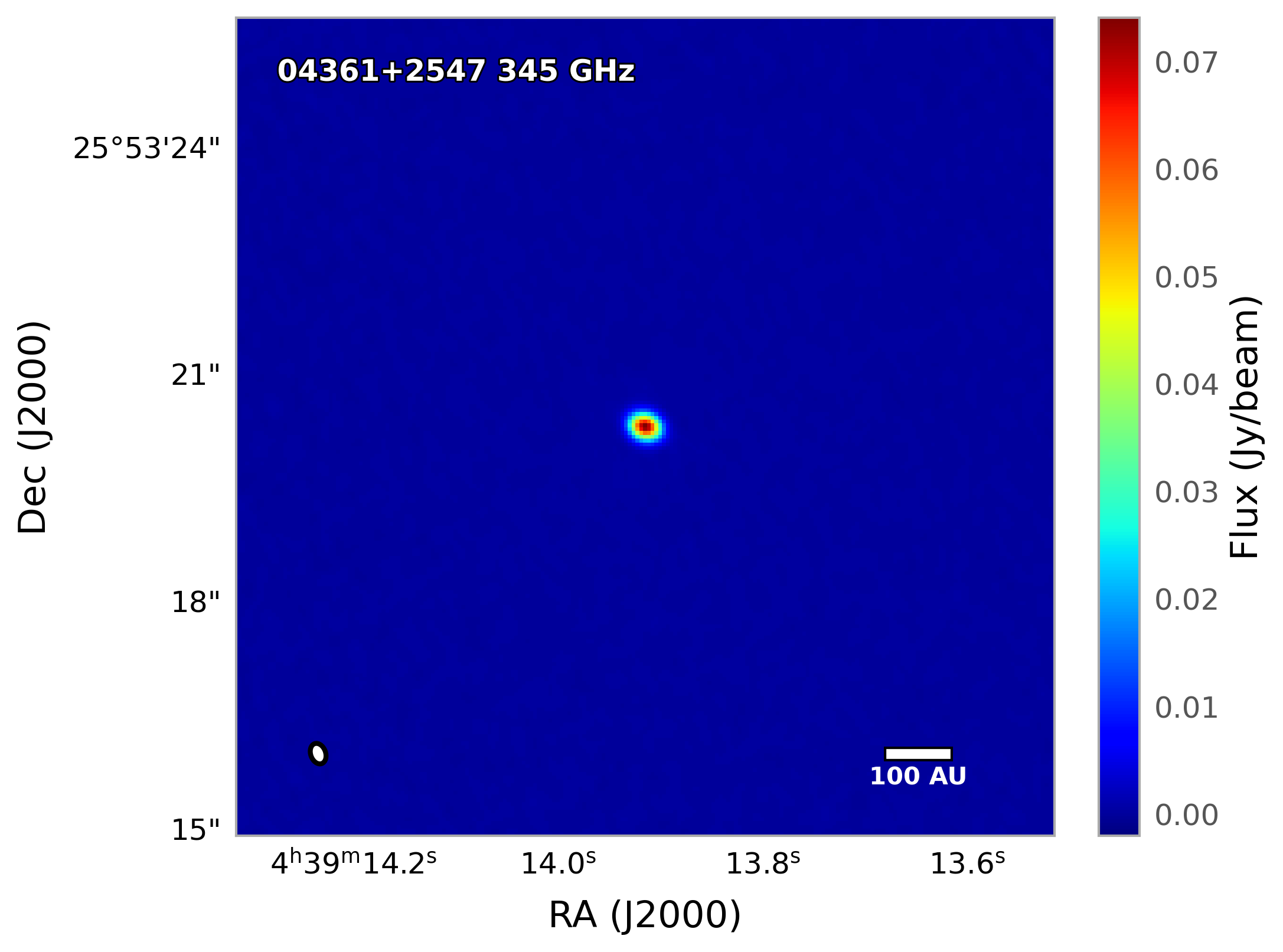}
\figsetplot{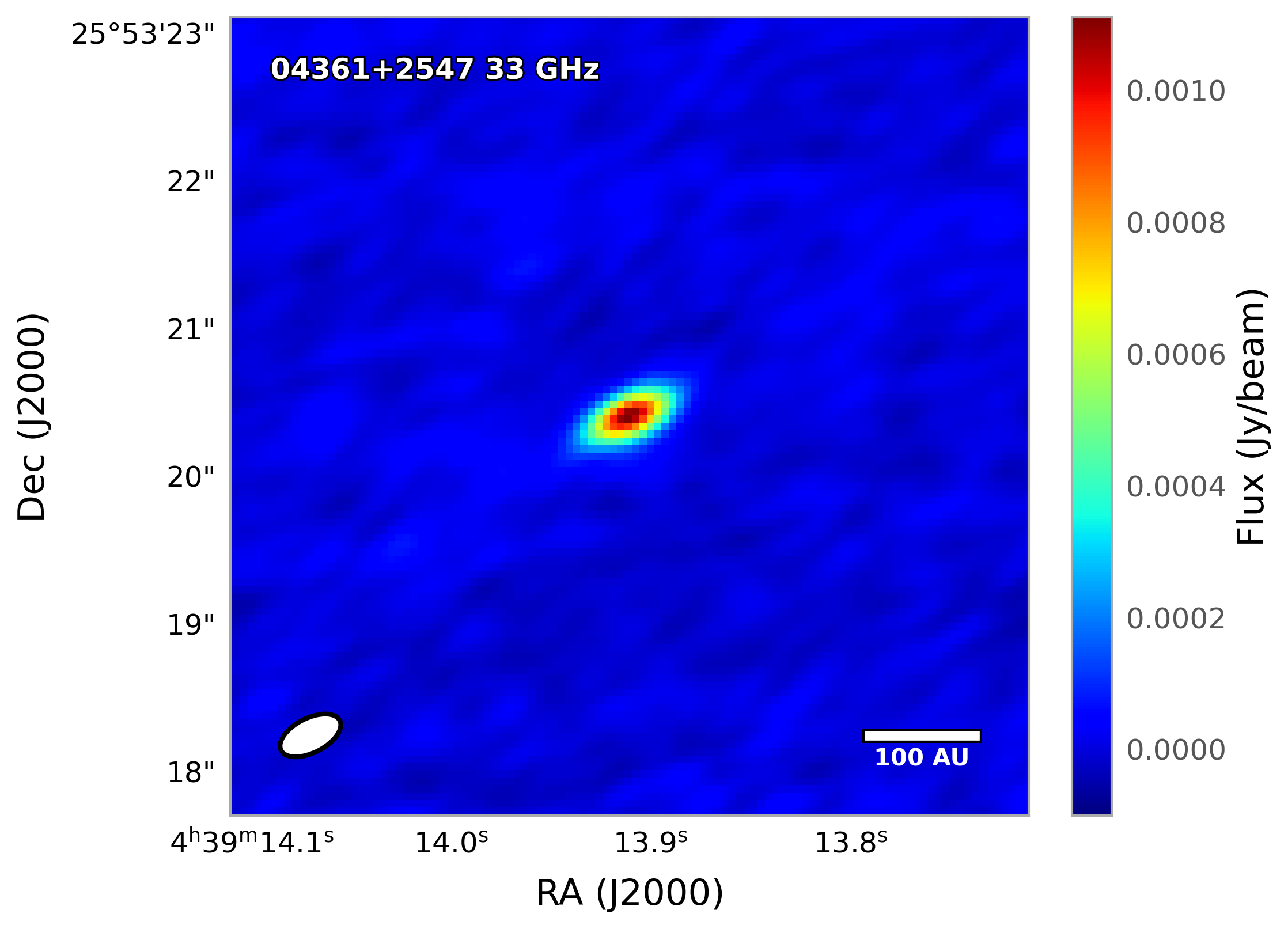}
\figsetgrpnote{ALMA 345~GHz and VLA 33~GHz continuum images of 04361+2547. The images are shown in units of Jy~beam$^{-1}$ in celestial coordinates (R.A./Decl.; J2000). A synthesized beam is shown in white in the lower left of each panel, and a 100~AU scale bar is shown in white in the lower right. Both images use a linear intensity scale.}
\figsetgrpend

\figsetgrpstart
\figsetgrpnum{A1.21}
\figsetgrptitle{04365+2535}
\figsetplot{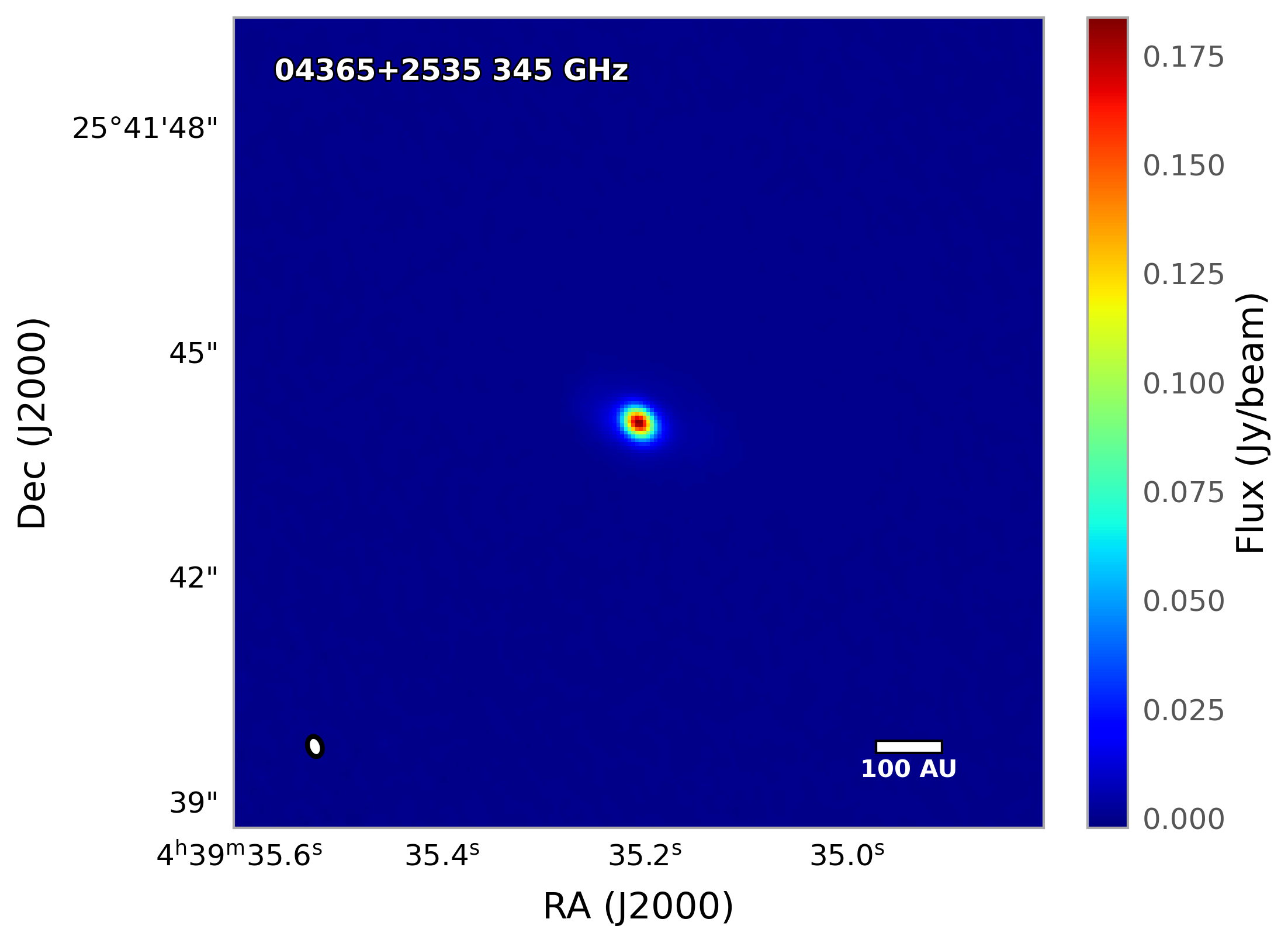}
\figsetplot{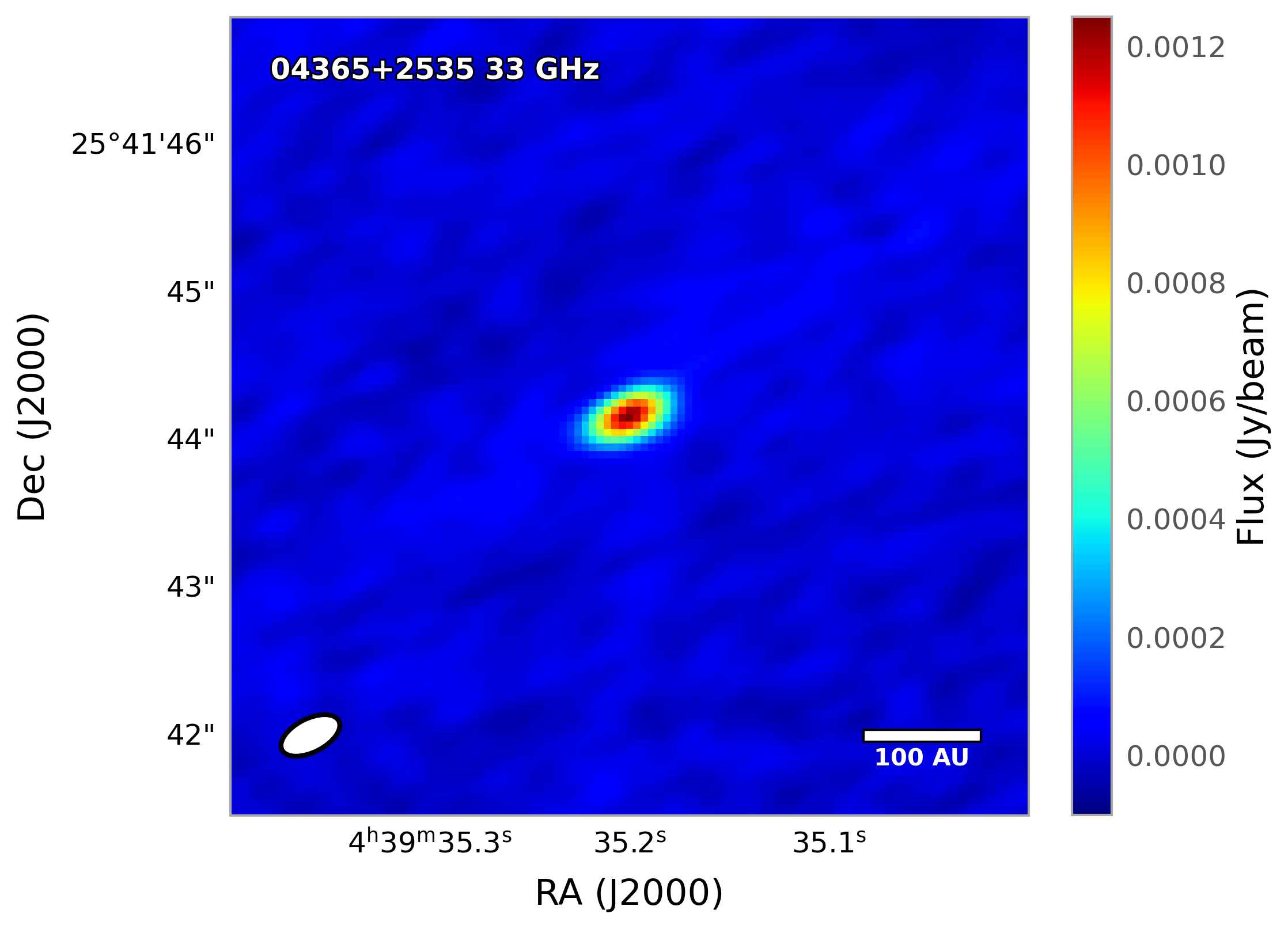}
\figsetgrpnote{ALMA 345~GHz and VLA 33~GHz continuum images of 04365+2535. The images are shown in units of Jy~beam$^{-1}$ in celestial coordinates (R.A./Decl.; J2000). A synthesized beam is shown in white in the lower left of each panel, and a 100~AU scale bar is shown in white in the lower right. Both images use a linear intensity scale.}
\figsetgrpend

\figsetgrpstart
\figsetgrpnum{A1.22}
\figsetgrptitle{04368+2557}
\figsetplot{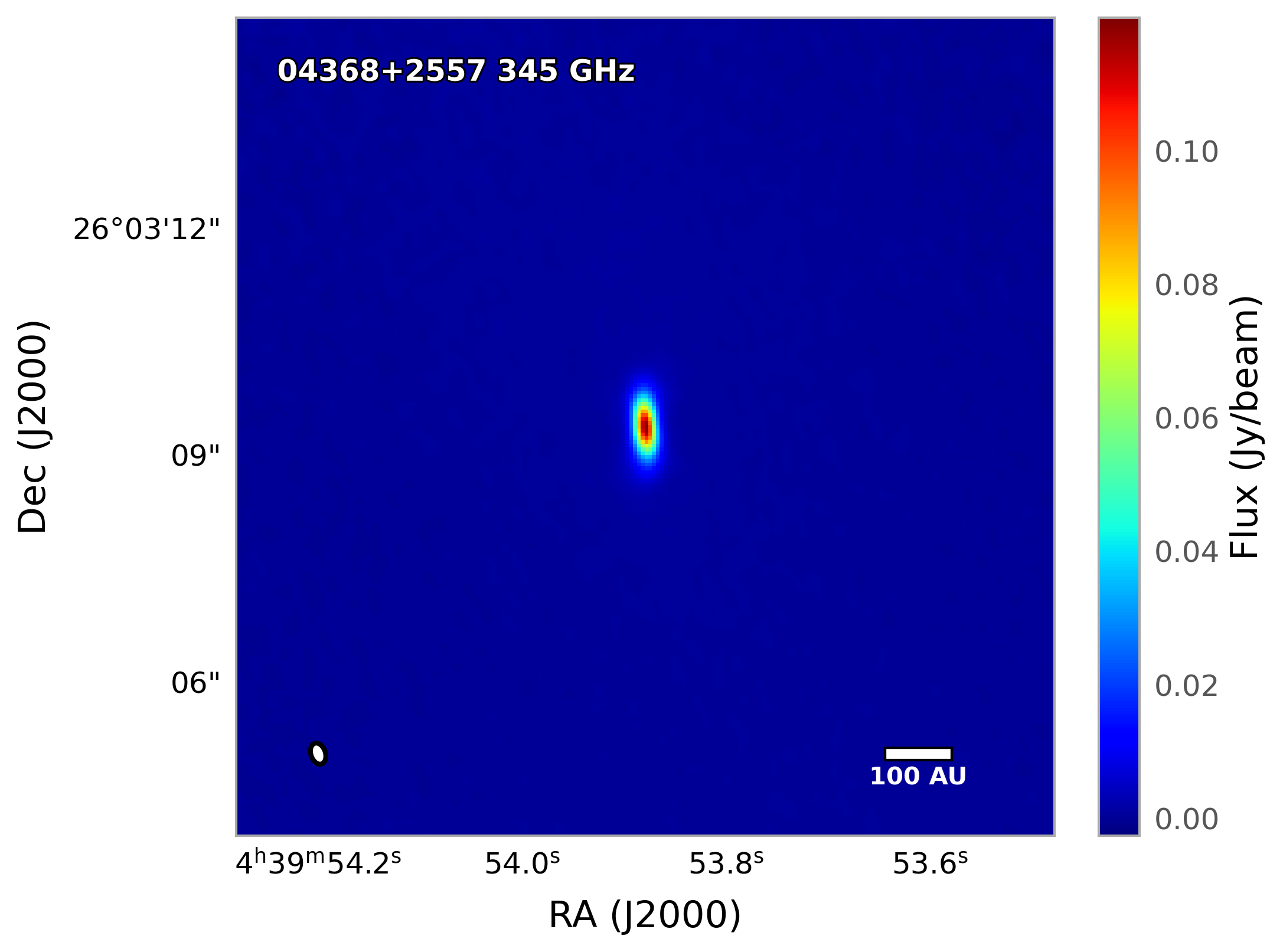}
\figsetplot{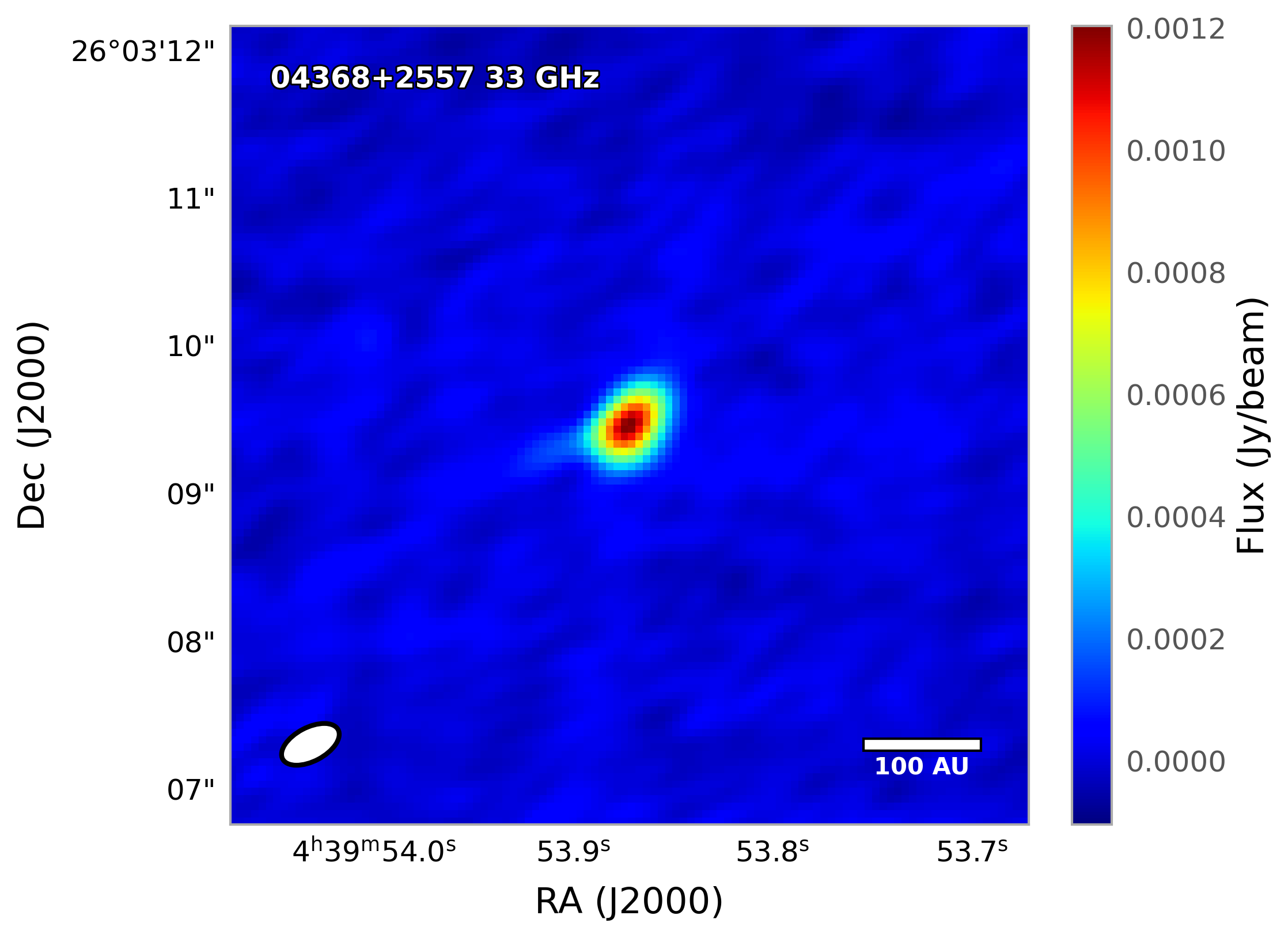}
\figsetgrpnote{ALMA 345~GHz and VLA 33~GHz continuum images of 04368+2557. The images are shown in units of Jy~beam$^{-1}$ in celestial coordinates (R.A./Decl.; J2000). A synthesized beam is shown in white in the lower left of each panel, and a 100~AU scale bar is shown in white in the lower right. Both images use a linear intensity scale.}
\figsetgrpend

\figsetgrpstart
\figsetgrpnum{A1.23}
\figsetgrptitle{04381+2540AB}
\figsetplot{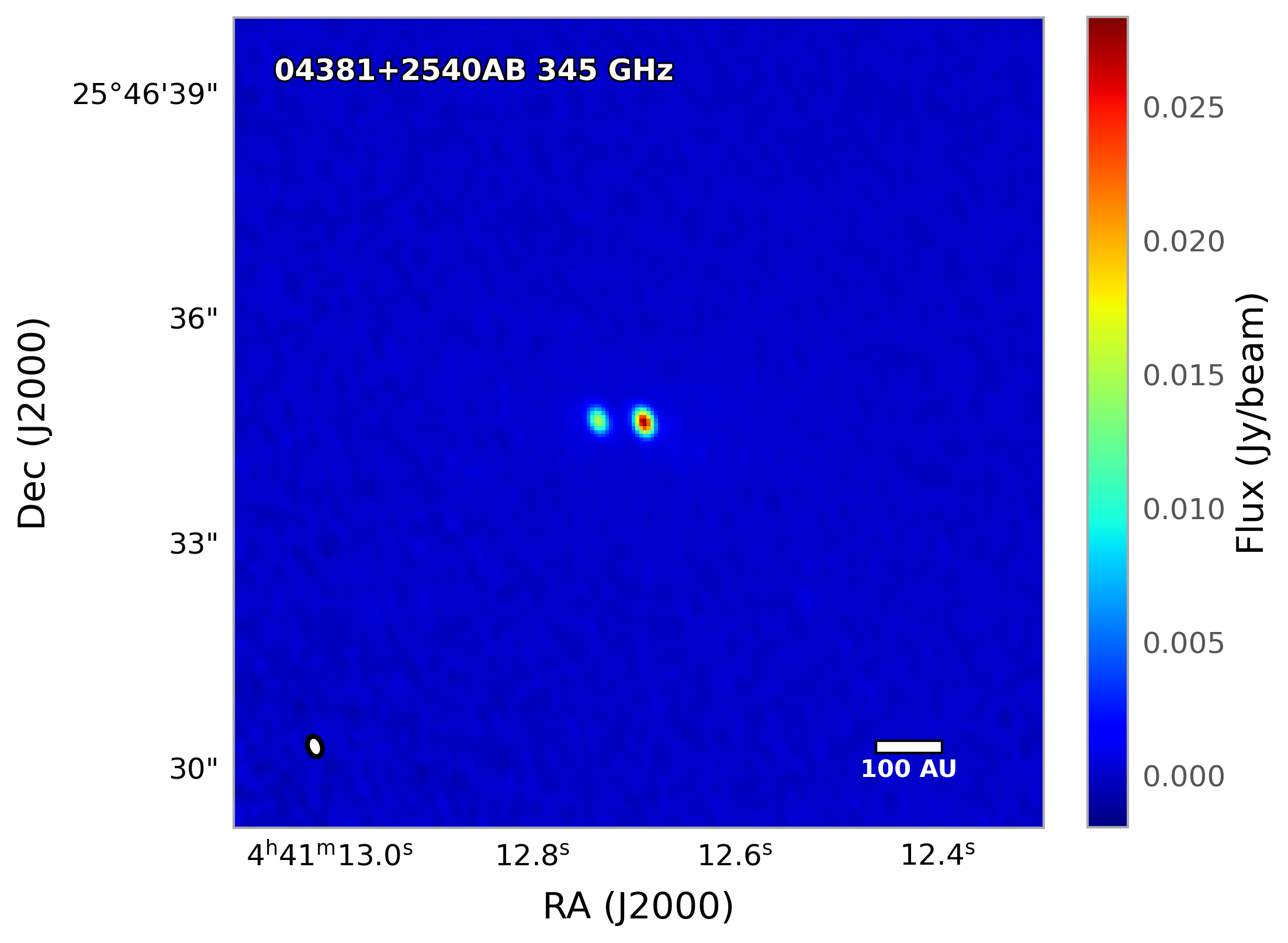}
\figsetplot{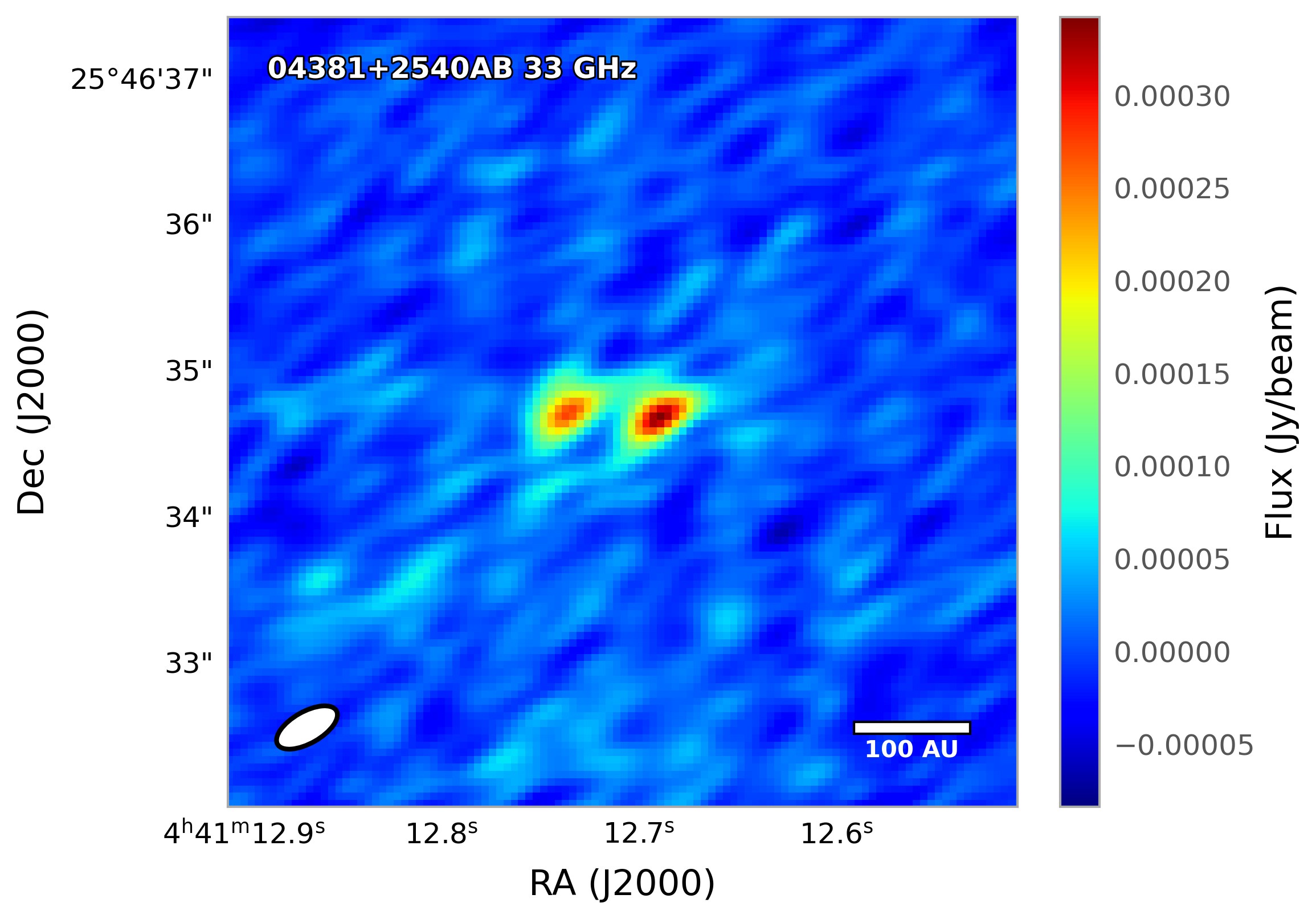}
\figsetgrpnote{ALMA 345~GHz and VLA 33~GHz continuum images of 04381+2540AB. The images are shown in units of Jy~beam$^{-1}$ in celestial coordinates (R.A./Decl.; J2000). A synthesized beam is shown in white in the lower left of each panel, and a 100~AU scale bar is shown in white in the lower right. Both images use a linear intensity scale.}
\figsetgrpend

\figsetgrpstart
\figsetgrpnum{A1.24}
\figsetgrptitle{04385+2550}
\figsetplot{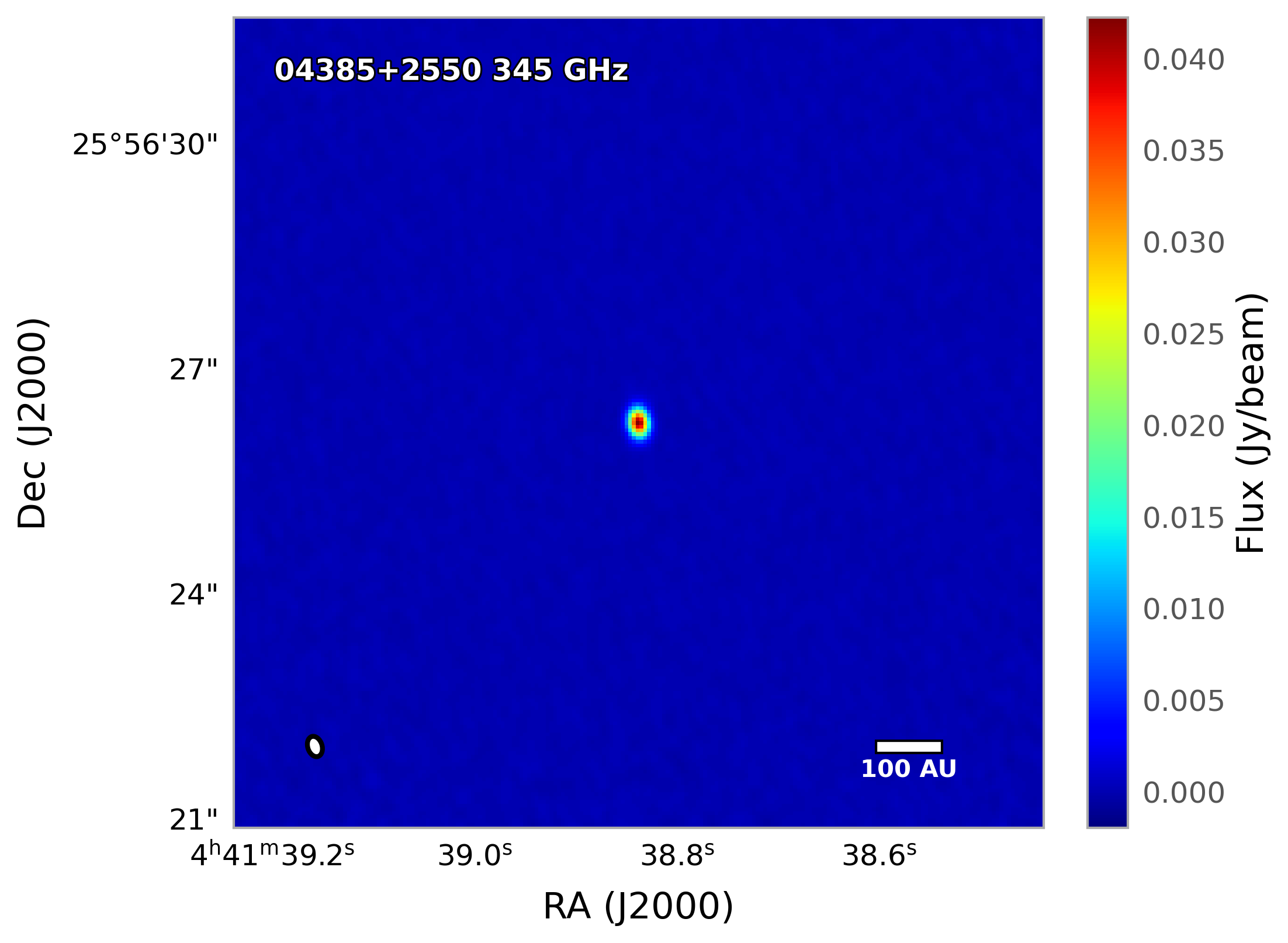}
\figsetplot{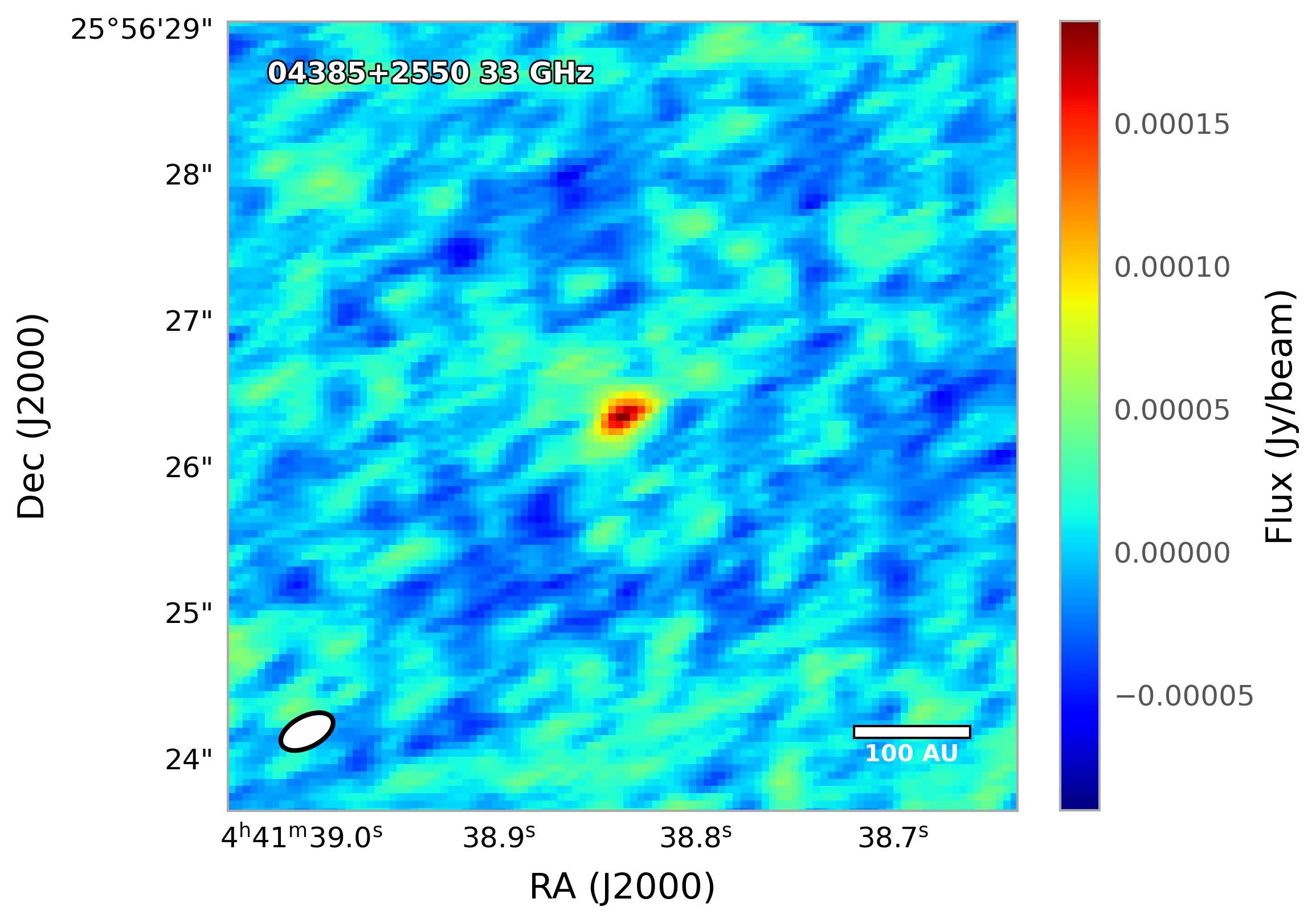}
\figsetgrpnote{ALMA 345~GHz and VLA 33~GHz continuum images of 04385+2550. The images are shown in units of Jy~beam$^{-1}$ in celestial coordinates (R.A./Decl.; J2000). A synthesized beam is shown in white in the lower left of each panel, and a 100~AU scale bar is shown in white in the lower right. Both images use a linear intensity scale.}
\figsetgrpend

\figsetgrpstart
\figsetgrpnum{A1.25}
\figsetgrptitle{04489+3042AB}
\figsetplot{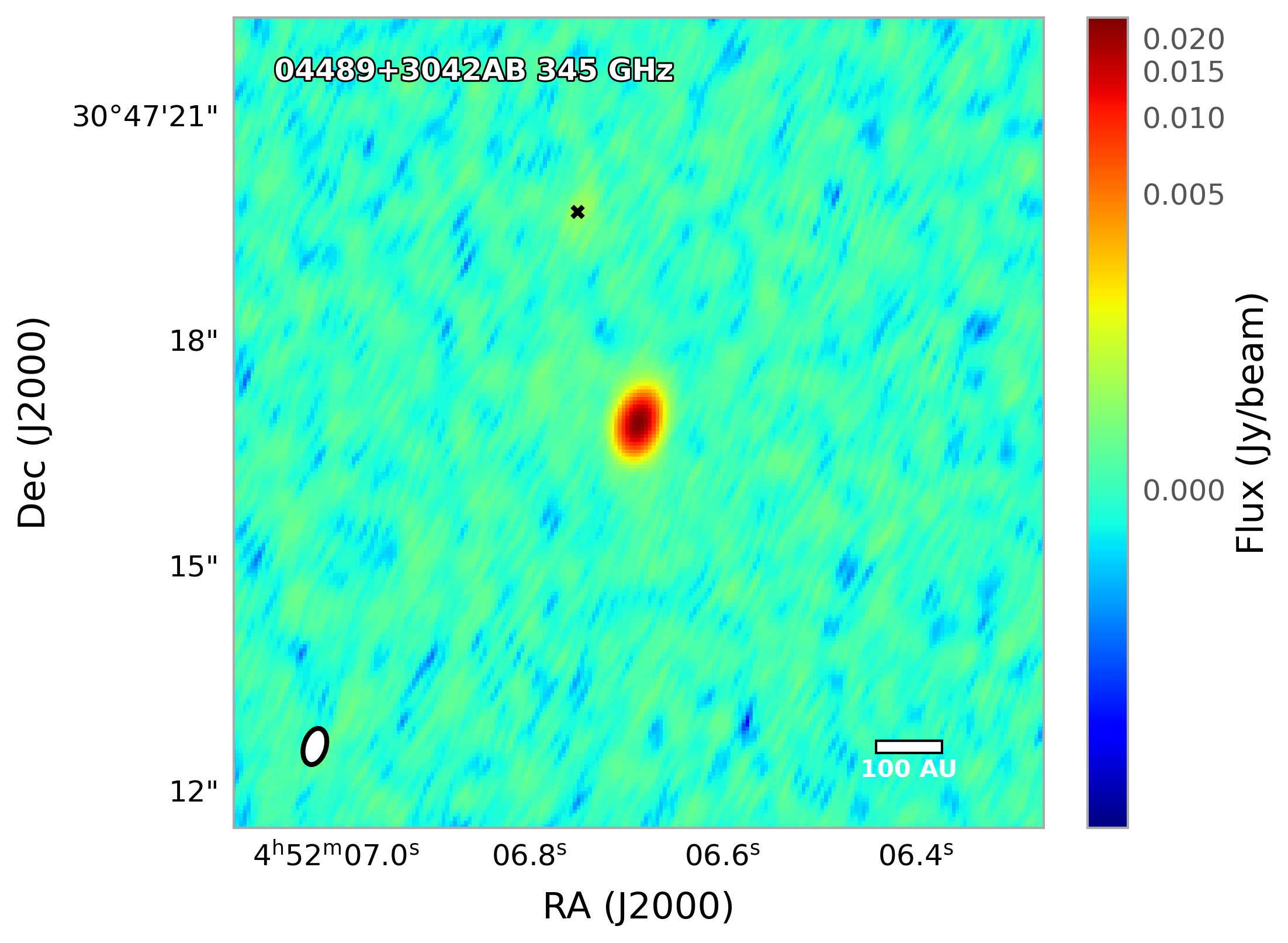}
\figsetplot{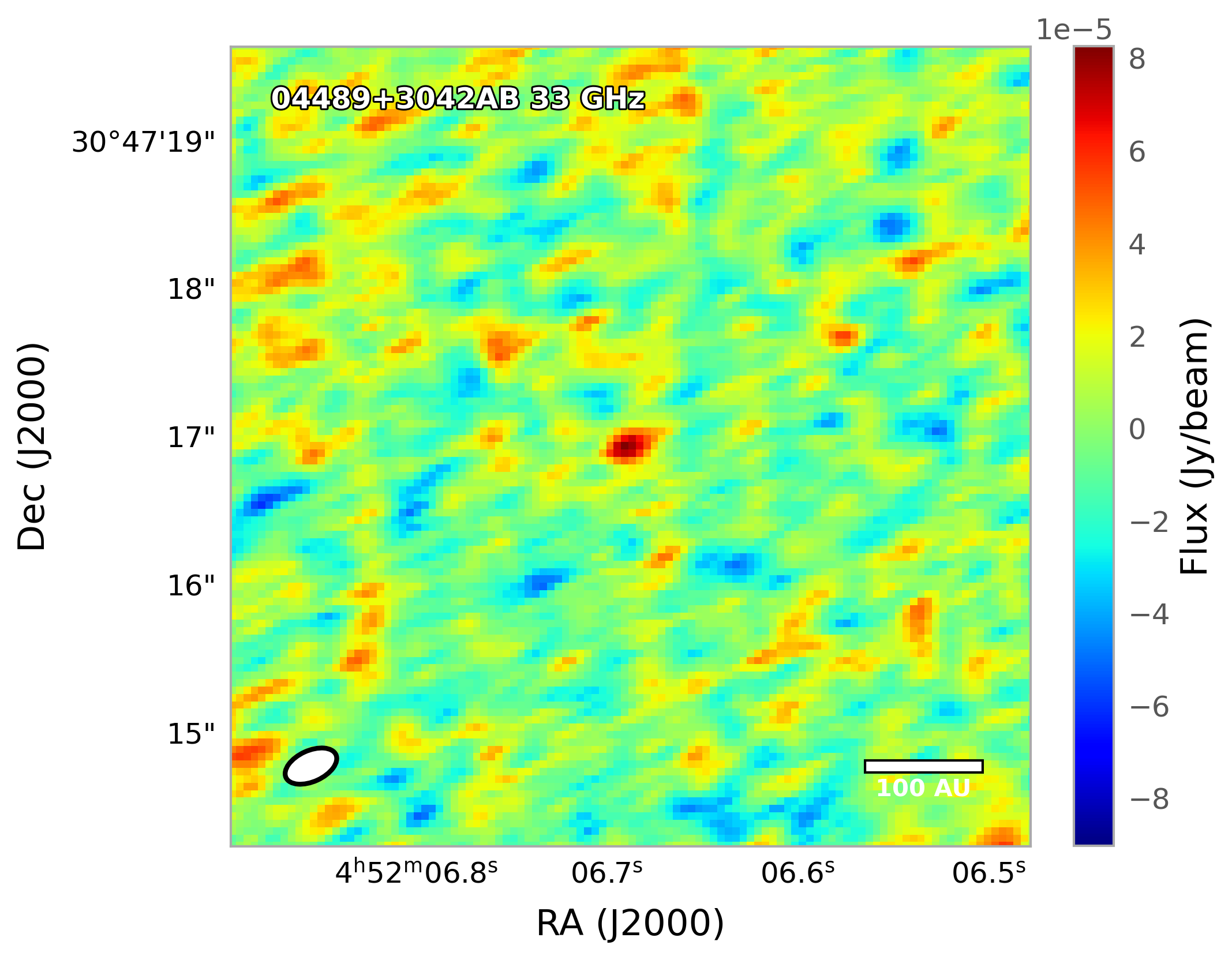}
\figsetgrpnote{ALMA 345~GHz and VLA 33~GHz continuum images of 04489+3042AB. The images are shown in units of Jy~beam$^{-1}$ in celestial coordinates (R.A./Decl.; J2000). A synthesized beam is shown in white in the lower left of each panel, and a 100~AU scale bar is shown in white in the lower right. The ALMA image is displayed with a logarithmic intensity scale to highlight faint emission from the companion 04489+3042B, whose position is marked with a small black X; the VLA image uses a linear intensity scale.}
\figsetgrpend

\figsetgrpstart
\figsetgrpnum{A1.26}
\figsetgrptitle{DG~Tau~B}
\figsetplot{alma_DG_TauB.png}
\figsetplot{vla_DG_TauB-log.png}
\figsetgrpnote{ALMA 345~GHz and VLA 33~GHz continuum images of DG~Tau~B. The images are shown in units of Jy~beam$^{-1}$ in celestial coordinates (R.A./Decl.; J2000). A synthesized beam is shown in white in the lower left of each panel, and a 100~AU scale bar is shown in white in the lower right. The ALMA image uses a linear intensity scale, whereas the VLA image is displayed with a logarithmic intensity scale to highlight faint emission.}
\figsetgrpend

\figsetgrpstart
\figsetgrpnum{A1.27}
\figsetgrptitle{HH~30}
\figsetplot{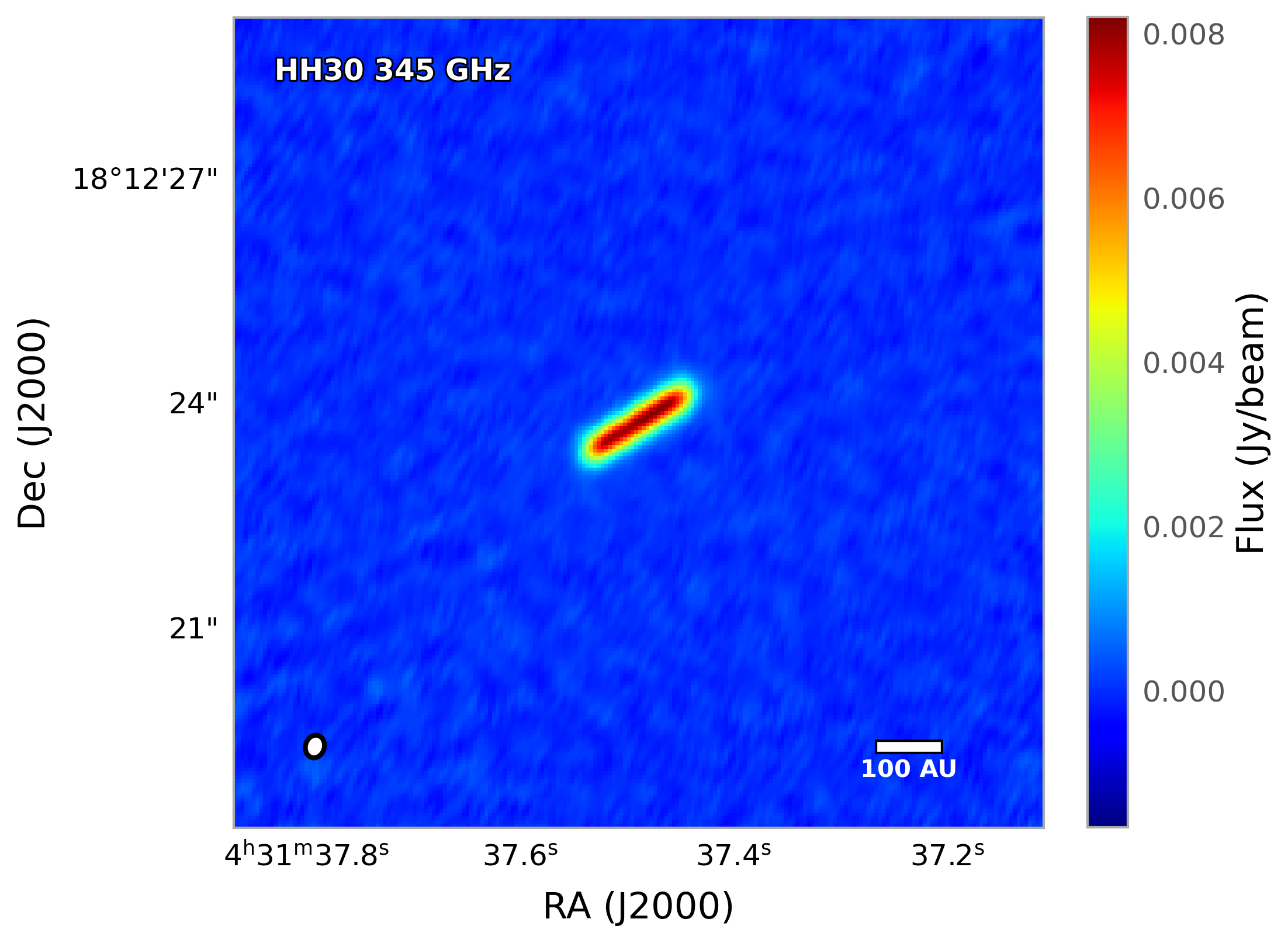}
\figsetplot{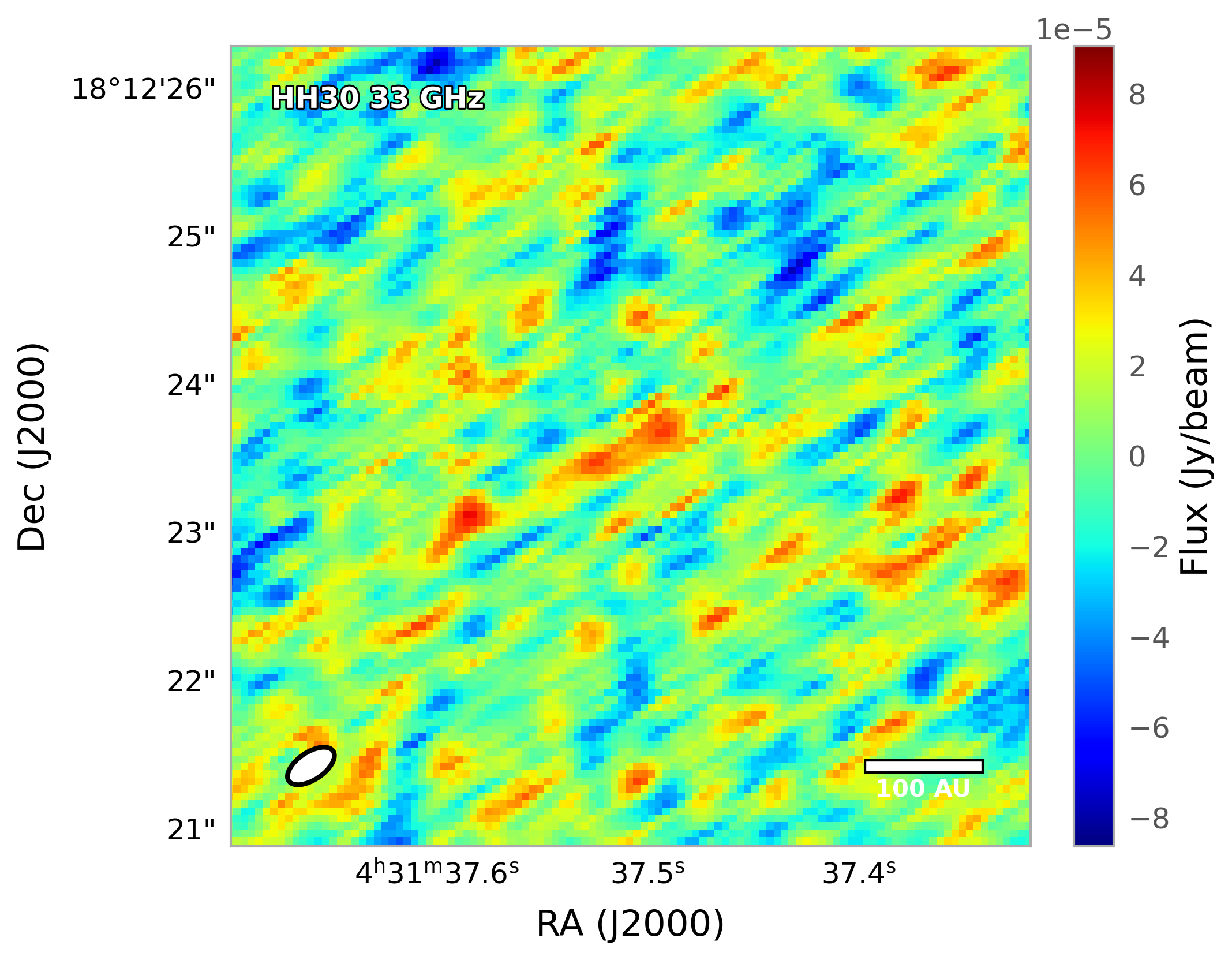}
\figsetgrpnote{ALMA 345~GHz and VLA 33~GHz continuum images of HH~30. The images are shown in units of Jy~beam$^{-1}$ in celestial coordinates (R.A./Decl.; J2000). A synthesized beam is shown in white in the lower left of each panel, and a 100~AU scale bar is shown in white in the lower right. Both images use a linear intensity scale.}
\figsetgrpend

\figsetgrpstart
\figsetgrpnum{A1.28}
\figsetgrptitle{IRAM~04191}
\figsetplot{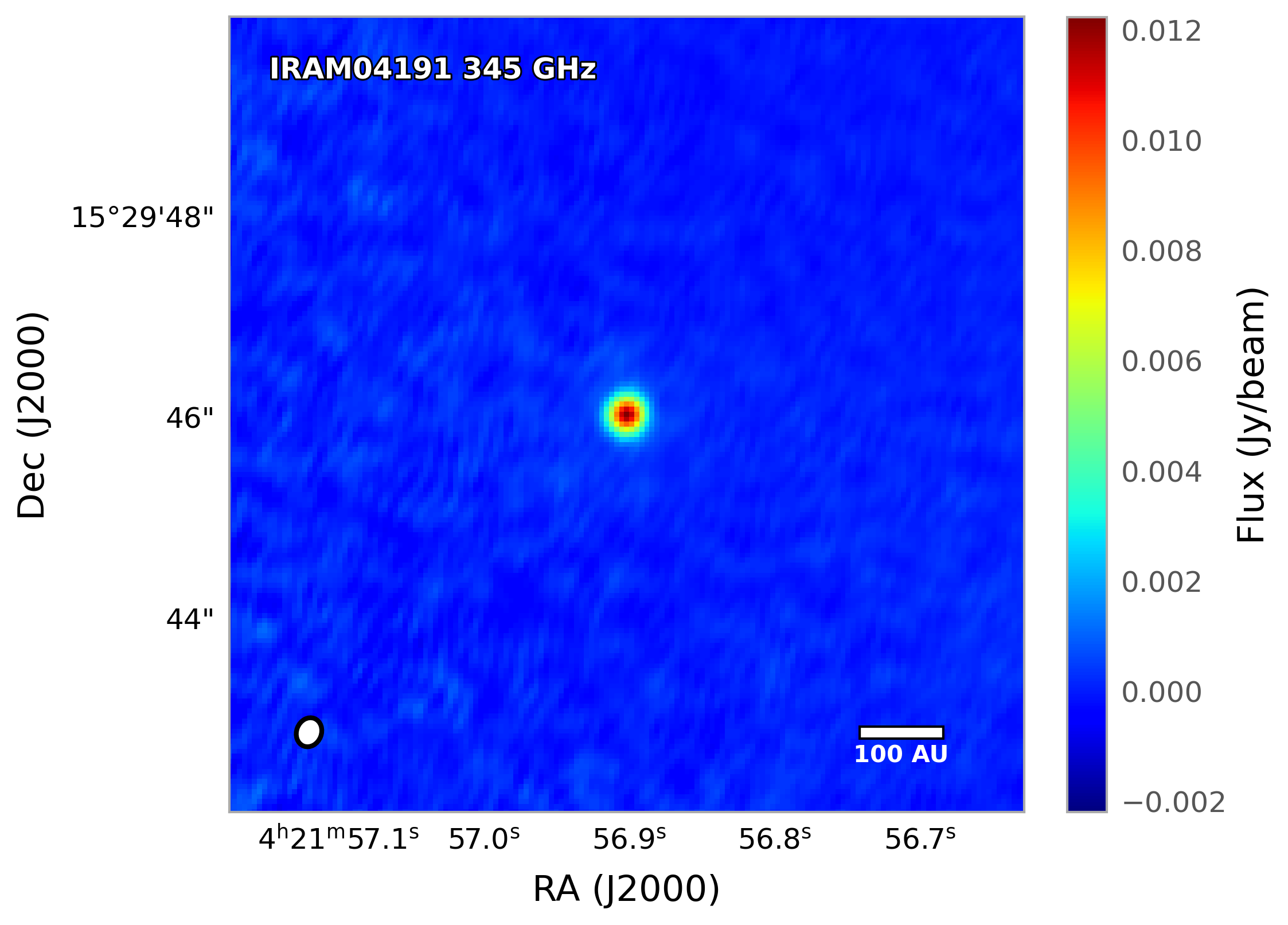}
\figsetplot{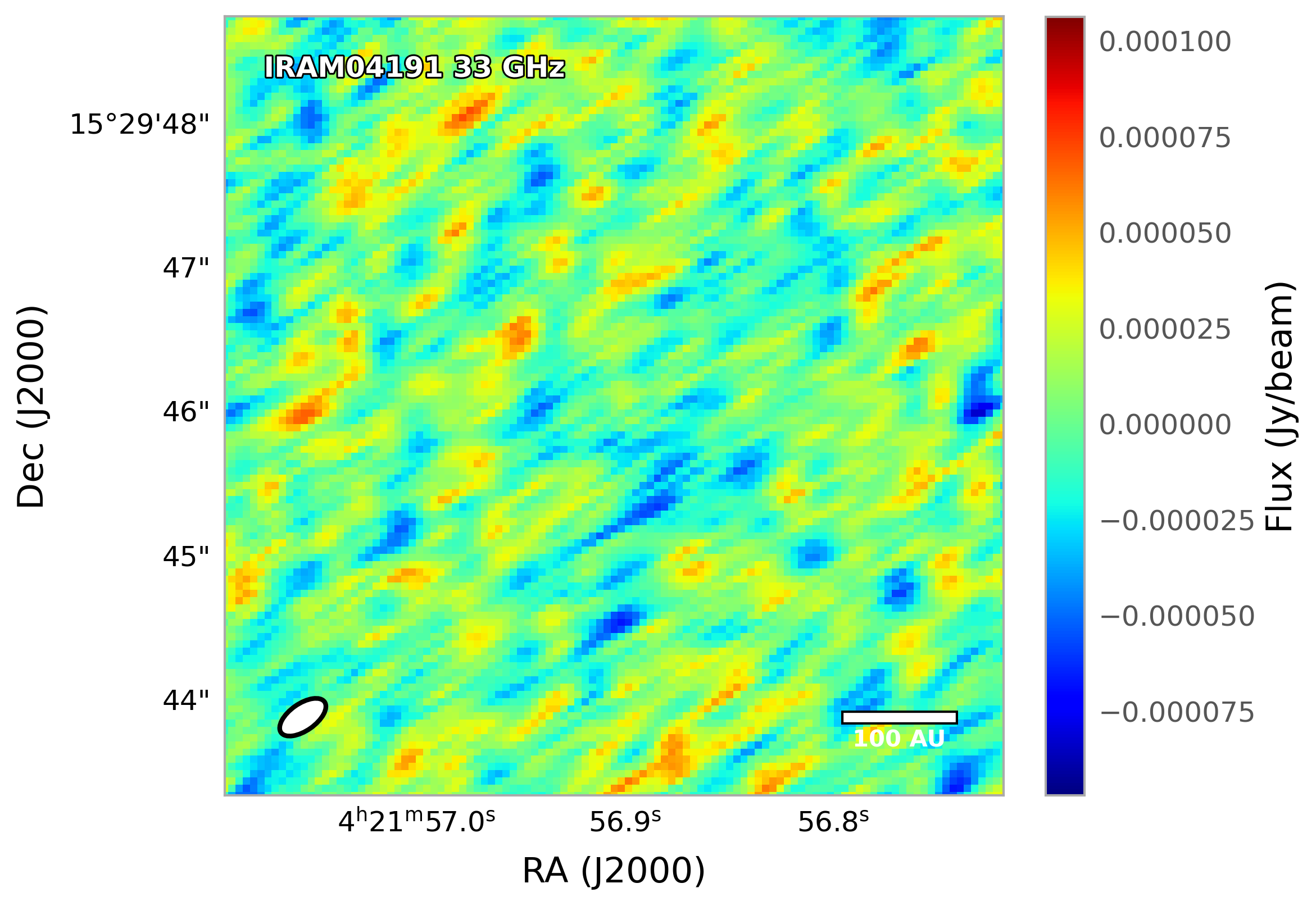}
\figsetgrpnote{ALMA 345~GHz and VLA 33~GHz continuum images of IRAM~04191. The images are shown in units of Jy~beam$^{-1}$ in celestial coordinates (R.A./Decl.; J2000). A synthesized beam is shown in white in the lower left of each panel, and a 100~AU scale bar is shown in white in the lower right. Both images use a linear intensity scale.}
\figsetgrpend

\figsetend

\begin{figure}
    \centering
    \includegraphics[width=0.48\linewidth]{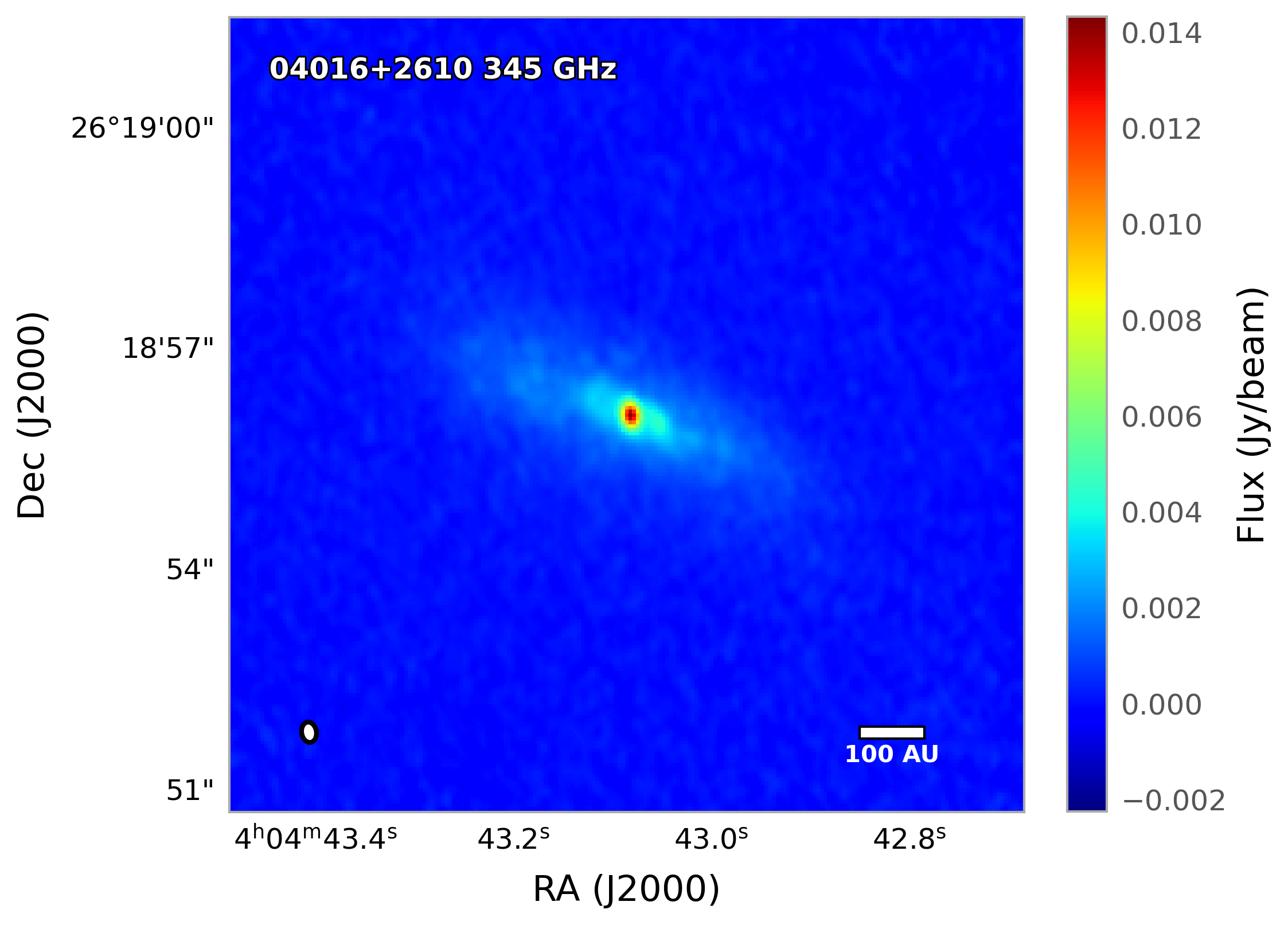}
    \includegraphics[width=0.48\linewidth]{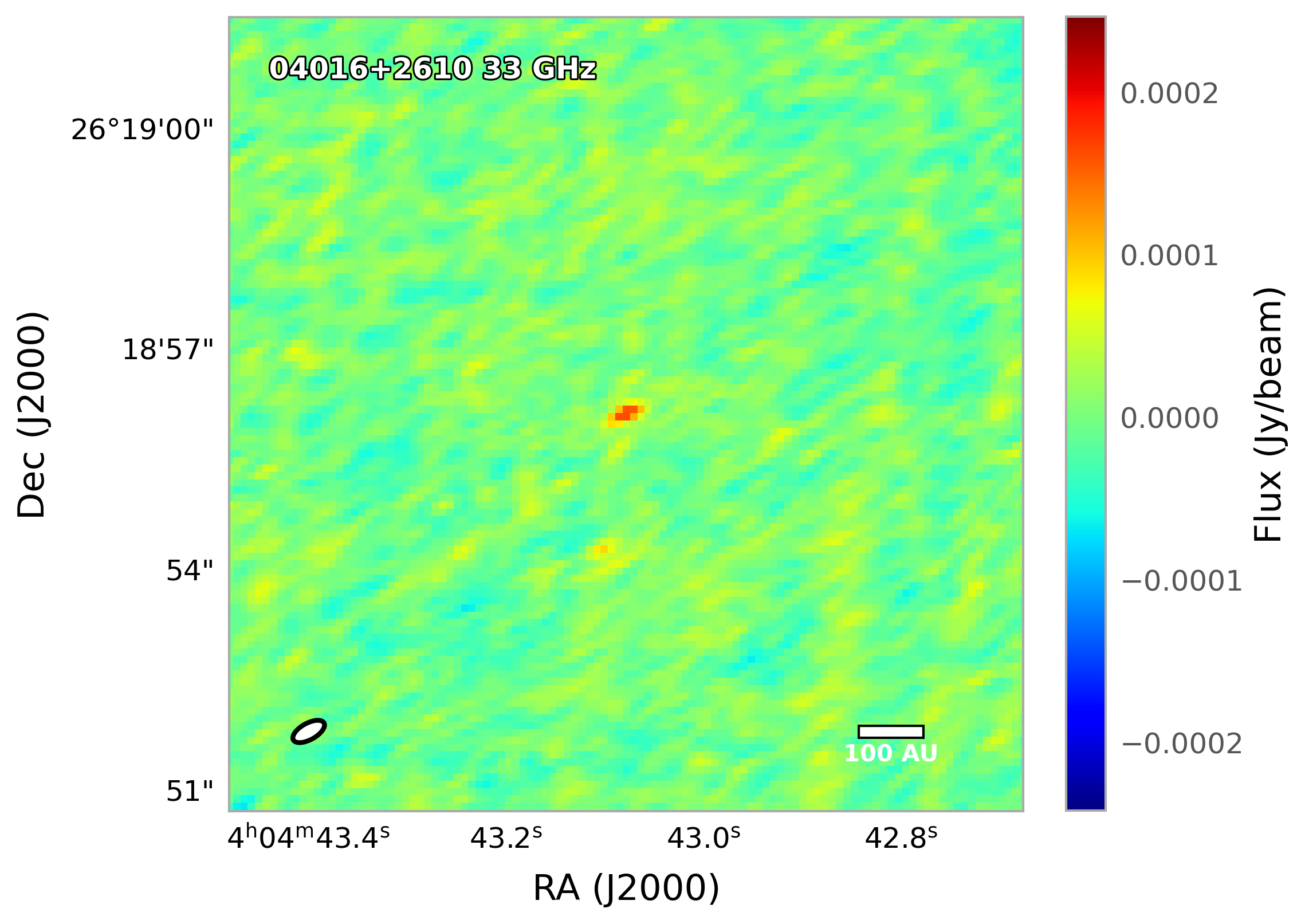}
    \caption{Continuum images of 04016+2610 observed with ALMA at 345~GHz (left) and the VLA at 33~GHz (right). The complete figure set containing continuum images of all sources is available in the online journal.}
    \label{fig:continuum_images}
\end{figure}

\section{$^{13}$CO and C$^{18}$O Channel Maps} \label{app:channel_maps}
In this appendix, we present the $^{13}$CO and C$^{18}$O channel maps for all sources as a figure set. Each panel is labeled by the channel velocity relative to the systemic velocity. The synthesized beam is shown in the lower left corner, and a 100~AU scale bar is shown in the lower right. The black contour shows the Keplerian mask used in the analysis. For clarity, we show only the channels spanning the masked range, beginning two velocity channels before the first masked channel and ending two velocity channels after the last masked channel. Details of the Keplerian mask construction are given in Section~\ref{subsec:line-analysis}.\\

\figsetstart
\figsetnum{A2}
\figsettitle{$^{13}$CO and C$^{18}$O channel maps for all Taurus Class~0/I sources}

\figsetgrpstart
\figsetgrpnum{A2.1}
\figsetgrptitle{04016+2610}
\figsetplot{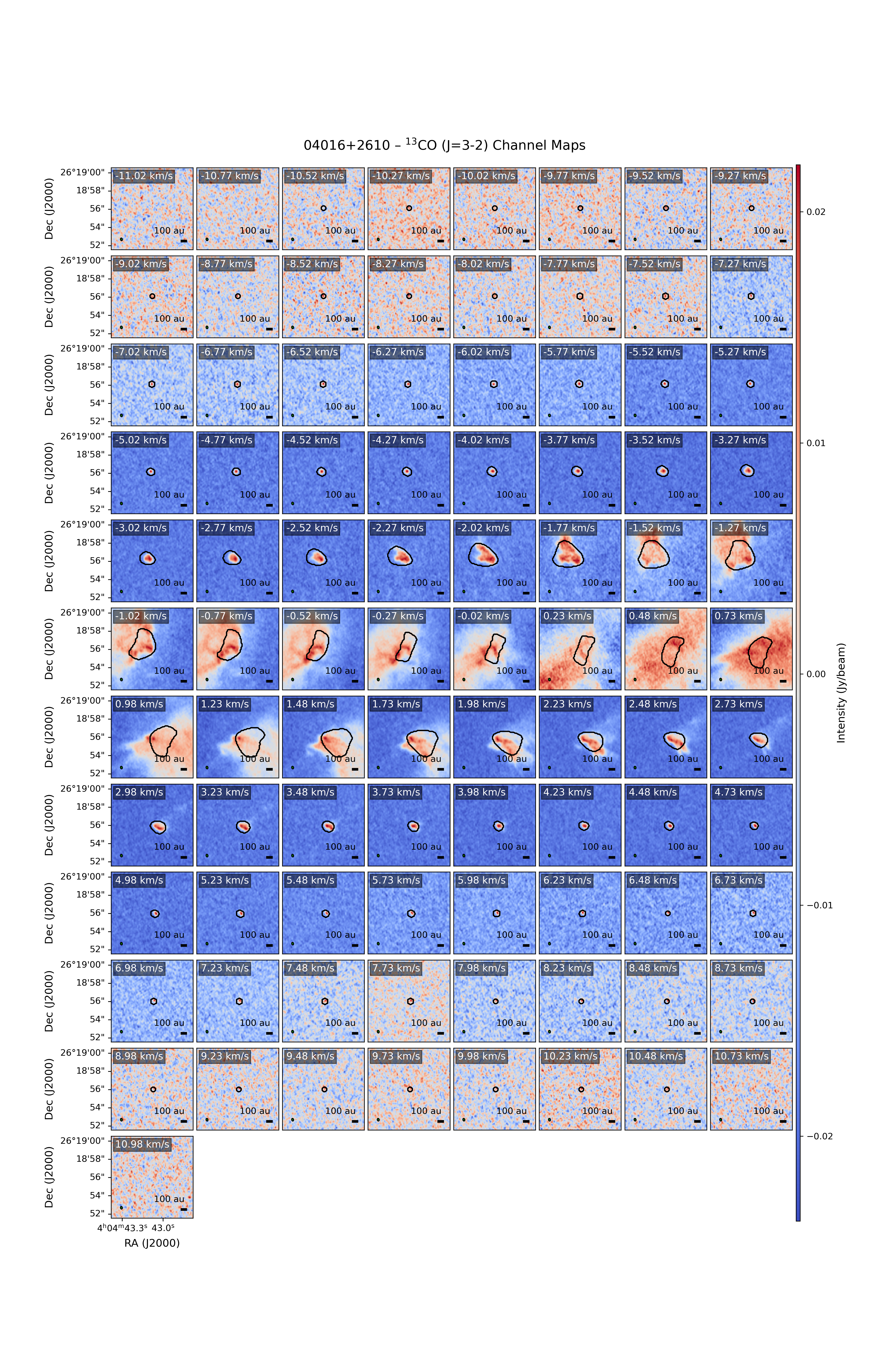}
\figsetplot{04016+2610_C18O_channel_maps.png}
\figsetgrpnote{$^{13}$CO and C$^{18}$O line images for 04016+2610.}

\figsetgrpstart
\figsetgrpnum{A2.2}
\figsetgrptitle{04108+2803B}
\figsetplot{04108+2803B_13CO_channel_maps.png}
\figsetplot{04108+2803B_C18O_channel_maps.png}
\figsetgrpnote{$^{13}$CO and C$^{18}$O line images for 04108+2803B.}

\figsetgrpstart
\figsetgrpnum{A2.3}
\figsetgrptitle{04158+2805}
\figsetplot{04158+2805_13CO_channel_maps.png}
\figsetplot{04158+2805_C18O_channel_maps.png}
\figsetgrpnote{$^{13}$CO and C$^{18}$O line images for 04158+2805.}

\figsetgrpstart
\figsetgrpnum{A2.4}
\figsetgrptitle{04166+2706}
\figsetplot{04166+2706_13CO_channel_maps.png}
\figsetplot{04166+2706_C18O_channel_maps.png}
\figsetgrpnote{$^{13}$CO and C$^{18}$O line images for 04166+2706.}

\figsetgrpstart
\figsetgrpnum{A2.5}
\figsetgrptitle{04169+2702}
\figsetplot{04169+2702_13CO_channel_maps.png}
\figsetplot{04169+2702_C18O_channel_maps.png}
\figsetgrpnote{$^{13}$CO and C$^{18}$O line images for 04169+2702.}

\figsetgrpstart
\figsetgrpnum{A2.6}
\figsetgrptitle{04181+2654A}
\figsetplot{04181+2654A_13CO_channel_maps.png}
\figsetplot{04181+2654A_C18O_channel_maps.png}
\figsetgrpnote{$^{13}$CO and C$^{18}$O line images for 04181+2654A.}

\figsetgrpstart
\figsetgrpnum{A2.7}
\figsetgrptitle{04191+1523A}
\figsetplot{04191+1523A_13CO_channel_maps.png}
\figsetplot{04191+1523A_C18O_channel_maps.png}
\figsetgrpnote{$^{13}$CO and C$^{18}$O line images for 04191+1523A.}

\figsetgrpstart
\figsetgrpnum{A2.8}
\figsetgrptitle{04191+1523B}
\figsetplot{04191+1523B_13CO_channel_maps.png}
\figsetplot{04191+1523B_C18O_channel_maps.png}
\figsetgrpnote{$^{13}$CO and C$^{18}$O line images for 04191+1523B.}

\figsetgrpstart
\figsetgrpnum{A2.9}
\figsetgrptitle{04239+2436AB}
\figsetplot{04239+2436AB_13CO_channel_maps.png}
\figsetplot{04239+2436AB_C18O_channel_maps.png}
\figsetgrpnote{$^{13}$CO and C$^{18}$O line images for 04239+2436AB.}

\figsetgrpstart
\figsetgrpnum{A2.10}
\figsetgrptitle{04248+2612AB}
\figsetplot{04248+2612AB_13CO_channel_maps.png}
\figsetplot{04248+2612AB_C18O_channel_maps.png}
\figsetgrpnote{$^{13}$CO and C$^{18}$O line images for 04248+2612AB.}

\figsetgrpstart
\figsetgrpnum{A2.11}
\figsetgrptitle{04260+2642}
\figsetplot{04260+2642_13CO_channel_maps.png}
\figsetplot{04260+2642_C18O_channel_maps.png}
\figsetgrpnote{$^{13}$CO and C$^{18}$O line images for 04260+2642.}

\figsetgrpstart
\figsetgrpnum{A2.12}
\figsetgrptitle{04263+2426A}
\figsetplot{04263+2426A_13CO_channel_maps.png}
\figsetplot{04263+2426A_C18O_channel_maps.png}
\figsetgrpnote{$^{13}$CO and C$^{18}$O line images for 04263+2426A.}

\figsetgrpstart
\figsetgrpnum{A2.13}
\figsetgrptitle{04263+2426B}
\figsetplot{04263+2426B_13CO_channel_maps.png}
\figsetplot{04263+2426B_C18O_channel_maps.png}
\figsetgrpnote{$^{13}$CO and C$^{18}$O line images for 04263+2426B.}

\figsetgrpstart
\figsetgrpnum{A2.14}
\figsetgrptitle{04264+2433AB}
\figsetplot{04264+2433AB_13CO_channel_maps.png}
\figsetplot{04264+2433AB_C18O_channel_maps.png}
\figsetgrpnote{$^{13}$CO and C$^{18}$O line images for 04264+2433AB.}

\figsetgrpstart
\figsetgrpnum{A2.15}
\figsetgrptitle{04287+1801}
\figsetplot{04287+1801_13CO_channel_maps.png}
\figsetplot{04287+1801_C18O_channel_maps.png}
\figsetgrpnote{$^{13}$CO and C$^{18}$O line images for 04287+1801.}

\figsetgrpstart
\figsetgrpnum{A2.16}
\figsetgrptitle{04288+1802}
\figsetplot{04288+1802_13CO_channel_maps.png}
\figsetplot{04288+1802_C18O_channel_maps.png}
\figsetgrpnote{$^{13}$CO and C$^{18}$O line images for 04288+1802.}

\figsetgrpstart
\figsetgrpnum{A2.17}
\figsetgrptitle{04295+2251}
\figsetplot{04295+2251_13CO_channel_maps.png}
\figsetplot{04295+2251_C18O_channel_maps.png}
\figsetgrpnote{$^{13}$CO and C$^{18}$O line images for 04295+2251.}

\figsetgrpstart
\figsetgrpnum{A2.18}
\figsetgrptitle{04302+2247}
\figsetplot{04302+2247_13CO_channel_maps.png}
\figsetplot{04302+2247_C18O_channel_maps.png}
\figsetgrpnote{$^{13}$CO and C$^{18}$O line images for 04302+2247.}

\figsetgrpstart
\figsetgrpnum{A2.19}
\figsetgrptitle{04325+2402A}
\figsetplot{04325+2402A_13CO_channel_maps.png}
\figsetplot{04325+2402A_C18O_channel_maps.png}
\figsetgrpnote{$^{13}$CO and C$^{18}$O line images for 04325+2402A.}

\figsetgrpstart
\figsetgrpnum{A2.20}
\figsetgrptitle{04325+2402B}
\figsetplot{04325+2402B_13CO_channel_maps.png}
\figsetplot{04325+2402B_C18O_channel_maps.png}
\figsetgrpnote{$^{13}$CO and C$^{18}$O line images for 04325+2402B.}

\figsetgrpstart
\figsetgrpnum{A2.21}
\figsetgrptitle{04361+2547}
\figsetplot{04361+2547_13CO_channel_maps.png}
\figsetplot{04361+2547_C18O_channel_maps.png}
\figsetgrpnote{$^{13}$CO and C$^{18}$O line images for 04361+2547.}

\figsetgrpstart
\figsetgrpnum{A2.22}
\figsetgrptitle{04365+2535}
\figsetplot{04365+2535_13CO_channel_maps.png}
\figsetplot{04365+2535_C18O_channel_maps.png}
\figsetgrpnote{$^{13}$CO and C$^{18}$O line images for 04365+2535.}

\figsetgrpstart
\figsetgrpnum{A2.23}
\figsetgrptitle{04368+2557}
\figsetplot{04368+2557_13CO_channel_maps.png}
\figsetplot{04368+2557_C18O_channel_maps.png}
\figsetgrpnote{$^{13}$CO and C$^{18}$O line images for 04368+2557.}

\figsetgrpstart
\figsetgrpnum{A2.24}
\figsetgrptitle{04381+2540AB}
\figsetplot{04381+2540AB_13CO_channel_maps.png}
\figsetplot{04381+2540AB_C18O_channel_maps.png}
\figsetgrpnote{$^{13}$CO and C$^{18}$O line images for 04381+2540AB.}

\figsetgrpstart
\figsetgrpnum{A2.25}
\figsetgrptitle{04385+2550}
\figsetplot{04385+2550_13CO_channel_maps.png}
\figsetplot{04385+2550_C18O_channel_maps.png}
\figsetgrpnote{$^{13}$CO and C$^{18}$O line images for 04385+2550.}

\figsetgrpstart
\figsetgrpnum{A2.26}
\figsetgrptitle{04489+3042A}
\figsetplot{04489+3042A_13CO_channel_maps.png}
\figsetplot{04489+3042A_C18O_channel_maps.png}
\figsetgrpnote{$^{13}$CO and C$^{18}$O line images for 04489+3042A.}

\figsetgrpstart
\figsetgrpnum{A2.27}
\figsetgrptitle{DG TauB}
\figsetplot{DG_TauB_13CO_channel_maps.png}
\figsetplot{DG_TauB_C18O_channel_maps.png}
\figsetgrpnote{$^{13}$CO and C$^{18}$O line images for DG TauB.}

\figsetgrpstart
\figsetgrpnum{A2.28}
\figsetgrptitle{HH 30}
\figsetplot{HH30_13CO_channel_maps.png}
\figsetplot{HH30_C18O_channel_maps.png}
\figsetgrpnote{$^{13}$CO and C$^{18}$O line images for HH 30.}

\figsetgrpstart
\figsetgrpnum{A2.29}
\figsetgrptitle{IRAM 04191}
\figsetplot{IRAM04191_13CO_channel_maps.png}
\figsetplot{IRAM04191_C18O_channel_maps.png}
\figsetgrpnote{$^{13}$CO and C$^{18}$O line images for IRAM 04191.}

\figsetgrpend
\figsetend

\begin{figure*}
\centering
\includegraphics[width=\textwidth,height=0.88\textheight,keepaspectratio]{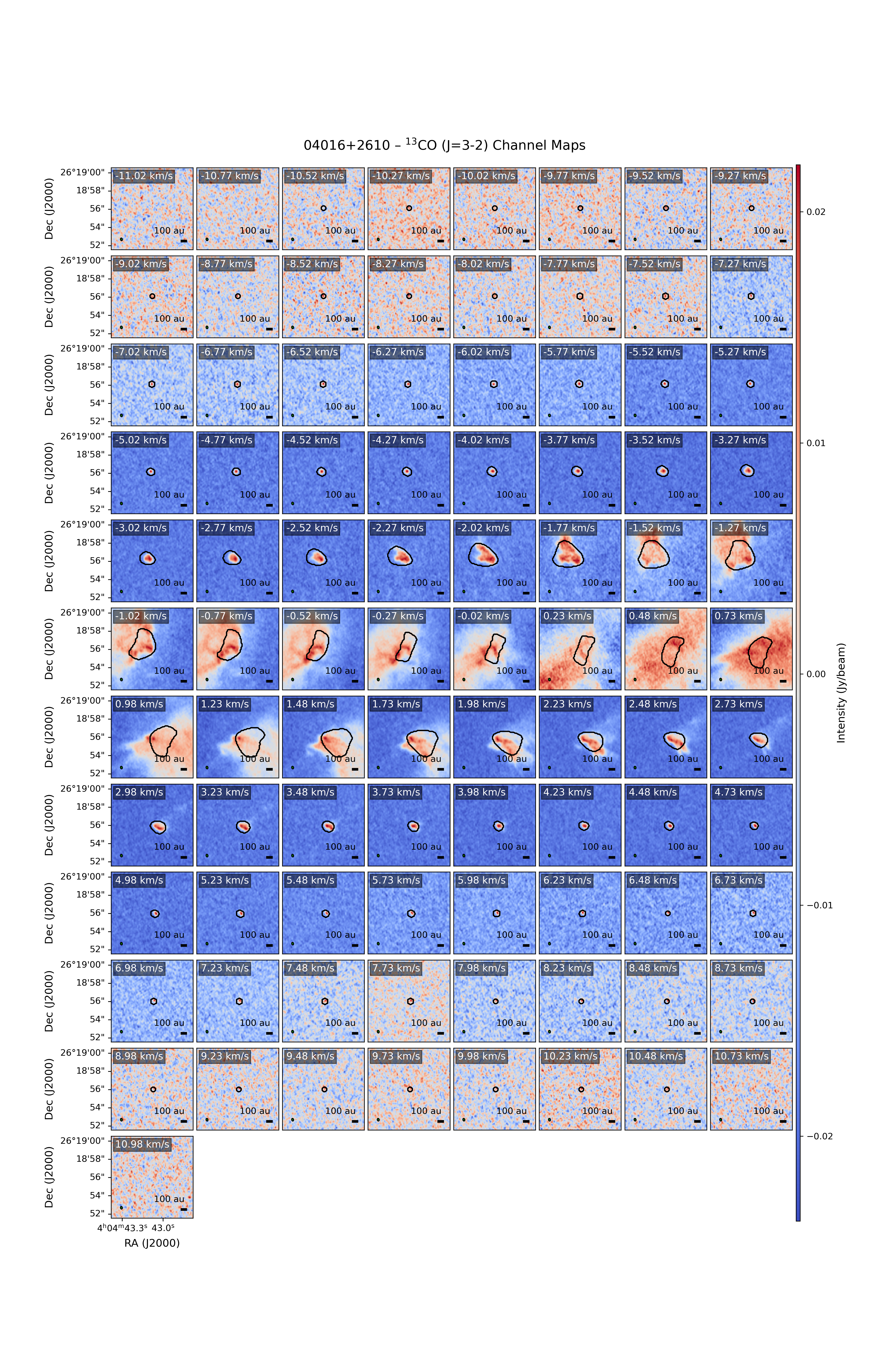}
\caption{Representative $^{13}$CO channel maps for 04016+2610. Each panel is labeled by the channel velocity relative to the systemic velocity, and the black contours show the Keplerian mask used in the analysis. The full figure set, which includes both $^{13}$CO and C$^{18}$O channel maps for all sources, is available in the online journal.}
\label{fig:channel_maps_rep_13co}
\end{figure*}

\bibliographystyle{aasjournal}


\end{document}